\documentclass[preprint,aps,amsmath,nofootinbib,tightenlines,superscriptaddress]{revtex4}

\usepackage{graphicx}
\usepackage{amsmath}
\usepackage{slashed}
\usepackage{bm}
\usepackage{xspace}
\usepackage{epsfig}
\usepackage{amssymb}
\usepackage{comment}
\usepackage[hyperfootnotes=false]{hyperref}

\newcommand{\nb}{\bar{n}}  
\newcommand{\nbs}{\slashed{\nb}}  
\newcommand{\ns}{\slashed{n}}  
\newcommand{\scetone}{\mathrm{SCET}_1}
\newcommand{\scettwo}{\mathrm{SCET}_2}
\newcommand{\scetthree}{\mathrm{SCET}_3}
\newcommand{\sceti}{\mathrm{SCET}_i}
\newcommand{\scetj}{\mathrm{SCET}_j}
\newcommand{\scetipone}{\mathrm{SCET}_{i+1}}

\newcommand{\scetn}{\mathrm{SCET}_N}
\newcommand{\mo}{\mathcal{O}}
\newcommand{\mb}{\mathcal{B}} 
\newcommand{\mcp}{\mathcal{P}}

\newcommand{\chib}{\bar{\chi}}
\newcommand{\nlosplit}{P^{(1)}_{qq}}
\newcommand{\losplit}{P^{(0)}_{qq}}
\newcommand{\moqq}{C^{(0)}_{1,\,\rm{LO}} \mo^{(0)}_{1}}
\newcommand{\moqqg}{C_{2, \mathrm{LO}}^{(1)} \mo^{(1)}_{2}(n_1, n'_1)}

\newcommand{\NLOa}{\mbox{NLO$(\alpha_s)$}\xspace}
\newcommand{\NLOl}{\mbox{NLO$(\lambda)$}\xspace}

\newcommand{\Dslash}{D\!\!\!\!\slash}
\newcommand{\bnP}{\bar {\cal P}}
\newcommand{\bnslash}{\bar n\!\!\!\slash}
\newcommand{\Bslash}{B\!\!\!\!\slash}

\newcommand{\beq}{\begin{equation}}
\newcommand{\eeq}{\end{equation}}

\newcommand{\OMIT}[1]{}

\newcommand{\eq}[1]{Eq.~(\ref{eq:#1})}
\newcommand{\eqs}[2]{Eqs.~(\ref{eq:#1}) and (\ref{eq:#2})}

\newcommand{\subsec}[1]{Sec.~\ref{subsec:#1}}
\newcommand{\app}[1]{App.~\ref{app:#1}}
\newcommand{\fig}[1]{Fig.~\ref{fig:#1}}

\newcommand{\tbl}[1]{Table~\ref{tbl:#1}}

  \newcount\hour \newcount\minute
  \hour=\time \divide \hour by 60 \minute=\time
  \newcommand{\todaytime}{\today \ -- \number\hour :\ifnum \minute<10 0\fi\number\minute}

\begin{document}
\setlength\baselineskip{15pt}

\preprint{ \vbox{ 
\hbox{MIT-CTP-4120}
} }

\title{\phantom{x}\vspace{0.5cm}
Systematic Improvement of Parton Showers with Effective Theory 
\vspace{0.6cm}
}

\author{Matthew Baumgart}
\email{baumgart@pha.jhu.edu}
\affiliation{Jefferson Physical Laboratory, Harvard University,\\ Cambridge, MA 02138, USA
\vspace{0.2cm}}
\affiliation{School of Natural Sciences, Institute for Advanced Study, \\ Princeton, NJ 08540, USA
\vspace{0.2cm}}
\affiliation{Department of Physics and Astronomy, Johns Hopkins University, \\ Baltimore, MD 21218, USA
\vspace{0.2cm}}

\author{Claudio Marcantonini}
\email{cmarcant@mit.edu}
\affiliation{Center for Theoretical Physics, Massachusetts
Institute of Technology,\\ Cambridge, MA 02139, USA
\vspace{0.2cm}\vspace{1cm}}

\author{Iain W. Stewart\vspace{0.5cm}}
\email{iains@mit.edu}
\affiliation{Center for Theoretical Physics, Massachusetts
Institute of Technology,\\ Cambridge, MA 02139, USA
\vspace{0.2cm}\vspace{1cm}}

\begin{abstract}
\vspace{0.3cm}

We carry out a systematic classification and computation of next-to-leading
order kinematic power corrections to the fully differential cross section
in the parton shower.  To do this we devise a map between ingredients in a
parton shower and operators in a traditional effective field theory framework
using a chain of soft-collinear effective theories. Our approach overcomes
several difficulties including avoiding double counting and distinguishing
approximations that are coordinate choices from true power corrections.
Branching corrections can be classified as hard-scattering, that occur near the
top of the shower, and jet-structure, that can occur at any point inside it.
Hard-scattering corrections include matrix elements with additional hard
partons, as well as power suppressed contributions to the branching for the
leading jet.  Jet-structure corrections require simultaneous consideration of
potential $1\to 2$ and $1\to 3$ branchings.  The interference structure induced
by collinear terms with subleading powers remains localized in the shower.

\phantom{x}
\vspace{0.5cm}
\noindent

\end{abstract}

\maketitle 

\tableofcontents

\newpage
\section{Introduction}

For scattering problems involving strongly-interacting particles, we are often
interested in final states with large multiplicities, sometimes including
thousands of hadrons.  To get to this level, we cannot rely solely on full
fixed-order calculations.  Tree-level event generators
\cite{Mangano:2002ea,Gleisberg:2008fv,Papadopoulos:2006mh, Kilian:2007gr} only
go up to $8$-$10$ external particles as Monte Carlo for higher multiplicity
phase space is increasingly intractable.  At one-loop, the frontier is 2
$\rightarrow$ 4 processes, which have been done at the level of differential
cross sections for $W+$ 3 jets \cite{KeithEllis:2009bu,Berger:2009ep} and
$t\bar{t}b\bar{b}$ \cite{Bredenstein:2010rs}.  At two-loops, there are 2
$\rightarrow$ 1 exclusive calculations for weak boson production by hadrons
followed by decay ($W$ and $Z$ \cite{Anastasiou:2003ds} and $W$
\cite{Catani:2010en} to leptons, and $H$ decaying to photons
\cite{Anastasiou:2005qj,Catani:2007vq}.).  Additionally, $e^+ e^- \rightarrow$ 3
jets to NNLO is known
\cite{GehrmannDeRidder:2007bj,GehrmannDeRidder:2007hr,Weinzierl:2008iv,Weinzierl:2009ms}.
In any case, a strict fixed order counting is not suitable for exclusive
observables with large multiplicities, nor for many inclusive observables where
certain regions of phase space receive kinematic enhancement by large
logarithms. If $Q$ is a hard scale in the process, then a subset of the
amplitude gets enhanced so that its coefficient is $(\alpha_s \ln^2(Q/p))^m$,
where $p \ll Q$ refers to a small scale that is induced by the choice of
observable or cuts.  Since we can resum these large logs by systematically
treating real radiation, we can give a leading log (LL) description of these
observables without performing multiloop computations.  The soft and collinear
limits that yield these large logs also allow us to simplify the amplitude.
Therefore, capturing the dominant contributions to these observables and
simulating processes with a large number of particles becomes feasible. This is
a main goal of parton Shower Monte Carlo (SMC).

A final state SMC is based on the ``strongly-ordered limit,'' which
describes the leading log contribution (accounting for soft emission by angular
ordering or other approximations).  In this kinematic configuration, each
radiated particle comes off much more collinear to its parent than the previous
one, a situation that can be formulated in terms of perpendicular momenta or
virtualities, $i.e.$
\begin{align} 
 & q_{0\perp} \gg q_{1\perp} \gg q_{2\perp} \gg \ldots, 
 & \text{or} &
 & q_0^2 & \gg q_1^2 \gg q_2^2 \gg \ldots \,.
 \label{eq:psLimit1}
\end{align}  
Furthermore, and important for practical computation, in this limit each
collinear emission is independent of the previous one.  Thus, if we have
calculated the differential cross section for $i$-parton emission, $d\sigma_i$,
then we can obtain the $(i+1)$-parton case as
\beq
d\sigma_{i+1} \,\propto\, \frac{P^{(0)}_{j\to kl}}{q_{i}^2}\, d\sigma_i,
\label{eq:splitFact}
\eeq
where $P^{(0)}$ is the leading order (LO) ``splitting function'' that
captures the probability for the $i^{th}$ emitted parton, of type $j$,
to split into two others, $kl$, and $q_i^2$ is its virtuality.  
We can therefore formulate the process in terms of a probabilistic Markov chain
of $i$ $1 \rightarrow 2$ particle splittings.  The probabilities
are determined by the functions $P^{(0)}_{j\to kl}$, which are the LO
Altarelli-Parisi kernels.  As an example, for $q\rightarrow qg$, after
averaging and summing over spins,
\beq
P_{q\to qg}^{(0)} \,=\, \frac{\alpha_s}{2\pi} \, C_F \frac{1+z^2}{1-z},
\label{eq:ap}
\eeq
where $z$ is the longitudinal momentum fraction of the daughter with
respect to the parent.  This classical, probabilistic process gives
rise to the SMC algorithms used by event generators such as
Pythia~\cite{Sjostrand:2006za,Sjostrand:2007gs} and
Herwig~\cite{Corcella:2000bw, Bahr:2008pv} to model radiation.
For a virtuality-ordered shower, such as the original version of Pythia, given some
initial offshellness, $q_0^2$, and an initial momentum fraction, $x_0$, SMCs
generate the virtuality and the momentum fraction of the daughter particle after
the spitting.  The former is determined by a Sudakov factor,
$\Delta(q^2,q_0^2)$, which gives the probability of a parton to evolve from
$q_0^2$ to $q^2$ without branching,
\beq
\Delta(q^2,q_0^2) \,=\, \exp \left[ -\int_{q_0^2}^{q^2} \frac{d q'^2}{q'^2} 
  \int\!\! dz\: \frac{\alpha_s}{2 \pi} P^{(0)}_{jk}(z)  \right].
\label{eq:sudakov}
\eeq
The traditional LL parton shower makes the multiplicity problem tractable, but it
has shortcomings related to the leading log approximation.  Even though
\eq{splitFact} is only correct in the collinear limit, the shower is used
everywhere in order to generate events that cover the full phase space.  In
addition, since each collinear emission is independent from the previous one in
the shower, the LL approximation does not include their spin or color
correlations, nor any of their interference.  The situation is different for
soft gluons where the inclusion of color effects
allows one to work in the simplifying limit of angular ordering.

The hierarchy of scales in the parton shower makes it amenable to an effective
field theory treatment. Since the shower regime occurs for particles in the soft
and collinear regions, we can describe it with Soft-Collinear Effective Theory
(SCET)~\cite{Bauer:2000ew, Bauer:2000yr, Bauer:2001ct, Bauer:2001yt}.  Like any
EFT, SCET comes with an expansion that allows, in principle, for systematic
improvement.  The first work on parton showers using SCET came in
\cite{Bauer:2006mk,Bauer:2006qp}, which we review in \subsec{bands}, where the
authors showed how the splitting functions and corresponding Sudakov factors,
along with the factorization of emissions emerge naturally.  Furthermore, they
could include virtual corrections by matching to QCD at higher order in
$\alpha_s$.  Unfortunately, in reproducing the LL shower in SCET, they
introduced many conventions whose extension to higher orders in the kinematic
expansion is unclear.  We therefore develop a modified approach to alleviate
these difficulties.

Before discussing our setup, we give an overview of advances in the parton shower
literature beyond the basic LL picture. The structure of these advances depends
on what aspect of the shower one aims to improve. Possible motivations
include accuracy at higher orders in $\alpha_s$, higher order in logs, and
higher order in powers of the kinematic expansions.  We first introduce some
terminology for higher order log resummation. If the resummation of large logs,
$L$, is at the cross section level
\begin{align}
  d\sigma \sim \Big[ \sum_k (\alpha_s L^2)^k \Big]_{\rm LL} 
    + \Big[ \sum_k \alpha_s L (\alpha_s L^2)^k \Big]_{\rm NLL} + \ldots
\label{eq:resumCross}
\end{align}
then we will refer to it as LL, NLL, etc, as indicated.  If the cross section
transformed to an appropriate set of variables has a resummation of logs in the
exponent
\begin{align}
  \ln d\sigma \sim L \Big[ \sum_k (\alpha_s L)^k \Big]_{\rm LL_{\rm exp}}
   +   \Big[ \sum_k  (\alpha_s L)^k \Big]_{\rm NLL_{\rm exp}} + \ldots
\label{eq:resumExp}
\end{align}
then we will attach a subscript ``exp'' to the orders to indicate this.  

A major concern with parton showers is how one handles the merging with matrix
element (ME) calculations that describe the initial underlying hard process.
One can consider a simple setup where one declares that a scale, $\mu_0$,
divides collinear from hard radiation.  Here, emissions above $\mu_0$
are described through tree-level ME calculations, and those beneath by running
SMC.  Each regime would get a reasonable treatment, but naively interfacing the
two leaves leading-log sensitivity to $\mu_0$.  This is because the LO (in
$\alpha_s$) result contains no Sudakov log resummation. Methods for carrying out
matrix element and parton shower merging including this information have been
considered in Refs.~\cite{Lonnblad:2001iq,Catani:2001cc,Caravaglios:1998yr} and
are referred to as CKKW-L and MLM.  In CKKW-L, one distributes the particles in
an event according to the probabilities given by the exact tree-level matrix
element, with $\mu_0^2$ as a lower cutoff related to the perpendicular momentum
between any two particles.  One then clusters the event using the $k_T$
algorithm \cite{Catani:1991hj} to determine the splitting scales, $q_{i\, T}^2$.
With these in hand, one reweights the event by multiplication by
appropriate Sudakov factors, as well as factors of $\alpha_s(q_{i\, T})/\alpha_s(Q)$,
where $Q$ is some hard scale.  We can then run a parton shower algorithm on
these squared amplitudes, vetoing any splitting $q_{i\, T}$ harder than $\mu_0^2$ to
avoid double counting. It was demonstrated that the $n$-jet rate depends on
$\mu_0$ only beyond NLL order, with the first missing term being $\alpha_s^2
\ln^2(Q/\mu_0)$.  CKKW-L has been built into Sherpa~\cite{Gleisberg:2003xi}.

Another important effect concerns soft gluons, which are also kinematically
enhanced.  Collinear emissions reinforce the picture of partonic radiation as an
isolated jet since they get distributed within some narrow cone about the
original hard parton. {\it A priori}, soft gluons have no preferred direction
and can communicate between elements of the shower.  Fortunately, 
wide-angle radiation only observes the net color charge
contained in the cone of emission.  Therefore, the pattern of soft radiation far 
from the collinear jet is not sensitive to splittings that have taken place within it.
This coherent branching and angular ordering
can be accommodated by methods such as evolving the shower by decreasing angle
monotonically, as is done in Herwig~\cite{Marchesini:1991ch}, or by enforcing it
with a veto in a virtuality-ordered shower (the rightmost expression in
\eq{psLimit1}), which is an option in Pythia~\cite{Sjostrand:2006za}.
Accounting for coherence properties leads to LL resummation for the soft
emissions~\cite{Mueller:1981ex,Ermolaev:1981cm,Bassetto:1982ma,
  Dokshitzer:1982xr,Dokshitzer:1982ia}.  Additional considerations treated in
shower programs include putting $\alpha_s$ at the $k_T$ scale of each splitting,
and encoding momentum conservation at each vertex, which give the parton shower
information beyond an analytic LO/LL calculation. These along with the overall
choice in evolution variable (mass, $k_\perp$, angle, {\it etc.}) are treated in
different fashions by different SMC codes.

There are of course further corrections to include to go to ${\rm NLO\; in\;}
\alpha_s$, denoted NLO$(\alpha_s)$, ${\rm NLL}$ in kinematic logs, and/or NLO in
power corrections to the strong ordering, denoted NLO$(\lambda)$.  The most
effort to date has gone to working out the \NLOa/LL contribution to incorporate
one-loop corrected amplitudes at the top of the shower.  Adding $\alpha_s$
corrections involves the numerical challenge of combining real and virtual
results which separately have IR divergences.  The basic resolution is to
extract the pole-portion of the real emission of $i$-partons and include it
along with the virtual contributions to the $i-1$ case.  Unfortunately, this
does not sum leading logs. One cannot blindly extend the CKKW procedure to
\NLOa/LL, as it leads to double-counting problems; the Sudakov factors in the
reweighting contain a portion of the one-loop contributions.  Separately adding
the full one-loop result would clearly overcount.

There are two main solutions to the \NLOa/LL merging problem in the context of
standard 1 $\rightarrow$ 2 splittings.  MC$@$NLO~\cite{Frixione:2002ik} works
by means of subtraction, finding the places where the Sudakovs will contribute
at \NLOa, and removing the splitting function contribution.  This approach is
conceptually clear.  Since the full amplitude and splitting function portions
are calculated separately before subtraction, the latter for each SMC program,
this is time-consuming.  Furthermore, since the subtractions occur for the
amplitude squared, one cannot guarantee positivity of the result and must
deal with negatively weighted events.  To avoid the computational difficulties
of process-by-process subtraction and negative weights, an alternative is the
POWHEG algorithm~\cite{Nason:2004rx}.  It keeps the IR-safe \NLOa cross section
manifest, and defines a Sudakov factor based on a modified splitting function to
handle LL$_{\rm exp}$ and a subset of NLL$_{\rm exp}$ resummation for the
hardest emission.  In this way, it makes use of quantities already obtained in
the fixed order \NLOa calculation, requiring fewer additional steps for its
implementation for each known process.  The conservation of probability obeyed
by the splittings and related Sudakov factors avoid double countings and give
back $\sigma_{\rm NLO}$ upon integration.

A separate set of approaches goes beyond the 1 $\rightarrow$ 2 formalism to
consider the radiation's effects on one or more ``spectators.''  The
consideration of an additional parton in the pre-emission configuration has led
to work known as dipole subtraction and dipole antennas.  The former was
initially developed in \cite{Catani:1996vz,Catani:2002hc}.  It explicitly
subtracts the IR divergence from real emission via a simplified ``dipole'' term.
Refs.~\cite{Schumann:2007mg,Dinsdale:2007mf} have proposed algorithms based on
these techniques.  There has also been development on the theoretical side of
subtractions by Nagy and
Soper~\cite{Nagy:2007ty,Nagy:2008ns,Nagy:2008eq,Soper:2008zp}, with the aim of
including spin and color effects, while improving the efficiency of
implementation \cite{Robens:2010zr}.  The original use of antennas came in the
ARIADNE program, which treats the 2 $\rightarrow$ 3 splitting as its basic
unit~\cite{Gustafson:1987rq,Andersson:1989ki,Pettersson:1988zu,Lonnblad:1992tz}
and allows for exact momentum conservation.  There have since been more
systematic attempts to extract the 2 $\rightarrow$ 3 ``antenna'' functions from
QCD and implement them in a shower, {\it e.g.} VINCIA \cite{Giele:2007di}.
Ref.~\cite{Larkoski:2009ah} even derives spin-dependent antenna functions,
though its SMC implementation is yet to appear.

A different approach is the GenEvA framework \cite{Bauer:2008qh,Bauer:2008qj}
which allows the issues of phase space double counting and combining matrix
elements and log resummation to be treated independently. This is done using
effective theory ideas for how to separate scales.  In this setup, one
manifestly avoids negative weights and double counting by using
multiplicative merging. For example, GenEvA yields a calculation that is
equivalent to POWHEG for the \NLOa/LL matching and at the same time a CKKW-L
type LO($\alpha_s$)/LL matching for additional emissions. In a similar
fashion, the power suppressed matrix element computations and subleading
no-branching probabilities derived here could be implemented in GenEvA, and work
in this direction is commencing.

Another approach to go beyond LL is to incorporate the contribution of
the $\mo(\alpha_s^2)$ corrections to the Altarelli-Parisi splitting
kernels, $\nlosplit$.  This was done to resum soft logs to NLL for
semi-inclusive variables in DIS and Drell-Yan~\cite{Catani:1990rr}.
In order to conserve probability, these corrections must be correctly
accounted for in both the probability for real emission in \eq{ap}, as
well as no-branching branching probabilities.  This is related to why
POWHEG only implements them for the hardest splitting, where they have
information from the full fixed-order computation.  The KRKMC group
incorporates the subleading real emission contributions into fully
exclusive partonic configurations in SMC
\cite{Jadach:2009gm,Skrzypek:2009jk, Jadach:2010ew}.  Some of the
subleading contributions take the form of $1 \to 3$ splittings,
requiring a modification of the usual $1\to 2$ algorithm.  Similar to
CKKW, the KRKMC groups corrections take the form of a multiplicative
reweighting.  For a particular configuration of partons in phase
space, they reweight by a factor that includes the insertion of $1\to
3$ ``defects'' and loop-corrected $1 \to 2$ splittings that account
for the effects of $\nlosplit$.  If $\rho$ is the fully differential
cross section, they define a corrected weight for $n$ partons, $w_n$
as:
\beq
w_n = \frac{\rho_{\rm LO}(k_1,\ldots,k_n) \,+\, \sum_{r=1}^{n/2} \rho_{\rm N^rLO}(k_1,\ldots,k_n)}{\rho_{\rm LO}(k_1,\ldots,k_n)},
\label{eq:jadach}
\eeq
where $r$ determines the number of defect insertions in any configuration. Since
this reweighting involves splitting probabilities and not subleading no-branching
probabilities, it does not clearly improve the level of log resummation.

In this work we set up an EFT framework to classify and study perturbative
$\alpha_s$ corrections, higher order log resummation and/or kinematic power
corrections to parton showers. While the ultimate goal is to facilitate the
implementation of a NLL/\NLOa parton shower algorithm accounting for the leading
deviations from strong ordering, our task here is much more modest.\footnote{ In
  particular we note that soft NLL resummation may only be feasible at leading
  orders in 1/$N_c$ \cite{Bonciani:2003nt,Frixione:2007vw}.}  We focus primarily
on kinematic power corrections in the fully differential cross section for an
arbitrary number of final state emissions. That is, our main goal is to compute
\begin{align}
  \frac{d\sigma^{\rm LO}}{d\vec p_1^{\,3}\cdots d\vec p_n^{\,3}}
  + \frac{d\sigma^{\rm NLO(\lambda)}}{d\vec p_1^{\,3}\cdots d\vec p_n^{\,3}}
 \,.
\label{eq:goalDef}
\end{align} 
Here \NLOl is the next-to-leading order power correction in the cross section,
which involves terms that are \NLOl and NNLO$(\lambda)$ in the amplitude.
Similarly to \cite{Bauer:2006mk,Bauer:2006qp}, we use an operator approach based
on SCET.  A main issue to resolve is taking into account different possibilities
for the kinematic configurations of subsequent emissions, to go beyond the
strong ordering described in Eq~(\ref{eq:psLimit1}).  The
hierarchy between regions is expressed by the power counting parameter $\lambda \ll 1$.
We overcome this issue by setting up a tower of related soft collinear effective theories, 
called SCET$_i$, which also helps us deal with several technical obstacles.  We
formulate the shower description as a standard matching procedure between
operators in different SCET$_i$. Power corrections are encoded by performing
matching computations at subleading order in the kinematic expansion.  
These corrections modify the processes that initiate the shower, modify
certain early branching probabilities, and open up the $1\to 3$ splitting
channel.  Virtual perturbative $\alpha_s$ corrections are included by performing
matching calculations beyond tree level between $\sceti$ theories.  Finally,
corrections to the Sudakov no-branching probabilities are encoded through
anomalous dimensions of leading and subleading operators at the appropriate
order within different $\sceti$'s.  When we refer to a parton shower in the
context of our calculations, we mean an explicit amplitude formula that would
agree numerically with a corresponding shower algorithm.  We will carry out the
necessary computations for the power corrected matching equations, and a subset
of the required calculations for anomalous dimensions occurring for operators
beyond the LL shower.  This analysis includes the leading corrections to
the shower from interference and from spin correlations.  As much as possible,
we attempt to give pointers for additional computations that are needed in
places where our analysis is incomplete. For example, to simplify things we have
not treated color correlations since doing so increases the basis of operators
and the number of computations, but does not change the conceptual setup.

The outline of our paper is as follows.  We present a brief overview of SCET in
Section~\ref{subsec:introSCET}. We review the Bauer-Schwartz SCET shower method
in Section~\ref{subsec:bands} and discuss the technical obstructions to
extending it to include power corrections. In Section~\ref{subsec:sceti}, we
present our $\sceti$ framework to resolve these issues. In Section \ref{sec:LO},
we analyze the LL shower in the $\sceti$ framework, and show that the transition
between SCETs, $\sceti \to \scetipone$, can be encoded by operator replacement
rules on single parton collinear fields.  Soft emissions in $\sceti$ are
discussed, and we summarize the correspondence between $\sceti$ objects and LL
shower ingredients.  In Section \ref{sec:nlo}, we use the $\sceti$ formulation
to classify and compute various corrections to the shower to $\mo(\lambda^2)$ in
the cross section. Two main categories of branching corrections emerge, which we
refer to as ``hard-scattering'' and ``jet-structure.'' We also discuss
ingredients needed for renormalization group evolution corresponding to
no-branching probabilities, derive all the LL anomalous dimensions for our
subleading operators.  Additionally, we mention the issues involved in obtaining
NLL$_{\rm exp}$ resummation from our results.  A summary of corrections in the
$\sceti$ framework is presented as a table in section~\ref{subsec:nlomap},
including the type of corresponding ingredients needed in a subleading shower.
We present in Eqs.~(\ref{eq:weight})-(\ref{eq:weightDefs}) a parton shower reweighting factor that
should allow one to implement our corrections.  We also discuss the
correspondence of these corrections with those currently included in other Monte
Carlos.  Conclusions are given in Section \ref{sec:conclusion}. At the present
time, we do not have an algorithmic implementation of our power suppressed
shower results, but work in this direction is in progress.

Many details are relegated to the Appendices.  Further details about SCET can be
found in Appendix \ref{app:scet}.  We describe finite reparametrization
transformations in Appendix \ref{app:rpi}, which is an important symmetry that
we use in our matching computations to disentangle kinematic coordinate
conventions from kinematic power corrections.  Details on the matching of QCD
$\to \scetone$, $\scetone\to \scettwo$, and $\scettwo\to \scetthree$ can be
found in Appendices \ref{app:QCD/SCET1}, \ref{app:scet1/scet2}, and
\ref{app:scet2/scet3}, respectively.  A complete list of the operators needed
to compute \eq{goalDef} in  $\scetn$ is given in \app{scet2/scet3}.
Appendix \ref{app:nloSplit} contains a
cross-check on our results, where we integrate a subset of our power suppressed
terms to rederive the abelian terms in $P_{q\to qg}^{(1)}$, namely the
$\mo(\alpha_s)$ correction to the $q\rightarrow qg$ splitting
function~\cite{Curci:1980uw}.

Those readers looking to find a quick summary of our results should look in 
Secs.~\ref{subsec:Summ LL} and \ref{subsec:nlomap}.

\section{Obtaining the Parton Shower with SCET}
\label{sec:SCETshower}

\subsection{SCET Basics} 
\label{subsec:introSCET}

Soft-Collinear Effective Theory is an effective field theory of QCD that
describes the interactions of collinear and soft particles \cite{Bauer:2000ew,
  Bauer:2000yr, Bauer:2001ct, Bauer:2001yt}.  We present here the basic ideas
needed for our analysis of the parton shower, including how collinear sectors
are organized into equivalence classes by the power counting parameters. Further
SCET concepts are reviewed in Appendix \ref{app:scet}.

The momentum, $p$, of any particle can be decomposed along two light-cone
vectors, $n$ and $\bar{n}$, with $n^2=0$, $\bar{n}^2=0$ and $n\cdot \bar{n}=2$,
as
\begin{align}
p^{\mu}&= \bar{p} \,\frac{n^\mu}{2}+p^\mu_\perp+n\!\cdot\!p \frac{\bar{n}^\mu}{2}\, ,
\end{align} 
where $\bar{p}=\bar{n} \cdot p$ and the particle's invariant mass is 
$p^2= n \cdot p \, \bar p +p_\perp^2$.  We use a Minkowskian notation for 
$p_\perp^2 = -\vec p_\perp^{\,2}$, where $\vec p_\perp$ is Euclidean. SCET's 
degrees of freedom include $n_i$-collinear fields for a set of distinct directions 
$\{n_i\}$, and soft fields.\footnote{Our primary interest here is the perturbative structure of
  jets, so we use ${\rm SCET}_{\rm I}$ theories with collinear and ultrasoft
  modes. For simplicity we will always use the phrase soft in place of
  ultrasoft.}  
A particle is collinear to a direction $n$ if its momentum scales
as:
\begin{equation}
(n \cdot p, \,\bar{p},\, p_\perp)\sim (\lambda^2, \,1, \,\lambda)\,\bar p\, ,
\label{eq:powercountingSCET}
\end{equation}
where $\bar p\sim Q$ is some hard scale in the process, and $\lambda \ll 1$ is the SCET
power counting parameter.  A particle is soft if
\begin{equation}
(n \cdot p, \,\bar{p},\, p_\perp)\sim (\lambda^2, \, \lambda^2, \,\lambda^2)\,Q\, .
\label{eq:powercountingSCETsoft}
\end{equation}
Collinear and soft fields have virtuality $\sim \, Q^2\lambda^2$ and $Q^2\lambda^4$,
respectively.  We obtain SCET from QCD by expanding in powers of $\lambda$,
integrating out hard modes, and dividing the remaining ones into collinear and
soft fields. Our collinear and soft degrees of freedom also contain all the IR
regions that can be obtained by a rescaling of $\lambda\to \lambda^i$, for
$i>1$. The leading order SCET Lagrangian is
\begin{align}
  {\cal L}_{\rm SCET}^{(0)} = {\cal L}_s^{(0)}+ \sum_{n\in \{n_i\}}
  \mathcal{L}_n^{(0)} \,,
\end{align}
where $\mathcal{L}_n^{(0)}$ is defined in Eq.~(\ref{eq:LagSCET}) and 
has only interactions among particles
collinear to the same $n$.  ${\cal L}_s^{(0)}$ is the Lagrangian for soft interactions 
discussed further in \app{scet}.
Particles collinear to different directions can
interact either by the exchange of soft modes, or from
their coupling to other sectors in external operators.  Two collinear sectors in
SCET, $n_1$ and $n_2$, are distinct if~\cite{Bauer:2002nz}:
\begin{align}
n_1\!\cdot\! n_2 \gg \lambda^2 \,,
\end{align}
so any particle is collinear to at most one direction within a given
SCET. The collinear sectors $\{n_i\}$ in SCET are really sets of
equivalence classes of null vectors, $\{[n_i]\}$, where the
equivalence class is $[n_j]=\{ n\in [n_j] |\, n\cdot n_j \lesssim
\lambda^2\}$. A class $[n_j]$ consists of all light-like vectors
connected to $n_j^\mu$ by a type-I reparametrization invariance (RPI)
transformation, $n_j^\mu \to n_j^\mu + \Delta_{n_j\perp}^\mu$, where
the scaling of the transformation parameter is
$\Delta_{n_j\perp}^\mu\sim \lambda$ (see \app{rpi} for a detailed
discussion of RPI). Physically, the class $[n_j]$ corresponds to
light-like vectors for particles whose momenta is in a cone centered
on $\vec n_j$ with an opening angle $\sim \lambda$ ({\it cf.}
Fig.~\ref{fig:singlecone}).

Thus, the defining concepts of a SCET-theory are its hard-scale $Q$, its
collinear sectors $\{[n_i]\}$, and its power counting parameter $\lambda$ which
governs the importance of operators and the size of the collinear sectors in phase
space.

Most of our discussion will involve interactions with collinear fields,
and we use the notation $\chi_n$ for quarks and
$\mathcal{B}_{n\perp}^\mu$ for gluons (definitions of these fields can
be found in Eq.~(\ref{eq:collFieldsDef}), and they incorporate
collinear Wilson lines built out of $\nb\cdot A_n$ fields).  We can
match QCD onto a series of SCET operators organized by powers of
$\lambda$.  The key building blocks are: $\chi_n$, $\mb_{n\perp}^\mu$,
and $\mathcal{P}_{n\perp}^\mu$ (a type of derivative operator that
yields the perpendicular momentum of an $n$-collinear field), each of
which scale as $\lambda$ in the kinematic power counting.  A general
notation for the $i$-parton operators we will consider is:
\beq
\mo^{(j,k,\ell)}\big(n_1^{[\ell_1]},\ldots,n_{j+k}^{[\ell_{j+k}]}\big) \,=
 \bigg[ \prod_{a=1}^{j/2} ({\cal P}_{n_a\perp})^{\ell_a} \chi_{n_a} 
  \bigg] 
 \bigg[ \prod_{b=j/2+1}^{j} ({\cal P}_{n_b\perp})^{\ell_b} \bar\chi_{n_b} 
  \bigg]  
 \bigg[ \prod_{c=1}^{k} ({\cal P}_{n_c\perp})^{\ell_c} g \mb_{n_c\perp} 
  \bigg], 
\label{eq:opDef}
\eeq
where the number of partons is the sum of quarks and gluons, $j+k =
i$, and the total number of $\perp$ derivatives is
$\ell=\sum_{m=1}^{j+k}\ell_m$.  In the operator argument, we list the
index labels, $n_d$, of the parton fields on the RHS.  The
superscripts in the argument on the LHS denote the number of
derivatives acting on the field with the corresponding direction.
There may be a degeneracy among the index labels, $n_d$, and so the
operator has at most $i$ distinct collinear directions.  The scaling
of these operators is ${\cal O}^{(j,k,\ell)}\sim \lambda^{j+k+\ell}$.
They are tensors in the space of spinors and Lorentz vectors, and the
indices get contracted with structures contained in the Wilson
coefficient $C$ for the operator. If $C {\cal O}$ is a Lorentz scalar,
then $j$ is even. Since the collinear fields carry a label referring
to a specific light-cone vector, these operators describe particles in
a specific region of phase space.  SCET therefore distinguishes
situations with the same particle content, but different kinematics,
in a straightforward way.
\begin{figure}[t!]
\centering
\includegraphics[width=0.9\textwidth]{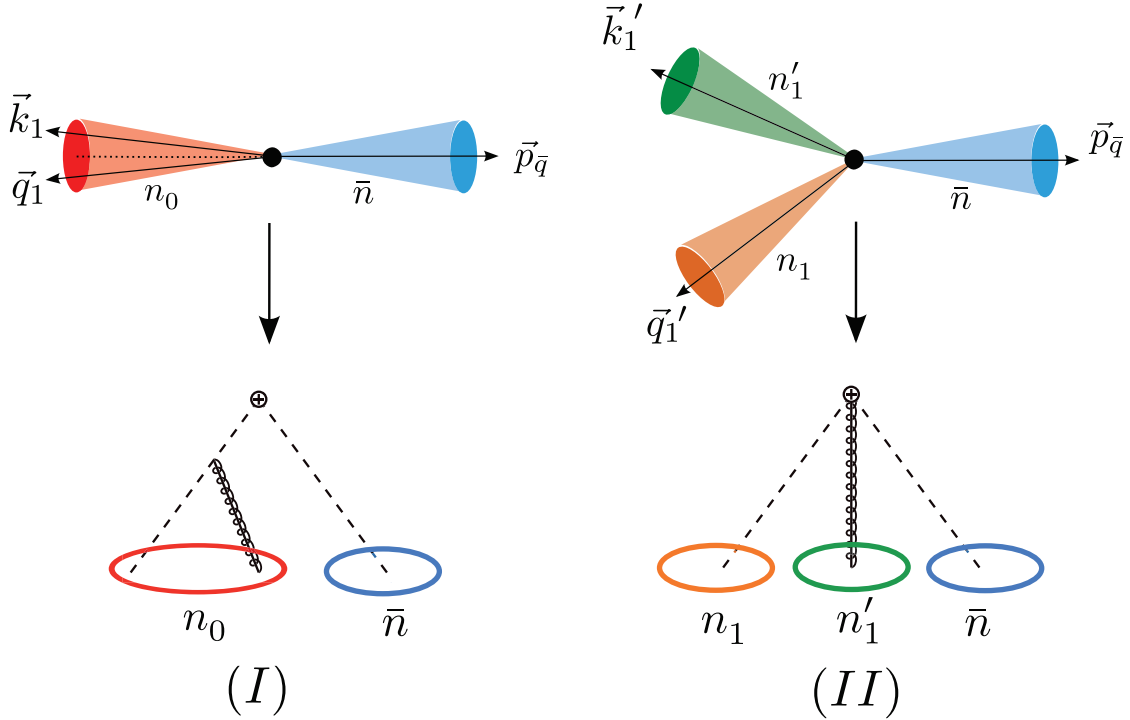} 
\caption{Different kinematic configurations of a final state with a quark,
  antiquark, and gluon are described by
  different SCET operators.  In ({\it I}), the quark and the gluon are collinear
  to the direction $n_0$, represented by their sharing a common cone.  In ({\it
    II}), the vectors $q'_1$ and $k'_1$ are too far apart to be collinear.  The
  Feynman diagrams show that collinear particles can come from Lagrangian
  insertions, whereas non-collinear ones arise exclusively from
  higher-multiplicity operators.  The Feynman diagram in ({\it I}) only depicts
  the first term on the RHS of Eq.~(\ref{eq:AI}).}
\label{fig:cone3aC}
\end{figure}

For example, one can take an amplitude for three external particles: a quark,
gluon, and antiquark.  We can consider two different
configurations, $|q_{n_0} g_{n_0}\bar{q}_{\bar n} \rangle$ and
$|q_{n_1} g_{n_1'}\bar{q}_{\bar n} \rangle$. In the first, shown in
\fig{cone3aC}($I$), the quark and the gluon are $n_0$-collinear, and the
antiquark is collinear to a different direction, $\bar{n}$. Here the amplitude
is described by operators with two distinct directions, say
\begin{align}
 {\cal O}^{(2,0,0)}(n_0,\nb) &=\bar{\chi}_{n_0}
\Gamma \chi_{\bar{n}}\sim \lambda^2\,,
 &{\cal O}^{(2,1,0)}(n_0,n_0,\nb) &=\bar{\chi}_{n_0}g  \mb_{n_0 \perp}^\mu
\Gamma^\prime \chi_{\bar{n}}\sim \lambda^3 \,,
\end{align}
where the form of the Dirac structures $\Gamma$ and $\Gamma^\prime$ are not
central to our discussion here.  ${\cal O}^{(2,0,0)}$ can emit $\nb\cdot
A_{n_0}$ gluons from the Wilson line in $\chi_{n_0}$, but requires a Lagrangian
insertion to emit an $A_{n_0}^\perp$ gluon.  Schematically, the amplitude
for a transverse gluon has contributions:
\begin{align} \label{eq:AI}
A^I = \int\!\!dx\: \big\langle 0\big| 
 T\: \{ \mathcal{L}^{(0)}_{n_0}(x) \, \bar{\chi}_{n_0} \Gamma \chi_{\bar{n}}(0) \}
   \big|q_{n_0} g_{n_0} \bar{q}_{\nb} \big\rangle 
 + \big\langle 0 \big| \bar{\chi}_{n_0}g \mb_{n_0 \perp}^\mu
\Gamma^\prime \chi_{\bar{n}}(0) \big|q_{n_0} g_{n_0} \bar{q}_{\nb} \big\rangle 
\,.
\end{align}
In \fig{cone3aC}($II$), each of the particles is collinear to a distinct
direction, so no cone of size $\sim \lambda$ fits two of the momenta.  In this
case, the amplitude can only come from an operator with three distinct
labels, such as $\bar{\chi}_{n_1} \mb_{n_1' \perp}^\mu \Gamma^{\prime\prime}
\chi_{\bar{n}}$:
\begin{align} \label{eq:AII}
A^{II} = \langle 0| \bar{\chi}_{n_1}g \mb_{n_1' \perp}^\mu \Gamma^{\prime\prime}
\chi_{\bar{n}} |q_{n_1} g_{n_1'} \bar{q}_{\nb} \rangle \, .
\end{align}
The ability of SCET to cleanly separate contributions such as those in
\eqs{AI}{AII} will be useful for formulating a complete set of power suppressed
corrections to the parton shower.

\subsection{Bauer-Schwartz Method} 
\label{subsec:bands}

The original application of SCET to study and improve the parton shower was
carried out in \cite{Bauer:2006mk,Bauer:2006qp} by Bauer \& Schwartz.  The
main reasons why SCET is useful for this are:
\begin{itemize}
\item{ The SCET fields, soft and collinear quarks and gluons, 
		have support in the infrared exactly where the parton shower 
		amplitudes have their dominant contributions in phase space.  }
\item{Since SCET is improvable order-by-order in the kinematic expansion
    parameter, $\lambda$, one has the potential to systematically correct the
    shower.  }
\end{itemize}
We will give a short overview of the Bauer-Schwartz approach, and then discuss
the complications that arise when trying to extend the analysis to NLO in the
$\lambda$ expansion, namely \NLOl.  In this section we will use notation that is
not found elsewhere in the paper to retain consistency with
\cite{Bauer:2006mk,Bauer:2006qp}.

The procedure of \cite{Bauer:2006mk,Bauer:2006qp} starts by constructing
$i$-parton operators, $\mo_i$, through matching SCET to QCD at a hard scale.
For example, their $\mo_2$ will equal $\mo^{(2,0,0)}(n_1,\, n_2)$ in the
notation of \eq{opDef}, and $\mo_3$ will be $\mo^{(2,1,0)}(n_1,\, n_2,\, n_3)$.
As we run $\mo_i(\mu)$ down, the leading log renormalization group evolution (LL
RGE) does not mix operators and the exponential evolution kernel encodes the
no-branching probability.  The evolution continues until another parton becomes
apparent at a scale $\mu=p_T$. 

If we have an $i$-parton operator, ${\cal O}_i=\mo^{(j,i-j,0)}(n_1,\ldots,n_i)$
with all $n$'s distinct, then it has the RG solution ${\cal O}_i(\mu)=
U^{(j,i-j,0)}(Q,\mu) {\cal O}_i(Q)$ with
\beq
U^{(j,i-j,0)}(Q,\,\mu) \,=\, \exp \left[ - \int_\mu^Q \frac{d\mu'}{\mu'} \gamma^{(j,i-j,0)} (\mu')  \right],
\label{eq:rgKernel}
\eeq
where $\gamma^{(j,i-j,0)}$ is the operator's anomalous dimension.  The
leading-log resummation effects of the Sudakov factor in the PS enter through
one-loop operator running in SCET, as dictated by the cusp anomalous dimension.
The one-loop cusp portion is especially easy to calculate in SCET as it depends
solely on the number of collinear fields, even though the calculations have
loops involving soft ones as well~\cite{Bauer:2006mk,Bauer:2006qp},
\beq
\gamma^{(n_q,n_g,0)}_{\rm{LL}} (\mu) \,=\, -\frac{\alpha_s}{\pi} 
\left[ \frac{n_q}{2}C_F + \frac{n_g}{2}C_A \right] \log \frac{\mu^2}{Q^2} \,.
\label{eq:oneLoopCusp}
\eeq 
This form of the kernel gives a product of Sudakov factors which are the
no-branching probabilities for each parton in the operator:
\beq
U^{(j,i-j,0)} _{\rm{LL}}(Q,\,\mu)\,=\, \Delta_q^{\frac{j}{2}}(Q,\,\mu)
\Delta_g^{\frac{i-j}{2}}(Q,\,\mu).
\label{eq:runningAsSudakov}
\eeq
Here, as in ~\cite{Catani:2001cc}, one accounts for leading-log effects for
any particle multiplicity by simply multiplying matrix elements by appropriate
Sudakov factors.

As we run $\mo_i(\mu)$ down, another parton becomes apparent at a scale
$\mu=p_T$.  To account for this, Bauer-Schwartz devised a ``threshold
matching'' of $\mo_i$ to a new, higher multiplicity operator, $\mo^{(i)}_{i+1}$,
where the subscript still denotes the number of partons in the operator and the
superscript tracks the parent operator.  The general threshold matching equation
is
\begin{align} \label{eq:tmatch}
 \big[ C_n^{(j)} \langle {\cal O}_n^{(j)} \rangle \big]_{\mu=p_T+\epsilon} 
  = \big[ C_{n+1}^{(j)} \langle {\cal O}_{n+1}^{(j)} \rangle
  \big]_{\mu=p_T-\epsilon} 
  \,.
\end{align}
After further running and threshold matching, we eventually have $\mo^{(i)}_{n}$
for various $n>i$.  The $n-i$ particles emitted at increasingly lower scales by
this process correspond to the parton showering of the original fields created
at the hard scale by $\mo_i$.  Additionally, they also showed that an
appropriate list of SCET operators ($\mo_i$'s and $\mo_i^{(n)}$'s) can
interpolate between fixed-order QCD and parton shower (PS) calculations of
IR-safe observables.  Furthermore, they derived the $\mo(\alpha_s)$ effects from
matching QCD to SCET at one-loop for ${\cal O}_3$.

That subsequent emissions reproduce the usual parton shower splitting function
emerges easily from SCET. Consider an operator $\mo_i \,=\,
\bar{\chi}_{n_0}\Omega$, where $\Omega$ is arbitrary and we have made explicit a
single collinear quark field, $\bar{\chi}_{n_0}$.  If we emit a collinear gluon
from this quark, $q(q_0^\mu)\to q(q_1^\mu) g(k_1^\nu)$, the amplitude for the
process is
\beq \label{eq:AXqg}
A^{X+qg}_{LO} \,=\, \bar{u}_{n_0}(q_1) \rho^{\alpha} \frac{\bar{q}_0}{q_0^2} \Omega,
\eeq
where $u_{n_0}$ is the collinear quark spinor, and $\rho^{\alpha}$ is the
combination of the SCET single gluon emission Feynman rule plus the $\bar
\chi_{n_0}$ Wilson line emission (the quark ${\cal L}_n^{(0)}$ can be found in
Eq.~\ref{eq:LagSCET}),
\beq \label{eq:rhoDef0}
\rho^\alpha = n_{0}^\alpha 
 + \frac{(\slashed{q}_1)_{n_0 \perp} \gamma_{n_0\perp}^ \alpha}{\bar{q}_1} 
 + \frac{\gamma_{n_0\perp}^ \alpha (\slashed{q}_0)_{n_0\perp} }{\bar{q}_1} 
 - \frac{\nb^\alpha}{\bar q_0} \bigg[ \frac{q_0^2}{\bar k_1}  
  + \frac{(\slashed{q}_1)_{n_0\perp} (\slashed{q}_0)_{n_0\perp}  }{\bar{q}_1}
  \bigg] . 
\eeq
Note that $\rho^\alpha$ in SCET comes entirely from $\bar\chi_{n_0}$ without
reference to anything residing in $\Omega$.  The subscript $(n_0\!\perp)$ refers
to components perpendicular to $n_0^\mu$ and $\nb^\mu$, which we denote by
$\perp$ for the remainder of this computation. The amplitude in \eq{AXqg} is
gauge invariant and $k_1^\alpha \rho_\alpha=0$.  Squaring $A_{LO}^{X+qg}$ and
summing over spins we have $\sum_{spin} \bar u_{n_0}(q_1) u_{n_0}(q_1) = \bar q_1
\ns_0/2$, and the gluon polarization sum denoted $\sum_{spin}
\epsilon_\alpha \epsilon^*_\beta = d_{\alpha\beta}$.  Since $\rho^{\alpha}$
commutes with $\ns_0$, we get an answer proportional to $\rho^{\alpha}
\rho^{\dagger \beta} \, d_{\alpha \beta}$, where without loss of generality we
can use a light-cone gauge, $d_{\alpha \beta} =\, -g_{\alpha \beta} +
(\nb_{\alpha} k_{1\beta} +k_{1\alpha} \nb_{\beta})/\bar{k}_1$.  Crucially, this
is a Dirac scalar:
\beq \label{eq:rhoSq0}
\rho^{\alpha} \rho^{\dagger \beta} \, d_{\alpha \beta} \,\equiv\, |\rho|^2 \,=\, 
2 \Big( \frac{2\, q_0^2}{\bar{k}_1\bar q_0 }  
 - \frac{q_{1\perp}^2}{\bar{q}_1^2}
 + \frac{2 q_{0\perp} \cdot q_{1 \perp}}{\bar{q}_0 \bar{q}_1}
 - \frac{q_{0\perp}^2}{\bar{q}_0^2}
  \Big) 
\times \boldsymbol{\mathbb{ I}}_4,
\eeq
where we have used the on-shell conditions $q^2_1=0$ and $k_1^2=0$.

In a frame where $q_0^\perp=0$ we have $q_{1\perp}=-k_{1\perp}$ and $\bar
q_0/q_0^2 = 1/(n_0\cdot q_0)$. Here $n_0\cdot q_0 = n_0\cdot k_1 + n_0\cdot q_1
= -k_{1\perp}^2/[\bar q_0\, z(1-z)]$, where $z \equiv \bar{q}_1/\bar{q}_0$. Thus
we have the simpler expression
\beq \label{eq:rhoDef}
\rho^\alpha 
= n_{0}^\alpha 
 + \frac{(\slashed{q}_1)_{n_0 \perp} \gamma_{n_0\perp}^ \alpha}{\bar{q}_1}
 , 
\eeq
which we have written in light-cone gauge without the Wilson line contribution 
($\propto \nb^\alpha q_0^2$), and 
\beq \label{eq:rhoSq}
\rho^{\alpha} \rho^{\dagger \beta} \, d_{\alpha \beta} 
 \,=\,  2 \Big( \frac{2\, n \cdot k_1}{\bar{k}_1} -
\frac{2 q^2_{1 \perp}}{\bar{q}_1 \bar{k}_1} - \frac{q_{1\perp}^2}{\bar{q}_1^2} \Big) 
\times \boldsymbol{\mathbb{ I}}_4
 = - \frac{2\, k_{1\perp}^2}{\bar q_0^2}\: \frac{(1+z^2)}{z^2(1-z)^2} \,.
\eeq
Putting these properties together in the full amplitude squared we get
\begin{align}  
|A^{X+qg}_{LO}|^2 
 &= \frac{g^2 C_F}{(n_0 \cdot q_0)^2} \frac{\bar{q}_1}{2} \, 
{\rm Tr} \left[ \ns_0 \rho^{\alpha} \Omega \Omega^{\dagger}
 \rho^{\dagger \beta} \right] d_{\alpha \beta} 
= \frac{g^2 C_F\,\bar{q}_1}{(n_0 \cdot q_0)^2}  |\rho|^2 \, 
{\rm Tr} \left[ \frac{\ns_0}{2}  \Omega 
 \Omega^{\dagger} \right] \nonumber \\ 
&= g^2 C_F\, 2 z\, \frac{(1+z^2)}{|k_{1\perp}|^2} \,
 {\rm Tr} \left[ \bar{q}_0\frac{\ns_0}{2} \Omega \Omega^{\dagger} \right]
 \,. 
\label{eq:loAmpFact} 
\end{align}  
Thus, all information about the emission factors out to the front and is
independent of the rest of the process encoded by $\Omega$. Since the power
expansion is built into SCET, there was no need to expand terms in the amplitude
to obtain this result (unlike the analogous computation in full QCD).  In order
to recover \eq{ap}, we still need to include the $z$-dependence from phase
space, since $P^{(0)}_{jk}(z)$ operates at the level of the cross section. Using
$d^3k/(2E_k) = d\bar{k} d^2k_\perp/(2\bar{k})$, for $q_1$ and $k_1$ we have
\begin{align}
 \frac{d\bar{q}_1 d^2q_{1\perp}}{2 \bar q_1} \frac{d\bar{k}_1 d^2k_{1\perp}}{2\bar k_1}
 & \rightarrow \frac{d\bar{q}_0 d^2q_{0\perp}}{2 \bar q_0} \frac{dz\, d^2k_{1\perp}}{2z(1-z)},
\label{eq:spPhaseSpace}
\end{align}
where the arrow means that we insert
$d^4 q_0\, \delta^{(4)}(q_0 - q_1 - k_1)$
and integrate $d^3 q_1$ along with $d(n_0 \cdot q_0)$.
Thus, we recover the expected $1/(1-z)$ dependence from the measure.  Combining
pieces and performing the trivial azimuthal integral $d\phi_{k_1}$, we get the
expected expression:
\beq
d\sigma_{X+qg} \,=\, dz\,\frac{d k_{1\perp}^2}{k_{1\perp}^2} 
 P^{(0)}_{q\to qg}(z) \, d\sigma_{X+q},
\label{eq:splitCross}
\eeq
where $P^{(0)}_{q\to qg}(z)$ is the quark splitting function in \eq{ap}.  Here
$d\sigma_{X+q}$ is the cross section for the rest of the process with emission
of a momentum $q_0$ quark, and the corresponding amplitude squared is ${\rm Tr}
\left[\frac{\bar{q}_0}{2} \ns_0 \Omega \Omega^{\dagger} \right]$.
Whether $\Omega$ represents a simple hard current or an entire chain of
collinear splittings, we see that the $q\to q g$ emission factors out with the
expected soft-collinear double pole, as in \eq{splitFact}.

In order to obtain their results, Bauer-Schwartz introduced choices and
approximations at several points which obscure the path toward systematically
computing \NLOl corrections.
Indeed, they concluded that obtaining these corrections  may be
prohibitively difficult~\cite{Bauer:2006qp}.  Some of the issues one
encounters trying to work at higher orders are:
\begin{enumerate}
\item{At \NLOl, it becomes crucial to distinguish which simplifications correspond
    to approximations with power corrections, and which involve a choice of
    coordinates where a symmetry makes the final answer coordinate independent.  For example,
    a collinear state typically has nonzero momentum components
    perpendicular to the index $n$ of the field that annihilates it.
    Refs.~\cite{Bauer:2006mk,Bauer:2006qp}, however, dictated that collinear
    SCET fields in their operators only create particles whose momenta perfectly
    align with their index direction, $n$:
    \begin{eqnarray} \label{eq:condOne} 
    \chi_n|q \rangle &=& \delta_{n,n_q}, 
    \ \ \mathrm{where} \:\: n_q^\mu = q^\mu/E_q,
    \end{eqnarray}
    leaving it ambiguous what amount of symmetry protects this choice.
    \eq{condOne} enforces certain kinematical restrictions on final state
    particles, and requires that fermion fields be rotated to an appropriate
    $n_q^\mu$ via $\xi_n\to (\ns\nbs/4) \xi_{n_q}$.  }
 \item{At LO, it was possible to avoid a potential double counting between
     collinear and soft fields by dropping soft emission and Wilson line
     emission, and taking only collinear emissions with transverse polarization.
     The threshold matching procedure is designed to avoid double counting of
     collinear operators, such as a Lagrangian emission from ${\cal O}_2$ and
     direct emission from ${\cal O}_3^{(2)}$, since only one of these is allowed
     to operate at a time. However, the threshold matching in \eq{tmatch} makes
     the technical procedure for incorporating power corrections unclear.
   }
 \item{ Threshold matching contains another impediment to systematic
     improvement.  Through this procedure, the initial operator $\mo_2$ has
     nonzero projection onto Fock states of any multiplicity, but the number of
     particles created by an operator is a scale-dependent question. The
     matching scales are determined by the strong ordering kinematics,
     $p_{1\perp} \gg \ldots \gg p_{m\perp}$. At the scale of an emission, say
     $p_{1\perp}$, one threshold matches to the operator $\mo_3^{(2)}$, which only adds
     one parton at a time.  However, going to higher orders in the shower
     necessitates more general configurations. }
\end{enumerate}
In carrying out their method, Bauer-Schwartz carefully enumerated the above
approximations.  They affect the ability to include corrections in $\lambda$,
but do not impact the terms necessary for a LL shower. 

Building on the work of Refs.~\cite{Bauer:2006mk,Bauer:2006qp}, the main goal of
the framework we develop in the next section is to overcome this list of issues
so that we can determine power corrections to the shower using SCET.

\subsection{Using SCET${}_i$}
\label{subsec:sceti}

The main feature of the parton shower is the ability to capture the dominant
physics of particles emitted in kinematically hierarchical regions of phase
space. Our goal is to formulate the SCET interface with the shower using
a standard sequence of matching and running steps in different versions of SCET,
\begin{align} \label{eq:scetichain}
  \text{QCD} \to \text{SCET}_1 \to \text{SCET}_2 \to \cdots \to \text{SCET}_N\,.
\end{align}
We refer to this as the $\sceti$ procedure. The key distinction between a SCET
at one stage and the next is the definition of the corresponding resolution
parameters $1\gg \lambda_1 \gg \lambda_2 \gg \cdots \gg \lambda_N$, where
$\lambda_i$ sets the power counting for SCET$_i$. As we move down the chain, the
corresponding SCET resolves smaller $\sim (Q\lambda_j)^2$ invariant masses and
relative squared perpendicular momenta, and has a different meaning for its
collinear sectors $\{[n_i]\}_{{\rm SCET}_j}$. To keep track of this, we will
attach a subscript to the operators to denote the $\sceti$ in which its fields
live,
\begin{align}
 {\cal O}^{(j,k,\ell)}_i(n_1,\ldots,n_{j+k}) \,.
\end{align}
Effectively with \eq{scetichain}, we partition the momenta of partons in the
shower history into classes,
\begin{align}
\Omega_0 \supset \Omega_1 \supset \ldots \supset \Omega_N \,,
\label{eq:omegas}
\end{align}
where $\Omega_j$ contains the momenta of all propagators having
$p^2\sim (Q\lambda_j)^2$ or smaller, or an equivalent condition on
relative perpendicular momenta.  The allowed momenta in $\Omega_i$
correspond to the collinear modes of SCET$_i$.  The sequence of
$\sceti$'s is truncated when we resolve a scale of order the parton
shower cutoff, $Q\lambda^{N} = p_T^{\rm cut} \simeq 1\,{\rm GeV}$,
that is in $\text{SCET}_N$.  

Note that we do not associate a large hierarchy to the hard scales $\bar{p}_i$
between $\text{SCET}_j$ and $\text{SCET}_{j+1}$. That is to say we do not
associate the energy loss due to splitting with a power of $\lambda_j$.  Instead
if $Q$ is the scale of the primary hard interaction then we consider
$\bar{p}_i\sim \eta^j \, Q$ in $\text{SCET}_j$, where $\eta\gg \lambda_j$ and
for numerical estimates we can take $\eta \sim \frac{1}{2}$.  (For each
branching the geometric mean of the two daughters' $\bar{p}$ fraction averages
to $0.4$ which is roughly one half.)  Parametrically, the decrease in the
parton energy is not as rapid as that for the perpendicular momenta encoded in
the power counting parameter $\lambda_j$.  In principle, we can account for
$\eta$ as a separate factor.  In practice, we will be most interested in tracking
powers of $\lambda_j$ and will only include $\eta$ factors in places where the
corresponding powers of two have a numerical impact on the implementation, or if
we wish to disentangle the changes in offshellness due to strong-ordering
effects and those coming from the more modest decrease in $\bar{p}_i$.

The strongly ordered configuration of partons in \eq{psLimit1} corresponds to
removing a single $q_j^2$ in $\Omega_j$ as we pass from $\Omega_j\to
\Omega_{j+1}$.  However, with \eq{omegas}, nothing stops us from having multiple
emissions at a single scale.  If two mother particles, with $q_j^2$ and
$q_{j+1}^2$, are associated to the same $\Omega_k$, then when we integrate out
that scale in SCET$_{k+1}$ this configuration just contributes to an operator
with a different parton multiplicity from the strongly ordered one.  Thus, with
\eq{scetichain} there is no obstacle to considering corrections from an
arbitrary assignment of $q_j^2$'s to $\Omega_k$'s. This resolves issue 3.~of
\subsec{bands} since we can treat emissions where the shower tree has momenta
with the same parametric scaling in $\lambda$.

To carry out calculations in the $\sceti$ framework, it is convenient and
sufficient to take a specific definition of the power counting parameters,
$\lambda_i = (\lambda)^i$.  We want the hierarchy between neighboring splittings
to stay the same throughout the shower so as not to privilege any portion of it.
We will see in \subsec{ampSq} that this democratic setup allows us to interpret
part of our $\mo(\lambda)$ corrections to $i$-parton amplitudes as universal
corrections to the splitting probability, given at LO by \eq{ap}.  As we go to
lower scales, our definition of collinearity also changes, and by analogy to
\eq{powercountingSCET}, fields collinear to $n$ within $\Omega_i$ have:
\beq \label{eq:omegaCond}
 (n \cdot q_i, \,\bar{q}_i,\, q_{i \perp}) \sim
(\lambda^{2i}, \,1, \,\lambda^i)\, \bar q_i, 
\eeq 
and virtuality $\sim\, (\bar q_i)^2 \lambda^{2i}$.  In $\sceti$, ${\cal
  L}_n^{(0)}$ again only couples collinear fields in the same direction $n$.
Since different $\sceti$'s have different definitions of collinearity, our
description of identical physical processes changes when we switch to a theory
with a lower scale. For convenience, we will use the same auxiliary vector
$\nb^\mu$ for any $n_j$-collinear field in any $\sceti$.  If $\nb$ is a valid
auxiliary vector for $n$-collinear fields in $\text{SCET}_1$, then it is readily
apparent that it will be a valid choice for all subsequent collinear fields in
$\sceti$'s that descend from an $n$-collinear mother in $\text{SCET}_1$. Our
default choice is stronger: given a set of light-like vectors in $\{n_j\}$ in
$\text{SCET}_1$ we take a light-like $\nb$ that is parametrically close or
aligned with the antiquark direction. We then adjust the magnitude of $n_j^0$
and of $\vec n_j$ so that $n_j^2=0$ and $\nb\cdot n_j=2$ (for a related
discussion based on RPI see Appendix~\ref{app:QCD/SCET1}).
\begin{figure}[t!]
\centering
\includegraphics[width=0.9\textwidth]{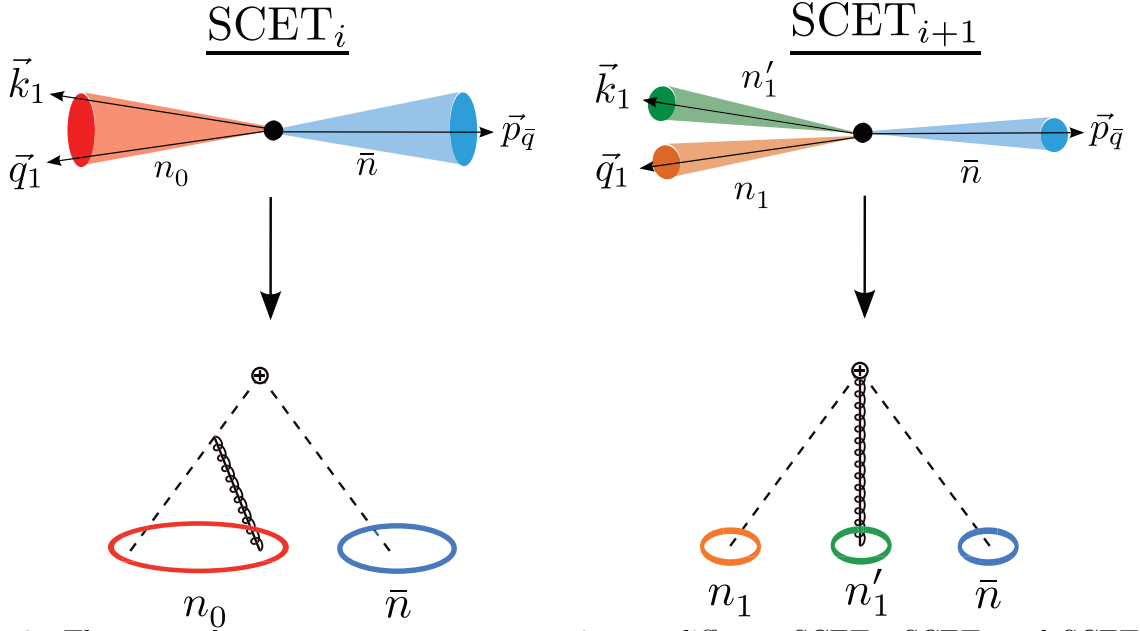} 
\caption[SCET$_i$ operators]{The same three-parton process as seen in two
  different SCETs, $\sceti$ and $\scetipone$.  Above: Kinematic configuration
  of the quarks and gluon.  The solid cones represent the regions considered
  collinear to the vectors drawn.  Below: Feynman diagrams for the corresponding
  amplitude.  Note that in $\scetipone$ we have removed a degree of freedom that
  propagates in $\sceti$.  The amplitude thus comes from a higher dimension
  operator $\mo^{(1)}_{i+1}$, rather than from a time-ordered product of
  $\mathcal{L}_{\sceti}$ with $\mo^{(0)}_{i}$, as it did in $\sceti$.}
\label{fig:figureTprC}. 
\end{figure}

We depict the different descriptions of the same physical configuration in
\fig{figureTprC}, where the left panel is in $\sceti$ and the right panel is in
$\scetipone$.  In $\sceti$, the quark ($\vec{q}_1$) and gluon ($\vec{k}_1$) are
$n_0$-collinear.  This means that at LO they are emitted from a $qqg$ vertex in
the LO $\sceti$ Lagrangian (or a Wilson line interaction).  Schematically, the
amplitude for a $\perp$-polarized gluon looks like\footnote{ From here on, we
  will drop the superscript $(0)$ and the subscript $n$ from the collinear
  Lagrangian.}
\beq
A^{q\bar{q}g} \,=\,
C^{(2,0,0)}\int\!dx \langle 0 | T \{ \mathcal{L}_{\sceti}(x) \mo^{(2,0,0)} \}
|q\bar{q}g \rangle, 
\eeq
namely like the first term in \eq{AI}.  The right-hand
panel of \fig{figureTprC} denotes the same configuration as seen by
$\scetipone$.  The scale of this theory is lower and the definition of
collinearity stricter, so the quark and gluon are not collinear here.
Therefore, the amplitude now comes from a three-parton operator,
\beq
A^{q\bar{q}g} \,=\, C^{(2,1,0)}\langle 0 | \mo^{(2,1,0)} |q\bar{q}g
\rangle, 
\eeq
as in \eq{AII}.  We match $\sceti \,\rightarrow\, \scetipone$ to calculate
$C^{(2,1,0)}$.  

Given the above conventions and with the notation in
Fig.~\ref{fig:figureTprC} at hand, it is worth stating some simple
kinematic relations that we will use later on.  Take an
$n_0$-collinear mother particle of momentum $q_0^\mu = q_0^+
(\nb^\mu/2) + \bar{q}_0 (n_0^\mu/2)$. Let $q_0$ decay to two onshell
massless daughters, $k_1$ and $q_1$, with momentum fractions $x$ and
$(1-x)$, back-to-back $\perp$-momenta $\vec k_\perp$, and light-like
directions $n_1$ and $n_1'$, then
\begin{align}
  k_1^\mu &= k_1^+ \frac{\nb^\mu}{2} + \bar{k}_1 \frac{n_0^\mu}{2} + k_\perp^\mu =
  \bar{k}_1 \frac{n_1^{\prime\,\mu}}{2} \,,
 & q_1^\mu &= q_1^+ \frac{\nb^\mu}{2} + \bar{q}_1 \frac{n_0^\mu}{2} - k_\perp^\mu =
  \bar{q}_1 \frac{n_1^\mu}{2} \,.
\end{align}
Note that our convention of using the same $\nb^\mu$ auxillary vector ensures
that in these decompositions the momentum multiplying $n_0^\mu$ is the same as
the momentum multiplying $n_1^{(\prime)\mu}$.  The collinearity of $k_1$ and
$q_1$ can be determined by the size of $k_\perp^2$, $q_0^2$, or $n_1\cdot n_1'$,
and the relation between these three choices is
\begin{align} \label{eq:n1dotn1p}
  n_1 \cdot n_1' = \frac{ 2\, \vec k_\perp^2}{(\bar{q}_0)^2\, x^2(1-x)^2} = \frac{2\,
    q_0^2}{(\bar{q}_0)^2\, x(1-x)}\,. 
\end{align}
Since we take $\vec k_\perp^2 / (\bar{q}_0)^2 \sim \lambda_i^2$ in $\sceti$, we have
$q_0^2/(\bar{q}_0)^2 \sim \lambda_i^2/\eta^2$ and $n_1\cdot n_1'\sim
\lambda_i^2/\eta^4$. Thus, all three choices are equivalent for counting powers
of $\lambda_i$, but differ with respect to how powers of the energy loss
parameter $\eta \sim 1/2$ appears.

After this introduction to $\sceti$, we now list some technical advantages of
this framework for our analysis:
\begin{enumerate}
\item{ Collinear fields in SCET with different $n$-labels, as well as soft
    fields, do not overlap in Hilbert space.  This allows us to separate an
    $i$-jet process with $i$ distinguished partons, from an $(i-1)$-jet process
    with $i$ partons, where two are collinear and unresolved. Lower-scale
    $\sceti$'s distinguish configurations more finely based on their stricter
    definition of collinearity.  This resolves issue 2, avoiding the
    double-counting of similar configurations, from \subsec{bands}.  This SCET
    property also illuminates simplified structures in the power corrections,
    such as the form of the amplitude interference (cf.~section
    \ref{subsec:ampSq}).  }
\item{ Soft modes communicate between collinear sectors and threaten the
    factorization of different jets.  Fortunately, SCET constrains the
    interactions they have with collinear fields.  In fact, one can decouple
    them using soft Wilson lines in the LO SCET Lagrangian. At LO, using the
    $\sceti$ soft Wilson lines, we maintain factorization, obtain angular
    ordering, and rederive the coherent branching of soft emissions (cf.~section
    \ref{subsec:softemissions}).  Soft interactions which are power suppressed
    can also be systematically studied in SCET with Lagrangians available in the
    literature~\cite{Bauer:2003mga,Beneke:2002ph,Beneke:2002ni},
    which we give in \eq{Lxxnew}.  }
\item{ 
    In $\sceti$, we have a symmetry group ${\rm RPI}_i$ which corresponds to
    coordinate choices.  In $\scetipone$, only a subset of this, ${\rm RPI}_{i+1}
    \,\subset\, {\rm RPI}_i$, remains a symmetry of the new theory.  The
    kinematics in the coset portion ${\rm RPI}_i/{\rm RPI}_{i+1}$ within
    $\sceti$ give a set of higher-dimension operators in $\scetipone$, and
    describe configurations which would not otherwise be contained in the
    $\scetipone$ Lagrangian (cf.~section \ref{sec:LO} and Appendix
      \ref{app:rpi}).  This resolves issue 1.~from \subsec{bands}
       making the difference clear between approximations and conventions
       chosen for simplicity.  
}
  \item{ In matching between $\sceti$ and $\scetipone$, suppressed operators
      in the lower-scale theory are needed to reproduce the physics of the
      higher one.  It can be proven that all higher order purely collinear
      operators can be built from quark fields ($\chi_n$), perpendicular gluon
      fields ($\mb_{\perp \, n}$), and the perpendicular momentum operators
      ($\mcp_{\perp \, n}$)~\cite{Marcantonini:2008qn}.  Thus the symmetries and
      equations of motion of SCET greatly simplify the operator basis one needs
      to consider at each order in $\lambda$ (cf.~section \ref{sec:nlo} and
      Appendices \ref{app:QCD/SCET1}, \ref{app:scet1/scet2}, and \ref{app:scet2/scet3}). }
\end{enumerate}

The final $\text{SCET}_N$ corresponds to the scale where the shower stops, {\it
  i.e.}  where $Q\eta^N\lambda^N \sim p_{T}^{\rm cut}$. In $\text{SCET}_N$, we only
need the coefficients of the operators where all collinear partons have distinct
$n$-labels, and which have no ${\cal P}_{n\perp}$'s, $C^{(j,k,0)}_N {\cal
  O}^{(j,k,0)}_N$.  Once we reach the physical resolution scale, it is only
meaningful to have one collinear parton in each distinguished block of phase
space.  Using RPI$_N$, we can set $n_j^\mu = p_j^\mu/p^0_j$.  This is as in
\eq{condOne}, but we only do this when we run up against the physical limit that
requires just one parton per equivalence class.  At intermediate stages, we
allow different fields to share $n$-labels, which also results in operators
containing $\mcp_{n\perp}$.  The coefficients $C_N^{(j,k,0)}$ encode the history
of the shower. They can be written entirely in terms of: dot products $n_i\cdot
n_{i'}$, equivalent to products of final parton momenta, which carry the scaling
in $\lambda$; hard momenta $\bar p_i$, the renormalization scale $\mu$, and
collinear cutoff parameters encoded in $\Theta$-functions.

As far as the shower is concerned, $\lambda$ is merely a bookkeeping device
which determines what pieces are needed beyond LO.  One could try defining
$\lambda_1= k_{1\perp}/Q$, $\lambda_2=k_{2\perp}/Q$, etc., but this is not ideal
since there is a chance for events where $k_{1\perp} \sim Q$ or $k_{1\perp} \sim
p_{T}^{\rm cut}$. The organization in \eq{omegas} instead exploits the fact that
{\it on average} showers are strongly-ordered.  Our expansion in $\lambda$ will
then {\it on average} give a description of the most likely deviations from
strong-ordering. Our goal in using the $\sceti$ framework is to extract an
amplitude suitable for reweighing the parton shower to this level of
accuracy.\footnote{As a well-defined EFT, one certainly could also do standard
  factorized cross section computations in any $\sceti$ if one wanted.}  
From the SCET side, we pass to the shower weights built from ${\rm SCET}_N$
squared amplitudes ({\it cf.} Eqs.~\ref{eq:weight}-\ref{eq:wjk}).  
They contain the information needed to describe a strongly-ordered 
shower and its leading kinematic corrections.

Before proceeding to our computations, it is worth commenting explicitly on
which shower ingredients we do not compute.  We only treat the case of a
showering quark $q \to qg$ and in general take the abelian limit of QCD ($C_A =
0$).  We have left out gluon splittings, $g\to q\bar{q}$ and $g\to gg$, from
this analysis, though we expect that the extension to these cases should be
straightforward.  We have also not determined the effect of \NLOl power
corrections from subleading soft interactions, although we briefly examine the
factorized structure of LO softs in section~(\ref{subsec:softemissions}).  These
items are all left to future investigations.

\section{Parton Shower in SCET via Operator 
Replacement}
\label{sec:LO}

%
\begin{figure}[t!]  
\centering
\includegraphics{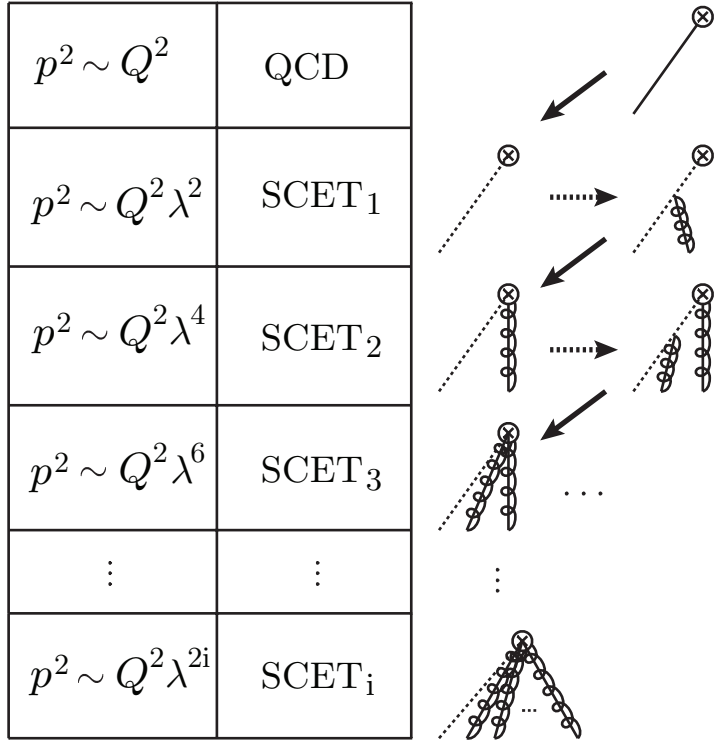}
\caption[SCET$_i$ matching]{Operators that reproduce strongly-ordered gluons are
  constructed through a series of matching computations with emissions in
  different $\scetj$. The horizontal dashed arrows refer to the radiation of a
  gluon from a time-ordered product of the $\scetj$ Lagrangian with the operator
  creating fields at the point marked by $\otimes$.  The diagonal solid arrows
  denote the matching onto a higher multiplicity operator in
  $\mathrm{SCET}_{j+1}$.}
\label{fig:figureLOC}
\end{figure}

In the previous section, we presented our approach of using a series of EFTs, 
the $\sceti$, to handle processes with a hierarchy of many scales.  We will now use 
this technique to calculate the leading contribution to a series of collinear emissions, 
as occurs in the parton shower.  Our ultimate goal is to incorporate corrections,
but as a starting point we want to easily reproduce the strongly-ordered configuration 
of Eq.~(\ref{eq:psLimit1}). We can do this if we declare that in a shower, 
the $i^{th}$ particle decomposes as:
\beq (n \cdot q_i, \,\bar{q}_i,\, q_{i \perp})\sim (\lambda^{2i}, \,1,
\,\lambda^i)\,\bar q_i, \eeq
and therefore has virtuality $q_i^2 \,\sim\, (\bar q_i)^2 \lambda^{2i}$
({\it cf.} Fig.~\ref{fig:LOPartonShower}).  This is exactly the same
condition as Eq.~(\ref{eq:omegaCond}), which we used to define the
EFT, $\sceti$. 

\begin{figure}[t!]
\centering
\includegraphics{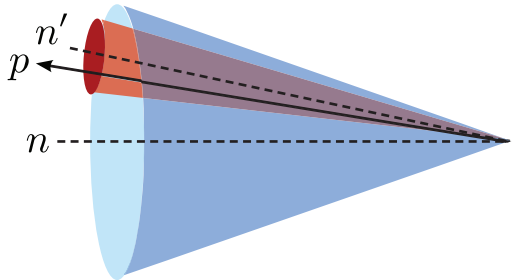}
\caption[Finite RPI]{The opening angle of the light grey (blue) cone is
  $\sim\lambda^{2i}$, and the opening angle of the dark grey (red) one is
  $\sim\lambda^{2(i+1)}$.  The particle with momentum $p$ is collinear to both
  $n$ and $n'$ in $\sceti$, but only to $n'$ in $\scetipone$.  RPI$_i$ allows us
  to move the field label, $n$, to any location inside the appropriate cone for
  $\sceti$ while keeping the theory invariant.}
\label{fig:ConeC}
\end{figure}

To calculate the operators that describe $i$ emissions in the strongly-ordered
limit, we will perform a series of matchings $\sceti \rightarrow \scetipone$.
We will find that the most efficient way to describe the process at LO in
$\lambda$ is to be in $\scetipone$ for $i$-parton radiation.  Thus, we emit and
match $i$-times in series, as shown by Fig.~\ref{fig:figureLOC}.  At LO, we
will show that one can implement this using an operator replacement rule.  In
the case of $q \rightarrow qg$ emission, it takes the form:
\begin{equation}
\chi_{n_1} \rightarrow c\, g \mathcal{B}_{n_3 \perp }^\alpha \,\chi_{n_2} \, ,
\label{eq:replaceRule}
\end{equation}
where $\chi_n$ and $\mathcal{B}_{n\perp}^\alpha$ are the SCET fields associated
with collinear quarks and gluons, respectively, and $c$ is the Wilson
coefficient whose spin and color indices are suppressed.  Though we do not
compute them, there are similar $\mb_{n_1 \perp}^\alpha \to c' \,
\chib_{n_2} \chi_{n_3}+ c''\, \mb_{n_2 \perp}^\beta \mb_{n_3 \perp}^\gamma$
rules as well.  In SCET, each collinear field carries the label $n$, which gives
its direction of collinearity.  Note that the quark field on the LHS of
(\ref{eq:replaceRule}) has a different one from those on the RHS.  This relates
to the stricter definition of collinearity in $\scetipone$ shown in
Fig.~\ref{fig:ConeC}.
In order to perform the matching, we will make use of the reparametrization
invariance (RPI) discussed in point 3.~of \subsec{sceti} to change fields' $n$-labels.

\subsection{Leading Shower Revisited}   
\label{subsec:loShowerRe}

We first want to reproduce the strongly-ordered contribution to $i$-gluon
radiation from the quark in an initial $\gamma^*\to q\overline{q}$ pair
production.  Our iterative matching procedure for multiple EFTs takes a
particularly simple form at LO in $\lambda$.  For our standard example, we take
the process $e^+e^- \rightarrow$ jets.  Starting in QCD, we couple the quarks to
another sector via the operator, $J^\mu_{\rm QCD} \,=\, \bar{q} \,\Gamma^{\mu}
q$.  This allows us to avoid complications that come from the initial state such
as backward evolution.  In $\scetone$ (which is equivalent to the usual SCET),
matching to QCD at tree-level converts the quark coupling to the following
operator at LO: $\bar{\chi}_{n_0} \Gamma^{\mu} \chi_{\bar{n}}$, which produces
$q$ and $\bar{q}$ in different collinear directions.  Details on the matching of
QCD to SCET$_1$ are given in App.~\ref{app:QCD/SCET1}.  Using the notation in
Eq.~(\ref{eq:opDef}), we write the $\scetone$ operator in the following way:

\begin{align}
\bar{\chi}_{n_0} \Gamma^{\mu} \chi_{\bar{n}}= \left(C^{(2,0,0)}_{1,\,\rm{LO}} \right)_{ij} 
\left(\mo^{(2,0,0)}_{1}(n_0, \nb) \right)_{ij} \, ,
\label{eq:LO0em}
\end{align}
where
\begin{align}
\label{eq:loDiquark}
 \left(\mo^{(2,0,0)}_{1}(n_0, \nb) \right)_{ij} &=(\bar{\chi}_{n_0})_i (\chi_{\nb})_j \, , \\
 \left(C^{(2,0,0)}_{1,\,\rm{LO}} \right)_{ij}&=(\Gamma^\mu)_{ij} \, , \nonumber
\end{align}
and $i$ and $j$ are spinor indices. The subscripts 1 in Eq.~(\ref{eq:loDiquark})
indicate that the fields are defined in $\scetone$.  Our focus is on gluon
emissions from the quark, and we always take the antiquark in the same
direction, $\nb$, therefore we drop it from the list of $n$-labels. Also, we
will use the following shorthand notation for the most common operator,
\begin{align}
\label{eq:Oi}
\mo^{(2,k,0)}_{i}(n_1,n'_1,\dots, n'_k, \nb) \equiv \mo^{(k)}_{i}(n_1,n'_1,\dots, n'_k) \, ,
\end{align}
where the subscript marks these as being in SCET$_i$.
In the rest of the paper, we will often drop the spinor indices.
Using the above convention, we write the operator in Eq.~(\ref{eq:LO0em}) as:
\begin{align}
\bar{\chi}_{n_0} \Gamma^{\mu} \chi_{\bar{n}}=  C^{(0)}_{1,\,\rm{LO}} \mo^{(0)}_{1}(n_0) \, .
\label{eq:simplerDef}
\end{align}
The LO derivations are independent of the exact structure of $\Gamma^{\mu}$.  In fact,
even the antiquark is a spectator, and we could just as easily use $\mo^{(q)}
\,=\, \bar{\chi}_{n_{0}}\Omega$, where $\Omega$ is arbitrary.
However, as we will discuss in Sec.~\ref{sec:nlo}, matching QCD to $\scetone$ at higher
orders requires us to specify $\Omega$.  

\begin{figure}[t!]
 \centering
 \includegraphics{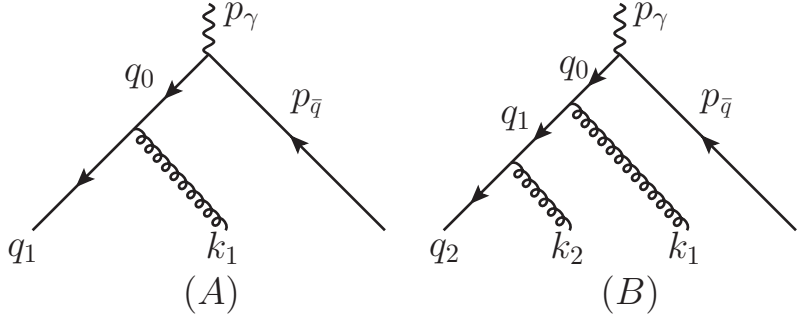}
 \caption{Momentum labels for single ($A$) and double ($B$) gluon emission.} 
 \label{fig:kin1E}
 \end{figure}

To calculate operators in $\scettwo$, we start with single gluon
radiation.  In this case, shown in Fig.~\ref{fig:kin1E}, the emission
amplitude is:\footnote{All the amplitudes we write in this work refer
only to the hadronic part of $e^+e^- \,\rightarrow$ jets, thus
$A^{q\bar{q}g}_{\mathrm{LO}}$ is the amplitude of $\gamma^{*}
\rightarrow q\bar{q}g$.}
\begin{eqnarray}
A^{q\bar{q}g}_{\mathrm{LO}} &=&  C^{(0)}_{1,\,\rm{LO}} \langle 0|\int \!\!d x\, T\{\mathcal{L}^{\scetone} (x)
\mo^{(0)}_{1}(n_0) \}  |q_{n_0}   g_{n_0} \bar{q} _{\nb} \rangle  \label{eq:singleGluOp} \\
&=&g\, \bar{u}_{n_0}(q_1) \left( n_{0}^\alpha + \frac{(\slashed{q}_1)_{n_0 \perp}
\gamma_{n_0\perp}^ \alpha}{\bar{q}_1} \right)\frac{\bar{q}_0}{q_0^2}
\Gamma^{\mu} v_{\nb}(p_{\bar{q}}),
\label{eq:singleGlu}
\end{eqnarray}
where we have labeled the collinear directions of the particles in the state
$|q_{n_0} g_{n_0} \bar{q} _{\nb} \rangle$ for later convenience.  The $\scetone$ 
Lagrangian is given in Eq.~(\ref{eq:LagSCET}).  Here we study the process in 
the center of mass frame with $p_\gamma  =(Q,0,0,0)$ and the quark ($q_0$) and antiquark ($p_{\bar{q}}$) along the 
directions $n_0=(1,0,0,1)$ and $\nb=(1,0,0,-1)$, respectively:
\begin{align}
p_\gamma^\mu & = \frac{Q}{2} n_0^\mu + \frac{Q}{2} \nb^\mu\, \nonumber \\
p_{\bar{q}}^\mu &= \frac{n_0 \cdot p_{\bar{q}}}{2} \bar{n}^\mu,\, \nonumber \\ 
q_0^\mu & = \frac{\bar{q}_0}{2}n_0^\mu + \frac{n_0 \cdot q_0}{2} \bar{n}^\mu \,.
\end{align}
We decompose the emitted quark ($q_1$) and gluon ($k_1$) along the directions $(n_0, \bar{n})$,
\begin{align}
q_1^\mu & = \frac{\bar{q}_1}{2}n_0^\mu +({q}_1)_{n_0 \perp}^\mu+ \frac{n_0 \cdot q_1}{2} \bar{n}^\mu \,,\\
k_1^\mu & = \frac{\bar{k}_1}{2}n_0^\mu +({k}_1)_{n_0 \perp}^\mu+ \frac{n_0 \cdot k_1}{2} \bar{n}^\mu \,. \nonumber
\end{align}
The variables are illustrated in Fig.~\ref{fig:kin1E}. By momentum conservation we
have $({k}_1)_{n_0 \perp}=-({q}_1)_{n_0 \perp}$,
$Q=\bar{q}_0=\bar{k}_1+\bar{q}_1$ and $n_0 \cdot p_{\bar{q}} =Q -n_0 \cdot q_1-
n_0 \cdot k_1$. We take all the external particles on-shell, thus $n_0 \cdot
q_1 = - ({q}_1)_{n_0 \perp}^2/\bar{q}_1$ and similarly for $n_0 \cdot k_1$.  As
we discussed in Section \ref{subsec:bands}, \cite{Bauer:2006mk,Bauer:2006qp}
showed that single gluon emission in SCET reproduces the splitting function,
Eq.~(\ref{eq:ap}), and factorization behavior, Eq.~(\ref{eq:loAmpFact}), of the
standard parton shower.  This simple behavior for a single radiation will
reproduce the shower for an arbitrary number of gluons.

\begin{figure}[t!] 
\centering
\includegraphics{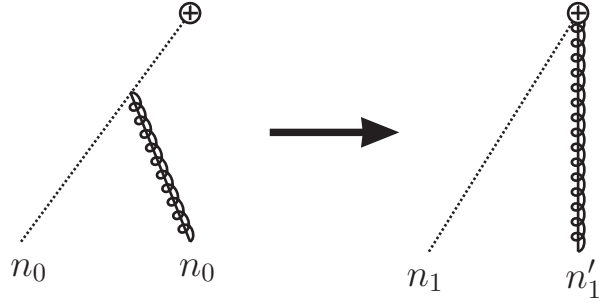}
\caption[Matching SCET$_1$ to SCET$_2$]{(Left panel) Single gluon
emission in $\scetone$ comes from the time-ordered product of the
Lagrangian with a quark-creating operator, $A = \langle 0|T\{
\mathcal{L}_{\scetone} \mo^{(0)}_1\}|qgX \rangle $. (Right panel)  For
parent quarks with virtuality $\gg Q^2\lambda^4$, the gluon comes
from the central vertex in $\scettwo$ via a higher-dimensional operator, $A =
\langle 0| \mo^{(1)}_2|qgX \rangle $.}
\label{fig:oneGluSCET2}
\end{figure}

We now want to match the single emission to $\scettwo$ 
({\it cf.}~Fig.~\ref{fig:oneGluSCET2}).  There is a slight technical complication
due to the different definitions of collinearity in the two theories,
as illustrated by \fig{ConeC}.  In $\sceti$, a collinear field with
label $n$ can annihilate a state containing a particle whose momentum
vector lies anywhere in a cone with angle $\sim \lambda^i$ about $n$.
When we change to a lower-scale theory in a matching equation, we have
to take care that the operators' $n$-labels are appropriate for the
desired amplitude.  Using the terminology of \fig{ConeC}, while any
label vector in the light grey (blue) cone is sufficient for a
particle with momentum $p$ in $\sceti$, for $\scetipone$ we need one
in the dark grey (red) cone.  This is where RPI$_i$ comes in, as
mentioned in \subsec{sceti}.  We use it in $\sceti$ to transform all
quantities in the amplitude (spinors and vectors) that depend on the
label vectors, such that the label after rotation lies within a
collinear cone with angle $\sim \lambda^{i+1}$ about the particle
momentum.

The simplest convention is to choose the $n$-label to align perfectly with the particle.
If desired, we could make any choice consistent with RPI$_{i+1}$ transformations.
For the process under consideration, we define labels, $n_1,\, n'_1$ such that,
\begin{align}
\label{eq:q1k1}
  q_1 &= \bar{q}_1\, \frac{n_{1}}{2} \, , \nonumber \\
  k_1 &= \bar{k}_1\, \frac{n'_{1}}{2} \, .
\end{align}
In $\scetone$, we are free to use $n_0$ or $n_1$ to describe the $q_1$
quark and $k_1$ gluon because of the RPI$_1$ symmetry.  Since $n_1$ is
a valid index for the quark field in $\scettwo$, we do the matching
computation using the same spinor, $u_{n_1}(q_1)$, in both theories.
In App.~\ref{app:rpi}, we derive the RPI transformations we use here
and other rotation formulas.  For now, we quote the results we need:
\begin{align}
 \label{eq:SpinorRot1} 
u_{n_0} &= \frac{\ns_0 \nbs}{4} u_{n_1},\\
n_1^{\alpha} &= n_0^{\alpha} + \frac{2{(q_1)}_{n_0\perp}}{\overline{q}_1} - \frac{{(q_1)}_{n_0\perp}^2}{\bar{q}_1^2}\nb^{\alpha}\,, \nonumber \\
n_1^{\prime\alpha} &= n_0^{\alpha} + \frac{2{(k_1)}_{n_0\perp}}{\overline{k}_1} - \frac{{(k_1)}_{n_0\perp}^2}{\bar{k}_1^2}\nb^{\alpha}. \nonumber 
\end{align}
As required, the two different $n_i$-vectors' directions lie within cones of size $\lambda$
about $n_0$.  It is simple to check that in the new basis, 
$(q_1)_{n_1\perp} \,=\, q_1 - (n_1 \cdot p)\bar{n}/2 - \bar{q}_1 \, n_1/2 \,=\, 0$ and similarly 
for $(k_1)_{n'_1\perp}$.  Acting on Eq.~(\ref{eq:singleGlu}), we get:
\beq
A^{q\bar{q}g}_{\rm{LO}} = g \frac{\bar{q}_0}{q_0^2} \bar{u}_{n_1} \left( n_0^{\alpha} + 
\frac{(\slashed{q}_1)_{n_0 \perp}\gamma^{\alpha}_{n'_1 \perp}}{\bar{q}_1} \right)  
\frac{\nbs\ns_{0}}{4} \Gamma^{\mu} v_{\bar{n}}\, ,
\label{eq:loRepS2}
\eeq
where $q_0 = q_1 + k_1$.  Having changed bases, we can easily write the
$\scettwo$ operator that reproduces \eq{loRepS2}, $C_{2, \mathrm{LO}}^{(1)}\,
\mo^{(1)}_{2}(n_1, n'_1)$, where:\footnote{ See Appendix \ref{app:scet1/scet2}
  for more detail on this matching.  Though we have written \eq{loRep} to look
  as much like the $\scetone$ amplitude as possible, we can rewrite it purely in
  terms of external momenta, as in \eq{C1S2}.}
\begin{align}
\mo^{(1)}_{2}(n_1, n'_1) =& \left(\bar{\chi} _{n_{1}}\right)_j g \mb^{\alpha}_{n'_1 \perp } \left(\chi _{\bar{n}}  \right)_k\, , 
\nonumber \\
C^{(1)}_{2,\,\rm{LO}}(n_1, n'_1)  = &\, U^{(2,0,0)}_{\rm{LL}}(n_0;\,Q,\mu_1) 
\left[ \frac{\bar{q}_0}{q_0^2} \left( n_{0}^{\alpha} + 
\frac{(\slashed{q}_1)_{n_0 \perp}\gamma^{\alpha}_ {n'_1 \perp}}{\bar{q}_1} \right) 
\frac{\nbs \ns_{0}}{4} \Gamma^{\mu} \right]_{jk} \Theta_{ \delta_2} [n_1 \cdot n'_1 ]  \,.
\label{eq:loRep}
\end{align}
We note that we have also given the Wilson coefficient the $n$-labels
of the operator it multiplies.  In cases where it is clear, we will
only explictly label one of $C$ or $\mo$.  In addition to the expected
tree-level amplitude term in brackets, we also give the RG kernel,
$U^{(2,0,0)}_{LL}$, and an angular phase-space cutoff,
$\Theta_{\delta_2}$.  We discuss each of them in turn.  

The former comes from running the SCET$_1$ operator $\mo^{(0)}_1$ from
$Q$ to the scale $\mu_1\sim \lambda Q$.  When $U_{\rm{LL}}$ refers to
an operator where all collinear directions are distinct, we will drop
$n$'s from the notation.  From \eq{runningAsSudakov}, we have
\begin{align}
U^{(2,0,0)}_{\rm{LL}}(n_0;\,Q,\mu_1)=\Delta_q(Q, \mu_1)\, ,
\end{align}
where LL refers to the fact that we take the one-loop cusp anomalous dimension,
which resums the leading logs of this running.  As mentioned in \subsec{bands},
\cite{Bauer:2006mk,Bauer:2006qp} showed this resummation to be equivalent to
that of no-branching Sudakov factors of CKKW-L.  We discuss the running of our
operators in more detail in \subsec{opRunning}.

The phase-space cutoff $\Theta_{ \delta_2} [n_1 \cdot n'_1 ]$ encodes that $n_1
\cdot n'_1 \lesssim \lambda^2/\eta^4$ (the power of $\eta^{-4}$ was discussed in
\subsec{sceti}). The SCET$_2$ operator, $\mo^{(1)}_{2}(n_1, n'_1)$, can only
distinguish that the quark and gluon are not collinear in SCET$_2$, but does not
know that they {\it were} collinear in SCET$_1$.  Thus, we put a cutoff on how
far apart they are using $n_1\cdot n_1'$ to ensure that this $\scettwo$ operator
cannot create them in a region of phase-space where they would have been
non-collinear, even in $\scetone$.
As an example, we could choose $\Theta$ to be the usual step-function
\begin{align}
  \Theta_{\delta _k}[n_i \cdot n_j] &= \begin{cases}
  1 & n_i \cdot n_j \,\leq\, \delta _k \\
  0 & n_i \cdot n_j \,>\, \delta _k
  \end{cases} \,, \nonumber \\
 \tilde\Theta_{\delta k} &= 1 - \Theta_{\delta_k} \,.
\label{eq:thetaExpl}
\end{align}
In practice we will use a smooth version of the above step.
For later convenience, we defined the complement, $\tilde{\Theta}_{\delta_k}$.
In working with $\sceti$ operators, we relate $\delta_k$ to $\lambda$.  In
general, the Wilson coefficient in $\sceti$ has to encode whether $ n_i \cdot
n_j \,\leq\, \lambda^{2( i-1)}/\eta^4$ or $ n_i \cdot n_j \,>\, \lambda^{2(
  i-1)}/\eta^4$, in order to do it we will set $\delta _i = \lambda^{2
  i-3}/\eta^4$. This satisfies the necessary criteria since $\lambda^{2i-2} \ll
\lambda^{2i-3} \ll 1$ (and recall that $\eta$ is the parameter that accounts for
the decrease in $\bar{p}$ of a daughter relative to its mother).  For $C_{2,
  \mathrm{LO}}^{(1)}$ above, this means $\delta_2 = \lambda/\eta^4$.  At the end
of Sec.~\ref{subsec:sceti}, we discussed how $\lambda$ gives us a way to
parametrize strong-ordering and deviations from it.  To this end, we did not
need to assign it a numerical value beyond $\lambda \ll 1$.  Here for the
implementation, we do have to make an explicit choice as to where our $\Theta$
functions turn over, and for this purpose we will use fixed values such as
$\lambda=0.1$ and $\eta=1/2$. This means $\delta_2=1.6$ and since the
$\eta^{-4}$ is a common overall factor that all $\delta_{k\ge 3} \le 0.16$. The
smoothness of both $\Theta$ and our physical processes gives us great leeway in
the choice for $\lambda$, and we expect that any $\lambda \simeq 0.1$ will
suffice ({\it cf.}  Fig.~\ref{Merge2J-3JB}).

\begin{figure}[t!]
\centering
\includegraphics[width=0.75\textwidth]{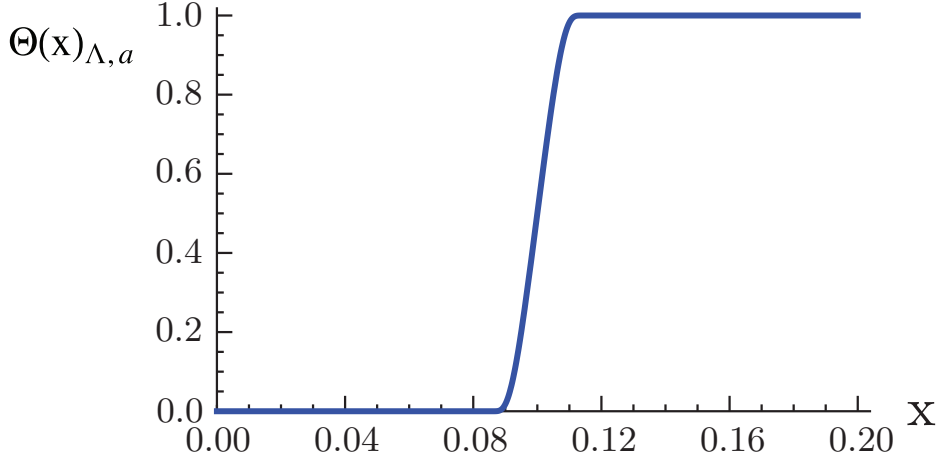}
\caption{Plot of the smoothed $\Theta$-function, $\Theta(x)_{\Lambda,a}$,
  defined in Eq.~(\ref{eq:thetaEx}), taking $\Lambda=0.1$ and $a=0.016$.  The
  parameter $\Lambda$ determines the value of $x$ where the function switches
  from 0 to 1, and $2a$ is the range in $x$ over which the transition is made.
  Comparing this smoothed $\Theta$-function to $\Theta_{\delta}$ in
  \eq{thetaExpl}, we have parametrically $\Lambda\simeq \delta_k$ and $a \ll
  \delta_k$. This plot is for the case $\delta_3=0.1$.}
\label{PlotTheta-3JB}
\end{figure}

Once we square and integrate our operators, we have certain practical
considerations to take into account.  For example, it is better to use a
smoothed step.  We give an example of such a function in \eq{thetaEx}, and plot
it in Fig.~\ref{PlotTheta-3JB}.
where we choose an appropriate numerical value for $\delta_k$.  If one only
wishes to recover the LL shower, then one should use $\Theta = 1$, as the errors
induced by this do not affect the leading resummation. Furthermore, taking
$\Theta=1$ ensures that the LL shower can cover all of phase space.  Once we
include corrections, though, then it is important to keep different types of
collinearity distinct and include non-trivial $\Theta$'s. In the presence of
corrections, there will always be amplitudes with a $\Theta$ and others with a
$\tilde\Theta$, which together cover all of phase space (see also
Fig.~\ref{Merge2J-3JB}).

Unlike standard SCET, where all the coefficients are of order
$\lambda^0$, $C_{2, \mathrm{LO}}^{(1)}$ has an overall weight of
$\lambda^{-1}$.  We get $\lambda^{-2}$ from the $\scetone$ propagator,
$1/q_0^2$.  The numerator is proportional to $\lambda$ and comes from
the vertex: $\left( n_0^{\alpha} + (q\!\!\!\slash_1)_{n_0
\perp}\gamma^{\alpha}_{n'_1 \perp}/\bar{q}_1 \right)$.  The second
term is straightforwardly $\mo(\lambda)$ from $(\slashed{q}_1)_{n_0
\perp}$.  Since $n_{0}^{\alpha}$ gets contracted with
$\mb^{\alpha}_{n'_1 \perp}$, it only contributes its perpendicular
component in the $n'_1$ frame.  From Eq.~(\ref{eq:SpinorRot1}), we see
that $(n_0)_{ n'_1\perp} \,\sim\, n_0 - n'_1 \,\sim\,
(k_1)_{n_0\perp}/\bar{k}_1 \,\sim\, \lambda$.  

$C_{2, \mathrm{LO}}^{(1)}$ is gauge invariant despite the presence of the
$\Theta_{\delta_k}$ function.  This follows from writing Eq.~(\ref{eq:loRep})
only in terms of scalar products of $n$ vectors, ({\it cf.}  Eq.~\ref{eq:C1S2}),
since collinear directions are invariant under collinear gauge transformations
\cite{Bauer:2001yt}.

We note that we can obtain $C_{2, \mathrm{LO}}^{(1)}
\mo^{(1)}_{2}(n_1, n'_1)$ from the original two-parton operator,
$C_{1, \mathrm{LO}}^{(0)} \mo^{(0)}_{1}(n_0)$, in two steps:
first we multiply it by the running factor
\begin{align}
U^{(2,0,0)}_{\rm{LL}}(n_0;\,Q,\mu_1)=\Delta_q(Q, \mu_1)\, ,
\end{align}
where the formulas for $U^{(0)}_{\rm LL}$ are given in 
Eqs.~(\ref{eq:rgKernel}-\ref{eq:oneLoopCusp}).
Secondly, we apply the replacement rule
\beq
(\chib_{n_{0}})_i \rightarrow  \, (c_{\rm{LO}}^\alpha (n_0))_{ji} \, (\bar{\chi}_{n_{1}})_j g\mb^{\alpha}_{n'_1 \perp}, 
\label{eq:loArrow}
\eeq 
where $c_{\rm{LO}}^\alpha$ is: 
\begin{align}
c_{\rm{LO}}^\alpha (n_0) &= \frac{\bar{q}_0}{q_0^2}\, 
\left( n_{0}^{\alpha} + \frac{ (q_1)_{n_0 \perp}^\mu\gamma^{\alpha}_{n'_1 \perp}}{\bar{q}_1} \right) 
\frac{\nbs \ns_{0}}{4} \, \Theta_{ \delta_2} [n_1 \cdot n'_1 ] \,.
\label{eq:loCoeff}
\end{align}
The relation (\ref{eq:loArrow}) is the operator statement of splitting
in the parton shower.  The scale $\mu_1$ defines the endpoint of
running in the UV theory.  As we evolve down, more partons become
apparent.  We can see this here by the presence of two fields where
there had been one.  It makes the basic aspects of the shower
manifest.  The replacement rule affects the quark alone, and so we see
that the amplitude for splitting factorizes off from the rest of the
process.  The RG kernel reflects the no-branching probability.
Lastly, we can interpret the vertex portion of $c^{\alpha}_{\rm LO}$
as the ``square root'' of the splitting function.  The spinor
projector ($\nbs \ns_{0}/4$) in \eq{loCoeff} rotates the spin-sum from
$\ns_{1}$ to $\ns_{0}$ in accordance with \eq{splitFact}.  The
remaining part of $c^{\alpha}_{\rm LO}$ after stripping off the
$\Theta_{\delta_2}$ is:
\begin{align}
  P_{\alpha}\,\equiv\,  \frac{\bar{q}_0}{q_0^2} \left(
  n_{0}^{\alpha} + \frac{ (\slashed{q}_1)_{n_0 \perp} \gamma^{\alpha}_{n'_1
  \perp} }{ \bar{q}_1 } \right), 
  \label{eq:vertexRule}
\end{align}
which squares to a trivial Dirac structure.  Furthermore, even though
$\rho_{\alpha} (\bar{q}/q_0^2) \neq P_{\alpha}$ because of the RPI rotations we
performed (where $\rho$ is defined in Eq.~\ref{eq:rhoDef}), we have $|\rho|^2
(\bar{q}_0/q_0^2)^2 \,=\, |P|^2$ with respect to the gauge polarization sum,
$d_{\alpha\beta}$, so
\beq
|P|^2 \,=\, \frac{1+z^2}{k_{1\perp}^2}. 
\label{eq:pSq}
\eeq
Just as before, including the $z$-dependence from the measure and spin-sum, we
recover the the standard splitting function $\propto \, (1+z^2)/(1-z)$. Thus,
$c_{\rm{LO}}^\alpha$ weights the probability assigned to the expectation value
of $C_{2, \mathrm{LO}}^{(1)} \mo^{(1)}_{2}(n_1, n'_1)$ appropriately.

Having computed the LO result for a single gluon, it is
straightforward to proceed to an arbitrary number of emissions.  In
$\scettwo$, we know that a two-gluon process comes from the
$T$-product of the Lagrangian with $C_{2, \mathrm{LO}}^{(1)}
\mo^{(1)}_{2}(n_1, n'_1)$.  Similarly to before, the amplitude has the
contribution,
\begin{align}
A^{q \bar{q} g g }_{\rm LO}= C_{2, \mathrm{LO}}^{(1)} \langle 0| 
\int \!\!d x\, T \{\mathcal{L}_{\scettwo}(x) \mo^{(1)}_{2}(n_1, n'_1)\}|q_{n_1} g_{n_1} g_{n'_1}  \bar{q}_{\nb}  
\rangle \, .
\end{align}  
The vertex for gluon emission in the $\scettwo$ Lagrangian is identical to that
in $\scetone$.  Thus, integrating out the parent of the Lagrangian-emitted
gluon, we obtain a two-gluon $\scetthree$ operator, $C^{(2)}_{3,\,\rm{LO}}
\mo^{(2)}_3(n_2, n'_1, n'_2)$, similarly to before.  Also like in the matching
$\scetone \rightarrow \scettwo$, we can obtain $C^{(2)}_{3,\rm{LO}}
\mo^{(2)}_3(n_2, n'_1, n'_2)$ from the SCET$_2$, $\moqqg$, by multiplying it by
the running factor for $\mo^{(1)}_2$,
\begin{align}
U_{\rm{LL}}^{(1)}=\Delta_q(\mu_0,\mu_2)\,\Delta_g(\mu_1,\mu_2)^{1/2},
\end{align}
with $\mu _0\sim Q$ and applying the replacement rule:
\begin{align}
(\chib_{n_{2}})_i \rightarrow &  \, (c^{\alpha}_{\rm LO}(n_1))_{ji} \, (\bar{\chi}_{n_{3}})_j g \mb^{\alpha}_{n'_2 \perp}, 
\label{eq:loArrow2} \\
c^{\alpha}_{\rm LO}(n_1) \,=&\, 
\frac{\overline{q}_1}{q_1^2} \left( n_{1}^{\alpha} + 
\frac{(\slashed{q}_2)_{n_1 \perp } \gamma^{\alpha}_{n'_2 \perp}}{\overline{q}_2}  \right)
\frac{\nbs\ns_{1}}{4}  \, \Theta_{ \delta_{3}} [n_2 \cdot n'_2 ] \, , \nonumber
\end{align}
where $n_2$ and $n'_2$ are directions proportional to the quark and
second gluon momenta, defined in \eq{n2nq2def}, and $\delta_3 =
\lambda^3/\eta^4$.
\begin{figure}[t!]
\centering
\includegraphics{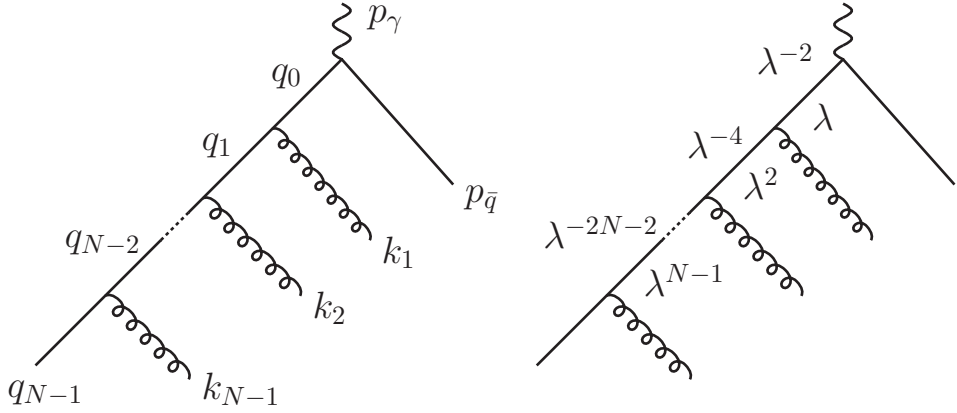} 
\caption[Parton showers]{ (Left panel) Our kinematic convention for a strongly
  ordered process.  Quark momenta are denoted by $q_i$ and gluon momenta by
  $k_i$. (Right panel) Power counting of the LO coefficient in SCET$_N$. The
  powers of $\lambda$ with negative exponents refer to the propagator
  contribution to the amplitude.  Those with positive exponents refer to the
  perpendicular momentum of the gluon with respect to its parent, which appears
  in the SCET vertex Feynman rule.  }
\label{fig:LOPartonShower}
\end{figure}  
One can iterate this procedure to obtain the LO result for $(N-1)$-gluon
emission.  If we use the replacement rule $N-1$ times we go down to the $\scetn$
operator $C^{(N-1)}_{N,\,\rm{LO}} \mo^{(N-1)}_{N} (n_{N-1}, n'_1,\dots,n'_{N-1})$,
after which Lagrangian emissions are no longer distinguished as separate
particles. We have:
{\allowdisplaybreaks
\begin{align}
\mo^{(N-1)}_{N}(n_{N-1}, n'_1,\dots,n'_{N-1}) =& \bar{\chi}_{n_{N}} \left( \prod_{k=1}^{N-1} 
g\mb_{\alpha_k}^{n'_k \perp} \right) \chi_{\nb}\, , 
\label{eq:loOp}\\
C^{(N-1)}_{N,\,\rm{LO}}(n_{N-1}, n'_1,\dots,n'_{N-1}) =& \left( \prod_{k=1}^{N-1} U^{(k-1)}_{\rm{LL}}(\mu_{k-1},\mu_{k}) \, 
c^{\alpha_k}_{\rm{LO}}(n_{k-1}) \right) \Gamma^{\mu} , 
\nonumber \\
c^{\alpha_{k}}_{\rm{LO}}(n_{k-1}) =\, &
\frac{\overline{q}_{k-1}}{q_{k-1}^2} \left( n_{{k-1}}^{\alpha_k} + 
\frac{( \slashed{q}_k)_{n_{k-1} \perp} \gamma^{\alpha_k}_{n'_k \perp}}{\overline{q}_k} \right)
\frac{\nbs\ns_{k-1}}{4}  \, \Theta_{ \delta_{k}} [(n_{k} \cdot n'_{k}) ]  \, , \nonumber \\
U^{(k-1)}_{\rm{LL}}(\mu_{k-1},\mu_{k}) =&\,\Delta_q (\mu_{k-1},\mu_{k})(\Delta_g(\mu_{k-1},\mu_{k}) )^{(k-1)/2}\, . \nonumber
\end{align}
}
The variables for $N-1$ emissions are illustrated in
Fig.~\ref{fig:LOPartonShower}, where $q_{k-1} = {\left( q_{N-1}\,+\,
    \sum_{j=k}^{N-1} k_j \right)^2}$ and $\delta_k = \lambda^{2k-3}/\eta^4$. From the
power counting one knows that $\mu_0=\,Q$, and $\mu_k\sim Q\lambda^{k}$, where
the latter scaling determines how $\mu_k$ depends on $p_\perp^j$ momenta, but
not how it depends on ratios of the large $\bar q_j$ momenta.  To sum LL$_{\rm
  exp}$ the approach taken by CKKW and elsewhere is to use $\mu_k^2=k_\perp^2$,
namely the transverse momentum squared of the
emission~\cite{Amati:1980ch,Bassetto:1984ik,Brown:1990nm,Catani:1991hj,Dokshitzer:1992ip}.
This accounts for soft interference effects and coherent branching, see
Ref.~\cite{Bassetto:1984ik} for a review. To investigate this scale choice in
the $\sceti$ framework requires an examination of the logs in the one-loop
matching computation for $c_{\rm LO}$, and consideration of soft gluons in
$\sceti$ and $\scetipone$. Having not carried out this computation ourselves, we
rely on the previous literature.  For our variables using \eq{n1dotn1p} this implies
\begin{align} 
\label{eq:muk2}
 \mu_k^2= \Big(\frac{\bar q_k\bar k_k}{\bar q_{k-1}}\Big)^2\ \frac{|n_k \cdot
   n'_k|}{2}
  \,.
\end{align}
(In contrast, the choice of invariant mass $q_{k-1}^2$ would have yielded
$\mu_k^2=(\bar q_k\bar k_k)|n_k \cdot n'_k|/2$, but this leads to incomplete
cancellations of soft divergences, and therefore problems with the resummation
of soft logs~\cite{Amati:1980ch}.)  The directions $n_k$ and $n'_k$ are aligned
with the external quark, $q_k$, and the gluon momenta, $k_k$.  They are related
to $n_{k-1}$ through an RPI$_k$ transformation.  We can extend the argument to
calculate the scaling of $C_{2, \mathrm{LO}}^{(1)}$ to the SCET$_N$ coefficient
in Eq.~(\ref{eq:loOp}).  Counting the contributions from the tree-level terms,
$c^{\alpha_k}_{\rm LO}$, $C_{N, \mathrm{LO}}^{(N-1)} \,\sim\,
\prod_i^{N-1}1/\lambda^{-i}=\lambda^{-N(N-1)/2}$, {\it cf.}
Fig.~\ref{fig:LOPartonShower}.

Similarly to the discussion above Eq.~(\ref{eq:pSq}), we can extract the vertex part of 
$c^{\alpha_k}_{\rm{LO}}$ to define $P^{\alpha_k}$.  We get that:
\beq
|P^{\alpha_k}|^2 \,=\, \frac{1+z_k^2}{(q^2_k)_{n_{k-1} \perp}},
\eeq
where $z_k \equiv \bar{q}_k/\bar{q}_{k-1}$.  Thus, the amplitude
squared goes like the factorized product of the appropriate 1
$\rightarrow$ 2 splitting functions.  Since $\mo^{(N-1)}_N (n_{N-1},
n'_1,\dots,n'_{N-1})$ is just built up from the repeated use of
Eq.~(\ref{eq:loArrow}), we see that it requires no added information
after we compute the first $q\to qg$ splitting.  Thus, what we need to
pass to a shower algorithm comes just from single real and single
virtual gluon computations, as we list below in \subsec{Summ LL} in
\tbl{map}.  The collinear splitting needed for a LL shower is entirely
handled by the replacement rule in \eq{loArrow}.\footnote{It is
  straightforward to see that we do not have additional contributions at
  LO in $\lambda$.  Firstly, consider the possibility of operators that
  do not take the form of a single-field replacement rule.  These would
  depend on the details of the hard process that produced the quark in
  the first place and could threaten the factorization of the shower.
  In fact, we will get such terms when we match QCD $\rightarrow \,
  \scetone$, but they are always suppressed, as we discuss in
  Sec.~\ref{sec:nlo}.  Returning to single-field replacement, let us
  consider matching $\scetone \,\rightarrow\, \scettwo$, as results in
  this case will generalize to all $\sceti$.  Rule (\ref{eq:loArrow})
  sends $\chi_{n_1} \rightarrow C \mathcal{B}_{n_3 \perp }^\mu
  \,\chi_{n_2}$.  At LO, we cannot get such a replacement involving
  multiple gluon fields, $\mathcal{B}_{n_j \perp }$, as this implies
  that we have integrated out multiple, hard ($\sim Q\lambda^2$)
  propagators.  Such a contribution would not be strongly ordered, and
  is suppressed.  In Sec.~\ref{sec:nlo}, we will also see that we do
  have such contributions at higher orders.  }

Lastly, we note that at higher orders in $\scetn$, we will only ever
need to compute the Wilson coefficient, $C^{(N-1)}_N$, of
$\mo^{(N-1)}_N$.  Since each field in this theory has its own
direction by the physical resolution constraint, we can use RPI$_N$ to
make all operators with $\mcp_{n\perp}$ equal to zero.

\subsection{Soft Emissions}
\label{subsec:softemissions}

SCET describes soft degrees of freedom using soft quark and gluon fields:
$q_s(x)$ and $A_s(x)$.  In this work, we focus on fully differential cross
sections where we can always distinguish collinear and soft modes.  In an
integrated cross section in SCET, we have to implement soft emissions with some
form of zero-bin subtractions~\cite{Manohar:2006nz} to avoid double counting
between soft and collinear radiation. (In the shower literature a proper
treatment of softs is also often implemented by subtraction
methods~\cite{Frixione:1995ms,Catani:1996vz,Catani:2002hc,Nagy:2007ty,Nagy:2008ns,Nagy:2008eq,Soper:2008zp}.)
The collinear sector and the soft sector couple through the covariant
derivative,
\begin{align}
i D^\mu_s= i \partial^\mu + g A_s^\mu\, ,
\end{align}
acting on the collinear fields.  At LO in $\lambda$, the collinear particles only couple to
the $n\cdot A_s$ component of the soft gluons and the soft-collinear
factorization guarantees that we can absorb this interaction into a Wilson line,
$Y(x)$, along the direction of the collinear particle,
\begin{align} \label{eq:Yn}
Y_n(x)=P \, \mathrm{exp}\Big[ ig \int_{-\infty}^0 \!\!\!\!\mathrm{d} s \, n \!\cdot\! A_s (x+ s n)  \Big] \, .
\end{align}
In SCET, this is accomplished by making field
redefinitions~\cite{Bauer:2001yt}, so that the new collinear fields no longer couple
to soft gluons through their kinetic term, as we review in \app{scet}. The
outcome for the composite fields considered here is that
\begin{align}
  \chi_{n}\to Y_n \chi_n\, , \qquad \mathcal{B}_{n}^{\mu} \to Y_n
  \mathcal{B}_n^{\mu} Y_n^{\dagger}\, .
\label{eq:softsub}
\end{align}
Note that here we consider nonabelian soft interactions, which is why the soft
Wilson lines do not cancel for the ${\cal B}_n^\mu$ field.

In matching $\sceti$ to $\scetipone$, we will only consider external soft modes
in $\scetipone$ with momenta $k \sim Q\lambda^{2(i+1)}$. These are contained as
a subset of the softs in $\sceti$. We do not consider particles with soft
momenta $k \sim Q\lambda^{2i}$ that could not be encoded by onshell modes in
$\scetipone$.  Such modes are forced to have larger momenta than the soft
fields in $\scetipone$, and they are not responsible for IR divergences. Any
contributions from momenta of this type can be encoded in the Wilson
coefficients of our $\scetipone$ operators.

In a given $\sceti$, after making the field redefinition, the effect of soft
gluons is encoded by Wilson lines $Y_n$ in the operators, with the form
\begin{align} \label{eq:OpLOsoft}
 \chib^{(0)}_{n_{N}} Y^\dagger_{n_{N}}\prod_{k=1}^{N} Y_{n'_k} 
\mb^{(0)\alpha_k}_{n'_k \perp}  Y_{n'_k}^{\dagger} \,  \Gamma_{\mu}Y_{\bar{n}} \chi_{\nb} \, .
\end{align}
The angular ordering property and the coherent parton branching formalism 
for soft emissions with multiple hard
partons emerge naturally from such operators in $\scetipone$.  
If we take the Fourier transform of $Y_n(x)$ we get
\begin{align}
Y=1+\sum_{m=1}^{\infty} \sum_{\mathrm{perms.}} \frac{(-g)^m}{m!} \frac{n \cdot A_s^{a_1} 
\cdots n \cdot A_s^{a_m}  }{n\cdot k_1 n\cdot (k_1+k_2) \cdots n\cdot(\sum_{i=1}^m k_i)} T^{a_m} \cdots T^{a_1}
\label{eq:Yexp}
\end{align}
where $k_1$, $k_2$, ... $k_n$ are the momenta of the gluon fields.  The eikonal
structure of (\ref{eq:Yexp}) leads to angular ordering. If a collinear particle
with momentum $q_i$ in the $n_i$ direction emits a soft gluon of momentum $k_s$,
the amplitude acquires a term proportional to
\begin{align}
 F_{\mathrm{soft}}=\frac{n_i\!\cdot\! \varepsilon_s}{n_i\!\cdot\! k_s}=\frac{q_i \!\cdot\! \varepsilon_s}{q_ i\!\cdot\! k_s} + O(\lambda) \, ,
\end{align}
where $\varepsilon_s $ is the polarization vector of the soft radiation and
$q_i^\mu=\bar{q} \,n_i^\mu/2$ up to power corrections.  If
$A_n(q_1,q_2,\cdots,q_n)$ is the amplitude to emit $n$ collinear particles with
momenta $q_1,q_2,\cdots,q_n$ and $A_{n+1}$ the amplitude with one more emission,
$k_s$, in the soft region, we get
$A_{n+1}(q_1,q_2,\cdots,q_n,k_s)\sim A_{n}(q_1,q_2,\cdots,q_n) \sum_{i=1}^nC_{i}\,
{q_i\cdot \varepsilon_s}/{q_i\cdot k}$,
where $C_{i}$ is a color factor. For the cross section this implies
\begin{align} \label{eq:sigmaSoft}
d \sigma_{n+1}=d\sigma_n \frac{d E_s}{E_s} \frac{d \Omega_s}{2 \pi} \frac{\alpha_s}{2 \pi} \sum_{i, j}C_{i, j}W_{i, j} \, ,
\end{align}
where $d \Omega_s$ and $E_s$ are the element of solid angle and the energy of
the emitted soft gluon, and $C_{i, j}$ is a color factor.
Here 
\begin{align} \label{eq:Wij}
 W_{i, j}= \frac{E_s^2 \, q_i \!\cdot\!
    q_j}{q_i\!\cdot\!k_s \,q_j\!\cdot\!k_s}
\end{align}
is known as the radiation function.  Without color weights, the integration of
$W_{i,j}$ over azimuthal angular variables would imply that soft gluons only
contribute when the gluon is confined to the cones centered in the directions of
particles $i$ and $j$, and are hence angular ordered.

To see how coherent branching emerges, we consider effects encoded by
operators with exactly the same collinear field content in $\sceti$
and $\scetipone$.  Graphs involving soft gluons will agree, and there is
no contribution to the matching. If we consider instead the collinear
calculations that lead to the LO replacement rule $\bar \chi_{n_0} \to
c_{\rm LO} \bar\chi_{n_1} {\cal B}_{n_1'}^{\perp}$, then the soft
gluons are encoded by
\begin{align}
&  \sceti:\ \  \bar\chi_{n_0} Y_{n_0}^\dagger \,,
& \scetipone: &\ \  c_{\rm LO}
\bar\chi_{n_1} Y_{n_1}^\dagger Y_{n_1'} {\cal B}_{n_1'}^{\perp} Y_{n_1'}^\dagger
\,.
\end{align}
For soft gluons at wide angles relative to $n_0$, $n_1$, and $n_1'$, the effect
of attachments to $Y_{n_1}^\dagger Y_{n_1'}$ are power suppressed because soft
emission from these two lines cancels up to terms that are power suppressed by
$n_1 \cdot n_1' \sim \lambda^{2i}/\eta^4$. The remaining attachment to
$Y_{n_1'}^\dagger$ looks the same as those to $Y_{n_0}^\dagger$ at leading
power, since $n_0\cdot n_1'\sim \lambda^{2i}/\eta^4$. Thus, wide angle soft gluons do not
resolve the substructure revealed by matching to $\scetipone$ and effectively
only couple to the overall color charge of the parent quark $\bar\chi_{n_0}$.
Soft radiation that is close in angle to $n_1$ and $n_1'$ resolves the split
into quark $\bar\chi_{n_1}$ and gluon ${\cal B}_{n_1'}^\perp$, compensating for the
$n_1\cdot n_1'$ suppression by additional collinear singularities in its
propagator factors. Thus, the coherent branching formalism for soft gluons emerges
naturally for amplitudes in our SCET$_i$ picture.

From the SCET point of view, it would be natural to distinguish soft and
collinear radiation in the shower and treat them independently, being careful
not to double count. For simplicity, all available shower codes treat them in a
simultaneous fashion.  Accounting for soft coherent branching in the
shower typically leads to modifications of the Sudakov probability factors (see
for example Ref.~\cite{Ellis:1991qj}), and affects the choice of evolution
variable or adds additional vetoes. In the context of SCET, the implications
of this were discussed recently in \cite{Bauertalk}.

\subsection{Summary for LO Parton Shower}
\label{subsec:Summ LL}

In Table~\ref{tbl:map}, we summarize results for the mapping between the LL
parton shower and our SCET$_i$ picture at LO in $\lambda$.  In the first column,
we put the elements needed for showering, and in the central column the
translation to elements in the $\sceti$ setup.  The usual splitting function is
related to our replacement rule $\bar\chi_{n_0}\to c_{\rm LO} \bar\chi_{n_1}
{\cal B}_{n_1'}^{\perp}$, that in turn is related to the SCET$_2$ coefficient of
the operator $\mo^{(1)}_2$.  The LL Sudakov comes from LL running factors
related to the one-loop cusp anomalous dimension as in \cite{Bauer:2006mk,Bauer:2006qp}.  At
leading order, soft emission in $\sceti$ is taken into account by adding soft
Wilson lines $Y_n$ into our operators. This leads to angular ordering and
coherent branching, which must be accounted for with modifications to the shower
to account for the soft singular regions.  Finally, showers are constructed with
different choices of evolution variables and the choice effects the structure of
power corrections.  In $\sceti$, we have seen that we can write all coefficients
in terms of the large momenta ($\bar{q}$) and dot product of $n$ vectors
($n_i\cdot n_j$), which are natural variables in the SCET$_i$ picture. One can
convert these variables to $k_T^2$, virtuality, or angles as desired. At LL this
translation is straightforward.
\begin{center}
  \begin{table}[ht!]
  \begin{tabular}{|c|c|c|}
  \hline
  Shower Concepts        & Quantity in $\sceti$       & Found In: \\[0pt] 
\hline\hline
  Splitting function        & Replacement rule      &         Eq.(\ref{eq:loArrow}) \\\hline
       
  LL Sudakov factor        & One-loop cusp    &     Eq.~(\ref{eq:runningAsSudakov})  \\
                           & anomalous dimension        &     \\ \hline      
Soft  emission  &  Soft amplitude              &    Eq.~(\ref{eq:OpLOsoft})      \\ \hline 
  \end{tabular}
\caption{Mapping between parton shower and $\sceti$ at 
  LO/LL.}
  \label{tbl:map}
  \end{table}
\end{center}

\section{SCET Power Corrections to the Shower}
\label{sec:nlo}

As we have seen in the previous section, we reproduce the usual parton shower by
matching collinear gluon emissions to increasingly lower-scale EFTs, the
$\sceti$.  Our goal is to catalog the leading power corrections (in $\lambda$)
to the differential cross section for the emission of an arbitrary number of
collinear gluons to a quark.  By this we mean all amplitude terms to
LO$(\lambda)$ and \NLOl, as well as those at NNLO$(\lambda)$ that can interfere
with LO$(\lambda)$.  As we will argue in Sec.~\ref{subsec:ampSq}, in most cases
of interest, there is no LO$(\lambda)$/NLO$(\lambda)$ interference, and so we
focus on the most important power suppressed terms which are
NLO$(\lambda)$$\times$NLO$(\lambda)$ and LO$(\lambda)$$\times$NNLO$(\lambda)$.
Just as in the strongly-ordered case, it is convenient to integrate down to
$\scetipone$ when describing the emission of $i$-gluons.  We obtain these
corrections by doing our matching computations at higher order.  We will show
that there are two distinct types of subleading matching, and they have a
different physical interpretation:
\begin{itemize}
\item{ One type originates in matching QCD $\rightarrow \, \scetone$ at higher
    orders.  This generates a set of subleading terms that remain suppressed as
    we move down to lower-scale $\sceti$'s.  We call them {\it hard-scattering}
    power corrections as they involve the details of the hard-scale process that
    created our original partons.  Also, they are most important for partons
    radiated closest to the hard vertex.  }
\item{ The other type comes from the subleading matching $\sceti
    \,\rightarrow\, \scetipone$.  They involve processes described by
    the $\sceti$ Lagrangian, but ones that get integrated out into
    higher dimension operators at lower scales.  These corrections are
    ubiquitous.  They do not depend on the hard-scattering details,
    and we can determine them for arbitrary $\sceti \,\rightarrow\,
    \scetipone$ once we have found them in $\scetone \,\rightarrow\,
    \scettwo$.  Furthermore, they relate to known $\mo(\alpha_s)$
    corrections to the $q \rightarrow qg$ splitting function, which
    exponentiate to sum part of NLL.  For this reason, we call them
    {\it jet-structure} corrections.  }
\end{itemize}
Determining the above to NLO$(\lambda)$ in the cross section will only involve
single and double gluon emission. Thus, we will never need to compute in a
lower-scale theory than SCET$_3$. We perform all the necessary
QCD$\rightarrow$SCET$_1$$\rightarrow$SCET$_2$$\rightarrow$SCET$_3$ matchings for
these amplitudes in Appendices \ref{app:QCD/SCET1}-\ref{app:scet2/scet3}. Below,
we discuss the final results for the corrections, with \subsec{qcdToSCET}
focusing on hard-scattering and \subsec{scetToSCET} on jet-structure.  For these
portions of the paper, the matching is only done at tree level, though formulas
in the Appendices include one-loop RG kernels.  We give the effects of LL
running on correction terms in \subsec{opRunning} along with a discussion of how
to include NLL resummation for the LO (in $\lambda$) Wilson coefficients.  In
\subsec{ampSq}, we will study the amplitude squared and will see there is a
great simplification of the interference structure in $\scetn$, and hence for
NLO$(\lambda)$ power corrections in a shower. Lastly, we give in \subsec{nlomap}
the NLO counterpart to our LO table in \subsec{Summ LL}.  We describe how our
corrections from subleading operators relate to improvement of the parton shower
with higher order resummation of logs, corrections at higher order in
$\alpha_s$, as well as corrections to spin correlations and interference.  These
effects are summarized in a shower reweighting formula, \eq{weight}.

\subsection{Hard-Scattering Corrections}
\label{subsec:qcdToSCET}

\begin{figure}[t!]
\centering
\includegraphics{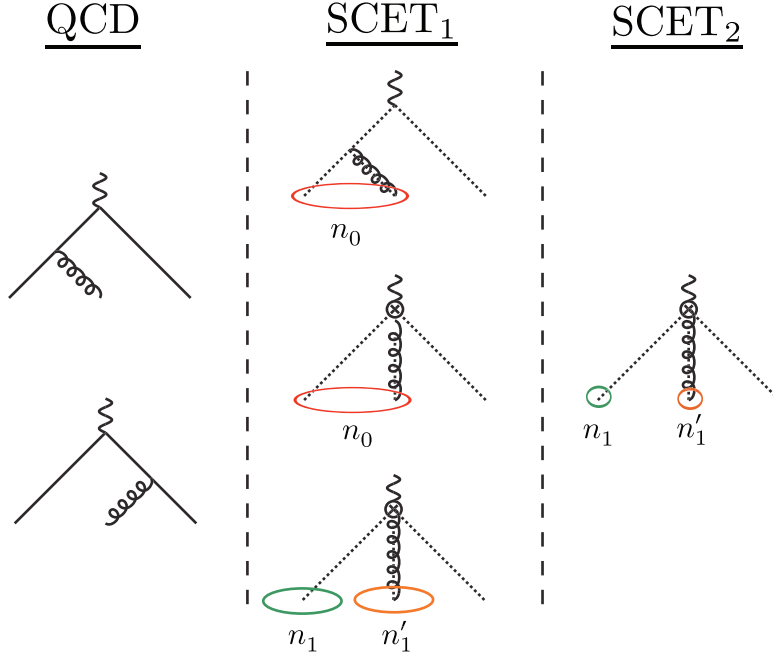}
\caption[Matching QCD to $\scetone$ to $\scettwo$ for one emission]{Matching QCD
  to $\scetone$ to $\scettwo$ for one gluon emission which is either collinear
  to the quark or is in its own direction (SCET graphs for emission collinear to
  the antiquark are not shown).  The figures represent operator structures that
  describe this process in each of the three theories.  The QCD contribution is
  standard.  In $\scetone$, we either emit a collinear gluon through the
  time-ordered product of the Lagrangian with an two-parton operator, or from
  three-parton operators.  In $\scettwo$, the emission relevant for us only
  arises from higher-dimension three-parton operators.}
\label{fig:figureNLO1C}
\end{figure}

Just as in \subsec{loShowerRe}, we begin by examining the matching
QCD$\rightarrow$SCET$_1$ for single gluon emission collinear to the
quark. For this case, all corrections are of the hard-scattering
type. Beyond LO, we can have dependence on the process that creates
the $\bar{q} q$ pair. For concreteness, we will consider the coupling
of QCD quarks to the vector current, $J_{{\rm QCD}}^{\mu} \,=\,
\bar{q} \gamma^{\mu} q$.  The matching is performed in the center of
mass frame with the initial virtual photon having momentum,
$p_\gamma=(Q,0,0,0)$.  The full details of this matching calculation
for QCD to SCET$_1$ are in Appendix \ref{app:QCD/SCET1}.  To reproduce
the full QCD current, $J_{{\rm QCD}}^{\mu}$, we need an infinite tower
of SCET$_1$ operators increasingly higher order in $\lambda$. However,
to get the required amplitude to NNLO$(\lambda)$, we only need four:
\begin{align}
\label{eq:S1amp0}
A_{\rm to \; NNLO}^{q\bar{q}g} &= C^{(0)}_{1,\,\rm{LO}}(n_0)  \int\!\! dx \langle 0| T \{ \mathcal{L}_{\scetone}(x)\mo^{(0)}_{1} \}  
|\, q_{n_0} g _{n_0}\bar{q}_{\nb} \rangle \nonumber \\
&+ C^{(1)}_{1,\,{\rm NLO}}(n_0, n_0 ) \langle 0| \mo^{(1)}_{1} |\, q_{n_0} g _{n_0}\bar{q}_{\nb} \rangle 
+ C^{(1)}_{1, \mathcal{T}}(n_0,n_0) \langle 0| \mathcal{T}^{(1)}_{1}|\, q_{n_0} g _{n_0}\bar{q}_{\nb} \rangle
\nonumber \\
&+C^{(1)}_{1}(n_1, n'_1 ) \langle 0| \mo^{(1)}_1 |\, q_{n_1} g _{n'_1}\bar{q}_{\nb} \rangle , 
\end{align}
where
\begin{align}
\mo^{(0)}_{1}(n_0)&= \bar{\chi}_{n_0}  \chi_{\bar{n}} \, , \nonumber \\
\mo^{(1)}_{1}(n_0, n_0 ) &=\,\chib_{n_0}\,g \mb^\alpha_{n_0\perp} \chi_{\nb}\, , \nonumber \\
\mathcal{T}^{(1)}_{1} (n_0, n_0 )&= 
\bar{\chi}_{n_0} \left[ \mathcal {P}_{n_0 \perp}^\beta \, g \mathcal{B}_{n_0 \perp}^\alpha  \right] \chi_{\bar{n}} \, , \nonumber \\
\mo^{(1)}_{1}(n_1,n'_1) &= \,\chib_{n_1}\,g \mb^\alpha_{n'_1\perp} \chi_{\nb} \, .
\label{eq:S1amp}
\end{align}
Here we introduced a short-hand for the notation established in \eq{opDef},
${\cal T}^{(1)}_1(n_0,n_0)= {\cal O}_1^{(2,1,1)}(n_0,n_0^{[1]})$. We give the
expression for $C^{(0)}_{1,\,\rm{LO}}$ in Eq.~(\ref{eq:loDiquark}).  The amplitude
from the operator $\mo^{(0)}_{1}(n_0)$ is shown in the first diagram in the
$\scetone$ column of Fig.~\ref{fig:figureNLO1C}, those from $
\mo^{(1)}_{1}(n_0,n_0) $ and $\mathcal{T}^{(1)}_{1} (n_0, n_0 )$ in the second,
and that for $\mo^{(1)}_{1}(n_1,n'_1)$ in the third. 

We call $ \mo^{(1)}_{1}(n_0,n_0) $ and $\mathcal{T}^{(1)}_{1} (n_0,
n_0 )$ ``two-jet'' operators as they are labeled with two distinct
collinear directions ($n_0$ and $\nb$) (we do not denote the antiquark
direction explicitly, following the convention in Eq.~\ref{eq:Oi}).
They describe a gluon collinear to the quark.  We obtain the
coefficients $C^{(1)}_{1}(n_0, n_0 )$ and $C^{(1)}_{1,
\mathcal{T}}(n_0,n_0)$ by expanding the QCD amplitude in the limit of
small gluon momentum transverse to the quark's direction with the
usual SCET proportionality: $(n_0 \cdot k_1, \bar{k}_1 , k_{1 n_0
\perp}) \sim (\lambda^2, 1,\lambda)Q$.  $C^{(1)}_{1}(n_0, n_0 )$ and
$C^{(1)}_{1}(n_1, n'_1 )$, are derived above Eq.~(\ref{eq:CoeS1}) and
given here:
\begin{align}
C^{(1)}_{1,\,{\rm NLO}}(n_0,n_0)&= \frac{1}{Q}  (n_{0}^{\mu}-\nb^{\mu} ) \gamma_{n_0 \perp}^\alpha \, , \nonumber \\
C^{(1)}_{1, \mathcal{T}}(n_0,n_0)  &= \frac{1}{\bar{q}_1 \bar{k}_1 }   
\gamma_{n_0 \perp}^\mu  \gamma_{n_0 \perp}^\beta \gamma_{n_0 \perp}^\alpha 
- \frac{2}{\bar{q}_1 Q} \,g^{\beta \mu} \gamma_{n_0 \perp}^\alpha \,.
\label{eq:singleGluonWilsonA}
\end{align}
We use the same kinematic variables as in Fig.~\ref{fig:kin1E}.  For
$C^{(1)}_{1}(n_0,n_0)$ and $C^{(1)}_{1, \mathcal{T}}(n_0,n_0)$, the
initial current is not a spectator, so neither term is simply
proportional to the $\gamma^\mu$ with which we started.  This
dependence on the details of the rest of process is a characteristic
feature of hard-scattering corrections.  There are an additional set
of two-jet configurations corresponding to the gluon collinear to the
antiquark.  These are trivial to obtain by charge conjugation.

The operator $\mo^{(1)}_{1}(n_1,n'_1)$ is a three-jet configuration,
as it describes three distinct directions.  Whenever we have an
operator where each field has its own index label, we can choose the
$n_i$ such that they are exactly aligned with the external particle
momenta.  We give the coefficient $C^{(1)}_{1}(n_1, n'_1 )$ in Eq.~(\ref{eq:C3j}).
 
Going to $\scettwo$ for single gluon emission is straightforward. The basis of
operators needed to reproduce the amplitude (\ref{eq:S1amp0}) is equal to
(\ref{eq:S1amp}), but with SCET$_2$ fields:
$\mo^{(0)}_{2}(n_0)$, $\mo^{(1)}_{2}(n_0, n_0 ) $, $\mathcal{T}^{(1)}_{2} (n_0,
n_0 )$, and $ \mo^{(1)}_{2}(n_1,n'_1) $.  As the computations get more
complicated with subsequent emissions, we wish to minimize our effort by only
including those terms necessary to give the corrections to a shower Monte Carlo.
This means we are only interested in the following:
\begin{enumerate}
\item We will need to keep those NNLO$(\lambda)$ contributions that can interfere
  with LO$(\lambda)$.  These give terms at the same order as an NLO$(\lambda)$ operator
  squared.  We do not compute NNLO$(\lambda)$ amplitude terms which have zero
  interference with the LO$(\lambda)$ amplitude. A list of the necessary
  computations is found in \app{scet1/scet2}.
  
\item Our ultimate goal is not a complete $\sceti$ theory from which one can do
  computations, but an improved shower algorithm.  In \tbl{map}, we give a list
  of those ingredients needed to construct a map between $\sceti$ and a LL
  parton shower.  We will augment the map with items needed for corrections
  (Eq.~\ref{eq:weight}, \tbl{map2}), but will not calculate contributions 
  which only contain redundant information for the shower amplitude. 
\end{enumerate}
The latter point has important implications for the sorts of operator
structures we need to consider.  If we wanted to do computations in
$\scettwo$, then we would need all operators and Wilson
coefficients to the order we are working.  However, single gluon
contributions in $\scettwo$ where the gluon and the quark are
collinear (inside a cone of angle $\sim \lambda^2$, {\it i.e.}
$\mo^{(1)}_{2}(n_0, n_0)$ or $\mathcal{T}_2^{(1)}(n_1,\, n_1)$)
correspond to a quark which does not split until after the scale of
matching $\scetone \,\rightarrow\, \scettwo$.  The corresponding
no-branching probability, however, is already determined in $\scetone$
from the one-loop RG kernel.  Thus, the coefficients of these
operators in $\scettwo$ are not required.  We only need to calculate
those single gluon contributions where each field has its own index
label in $\scettwo$, which means $C^{(1)}_{2}(n_1, n'_1 )$ for
$\mo^{(1)}_{2}(n_1,n'_1)$.

The matching equation for $C^{(1)}_{2}(n_1, n'_1 )$ in SCET$_2$ is:
\begin{align}
\label{eq:S1S2coeB}
& C^{(1)}_{2} (n_1,n_1^\prime)\langle 0 | \mo^{(1)}_{2} |
q_{n_1}\,g_{n^\prime_1}\,\bar{q}_{\nb} \rangle 
 \\
&= C^{(0)}_{1,\,\rm{LO}}(n_0) \int  d^4 x \langle 0|T \{ \mathcal{L}_{\scetone}(x) \mo^{(0)}_{1}  \}   | q_{n_1}\,g_{n^\prime_1}\,\bar{q}_{\nb} \rangle 
\nonumber \\
&\ \  + C^{(1)}_{1,\,{\rm NLO}} (n_0,n_0) \langle 0|   \mo^{(1)}_{1} |
q_{n_1}\,g_{n^\prime_1}\,\bar{q}_{\nb} \rangle 
 +  C^{(1)}_{1} (n_1, n'_1)  \langle 0|  \mo^{(1)}_1 | q_{n_1}\,g_{n^\prime_1}\,\bar{q}_{\nb} \rangle
\nonumber \\
&\ \  +   C^{(1)}_{1, \mathcal{T}} (n_0,n_0) \langle 0| \mathcal{T}^{(1)}_{1}
| q_{n_1}\,g_{n^\prime_1}\,\bar{q}_{\nb} \rangle 
\,. 
 \nonumber
\end{align}
It is convenient to decompose $C^{(1)}_{2}(n_1,n_1^\prime)$ as 
\begin{align} \label{eq:WCS21EB}
C^{(1)}_{2}(n_1,n_1^\prime) = C^{(1)}_{2,\,\rm{LO}}(n_1,n_1^\prime) +  
C^{(1) H,a}_{2,\,\rm{NLO}}(n_1,n_1^\prime) + C^{(1)H,b}_{2,\,\rm{NLO}}(n_1,n_1^\prime)  + 
C^{(1) H}_{2,\,\rm{NNLO}}(n_1,n_1^\prime)\,,
\end{align}
where the four terms on the RHS of \eq{WCS21EB} correspond to each of
the contributions on the RHS of \eq{S1S2coeB}.  We calculated in
$C^{(1)}_{2,\,\rm{LO}}$ in \eq{loRep} using RPI$_1$ to rotate objects
in the SCET$_1$ amplitude such that they can come from SCET$_2$
operators that annihilate the given external state.  The second
through fourth terms can be calculated in a similar manner.  Their
values are derived in Eqs.~(\ref{eq:C2S2})- (\ref{eq:S23j}):
\begin{align}
\label{eq:coeS2}
C^{(1)H,a}_{2,\,\rm{NLO}}(n_1,n_1^\prime) & =  
\frac{1}{Q}\Big(\frac{\bar{k}_1 n_1^{\prime\mu} + \bar{q}_1 n_1^\mu}{\bar{q}_0} 
 - \Big( 1+\frac{\bar{q}_1 \bar{k}_1}{2\, \bar{q}_0^2} (n_1\!\cdot\!n'_1 ) \Big)\nb^\mu \Big) \gamma_{n'_1 \perp}^\alpha
  \Theta_{ \delta_2} [n_1 \cdot n'_1 ]  \, , \\
C^{(1)H,b}_{2,\,\rm{NLO}}(n_1,n_1^\prime) & =  -\frac{2}{ (n_1\!\cdot\!n'_1) \bar{q}_1 \bar{k}_1} \gamma^\alpha \slashed{p}_\gamma \,\gamma_T^\mu \nonumber \\
&+ \Big[ \frac{1}{ (n\!\cdot\! p_{\bar{q}} )\bar{k}_1 } 
\Big( \gamma_T^\mu \, \slashed{p}_\gamma - \bar{q}_1\,n_{1T}^\mu \Big) 
+\frac{2  (n\!\cdot\! p_{\bar{q}} ) }{ (n_1\!\cdot\!n'_1) \bar{q}_1  \bar{k}_1} \nb_T^\mu \Big] \gamma^\alpha
    \tilde{\Theta}_{ \delta_2} [n_1 \cdot n'_1 ] \, ,\nonumber \\
 C^{(1)H}_{2,\,\rm{NNLO}}(n_1,n_1^\prime)  &= 
                           \Big(  \frac{1}{2 \,Q} \Big(\gamma^\mu_{n'_1 \perp} \sqrt{n_1 \cdot n'_1} \slashed{v}_1
                           + \nb^\mu \frac{\bar{q}_1}{Q} (n_1 \cdot n'_1) \Big)\gamma^\alpha_{n'_1 \perp}\nonumber \\
                          & +\frac{\bar{k}_1}{ Q^2} \Big(   \sqrt{n_1 \cdot n'_1} v_1^\mu 
- \bar{n}^\mu \,(n_1\!\cdot\! n'_1)\frac{  (\bar{k}_1^2 -\bar{q}_1^2)}{2\, Q^2} \Big)  \gamma^\alpha_{n'_1 \perp} \Big) 
 \Theta_{ \delta_2} [n_1 \cdot n'_1 ]  \,. \nonumber
\end{align} 
Here $n_1$ and $n'_1$ are aligned with the direction of the quark and
the gluon, and $v_1$ is defined in Eq.~(\ref{eq:v}).  In
Eq.~(\ref{eq:coeS2}), we have left off the running factors from
evolution of the SCET$_1$ operators.  The terms in
Eqs.~(\ref{eq:S1amp}) run differently.  In particular, the two-jet and
three-jet operators have different LL evolution.  Therefore, it is
important to decompose $C^{(1)}_{2}$ as in Eq.~(\ref{eq:WCS21EB}), so
that we can keep track of which SCET$_1$ evolution factor to include
for each. The running of these operators is discussed further in
Section~\ref{subsec:opRunning}.

\begin{figure}[t!]
\centering
\includegraphics[width=0.9\textwidth]{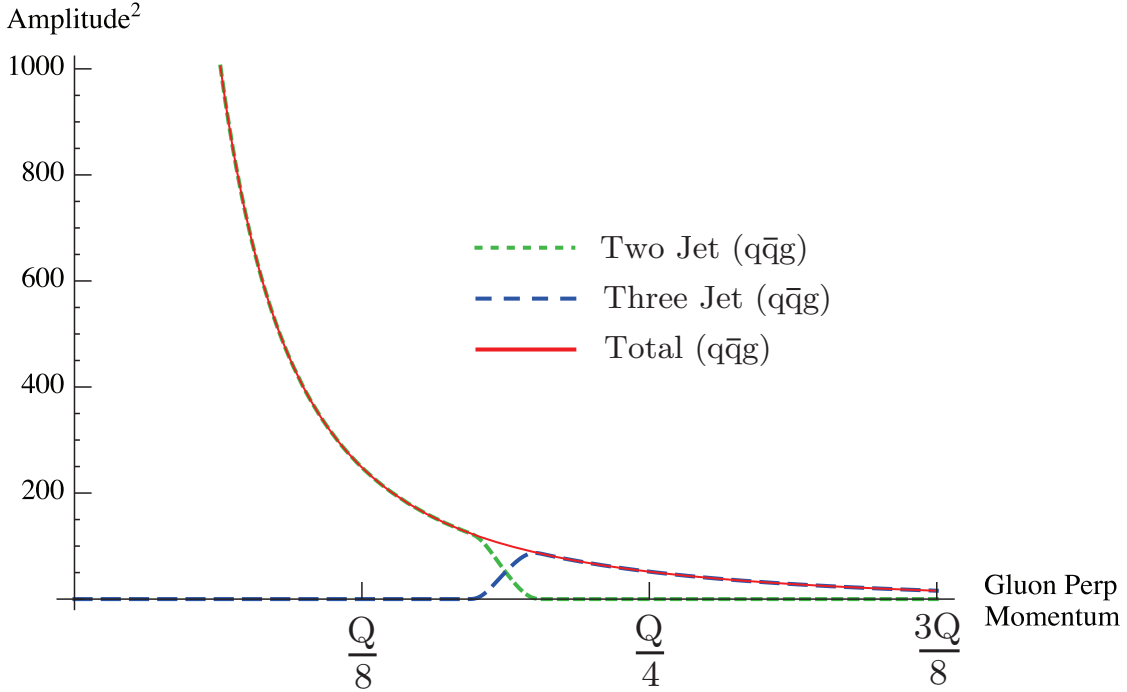}
\caption{Merging of the two-jet and and three-jet squared amplitudes using a
  smooth theta function for the $\gamma^* \rightarrow q \bar{q}g$ process.
  Plots of the amplitude squared components from $C_{2}^{(1)}
  \mo^{(1)}_2$:
  $|A^{q\bar{q}g}|_{\rm{LO}}^2+|A^{q\bar{q}g}|_{\rm{NLO},\, 2-jet}^2$ (short
  dashed green), $|A^{q\bar{q}g}|_{\rm{NLO},\,3-jet}^2$ (long dashed blue), and
  sum (solid red) versus $|k_1|_{n_0 \perp}$.  The amplitudes are evaluated
  without running coefficients, and taking $\bar{k}_1/\bar{q}_0=0.4$.  The $\delta_2$
  parameter in the $\Theta$-function is 1.2, which for the above $\bar{p}$ fraction
  corresponds to $\eta= 0.5$, and $\lambda = 0.08$.}
\label{Merge2J-3JB}
\end{figure}

We also note the different $\Theta$ dependence of the terms, where $\Theta$ and
$\tilde{\Theta}$ we introduced in \eq{thetaExpl} and the surrounding discussion.
We can read off from $C^{(1)H,b}_{2,\,\rm{NLO}}$ its origin as a three-jet term
in $\scetone$, while the others come from two-jet operators.  The $\Theta$
functions are necessary because without them $\scettwo$ operators, ({\it e.g.}
$\mo^{(2)}_2(n_1,\,n'_1)$) can only tell that the quark and gluon are not
collinear according to the SCET$_2$ definition.  By including these phase space
cutoffs, we can keep the distinct origins of different contributions manifest.
By adopting a smoothed step function, as suggested in \subsec{loShowerRe} and
given in \eq{thetaEx}, the amplitude squared for $C_{2}^{(1)} \mo^{(1)}_2$ will
be continuous despite having different supports in different parts of phase
space. An example of this is shown in Fig.~\ref{Merge2J-3JB}.  The full
expression for the plot is given in Eqs.~(\ref{eq:AmSqNLO}) and 
(\ref{eq:AmSqNLO2}).
\begin{figure}[t!]
\centering
\includegraphics[width=0.85\textwidth]{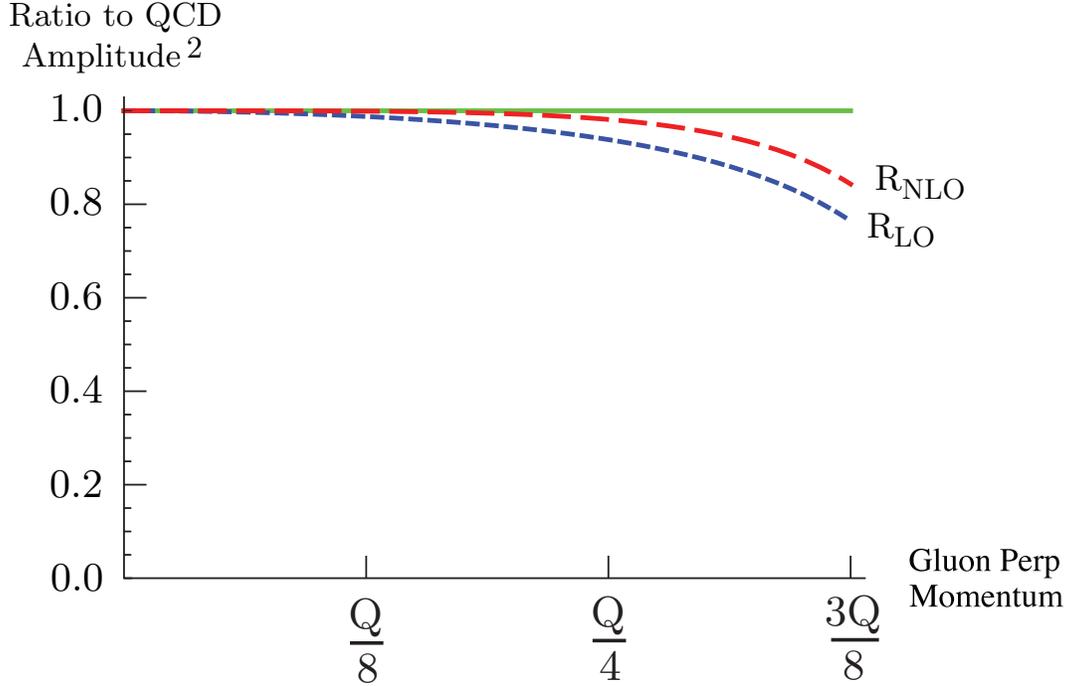}
\caption[Plot of the ratios of the QCD and SCET$_2$ amplitudes
  squared]{Plot of the ratios of the amplitudes squared for $\gamma^*
  \rightarrow q \bar{q}g$, namely $R_{\rm
  LO}=|A^{q\bar{q}g}|_{\rm{LO}}^2/|A^{q\bar{q}g}|_{\rm{QCD}}^2$ (blue short dashed)
  and $R_{\rm NLO}=(|A^{q\bar{q}g}|_{\rm{LO}}^2 +
  |A^{q\bar{q}g}|_{\rm{NLO},2-jet}^2)/|A^{q\bar{q}g}|_{\rm{QCD}}^2$
  (red long dashed) versus $|k_1|_{n_0 \perp}$, for $\bar{k}_1/\bar{q}_0=0.4$.
  The amplitudes are evaluated without running factors.}
\label{Plot:Plot_A22jvsQCDB}
\end{figure}
To illustrate the effects of including hard-scattering corrections, in
Fig.~\ref{Plot:Plot_A22jvsQCDB} we plot the ratios $R_{\rm LO}=|A^{q \bar{q} g}|
^2_{\rm{LO}}/|A^{q \bar{q} g}| ^2_{\rm{QCD}} $ and $R_{\rm NLO}=(|A^{q \bar{q}
  g}| ^2_{\rm{LO}}+|A^{q\bar{q}g}|_{\rm{NLO},2-jet}^2)/|A^{q \bar{q} g}|
^2_{\rm{QCD}}$ versus the gluon perp momentum. Here, $|A^{q \bar{q} g}|
^2_{\rm{QCD}}$ is the QCD amplitude squared for one-gluon emission, $|A^{q
  \bar{q} g}| ^2_{\rm{LO}}$ is the $\scettwo$ amplitude squared for one-gluon
emission from the LO coefficient $C_{2,\,\rm{LO}}^{(1)} \mo^{(1)}_2$ (from
Eq.~\ref{eq:loRep}), and $|A^{q\bar{q}g}|_{\rm{NLO},\,2-jet}^2$ is the
NLO$(\lambda)$ amplitude squared for one-gluon emission in the two-jet region
that comes from the coefficients $C^{(1)H,a}_{2,\,\rm{NLO}} $ and $
C^{(1)H}_{2,\,\rm{NNLO}}$ (given in Eq.~\ref{eq:coeS2}).  As we expect,
including corrections up to NNLO$(\lambda)$ in the amplitudes squared extends
the region where tree-level $\scettwo$ and QCD agree.  The advantage of using
the one-gluon $\scettwo$ amplitude over QCD comes from factorization properties
that effect interference as well as renormalization group evolution.  For
example the one-loop running in $\scettwo$ performs the LL Sudakov resummation.

With two-gluon emission, the $\scetone$ graphs will include
jet-structure corrections in addition to hard-scattering ones.  It is
straightforward to distinguish the types as the former result from
taking time-ordered products of the $\scetone$ Lagrangian with
operators generated by the LO replacement rule, \eq{loArrow}, while
the latter will come only from terms involving a power suppressed
$\scetone$ operator.  To fully identify the subleading contributions
to two-gluon emission, we must match down to $\scettwo$ where the LO
contribution is first uniquely identified.  We already know that it
comes from two applications of \eq{loArrow}.

\begin{figure}[t!]
\centering
\includegraphics{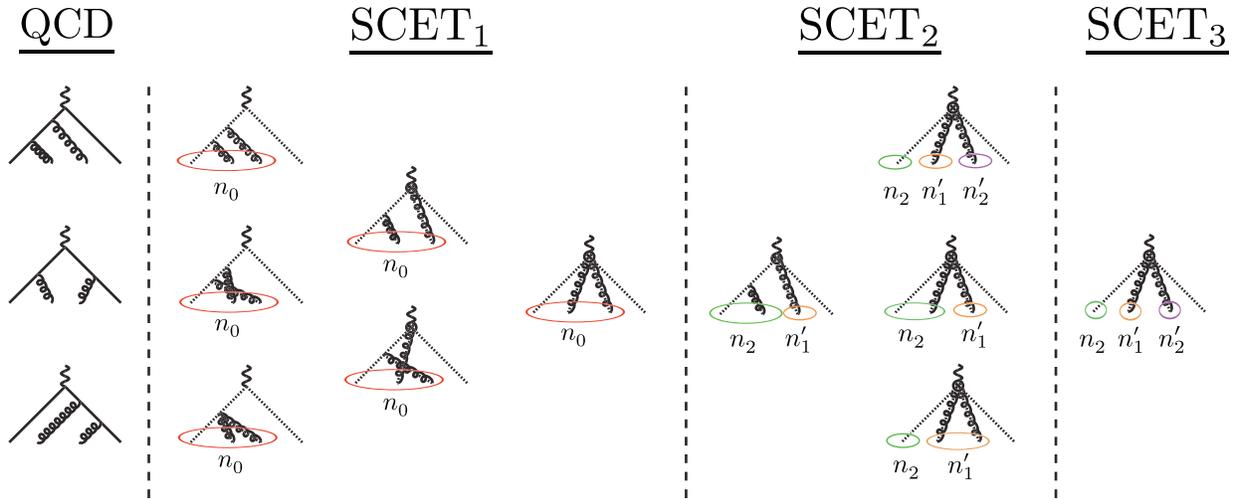}
\caption[Matching QCD to $\scetone$ to $\scettwo$ to $\rm{SCET}_3$ for two-gluon
emission] {Matching QCD to $\scetone$ to $\scettwo$ to $\rm{SCET}_3$ for
  two gluons emitted collinear to the quark direction (SCET graphs for other
  gluon kinematic configurations not shown).  Once again, we depict the operator
  structures that lead to this process in each of the theories.  Gluons drawn
  away from the central vertex are emitted by the leading order Lagrangian in
  that theory, while those coming from the vertex are due to higher dimension
  operators.}
\label{fig:figureNLO2C}
\end{figure}

In Fig.~\ref{fig:figureNLO2C},
we show the contributions to two-gluon emission in QCD, $\scetone$, $\scettwo$,
and $\scetthree$.  The first column in the $\scetone$ category corresponds to the
jet-structure corrections to be considered in the next section. In the second
column we have a set of hard-scattering corrections from taking the $T$-product
of the $\scetone$ Lagrangian with the suppressed single gluon operators we
calculated above in Eqs.~(\ref{eq:singleGluonWilsonA}), $C^{(1)}_{1}
\mo_1^{(1)}$ and $C^{(1)}_{1, \mathcal{T}} \mathcal{T}_1^{(1)}$.

In considering the basis of operators in $\scettwo$ we do not need operators
such as $\mathcal{T}_2^{(1)}(n_1,\, n'_1)$, since $\mcp_{n'_1 \perp} \,
\mb_{n'_1 \perp}$ = 0, with $n'_1$ lying along the gluon momentum.  We can use 
RPI$_2$ in $\scettwo$ to make a coordinate choice where they are not necessary.  As
mentioned above in the single gluon matching section, our interest is only in
calculating those terms needed to improve a shower algorithms, which precludes
us from considering operators such as $\mathcal{T}_2^{(1)}(n_0,\, n_0)$ or
$\mo_2^{(1)}(n_0,\, n_0)$, corresponding to an unbranched quark passing from 
$\scetone$ into $\scettwo$.  Therefore, for double gluon emission we only need to
calculate the coefficients of the following operators:
\begin{align} 
  \label{eq:OpS2}
  \mo^{(1)}_2 (n_1,\,n'_1)&=\bar{\chi}_{n_1} g \mb_{n'_1\perp}^\alpha  \chi_{\nb} \\
 \mo^{(2)}_2(n_2,n_2,n'_1)&=\bar{\chi}_{n_2} g \mb_{n_2\perp}^\alpha \, g \mb_{n'_1\perp}^\beta \chi_{\nb} \, ,  \nonumber\\
 \mo^{(2)}_2(n_2,n'_1,n'_1)&=\bar{\chi}_{n_2} g \mb_{n'_1\perp}^\alpha \, g \mb_{n'_1\perp}^\beta \chi_{\nb} \, , \nonumber  \\
   \mo^{(2)}_2(n_2,n'_1,n'_2)&=\bar{\chi}_{n_2} g \mb_{n'_1\perp}^\alpha \, g \mb_{n'_2\perp}^\beta \chi_{\nb} \, . \nonumber
\end{align}
Thus in $\scettwo$, we are interested in {\it two}-gluon operators where two
fields can have the same label.  When we pass to $\scetthree$, we can restrict
our interest to only $\mo^{(2)}_3(n_2,\,n'_1,\,n'_2)$. 

We already gave the coefficients of $\mo^{(1)}_2(n_1,\,n'_1)$ needed to compute
the leading power corrections in Eqs.~(\ref{eq:coeS2}) and
(\ref{eq:C2S2})-(\ref{eq:S23j}).  We get an NLO$(\lambda)$ contribution to the
two gluon amplitude by computing the matrix element, $ C^{(1)}_{2,\, \rm NLO}
\langle 0 | \mathcal{T}\{ \mathcal{L}_{\scettwo} \mo^{(1)}_2 \} | q\bar{q}gg
\rangle$ (first $\scettwo$ column in Fig.~\ref{fig:figureNLO2C}).  The
contribution receives no further suppression as the gluon from
$\mathcal{L}_{\scettwo}$ gives a tree-level vertex$\times$propagator factor of
$\lambda^{-2}$, just as with LO.  There are also coefficients we need from two-gluon matching
calculations for the operator $\mo^{(2)}_2$ (second $\scettwo$ column in
Fig.~\ref{fig:figureNLO2C}).  Putting in the index structures, these include
$C^{(2)J}_{2,\,\rm{NLO}} (n_2,\,n'_1,\,n'_2)$ for $\mo^{(2)}_2(n_2,n'_1,n'_2)$,
$C^{(2)J}_{2,\,\rm{NLO}} (n_2,\,n'_1,\,n'_1)$ for $ \mo^{(2)}_2(n_2,n'_1,n'_1)$
and $C^{(2)J}_{2,\,\rm{NLO}} (n_2,\,n_2,\,n'_2)+C^{(2)H}_{2,\, \rm{NNLO}}
(n_2,\,n_2,\,n'_2)$ for $ \mo^{(2)}_2(n_2,n_2,n'_1)$.  We include
NNLO$(\lambda)$ for the last one as only it interferes with LO$(\lambda)$.  In
the next subsection, we give the jet-structure corrections.  All hard-scattering
contributions to these structures just listed are beyond the order we need
except for $C^{(2) H}_{2,\,{\rm NNLO}} (n_2,\,n_2,\,n'_1)$, given by the
matching equation:
\begin{align}
\label{eq:C2HB}
& C^{(1)}_{1,\,{\rm NLO}}(n_0,n_0) \int \! \! dx \langle 0|   T \{ \mathcal{L}_{\scetone}(x) \mo^{(1)}_{1}(n_0, n_0) \} 
|q_{n_2} g_{n_2} g_{n_1^\prime} \bar{q}_{\nb} \rangle\\ 
&  - C^{(1)H,\,a}_{2,\,\rm{NLO}}(n_2, n'_1) \int \! \! dx \langle 0|T \{ \mathcal{L}_{\scettwo}(x)\mo^{(1)}_{2}(n_2, n'_1)\}
|q_{n_2} g_{n_2} g_{n_1^\prime} \bar{q}_{\nb} \rangle  \nonumber\\ 
& =  C^{(2)H}_{2,\,{\rm NNLO}}(n_2,n_2,n'_1)  \langle 0|\mo^{(2)}_2(n_2,n_2,n'_1) 
|q_{n_2} g_{n_2} g_{n_1^\prime} \bar{q}_{\nb} \rangle  \nonumber \,,
\end{align}
where we subtract the Lagrangian emission graph in $\scettwo$ from that in
$\scetone$ ($C^{(1)}_{1,\,{\rm NLO}}$ is given in
Eq.~\ref{eq:singleGluonWilsonA} and $C^{(1)H,\,a}_{2,\,\rm{NLO}}$ in
Eq.~\ref{eq:coeS2}).  The result for $C^{(2) H}_{2,\,{\rm
    NNLO}}(n_2,\,n_2,\,n'_1)$ is given in \eq{C2H221p}.  It is straightforward
to see why $\mo^{(2)}_2$ only gets hard-scattering at NNLO$(\lambda)$ and
higher.  By definition, hard-scattering has to involve a suppressed operator from
the QCD $\rightarrow \, \scetone$ matching, and so we begin at NLO$(\lambda)$ at
the lowest order.  Including a second gluon, but demanding that we cannot write
it as coming from a $\scettwo$ Lagrangian emission takes us to one order higher,
namely NNLO$(\lambda)$.

All the contributions we have discussed so far have come from the
hard-scattering, single-gluon, suppressed operators in $\scetone$.  There are
also those with two gluons.  That is to say a process where neither gluon comes
from the $\scetone$ Lagrangian, represented by the diagram in the third
$\scetone$ column in Fig.~\ref{fig:figureNLO2C}.  One example is double
$\perp$-gluon emission from the antiquark, as shown in the third QCD graph of
Fig.~\ref{fig:figureNLO2C}.  We know from applying Eq.~(\ref{eq:loArrow}) twice,
that LO for this process is at $\mo(\lambda^{-3})$, counting only the tree-level
vertex$\times$propagator factors, as these are all we need to compare different
$q\bar{q}gg$ processes.  We readily see that double antiquark emission is $\sim
\lambda^0$ as there are no small virtualities or emission angles for this term.
Thus, they are ${\rm N^3LO}$, and beyond this analysis.  Besides antiquark
vertices, we also have subleading emissions from the quark in QCD that arise
from the suppressed SCET-spinor portion of the QCD quark propagator ({\it cf.}
Appendix~\ref{app:scet}).  If both emissions come from the suppressed
propagator, once again, this is $\sim \, \lambda^0$ at lowest order, and so we
can neglect it.  Mixed antiquark/suppressed spinor contributions are also ${\rm
  N^3 LO}$.
  
Thus, we do not need corrections to double emission collinear to the quark if
they do not involve at least one $\scetone$ Lagrangian insertion.  We can extend
this argument further.  If there are no $\scetone$ Lagrangian insertions, then
the contribution goes like $\mo(\lambda^0)$, while LO goes like
$\mo(\lambda^{-\frac{i(i+1)}{2}})$.  Thus, to the order we are working, we only
need the single gluon hard-scattering corrections given by \eq{coeS2}, plus
Lagrangian insertions.

\subsection{Jet-Structure Corrections} 
\label{subsec:scetToSCET}
 
The {\it jet-structure} corrections only involve contributions from
the $\scetone$ Lagrangian.  These arise from the graphs in the first
$\scetone$ column in Fig.~\ref{fig:figureNLO2C}.  We specifically
designed our leading order replacement rule in Eq.~(\ref{eq:loArrow}),
so when used twice it only contains that part of double emission
corresponding to the leading strongly-ordered limit.  This occurs for
the gluons having collinearities $\sim \, \lambda, \, \lambda^2$,
respectively.  However, $\scetone$ describes other kinematic
situations and in this section we compute the corrections from them.

The prescription for obtaining two-gluon jet-structure corrections is to
compute the double gluon emission amplitude in $\scetone$ coming from two
Lagrangian insertions and take different limits on the relative collinearities
of $n_2,\,n'_2$, and $n'_1$, where these labels refer to the null vectors
exactly proportional the corresponding particle momenta.  We can define:
\beq
  A^{q\bar{q}gg}_{\mathrm{NLO}} \,=\, 
  C^{(0)}_{1,\,\rm{LO}}(n_0)\int\!\!dx_1 dx_2 \langle 0| T \{ \mathcal{L}_{\scetone}(x_1) \mathcal{L}_{\scetone}(x_2) \mo^{(0)}_{1}  \} 
  |q_{n_2}g_{n'_1} g_{n'_2}\bar{q}_{\nb} \rangle,
\eeq
and then calculate,
\begin{align}
  \lim_{n_2 \cdot n'_2 \sim \lambda^2} A^{q\bar{q}gg}_{\mathrm{NLO}} &=
  {C^{(2)J}_{ 2,\,\rm{NLO}}(n_2,\, n'_1,\, n'_2)} \langle 0| {\mo^{(2)}_{2}} 
  |q_{n_2}g_{n'_1} g_{n'_2}\bar{q}_{\nb} \rangle 
  \label{eq:originalReplaceOp} \,, \\ 
  \lim_{n'_1 \cdot n'_2 \sim \lambda^4} A^{q\bar{q}gg}_{\mathrm{NLO}} &=
  {C^{(2)J}_{ 2,\,\rm{NLO}}(n_2,\, n'_1,\, n'_1)} \langle 0| 
  {\mo^{(2)}_{2}}  |q_{n_2}g_{n'_1} g_{n'_2}\bar{q}_{\nb} \rangle 
  \label{eq:newSubleadingOp} \,, \\
  \lim_{n_2 \cdot n'_2 \sim \lambda^4} A^{q\bar{q}gg}_{\mathrm{NLO}} &= 
  {C^{(1)}_{2,\,\rm{LO}}}(n_2,n'_1) \langle 0| T \{ \mathcal{L}_{\scettwo} {\mo^{(1)}_{2}} \} 
   |q_{n_2}g_{n'_1} g_{n'_2}\bar{q}_{\nb} \rangle  \nonumber \\
  &+C^{(2)J}_{ 2,\,\rm{NLO}}(n_2,\, n'_1,\, n_2) \langle 0| 
  {\mo^{(2)}_{2}}  |q_{n_2}g_{n'_1} g_{n'_2}\bar{q}_{\nb} \rangle \,.
  \label{eq:loPlusSubleadingOp} 
\end{align}
We note a few things about the above equations.  Firstly, there is a 
correction to the LO Wilson coefficient obtained from the replacement rule
(Eq.~\ref{eq:loArrow}).
We cannot get it purely as a limit of
$A^{q\bar{q}gg}_{\mathrm{NLO}}$, so we need to subtract off the LO
contribution.  Secondly, the limit in \eq{originalReplaceOp} does not lead to
an expansion of any part of $A^{q\bar{q}gg}_{\mathrm{NLO}}$, as the scaling of
the $n$-indices' dot products is exactly that from $\scetone$.  Even though it
just gives back the same expression as the $\scetone$ amplitude,
$A^{q\bar{q}gg}_{\mathrm{NLO}}$, the $\scettwo$ result for ${C^{(2)J}_{
    2,\,\rm{NLO}}(n_2,\, n'_1,\, n'_2)}\mo^{(2)}_2$ tells us something more.
This Wilson coefficient is proportional to $\tilde{\Theta}_{\delta_2}[n'_1
\cdot n'_2] \tilde{\Theta}_{\delta_2}[n_2 \cdot n'_2]$, where the
$\tilde{\Theta}$'s only have support outside the phase space region of
\eq{newSubleadingOp}, as well as the strongly-ordered limit,
\eq{loPlusSubleadingOp}, (see Eqs.~(\ref{eq:thetaExpl}) and (\ref{eq:thetaEx})
for the definition of $\Theta$, $\tilde{\Theta}$).  The full results for the
Wilson coefficients shown in \eq{loPlusSubleadingOp} can be found in
Eqs.~(\ref{eq:C21p2p}), (\ref{eq:C11p1pa}), and (\ref{eq:Cn2n2n1p}).
At the amplitude level, given a particular phase space configuration for an
external state, we will only ever need one of these terms for double gluon
emission in $\scettwo$.  Squaring the result is straightforward as there will be
no interference between them.
 
We will now examine how to improve the matching of $\sceti$ to
$\scetipone$, and show that the jet-structure corrections computed
here generalize to that case.  We first notice that the first two
operators above do not interfere with the one giving LO, as they have
different index structures.  The subleading term in
\eq{loPlusSubleadingOp} does inhabit the strongly-ordered region of
phase space, but as we will argue in \subsec{ampSq}, LO$(\lambda)$/NLO$(\lambda)$
interference cancels out of most observables of interest.  Before
proceeding, we note that our description of corrections to two-gluon
emission gets even simpler when we match to $\scetthree$.  In
$\scetthree$, the only operator we need has distinct collinear
directions for all fields.  Thus, we can write all hard-scattering and
jet-structure corrections to two-gluon emission we have found in the
coefficient, $C^{(2)}_3$, for the operator ${\cal
O}_3^{(2)}(n_2,n_1',n_2')=\bar\chi_{n_2}g {\cal B}_{n_1'\perp}^\alpha
g{\cal B}_{n_2'\perp}^\beta \chi_{\nb}$, as we do in
Eqs.~(\ref{eq:CoeS3a}).  The same will hold for $i$-gluon emission in
$\scetipone$.  Our NLO$(\lambda)$ jet-structure operators therefore have the
following form:
\beq
C^{(2)J, I}_{3,\,\mathrm{NLO}}(n_2,\,n_1,\,n'_1) \mo^{(2)}_3  \,=\, h^{\alpha\beta}_I \, 
\chib_{n_2}g \mb^{\alpha}_{n'_1 \perp}g \mb^{\beta}_{n'_2\perp} \Gamma^{\mu} \chi_{\bar{n}} \, ,
\label{eq:twoGluNLO}
\eeq
where $h_I^{\alpha\beta}$ is given by \eq{hsc}. Here $I=\{1,2,3\}$, and we
distinguish the coefficients $C^{(2)J, I}_{3,\,\mathrm{NLO}}$ depending on
which SCET$_2$ operators they come from in order to properly account for their
RG evolution in $\scettwo$. 

When doing the LO matching for $\sceti$ to $\scetipone$, we found that the
replacement rule to go from $\scetone$ to $\scettwo$ generalized to the case of
$i$-gluon strongly-ordered emission.  Similarly, we can take the above operator,
\eq{twoGluNLO}, and recast it as a replacement rule for our original current
insertion, $C^{(0)}_{1 ,\,\rm{LO}}\mo^{(0)}_{1}$.  It takes the form of a $1\to
3$ replacement rule:
\beq
\chib_{n_0} \rightarrow h^{\alpha\beta}_I \, \chib_{n_2}g \mb^{\alpha}_{n'_1 \perp}g \mb^{\beta}_{n'_2 \perp},
\label{eq:nloReplace}
\eeq
with contributions from $I=1,2,3$.  

If we want to consider the NLO$(\lambda)$ radiation of $i$+1 gluons, we can
perform a very similar matching between $\sceti$ and SCET$_{i+2}$ to
the one above for $\scetone \rightarrow \scetthree$ to obtain an
operator $C^{(i+1) J}_{i+2,\,\mathrm{NLO}} \mo^{(i+1)}_{{i+2}}$.
Since the first $(i-1)$ emissions are strongly ordered, they
completely factor out.  Thus, the amplitude for the emission of the
final two gluons will be identical to that for simple two-gluon
emission.  We can therefore take the $(i-1)$ gluon LO operator, $
C^{(i-1)}_{i,\,\mathrm{LO}} \mo^{(i-1)}_{{i}} $, and use the
replacement rule in Eq.~(\ref{eq:nloReplace}), to obtain $
C^{(i+1)J}_{i+2,\,\mathrm{NLO}\, I} \mo^{(i+1)}_{i+2} $.  Our NLO$(\lambda)$
replacement rule corresponds to violating strong ordering at any
location in the shower, either by taking the $j^{\rm th}$ and
$(j+1)^{\rm th}$ gluons to have the same parametric collinearity with
respect to their parents, $k_{j+1\perp} \sim k_{j\perp}$
(Eqs.~\ref{eq:originalReplaceOp} and \ref{eq:newSubleadingOp}); or by
including the region of phase space where the propagator between them
is hard even in $\scetone$, and so we get no collinear divergence as
the quark and second gluon become collinear
(\ref{eq:loPlusSubleadingOp}).\footnote{At this point, one may ask why
we do not go farther and consider the case $k_{j+1\perp} \gg
k_{j\perp}$.  In fact, we do not have to.  Since the amplitude for
$i$-gluon emission has an underlying Bose symmetry, we are free to
partition phase space into $i!$ regions, each of which gives an
identical contribution to the cross section.  Thus, to get the final
answer, we only need to integrate over one of them.  While we can
choose this region such that $k_{j+1\perp} \gg k_{j\perp}$ never
occurs, we are forced to include $k_{j+1\perp} \sim k_{j\perp}$. If we
do not wish to partition phase space in this manner, then the Bose
symmetry implies that the result for $k_{j+1\perp} \gg k_{j\perp}$ can
be obtained from the configurations already discussed.}

It is not difficult to see that this gives an NLO$(\lambda)$ contribution for any
$j$.  If we have $i$-gluon strongly-ordered emission, the tree-level
factors, $c^{\alpha_{k}}_{\rm{LO}}(n_{k-1})$, ({\it cf.} Eq.~\ref{eq:loOp})
will go as $\lambda^{-i(i+1)/2}$, where the $j^{\rm th}$ gluon
contributes $\lambda^{-j}$.  If we violate strong ordering as we
mention above for any two gluons, the product of their vertices times
propagators goes like $\lambda^{-2j}$ instead of $\lambda^{-(2j+1)}$.
Thus, we can insert $\chib_{n_0} \rightarrow h_I \, \chib_{n_2} g
\mb_{n'_1 \perp} g \mb_{n'_2 \perp}$ instead of two successive
$\chib_{n_0} \rightarrow c_{\rm{LO}} \chib_{n_1} \mb_{n'_1\perp}$'s in
operator matching as a ``defect'' in strong ordering at any stage and
obtain an NLO$(\lambda)$ jet-structure correction.  The $\Theta$-functions
contained in the Wilson coefficients, $C^{(i+1)J}_{i,\,\mathrm{NLO}}$,
allow us to read off at which step in the shower we violated
strong-ordering.

In App.~\ref{app:nloSplit}, we show that an integrated version of
$h_I^{\alpha\beta}$ is related to the splitting function at NLO in $\alpha_s$,
which serves as a cross-check on our computations.

\subsection{Operator Running}
\label{subsec:opRunning}

Up until now, our discussion of matching has taken place mostly at tree-level.
Connecting to the no-branching probabilities and log resummation in the parton
shower however, requires that we include the anomalous dimensions needed for
running.  For this reason, our final expressions for Wilson coefficients in
Apps.~\ref{app:QCD/SCET1}-\ref{app:scet2/scet3} include the necessary notation
for evolution kernels.  Identifying the power suppressed amplitudes as
corresponding to perturbative corrections to more inclusive observables, it is
natural to take only LL$_{\rm exp}$ evolution for power suppressed or $\alpha_s$
suppressed corrections, and include NLL$_{\rm exp}$ evolution only for the
leading shower terms.  For the former, we assume (without carrying out the proof
in $\sceti$) that we must make the $k_T^2$ choice for the scales $\mu_k^2$ as in
\eq{muk2}, and that this accounts for the difference between LL and LL$_{\rm
  exp}$.  NLL$_{\rm exp}$ would require full one-loop, two-loop cusp, and NLL
$\alpha_s$ running, plus any modifications to the evolution induced by
subleading soft effects.  If subleading soft effects are neglected then in the
terminology of \cite{Bonciani:2003nt,Frixione:2007vw}, this gives the full {\it
  collinear} NLL$_{\rm exp}$ resummation.  The subleading logarithms coming from
pure soft effects involve the exponentiation of nonabelian matrices.  As
mentioned earlier, we do not compute the effects of subleading soft $\sceti$
operators here.  (In fact, for more than three hard, colored particles, the
problem is quite non-trivial~\cite{Bonciani:2003nt}.)

In this section, we determine the LL$_{\rm exp}$ running for our subleading
operators and discuss what is missing in our setup for a NLL$_{\rm exp}$
evolution kernel for emission anywhere in the shower.  To set the stage, we
consider $\scetone$ matched to QCD at the scale $Q$ for the first order power
corrections.  We then run down to $\mu$ in preparation for matching to
$\scettwo$.  The zero and single gluon operators in $\scetone$ acquire the
following running factors, $U$, ({\it cf.} the tree-level version in
Eq.~(\ref{eq:singleGluonWilsonA})):
\begin{align}
\label{eq:S1coeRun}
C^{(0)}_{0}(n_0)&=U^{(2,0,0)}(n_0;\,Q,\mu) \,  \gamma^\mu_{n_0 \perp}\nonumber  \\
C^{(1)}_{1,\,{\rm NLO}}(n_0,n_0)&=  U^{(2,1,0)}(n_0, n_0;\,Q,\mu)  \otimes\, \frac{n_{0}^{\mu}-\nb^{\mu} }{Q} 
\gamma_{n_0 \perp}^\alpha \, , \nonumber \\
C^{(1)}_{1,\,\mathcal{T}}(n_0,n_0)  &= U^{(2,1,1)}_{ \mathcal{T}} (n_0,n_0;\,Q,\mu)\otimes\,\frac{1}{\bar{q}_1 \bar{k}_1 }  
\Big( \gamma_{n_0 \perp}^\mu  \gamma_{n_0 \perp}^\beta \gamma_{n_0 \perp}^\alpha 
- \frac{2}{\bar{q}_1 Q} \,g^{\mu\beta} \gamma_{n_0 \perp}^\alpha \Big)\,, \nonumber \\
\ C^{(1)}_{1} (n_1, n'_1)&= -U^{(2,1,0)} (n_1, n'_1;\,Q,\mu) \, \Big( \ \frac{2}{ (n_1\!\cdot\!n'_1) \bar{q}_1 \bar{k}_1}
 \gamma_{n_0 \perp}^\alpha \,\slashed{p}_\gamma \,\gamma_T^\mu \nonumber \\
&+ \Big[ \frac{1}{(n\!\cdot\! p_{\bar{q}} )\bar{k}_1 } 
\Big( \gamma_T^\mu \, \slashed{p}_\gamma - \bar{q}_1\,n_{1T}^\mu \Big) 
+\frac{2  (n\!\cdot\! p_{\bar{q}} ) }{ (n_1\!\cdot\!n'_1) \bar{q}_1  \bar{k}_1} \nb_T^\mu \Big] \gamma_{n_0 \perp}^\alpha \Big) \, ,
\end{align} 
where the superscripts follow the convention in Eq.~(\ref{eq:opDef}).  We
inserted the symbol $\otimes$ in the second and third line of
Eq.~(\ref{eq:S1coeRun}) since an operator with multiple fields sharing the same
collinear direction can convolve the momentum fraction of $\bar{p}$ between the
corresponding RG kernel $U$ and momenta in the tree-level coefficient.  This is
because collinear fields that are in the same direction in SCET can exchange
momentum while running down from $Q$ to $\mu$.  The anomalous dimension of
an operator is independent of which $\sceti$ it is defined, but does depend on
the field content and in particular how many different collinear directions are
in the operator.  Thus, the RG-kernel for $\bar{\chi}_{n_0} g \mathcal{B}_{n_0
  \perp}^\alpha \chi_{\bar{n}}$ is different from that of $\bar{\chi}_{n_1}g
\mathcal{B}_{n_1^\prime\perp}^\alpha \chi_{\bar{n}}$.  In
Ref.~\cite{Bauer:2006mk,Bauer:2006qp}, the LL part of $U^{(j,i-j,0)}(Q,\,\mu)$ was related to
the Sudakov form factor, Eq.~(\ref{eq:runningAsSudakov}) (up to accounting for
the soft effects of angular ordering~\cite{Bauertalk}).  The cusp term in the
anomalous dimension resums the LL, and comes from soft and collinear one-loop
diagrams.  The result from the soft diagrams is constrained by that of the
collinear diagrams in order to cancel out infrared sensitivity that cannot be
absorbed in local counterterms at the hard scale. Here we will use this same
argument, but in reverse, in order to determine the LL$_{\rm exp}$ anomalous
dimension of various subleading operators.

Due to the soft-collinear factorization, the soft structure only depends on the
number of collinear directions.  After making the field redefinition, operators
like $\bar \chi_{n_0} \chi_{\nb}$ and $\bar{\chi}_{n_0}g \mathcal{B}_{n_0
  \perp}^\alpha \chi_{\nb}$ both have $Y_{n_0}^\dagger Y_{\nb}$, and so both
have the same soft divergences. Hence they have the same one-loop cusp term and
the same LL anomalous dimension from the sum of collinear and soft loops.  Thus,
the leading-log resummation only depends on the number of collinear index
directions in the operator, and not on the number of active partons. (At leading
power these concepts are the same, but it is not so for the power corrections.)
We therefore have
\begin{align}
U^{(2,0,0)}_{\rm{LL}}(n_0)= U^{(2,1,1)}_{\rm{LL}}(n_0, n_0) =
U^{(2,1,1)}_{\mathrm{LL}, \mathcal{T}} (n_0,n_0) \,,
\label{eq:kernelEq}
\end{align} 
where we give $U^{(2,0,0)}_{\rm{LL}}$ in Eqs.~(\ref{eq:rgKernel}) and (\ref{eq:oneLoopCusp}).
Thus, at LL order we have the full set of evolution kernels for subleading
collinear operators, and we account for these factors in the appendices.
Since this is a LL effect, we expect soft radiation
and angular ordering to be incorporated in a manner identical to the evolution
factor in the LL shower.

An important consequence of this result for the LL evolution is that it
justifies treating our hard-scattering corrections as improvements to the
fixed-order, matrix-element calculation that goes into a shower algorithm.
Correcting the two-jet amplitude with either $C^{(1)}_{1,\, \rm NLO}$ or
$C^{(1)}_{1,\, \mathcal{T}}$, we see that the LL resummation is the same as that
in the standard shower except that there is an extra parton already inside the
leading jet.  We thus get a shower correction just by using a matrix element
improved by including our hard-scattering terms.  This is unlike simply
running a LL shower on higher order matrix elements, as different anomalous
dimensions control different operators' evolution.  Some, like those just
mentioned with only $n_0$ and $\nb$ collinear directions, run like two-jet
configurations, that is with a quark-antiquark Sudakov.  Others, ({\it e.g.}
$C^{(1)}_1 \mo^{(1)}_1(n'_1,\,n_1)$) have three-parton running since they have
three distinct collinear directions.  This latter set corresponds to the usual
implementation of fixed order corrections in parton showers, but the former is a
novel type of shower improvement.

On the other hand, the effect of jet-structure corrections is not to modify the
initial scattering process, but to go hand in hand with the NLL change to the
leading operators' running. Similarly to \eq{loOp}, we might anticipate the
following Wilson coefficient for $\mo^{(N-1)}_N$ with evolution:
\begin{align}
C^{(N-1)}_{N,\,\rm{NLL,\,1}}(m) &=  \left[ \left( \prod_{k=1,\,k\neq m}^{N-1} U^{(k-1)}_{\rm{LL}}(\mu_{k-1},\mu_{k}) \, 
c^{\alpha_k}_{\rm{LO}}(n_{k-1}) \right) \right. \nonumber \\
 & \left. \times 
 U^{(m-1)}_{\rm{NLL}}(\mu_{m-1},\mu_{m})\, 
c^{\alpha_m}_{\rm{LO}}(n_{m-1})  \right] \Gamma^{\mu},
\label{eq:nllWils}
\end{align}
with a sum over all locations where the NLL evolution can be inserted:
\beq
C^{(N-1)}_{N,\,\rm{NLL,\,1}} = \sum_{m=1}^{N-1} C^{(N-1)}_{N,\,\rm{NLL,\,1}}(m).
\label{eq:nllWils2}
\eeq
One would expect to use $C^{(N-1)}_{N,\,\rm{NLL,\,1}}\mo^{(N-1)}_N$ along with
our real emission corrections (Eq.~\ref{eq:masterJet}) to correct a shower to resum at NLL the ratios of
all emission scales ({\it cf.} Eq.~\ref{eq:weight}).  The complication we face
for the calculation of $U^{(m-1)}_{\rm{NLL}}$ is that this correction to the
evolution kernel must, in principle, be carried out in the same scheme used to
distinguish the phase space regions for the jet-structure corrections, and hence
can depend on the choice for the $\Theta$ functions. In particular, we could have
non-trivial operator mixing on the edge where the cutoff makes a smooth
transition between operators with different numbers of jets, and we have not yet
performed the analysis that would determine whether this affects the resummation
at NLL$_{\rm exp}$ order. Furthermore, it is possible that power suppressed soft
effects will also have implications for the subleading evolution kernel, and may
make the nonabelian generalization of \eq{nllWils2} tricky. Our lack of an
appropriate NLL$_{\rm exp}$ evolution factor for the shower is due to these two
issues.

To setup the distinction between kinematic regions, we used Wilsonian
type $\Theta$ functions, but from the point of view of evolution
$\overline {\rm MS}$ would be simpler. Although it is only indirectly
relevant to our setup, it is nevertheless still interesting to
consider how the NLL evolution kernel would arise in $\overline {\rm
MS}$.  As we discuss below in \app{nloSplit}, when
integrated over phase space in dimensional regularization the
jet-structure corrections give the real emission portion of
$\nlosplit$, which is the $\mo(\alpha_s)$ correction to the
Altarelli-Parisi splitting kernel.  Combined with known SCET results
for single-emission at one-loop, we can recover all of the abelian
portion of $\nlosplit$.  Obtaining this expression is important
conceptually.  It validates our formal expansion in $\lambda$, showing
that corrections to $\mo(\lambda^2)$, along with a set of known
one-loop diagrams, capture contributions needed for collinear NLL$_{\rm exp}$
resummation. On the practical side, it provides a cross check on our
computations.

With $\nlosplit$ in hand, we can extend the argument of
\cite{Bauer:2006mk,Bauer:2006qp} that the Sudakov factor gives the LL part of
the the RG kernel $U^{(2,i,0)}(Q,\mu)$ (Eq.~\ref{eq:runningAsSudakov}) to the NLL
level, looking at $U^{(2,0,0)}(Q,\mu)$ for running of the operator $\moqq$.  Using
the Sudakov factor of \cite{Catani:2001cc} for quarks, we have:
\beq
\Delta_q(Q,\mu) \,=\, \exp \left\{ -\frac{C_F}{2\pi} \int^Q_\mu \frac{d\mu'}{\mu'} \alpha_s (\mu') 
\int_{\frac{\sqrt{\mu'}}{Q}}^{1-\frac{\sqrt{\mu'}}{Q}} dz \frac{1+z^2}{1-z} \right\},
\eeq
where we recognize $\losplit$, \eq{ap}.  Performing the $z$ integral and
expanding in the limit of large $Q$ gives:
\beq
\Delta_q(Q,\mu) \,\approx\, \exp \left\{ \frac{C_F}{\pi} \int^Q_\mu \frac{d\mu'}{\mu'} \alpha_s (\mu') 
\left[ \log \left( \frac{\mu'^2}{Q^2} \right) + \frac{3}{2} \right] \right\},
\label{eq:loSudakov}
\eeq
which is identical to $U^{(2,0,0)}(Q,\mu)$ at one-loop.  The term in the exponent
proportional to $\log (\mu'^2/Q^2 )$ sums the leading logs in the parton shower.
We also see that upon $\mu'$ integration, we get the double logarithm
characteristic of the soft-collinear divergence of collinear splitting.
Interpreting Eq.~(\ref{eq:loSudakov}) as an RG kernel, this $\log$ piece is
coming from the one-loop cusp anomalous dimension, $C_F$.  The factor of $3/2$
is the remaining part of the one-loop anomalous dimension, and it sums part of
the collinear NLL.\footnote{Since Eq.~(\ref{eq:loSudakov}) resums the NLL
  contributions expanded in the cross section ({\it cf.} Eq.~\ref{eq:resumCross}), 
  Ref.~\cite{Catani:2001cc} calls
  it the NLL Sudakov factor.}  In order to get the full collinear NLL$_{\rm exp}$ summation,
one also needs corrections corresponding to the two-loop cusp anomalous
dimension.  This is a known result in SCET for the operator $\chib_n
\chi_{\nb}$, which we can relate to $\nlosplit$, by adding the subleading
splitting function to the exponent of $\Delta_q(Q,\mu)$.  We wish to stress,
however, that the ultimate goal of improving parton showers through resummation
is to include all next-to-leading-logs.\footnote{At a practical level, while we
  see full collinear NLL$_{\rm exp}$ as coming from a straightforward extension of this
  work, pure soft NLL may only be possible at the leading orders in 
  $1/N_c$\cite{Bonciani:2003nt,Frixione:2007vw}.  }
In this paper, as mentioned previously we have not considered the effects of
soft NLL, nor those related to the two-loop running of $\alpha_s$, which will
affect collinear NLL.  Our formulas in
Apps.~\ref{app:QCD/SCET1}-\ref{app:scet2/scet3} include LL running for all
subleading operators. In App.~\ref{app:nloSplit} we discuss the relation of our
$1\to 3$ splitting amplitude with $P_{qq}^{(1)}$ in $\overline{\rm MS}$. The
collinear-NLL-improved Sudakov corresponding to this is
\beq
\Delta^{\rm NLL}_q(Q,\mu) \,=\, \exp \left\{ -\int^Q_\mu \frac{d\mu'}{\mu'}
\int_{\frac{\sqrt{\mu'}}{Q}}^{1-\frac{\sqrt{\mu'}}{Q}} dz\, 
\left[ \losplit(z,\alpha_s(\mu')) \,+\, \nlosplit(z,\alpha_s(\mu')) \right] \right\},
\eeq
where $\losplit$ given in Eq.~(\ref{eq:ap}) and $\nlosplit$
in~\cite{Curci:1980uw}.  Once again, we integrate in $z$, expanding in large $Q$
to get:
\begin{align}
\Delta^{\rm NLL}_q(Q,\mu) &= \exp \left\{ \int^Q_\mu \frac{d\mu'}{\mu'}
\left[ \frac{\alpha_s(\mu')}{\pi} C_F \left( \log \left( \frac{\mu'^2}{Q^2} \right) + \frac{3}{2} \right) \right. \right. 
\nonumber \\
 &+ \left. \left. \frac{\alpha^2_s(\mu')}{4\pi^2} C_F \left( C_g \left( \frac{67}{9} - \frac{\pi^2}{3} \right) 
-\frac{20}{9} C_F T_F\, n_F \right) \log \left( \frac{\mu'^2}{Q^2} \right) \right] \right\},
\end{align}
where the term $\propto \, \alpha^2_s$ reproduces the known result for
the two-loop cusp anomalous dimension.  While including this
$\overline {\rm MS}$ NLL effect for ``no-branching'' was already
possible, our result in \eq{masterJet} allows one to modify the
differential cross section for real emission to include the effects of
$\nlosplit$, as well.  Without including both, one does not have a
systematic improvement beyond LL. In~\cite{Catani:1990rr}, the authors
were able to get NLL$_{\rm exp}$ soft resummation by treating the
subleading real and virtual effects in semi-inclusive observables for
DIS and Drell-Yan.  A full implementation in our framework with more
exclusive observables must wait for computations that address the
missing NLL ingredients mentioned above.

\subsection{Squared Amplitudes and Interference Structures}
\label{subsec:ampSq}

As discussed previously, our series of matchings terminates with $\scetn$, where
each field has its own index direction.  Further Lagrangian emission from these
operators is physically meaningless, as the resolution scale is set $\sim
\mo({\rm GeV})$, below which we stop computing in perturbation theory and pass
to a hadronization routine.  Thus, we match everything to the single operator
$\mo^{(N-1)}_{N}(n_N,n'_1,\ldots,n'_{N-1})$ and all the information about the
shower at LO and NLO is encoded in the Wilson coefficients.  In this $\scetn$,
we square amplitudes and compute corrections to observables, as we detail in
\subsec{nlomap}.  As we saw in Secs.~\ref{subsec:qcdToSCET} and
\ref{subsec:scetToSCET}, for arbitrary $N$, we only needed one and two-gluon
computations to obtain leading corrections in $\lambda$ to the differential
cross section.  Using the LO replacement rule (Eq.~\ref{eq:loArrow}) will
account for the rest of the multiplicity.  Since the strongly-ordered emissions
it describes have trivial interference, we should expect that squaring our
results retains the simple picture we have for corrections at the amplitude
level.

\subsubsection{Interference for ${\rm LO^2}$ and for Jet-Structure Corrections}
\label{sec:InterJet}

It is a general statement about SCET fields with different $n$ index labels that
they have no overlap in Hilbert space.  As an example, we can take two different
operators, $\mo_{n_1}$ and $\mo_{n_2}$ where all the fields in $\mo_{n_1}$ and
$\mo_{n_2}$ are identical, except those labeled by $n_1$ and $n_2$ ({\it e.g.}
$\bar\chi_{n_1}$ versus $\bar\chi_{n_2}$).  For generality the field labeled by
$n_2$ may or may not be in the same equivalence class as $n_1$. We thus
have:\footnote{By RPI,
  $n_1$ and $n_2$ do not have to be exactly equal, but must concur up to an
  angle of $\mo(\lambda^i)$ in $\sceti$.}
\begin{align}
& \langle q_1,\, q_2,\, \ldots,\,q_m | \mo_{n_1}^\dagger | 0 \rangle \langle 0 | \mo_{n_2} | q_1,\, q_2,\, \ldots,\,q_m \rangle
\nonumber \\
& =\delta_{[n_1] ,\, [n_2]} \, \langle q_1,\, q_2,\, \ldots,\,q_m | \mo_{n_1}^\dagger | 0 \rangle \langle 0 | \mo_{n_2} | q_1,\, q_2,\, \ldots,\,q_m \rangle.
\label{eq:noOverlap}
\end{align}
This relation between $n_1$ and $n_2$ is simple when the difference is
encoded in the collinear fields in operators.  However, as discussed in
\subsec{qcdToSCET}, we also have to deal with situations where this
information ended up in Wilson coefficients when matching $\sceti$ to
$\scetipone$.  It is to guarantee a relation like \eq{noOverlap} that
our Wilson coefficients contain $\Theta$-functions ({\it cf.}
Eqs.~\ref{eq:thetaExpl} and \ref{eq:thetaEx}), which will cutoff the
overlap regions in phase space once we begin integrating.  The
amplitude squared is particularly simple in SCET$_N$, where we have only
the operator $\mo^{(N-1)}_{N}(n_N,n'_1,\ldots,n'_{N-1})$, and where
each particle is defined in a different collinear direction.

$\scetn$ (or $\sceti$, in general) easily distinguishes which
configurations are strongly-ordered by the structure of their Wilson
coefficients.  This means that we have no interference between
$C^{(N-1)}_{N,\, {\rm LO}} \mo^{(N-1)}_N$ and
$C^{(N-1)J}_{N,\,\mathrm{NLO}} \mo^{(N-1)}_N$ where $C^{(N-1)}_{N,\,
{\rm LO}}$ is the LO SCET$_N$ coefficient given in
Eq.~(\ref{eq:loOp}), and $C^{(N-1)J}_{N,\,\mathrm{NLO}}$ is in
Eq.~(\ref{eq:CNLOjet}).  Even though the $\mo$'s are the same, the
$\Theta$-functions in the $C$'s enforce different conditions, where
the former is strongly ordered, while the latter is not.  Thus, in the
analog of Eq.~(\ref{eq:noOverlap}), the Kronecker delta will give
zero.
\begin{figure}[t!]
\centering
\includegraphics{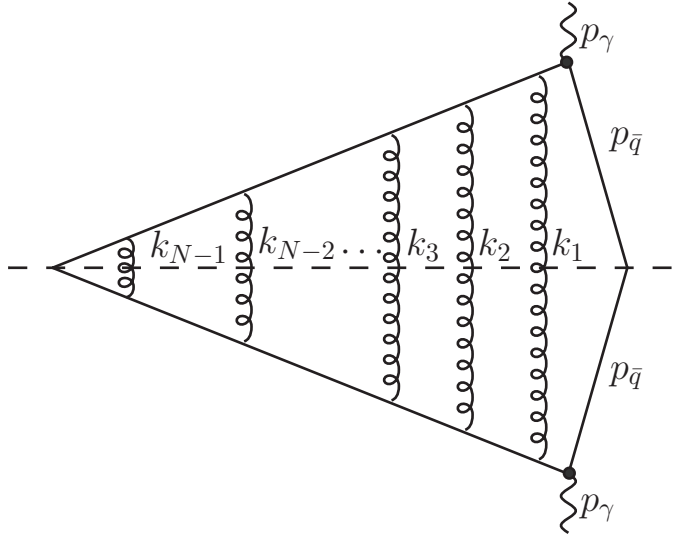}
\caption[Amplitude squared for the LO SCET$_N$ operator] {Amplitude squared for
  the LO SCET shower operator $C^{(N-1)}_{N,\, {\rm LO}} \mo^{(N-1)}_N$.  Rather
  than drawing the less intuitive squared amplitude in SCET$_N$, we
  illustrate the process here with a cut $\scetone$ Feynman diagram in order to
  emphasize the simple ladder structure.}
\label{fig:figureAmplSqrtLO1C}
\end{figure}

\begin{figure}[t!]
\centering
\includegraphics{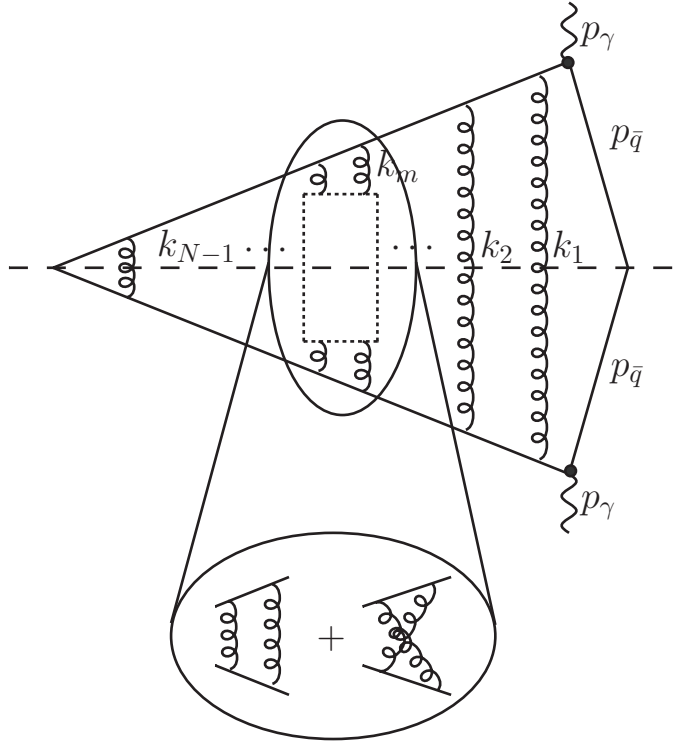}
\caption[Contribution to the amplitude squared of the jet-structure piece at
NLO] {Contribution to the amplitude squared of the jet-structure piece at NLO.
  We show a cut $\scetone$ Feynman diagram to emphasize that the square of the
  SCET$_N$ operator, $C^{(N-1)J}_{N,\,\mathrm{NLO}} \mo^{(N-1)}_{N}$, contains
  only a single deviation from the simple ladder structure appearing at LO
  in Fig.~\ref{fig:figureAmplSqrtLO1C}.}
\label{fig:figureAmplSqrtNLOC}
\end{figure}

We get a further simplification when we square the NLO$(\lambda)$ contributions.
Looking at $C^{(N-1)J}_{N,\,\mathrm{NLO}}$ in detail, we have:
\begin{align}
C^{(N-1) J}_{N,\,\rm{NLO}} =  \sum_{l=1}^{N-2} C^{(N-1) J}_{N,\,\rm{NLO}} (l) \, ,
\label{eq:masterJetPreEqn}
\end{align}
where
%
\begin{align}
C^{(N-1)J}_{N,\,\mathrm{NLO}}(l) =& 
\sum_{I=1}^3 \Big[ \Big( \prod_{k=1}^{l-1} U_{\rm{LL}}^{(k-1)}(\mu_{k-1} ,\,\mu_k)
 c^{\alpha_k}_{\rm{LO}}(n_{k-1})\Big)  U_{\rm{LL}}^{(l-1)}(\mu_{k-1} ,\,\mu_k) \otimes
 h^{\alpha, \beta}_I (n_{l+1}, n'_{l}, n'_{l+1})\nonumber \\
& \times \Big( \prod_{k=l+1}^{N-1} U_{\rm{LL}}^{(k-1)}(\mu_{k-1},\,\mu_k)
 c^{\alpha_k}_{\rm{LO}}(n_{k-1})\Big) \Big] \Gamma^{\mu} \,.
\label{eq:masterJetEqn}
\end{align}
%
In $C^{(N-1)J}_{N,\,\mathrm{NLO}}(l)$, we have made explicit that the
$l,(l+1)^{\rm th}$ gluons violate strong-ordering and come with the
factor $h^{\alpha \beta}$ of the subleading splitting rule,
Eq.~(\ref{eq:nloReplace}).  The sum in the last term over $I$ counts
the different types of NLO jet-structure terms given in
\eq{twoGluNLO}. The $c^{\alpha_k}_{\rm{LO}}$ are defined in
Eqs.~(\ref{eq:loOp}) and (\ref{eq:littlecDef}), and the $U$'s are
running factors given in
Eqs.~(\ref{eq:rgKernel})-(\ref{eq:runningAsSudakov}).  The complete
explanation of the symbols in Eq.~(\ref{eq:masterJetEqn}) can be found
in the discussion around Eq.~(\ref{eq:masterJet}).  The convolution
factor is explained below \eq{S1coeRun}.  Since different $l$
correspond to a violation of strong-ordering at different points in
the shower, each of the $C^{(i+1)J}_{i,\,\mathrm{NLO}}(l,l+1)$ encodes
a different $\Theta$ structure.  Therefore, there is no interference
for different values of $l$, and we have that the amplitude squared to
NLO$(\lambda)$ for jet-structure corrections (we call corrections of
$\mo(\lambda^2)$ at the amplitude squared level NLO$(\lambda)$) is just the sum
of squares of the individual operators:
\begin{align}
|A^{q(N-1)g\bar{q}\,J}|^2_{\rm to\, NLO} =|A^{q(N-1)g\bar{q}}|^2_{\rm{LO}} + |A^{q(N-1)g \bar{q}J}|^2_{\rm{NLO}} \, ,
\end{align}
where
\begin{align}
\label{eq:squareFact}
|A^{q(N-1)g\bar{q}}|^2_{\rm{LO}}&=\, |C^{(N-1)}_{N\, {\rm LO}}|^2 \, | \langle 0|\mo^{(N-1)}_N |q(N-1)g \bar{q} \rangle |^2 \, , \\
|A^{q(N-1)g\bar{q}\,J}|^2_{\rm{NLO}} &=\, \sum_{l=1}^{N-2} \, |C^{(N-1)J}_{N,\,\mathrm{NLO}}(l)|^2 \, | 
\langle 0| \mo^{(N-1)}_{N} |q(N-1)g \bar{q}\rangle|^2 , \nonumber
\end{align}
and $|q(N-1)g \bar{q} \rangle $ indicates the state with $N-1$ gluon
emission.  The simplification even extends inside each of the terms, since the
$j^{\rm th}$ gluon only gets contracted with itself.  Diagrammatically, this
means there are zero nearest-neighbor crossings in the $|{\rm LO}|^2$ diagram,
as we see in Fig.~\ref{fig:figureAmplSqrtLO1C}, and a maximum of {\it one} in the
$|{\rm NLO}|^2$, Fig.~\ref{fig:figureAmplSqrtNLOC}.
We thus only slightly modify the factorized emission formula, Eq.~(\ref{eq:splitFact}).  
Even for an arbitrary number of gluon emissions, we at most have to take into account a single 
defect that involves a full two-particle phase space.

We can see why terms that have non-trivial interference with more than
two gluons are suppressed by looking at the propagators in the
amplitude.  The amplitude for $i+1$ emissions has a factor $1/q^2_1
\times 1/q^2_2 \times \dots \times 1/q^2_i$.  The LO term comes form
the strong-ordered region where $q_1^2 \gg q_2^2 \gg \dots \gg q_i^2
$, Eq.~(\ref{eq:psLimit1}). The jet-structure NLO$(\lambda)$ is given when $q^2_j
\sim q^2_{j+1}$, which allows the two gluons $k_{j+1}$ and $k_{j+2}$
to share the same region of the phase space and therefore interfere.
To have an overlap of three or more gluons, we would need 
$q_j^2 \sim q_{j+1}^2 \sim \ldots \sim q_{j+k}^2$, which is clearly 
suppressed beyond NLO$(\lambda)$.

\subsubsection{Interference for Hard-Scattering Corrections}
\label{sec:InterHard}

The corrections to the differential cross section to $\mo(\lambda^2)$
involve squaring the subleading hard-scattering amplitudes as well.
Unlike the jet-structure case, these involve amplitude terms up to
NNLO$(\lambda)$.  As we argued above, they only modify the gluons closest to the
hard interaction.  Thus, we will not need to sum over many terms as we
do in Eq.~(\ref{eq:squareFact}).  In fact, for hard-scattering
corrections, we only need to worry about interfering $\sceti$
operators that arise from acting with the LO replacement rule
Eq.~(\ref{eq:loArrow}) on either $C^{(0)}_{1,\, {\rm LO}},\, C^{(1)}_{1,\, {\rm
NLO}}, \, {\rm and}\, C^{(1)}_{1,\, \mathcal{T}}$, given in
Eqs.~(\ref{eq:loDiquark}) and (\ref{eq:singleGluonWilsonA}), or
$C^{(2)H}_{2,\,{\rm NNLO}}$ given in \eq{S2SumE}.  Since the $3^{\rm
rd}$ through $i^{\rm th}$ gluons arise from the LO rule for all three
coefficients, they proceed as in the $|{\rm LO}|^2$ case.  The
interference to look at in detail is that of the first two gluons.  
In $\scetn$, we have:
\begin{align}
\label{eq:hsInt}
|A^{q(N-1)g\bar{q}\,H}|^2_{\rm{NLO}} \,=&\, \big(|C^{(N-1) \dagger}_{N,\,\rm{LO}}  C^{(N-1)H}_{N,\,\rm{NLO}} + C^{(N-1)H\dagger}_{N,\,\rm{NLO}}C^{(N-1)}_{N,\,\rm{LO}} | 
\,+\, |C^{(N-1)H}_{N,\,\rm{NLO}}| ^2\\
&+|C^{(N-1) \dagger}_{N,\,\rm{LO}}  C^{(N-1)H}_{N,\,\rm{NNLO}} + C^{(N-1)H\dagger}_{N,\,\rm{NNLO}}C^{(N-1)}_{N,\,\rm{LO}} | 
 \big)| \langle 0|  \mo^{(N-1)}_N
|q(N-1)g \bar{q}\rangle |^2 \nonumber 
\end{align}
The Wilson coefficients are found in Eqs.~(\ref{eq:loMaster}),
(\ref{eq:hardMasterNLO}), and (\ref{eq:hardMasterNNLO}), respectively.  
Nontrivial interference in \eq{hsInt} occurs between the first two-gluon
emissions.

The interference between LO and NLO$(\lambda)$ simplifies in many cases of interest. 
For example for one-gluon emission,
\beq
|A^{qg\bar{q}}_{\rm LO/NLO}|^2_{\mu\, \nu} \,=\, \frac{4 \bar{q}_1 \bar{p}_{\bar{q}} }{q_0^2}  k_{1\perp \, \nu}(n_{\mu} - \bar{n}_{\mu}).
\label{eq:lonloInt}
\eeq
If we can cleanly separate the initial and final states ({\it e.g.}  $e^+e^-
\rightarrow$ jets), then by a classic proof involving the Ward identity
(reproduced, for example, in \cite{Peskin:1995ev}), once we have integrated over
final state vector quantities (we can keep scalars such as $z_i$ unintegrated),
the resulting differential observable depends on $g^{\mu\nu} |A_{\rm
  LO/NLO}|^2_{\mu\, \nu}$, which for Eq.~(\ref{eq:lonloInt}) is zero.  This is
quite straightforward for leptonic initial states, and one may be able to extend
it to certain hadronic ones as well.  

One can account for these corrections by modifying the hard-scale matrix element
and then running a parton shower modified to include the different no-branching
probabilities for different phase space configurations of the same particle
content.  In the next section we discuss using a reweighting to implement these
corrections.

\subsection{Correction Summary at Subleading Order }
\label{subsec:nlomap}

In general, our corrections avoid double counting issues, because all
contributions, whether LO, hard-scattering, or jet-structure corrections are
kept separately with distinct $\Theta$ structures. Given the $\scetn$ amplitude
for $N+1$ final state particles with corrections implemented both for the
branching and for the no-branching, one can consider reweighting a LL shower in
order to implement our results.  For correcting the abelian emissions off a
single quark line, this weight factor would take the following form:
\begin{align}
w =& \frac{ \left[ J(N-1,\,0) \,+\, H (N-1,0)\right] }{  A(N-1,\,0) } \,,
\label{eq:weight}
\end{align}
where $A(N-1,0)$ is the LL amplitude squared for $N-1$ emissions from the quark
line, $J(N-1,0)$ includes the LL result along with power corrections and
subleading resummation associated with jet-structure corrections, and $H$
contains hard scattering corrections.  With our LO$(\lambda)$ result,
\begin{align} \label{eq:ANlo}
A(N-1,0) &=  |C^{(N-1)}_{N, {\rm LO}}|^2 | \langle 0|\mo^{(N-1)}_N |q(N-1)g
\bar{q} \rangle |^2 \,,  
\end{align}
but in general $A$ could be whatever amplitude squared a particular shower
algorithm has for a given configuration.  Eq.~(\ref{eq:weight}) then
reweights that particular shower to our \NLOl corrected result. An example of
shower Monte Carlo with an analytic expression for $A(N-1,0)$ is
GenEvA~\cite{Bauer:2008qh}. For a leading log shower without an explicit formula
for $A(N-1,0)$ one can use \eq{ANlo} with the understanding that it is likely a
good approximation to the shower output.  For the terms in the numerator of
\eq{weight} we have:
\begin{align} \label{eq:weightDefs} 
J(N-1,0) &= 
|C^{(N-1)J}_{N,\,\mathrm{NLL,\,1}}|^2 | \langle 0| \mo^{(N-1)}_{N} 
  |q(N-1)g \bar{q}\rangle|^2  +
|C^{(N-1)J}_{N,\,\mathrm{NLO}}|^2 | \langle 0| \mo^{(N-1)}_{N} |q(N-1)g
\bar{q}\rangle|^2  \nonumber \,, \\
H(N-1,0) &= 
|C^{(N-1)H}_{N,\,\mathrm{NLO}}|^2 | \langle 0| \mo^{(N-1)}_{N} |q(N-1)g
\bar{q}\rangle|^2 \nonumber \\ 
& \ \ +\, \Big(
 C^{(N-1)}_{N, {\rm LO}}\, C^{\dagger (N-1)H}_{N,\,\mathrm{NNLO}}   
\langle 0|\mo^{(N-1)}_N |q(N-1)g \bar{q} \rangle \langle q(N-1)g \bar{q} |
\mo^{\dagger (N-1)}_N | 0 \rangle + {\rm h.c.}\Big) ,
\end{align}
where we give formulas for $C^{(N-1)}_{N, {\rm LO}}$ and $\mo^{(N-1)}_N$ in
\eq{loOp}, $C^{(N-1)J}_{N,\,\mathrm{NLL,\,1}}$ is discussed near
\eqs{nllWils}{nllWils2}, $C^{(N-1)J}_{N,\,\mathrm{NLO}}$ is given in
\eqs{masterJetPreEqn}{masterJetEqn}, $C^{(N-1)H}_{N,\,\mathrm{NLO}}$ is given in
\eq{hardMasterNLO}, and $C^{(N-1)H}_{N,\,\mathrm{NNLO}}$ is given in
\eq{hardMasterNNLO}.  Our operators, $\mo^{(N-1)}_N$, describe a process with
$N-1$ emissions off the quark line.  The $A(N-1,0)$ amplitude squared in
\eq{ANlo} is contained within the first term in $J(N-1,0)$. As discussed in
\subsec{opRunning}, while we have worked out the real emission terms
($C^{(N-1)J}_{N,\,\mathrm{NLO}}$) completely, we have yet to determine the
subleading RG kernels needed for $C^{(N-1)J}_{N,\,\mathrm{NLL,\,1}}$.

We introduce the $A,\, J,\, H$ notation to describe more general abelian
processes.  $A(j,\,k)$ gives the amplitude squared necessary for the LL shower
of $j$ gluons collinear to the quark, and $k$ collinear to the antiquark.  The
correction, $J(j,\,k)$, contains the virtual and real corrections necessary for
NLL$_{\rm exp}$ resummation of collinear logs.  Since it contains an implicit
sum over insertions of a single defect, which can occur anywhere in the shower,
it depends on the total number of collinear emissions.  Including the
hard-scattering contributions to NLO$(\lambda)$ only requires modification of the first
two emissions, after which one simply uses the LO replacement rule,
\eq{loArrow}.  In the general case we denote it by $H'$, which differs from the
above by including corrections to antiquark emissions as well.  These are easily
obtained by charge conjugation.  Thus, an abelian two-jet process with $j+k$
gluons gets the following reweighting factor:
\beq \label{eq:wjk}
w(j,\,k) = \left[ J(j,\,k) \,+\, H'(j,\,k) \right] / A(j,\,k).
\eeq
These weight factors are positive definite. All contributing terms are squares
of amplitudes, except for ${\rm LO}(\lambda)\times {\rm NNLO}(\lambda)$
in $H(N-1,0)$.  This contributes in the same region of phase space as the
LO$(\lambda)$ amplitude squared, and the sum of these terms is positive. In the
full nonabelian case, with the presence of gluon splittings, one must sum over
possible shower histories in writing down the analog of \eq{wjk}.  Algorithms
for handling this complication can be found in
\cite{Bauer:2008qh,Mangano:2002ea}.

\begin{table}[t!]
\begin{tabular}{|c|c|c|c|}
  \hline
 Category &
  Shower Concepts        & Quantity in $\sceti$       & Found In: \\ 
\hline \hline
Hard Scattering
 & Hard matrix elements     & Wilson coeff. of $\bar\chi_{n_1} {\cal B}_{n'_1}^\perp
 \chi_{\bar n}$    & Eq.~(\ref{eq:coeS2}) \\
 &  with more partons          &   in $\scetone$   & \\
   \cline{2-4}
 & Power correction to initial branching  & Wilson coeff. of  $\bar\chi_{n_0} 
{\cal B}_{n_0}^\perp
 \chi_{\nb}$ & Eq.~(\ref{eq:coeS2}) \\ 
 &  within the leading jet & and $\bar\chi_{n_0} [{\cal P}_\perp {\cal B}_{n_0}^\perp] \chi_{\bar n}$ & \\  \cline{2-4}
 &  ${\cal O}(\alpha_s)$ hard virtual  &  One-loop matching for & See  \cite{Bauer:2006mk,Bauer:2006qp}
    \\ 
 &  correction  &  $\bar\chi_{n_1} \chi_{\bar n}$ & \\ 
\hline
Jet Structure 
 & $1\to 3$ Splitting functions   & Double gluon real     &  Eq.~(\ref{eq:twoGluNLO}) \\
 &                        & emission in $\scetone$           & \\ \cline{2-4}
& Combining $1\to 2$ splittings with  & Compute weights from SCET  & \eq{wjk} 
  \\ 
 & the various $1\to 3$ splittings & squared amplitudes&  \\ \cline{2-4}
  &  ${\cal O}(\alpha_s)$ virtual correction      & One-loop correction      &  Left for  \\
 & for LO $1\to 2$ splitting          & to $1\to 2$ replacement rule             &  future work   \\
 \cline{2-4}
  \hline 
  No Branching & NLL Sudakov factor for      &  NLL anomalous dimension  & See
  Sec.~\ref{subsec:opRunning},    \\
Probabilities & leading branching  &  for leading operators & Left for \\
 &    &    &   future work \\ 
 \cline{2-4}
    & LL Sudakovs for        &  LL anomalous dimensions for  &Eqs.~(\ref{eq:runningAsSudakov},\ref{eq:kernelEq})  \\
             &  subleading branching  & subleading operators &   \\ \hline
 Soft Emission & Subleading corrections   & Include effects of soft    &        Left for                \\
 &  from soft gluons           &  emission from subleading           &  future work     \\
 &                         &  SCET soft Lagrangians          &
 \\
 \hline
\end{tabular}
  \caption{Mapping between concepts in an NLO parton shower algorithm and
    computations in $\sceti$. For exclusive cross sections these ingredients
    would together yield results accurate to 
    NLO in the power expansion ($\lambda$), and with
    corresponding NLL resummation. }
\label{tbl:map2}
\end{table}
  
  In Table~\ref{tbl:map2} we list concepts that are addressed by our shower
  framework at subleading order, and associate these concepts with corresponding
  calculations in $\sceti$.  This table provides a summary of our results which
  appear in the weights given in \eq{wjk}, as well as pointers for future
  calculations. Since it is easier, in the table we use the language of
  $\scetone$ and $\scettwo$ to discuss the corrections, rather than referring to
  terms in the final $\scetn$.  In $\scetn$, the features of the $\scetone$
  operators that avoid double counting and allow the various contributions to be
  distinguished are encoded by $\Theta$ functions in the Wilson coefficients,
  and the operator language makes the discussion easier. For the total
  differential cross section, we found at NLO$(\lambda)$ two kinds of power
  corrections.  This includes a set of matrix-element corrections called
  hard-scattering corrections (\subsec{qcdToSCET}), and a set of contributions
  that improve double real emissions that we called jet-structure corrections
  (\subsec{scetToSCET}).

In the the hard-scattering category, we have overall three different kinds of corrections.
The first is due the the SCET$_1$ operator $\bar\chi_{n_1} {\cal B}_{n'_1}
\chi_{\bar n}$ that gives the SCET$_2$ coefficient $C^{(1)H,b}_{2,\,\rm{NLO}}$
in Eq.~(\ref{eq:coeS2}). This is an improvement of the hard matrix element that
takes into account the emission of an extra parton at the hard scale.  The
second is due to the SCET$_1$ operators $\bar\chi_{n_0} {\cal B}_{n_0}
\chi_{\nb}$ and $\bar\chi_{n_0} [{\cal P}_\perp {\cal B}_{n_0}] \chi_{\bar n}$
that give the SCET$_2$ coefficients $C^{(1)H,a}_{2,\,\rm{NLO}} $ and $
C^{(1)H}_{2,\,\rm{NNLO}} $ in Eq.~(\ref{eq:coeS2}). This correction also
accounts for more partons, but it describes a situation where they are initially
emitted close to the collinear quark. Therefore,
they are corrections which improve the description of the first branching within
the leading jet. It is important to note that because these two types of hard
corrections occur in different regions of phase space they have different
renormalization group evolution, and thus different Sudakov no-branching
factors. The required LL Sudakov factors were determined in our analysis.  For a
full \NLOa treatment we also need a third type of hard scattering correction,
the one-loop virtual corrections to the leading shower operator.  For the
required operator, $\bar{\chi}_{n_0} \chi_n$ these types of corrections were
discussed in Refs.~\cite{Bauer:2006mk,Bauer:2006qp}.

For the jet-structure corrections, there are several ingredients to consider. We
derived a replacement rule for two emissions $1 \rightarrow 3$,
Eq.~(\ref{eq:nloReplace}), that involved three different types of terms.  This
correction takes into account emissions in a region of the phase space that is
not strongly-ordered and automatically avoids double counting from multiple
$1\to 2$ emissions.\footnote{The method by which we avoid double counting for
  two gluon emission should be obvious, coming directly from our implementation
  of the $\Theta$ functions. Since $\Theta+\tilde\Theta=1$, the double $1\to 2$
  and $1\to 3$ together cover all of phase space without double counting. For
  three emissions we have either i) three $1\to 2$ emissions, ii) a $1\to 3$
  followed by a $1\to 2$ emission, or iii) a $1\to 2$ followed by a $1\to 3$
  emission.  Here there is an apparent combinatoric issue, as ii) and iii) both
  provide corrections for the middle gluon in i).  However they do so in
  nonoverlapping regions of phase space. The same is true for more than three
  emissions. We thank J.~Thaler for asking this question.}  In addition at
\NLOa/NLL we require the ${\cal O}(\alpha_s)$ virtual correction to the LO
splitting rule.  This would be derived from a one loop matching computation that
should be straightforward, but was not considered here.

We also discussed how no-branching Sudakov factors are associated with the
operator RG kernels, and by extension their anomalous dimensions (Section
\ref{subsec:opRunning}). To NLL, we need the NLL Sudakov factor for leading
branching and the LL Sudakovs for subleading branching. These are associated to
the full one-loop and two-loop cusp anomalous dimensions for the leading
operators, and one-loop cusp for the subleading ones.  At LL, we have determined
all the Sudakovs for subleading branching (Eqs.~\ref{eq:runningAsSudakov} and
\ref{eq:kernelEq}). We have not yet calculated the NLL Sudakov for leading
branching in the scheme with $\Theta$-functions that is needed for our setup, as
described in \subsec{opRunning}.

The last item in the table is the treatment of soft radiation at NLO. This can
be achieved by considering time-ordered products for the matching of QCD to
$\scetone$ and $\sceti$ to $\scetipone$ that involve soft gluons and subleading
soft Lagrangians that are known in
SCET~\cite{Bauer:2003mga,Beneke:2002ph,Beneke:2002ni} up to ${\cal
  O}(\lambda^2)$. For the terms involving collinear quarks they read
\begin{align} \label{eq:Lxxnew}
{\cal L}_{\xi\xi}^{(1)} 
  &=  \big(\bar \xi_n  W\big)\,  i\Dslash^\perp_{us} \frac{1}{\bnP} \big(W^\dagger 
  i \Dslash^\perp_n  \frac{\bnslash}{2} \xi_n \big)
  + \big( \bar \xi_n  i \Dslash^{\perp}_n W\big) \frac{1}{\bnP} i\Dslash^\perp_{us}
   \big( W^\dagger \frac{\bnslash}{2} \xi_n \big)\,, \\
  {\cal L}_{\xi\xi}^{(2)} 
  &=  \big( \bar \xi_n W\big)
 i\Dslash_{us}^\perp \frac{1}{\bnP} i\Dslash_{us}^\perp \frac{\bnslash}{2}
  \big( W^\dagger \xi_n\big) +
 \big( \bar \xi_n i\Dslash^\perp_n W\big) \frac{1}{\bnP^2} i\nb\!\cdot\! D_{us} 
  \frac{\bnslash}{2}\big( W^\dagger i\Dslash^\perp_n \xi_n\big) \,, 
\nonumber\\
{\cal L}^{(1)}_{\xi q} 
  &=   \bar\xi_n \: \frac{1}{i\nb\!\cdot\! D_n}\: 
 ig \Bslash_\perp^{\, n} W  q_{us} \mbox{ + h.c.}\,,
\qquad
{\cal L}^{(2b)}_{\xi q} 
  = \bar\xi_n \frac{\bnslash}{2} 
  i\Dslash_\perp^{\,c} \frac{1}{(i\nb\!\cdot\! D_n)^2}\:   ig\, \Bslash^{\,n}_\perp\, W 
  \,q_{us}   \mbox{ + h.c.}\,,
\nonumber \\
  {\cal L}^{(2a )}_{\xi q} 
  &= \bar\xi_n \frac{\bnslash}{2}
     \frac{1}{i\nb\!\cdot\! D_n}\: 
    [i\nb\cdot D_n, in\cdot D_n + gn\cdot A_{us} ]
    W \, q_{us} \mbox{ + h.c.} \,, 
   \nonumber
\end{align}
while the analogous pure glue Lagrangians can be found in
Ref.~\cite{Bauer:2003mga}.  Here the expressions are prior to the soft field
redefinition, and $ig \Bslash_\perp^{\,n} =[i\nb\cdot
D_n,i\Dslash_\perp^{\,n}]$.  One must then work out the effect that these NLO
soft amplitudes have on interference.  The associated soft calculations and
investigations have also been left for future work.

We also briefly comment on how the corrections in Table~\ref{tbl:map2} relate to
those already implemented in parton shower codes in the literature. In
most cases, the goal of these codes differed from the power suppressed
corrections considered here. This makes a strict association impossible, but
there is still a general correspondence that can be made.
CKKW~\cite{Catani:2001cc} is a LO($\alpha_s$)/LL procedure whose goal is to
merge matrix elements involving multiple partons with a parton shower in a
manner that avoids double counting.  In our language, this corresponds to the
real emission hard-scattering corrections in the first row of
Table~\ref{tbl:map2}.  The $\bar\chi_{n_0} \chi_{\bar n}$ and $\bar\chi_{n_1}
{\cal B}_{n_1'}^\perp \chi_{\bar n}$ operators describe processes with different
numbers of initial well-separated jets. In CKKW, a parameter $y_{\rm cut}$ is
used to separate the extra emission in the matrix element from emissions in the
shower. In our analysis, the contributions from showering $\bar\chi_{n_0}
\chi_{\bar n}$ does not interfere with the direct contribution from
$\bar\chi_{n_1} {\cal B}_{n_1'}^\perp \chi_{\bar n}$, and this is encoded by
$\Theta$ functions in the Wilson coefficient of $\scetn$. CKKW carries out this
procedure for several matrix element emissions, while we have only considered
one.

In MC$@$NLO~\cite{Frixione:2002ik} and POWHEG~\cite{Nason:2004rx}, virtual and
real matrix element corrections at \NLOa are incorporated into the shower, with
the goal of ensuring that it reproduces an associated cross section completely
at \NLOa. The implementation includes careful handling of the cancellation of
real and virtual IR divergences. Our goal was to implement corrections at
NLO($\lambda$) and we discussed NLL, but for all emissions from the shower
rather than just the first jet needed for the \NLOa cross section. At NLL, we
would have only terms up to ${\cal O}(\alpha_s\log)$ in the total cross section,
and hence this does not encode the entire NLO($\alpha_s$) result.  In our
language, the corrections that contribute to the \NLOa cross section correspond
to the hard scattering corrections in the first through third rows of
Table~\ref{tbl:map2}. In order to compute the \NLOa cross section it is not
necessary to distinguish between the terms in the first and second rows of the
table, and these terms are indeed considered simultaneously in MC$@$NLO and
POWHEG. The full NLO($\alpha_s$) virtual result are obtained in our language by
including the items mentioned in the 3$^{\rm rd}$ and 9$^{\rm th}$ rows of
Table~\ref{tbl:map2}.

The work of KRKMC~\cite{Jadach:2009gm,Skrzypek:2009jk, Jadach:2010ew}, on
the other side, aims to improve the shower algorithm taking into account an
exclusive version of the Altarelli-Parisi splitting function at \NLOa,
$P_{qq}^{(1)}$.  In our language, this corresponds to jet-structure
corrections and we show in Appendix \ref{app:nloSplit} how our replacement rule
in Eq.~(\ref{eq:nloReplace}) is also related to $P_{qq}^{(1)}$. Hence our $1\to
3$ emission corresponds to an exclusive version of $P_{qq}^{(1)}$, though in a
different scheme.  Part of the corrections in $P_{qq}^{(1)}$ involve order
$\alpha_s$ corrections to the $1\rightarrow 2$ splitting function, which are
taken into account by ${\cal O}(\alpha_s)$ virtual $1\to 2$ matching corrections
in our framework (6$^{\rm th}$ row of Table~\ref{tbl:map2}). In fact, in \subsec{ampSq},
we saw that $\sceti$ also leads one to view corrections to the shower as a
``defect'' insertion just as KRKMC.  In addition to these splitting corrections,
in our framework the amplitude also involves no-branching probabilities given by
evolution kernels that appear in the weight factors, which do not appear in the
KRKMC weights. Keeping track of the evolution also determines the appropriate
scale for evaluating $\alpha_s$.

\section{Conclusion} 
\label{sec:conclusion}

In this paper we developed a framework based on a tower of independent but
related EFTs, the $\sceti$, to study corrections to the parton shower.  The work
of \cite{Bauer:2006mk,Bauer:2006qp} showed how to formulate the LL parton shower
in terms of SCET, and how virtual corrections are straightforward to incorporate
by one-loop matching. Our $\sceti$ framework extends these ideas in a manner
that makes it easy to deal with: double counting, the issue of
disentangling coordinate choices from kinematic power corrections, and the
construction of a complete set of operators for corrections at a desired order.
The interference structures, and hence the leading corrections that give spin
and color correlations, also appear in a straightforward manner in
the $\sceti$ setup.

The $\sceti$ are iteratively used to integrate out the characteristic scale,
$Q\lambda^{i}$ for increasing $i$.  This approach allows us to perform a
systematic expansion which can correct both the hard-scale process that produces
partons to setup initial conditions for the shower algorithm and
the iterative shower itself. We described the parton shower through a set of
operators $\mo^{(j)}_i$ in ${\rm SCET}_i$, and used standard matching procedures
to make the transition from SCET$_i$ to SCET$_{i+1}$, where more partons become
apparent.  Performing the matching relied crucially on the RPI symmetry of SCET,
and we extended the usual infinitesimal version to carry out the finite
rotations that we needed.
At LO, a simple operator replacement rule generates the LL shower,
$\bar\chi_{n_0} \rightarrow c^\alpha_{\rm{LO}} \bar\chi_{n_1} g
\mathcal{B}_{n'_1\perp}^\alpha$, where $c_{\rm{LO}}$ is related to the standard
LO splitting-function. Also, angular ordering and coherent branching for LO soft
emissions emerge naturally in the $\sceti$ framework. A summary of ingredients
required for the shower with power corrections at \NLOl are given in
Table~\ref{tbl:map2}, including both calculations carried out here, as well as
those left for future work.  The main results of our paper are:
\begin{enumerate}
  
\item{At \NLOl we found two kinds of branching corrections: hard-scattering 
    and jet-structure. The hard-scattering corrections depend on the
    hard process and appear near the top of the shower tree. They came from
    matching QCD to SCET$_1$ at higher order. Since they only occur at the top
    of the shower, one can treat these as a modified form of matrix-element
    corrections.  A subset of these corrections correspond to the usual
    implementation of fixed-order matrix elements, while the remaining ones give
    power corrections to the initial branching in the LL shower. These two types
    require different Sudakov factors. This effect is apparent for the kinematic
    power corrections, but is beyond \NLOa for the fixed order counting.}
  
\item{The jet-structure corrections are independent from what happens at the
    hard scale, hence they are universal for any process we want to study. They
    come from matching SCET$_i$ to SCET$_{i+1}$ at higher order for any $i$.
    They can appear anywhere in the shower tree and they take into account
    emissions in regions of the phase space that are not strongly-ordered.  For
    these corrections we found that the \NLOl operators are related to the LO
    operator via a replacement rule for two emissions: $\bar\chi_{n_0}
    \rightarrow h_{I}^{\alpha \beta} \bar\chi_{n_2}g
    \mathcal{B}_{n'_1\perp}^\alpha g \mathcal{B}_{n'_2\perp}^\beta$. This \NLOl
    rule automatically avoids double counting with the iteration of two LO
    operator replacements. }
  
\item{The SCET$_i$ picture allowed us to easily take into account interference
    for the \NLOl power corrections.  Once we reach the final SCET$_N$ theory, all
    the fields are labeled in a different collinear directions.  Because in SCET
    we can only contract collinear fields that share the same collinear
    direction, in SCET$_N$ calculating the amplitude squared becomes very easy.
    Kinematic information that is encoded by the shower history from passing
    through earlier $\sceti$'s is encoded by $\Theta$ functions in the final
    $\scetn$ Wilson coefficients.  We demonstrated that when emitting an
    arbitrary number of partons, the non-trivial part of the amplitude squared
    involves at most four fields.}

\end{enumerate}
A comparison of how these $\sceti$ results relate to earlier parton shower
literature that goes beyond LL is given in Sec.~\ref{subsec:nlomap}.

The framework developed here allows for systematic improvement to arbitrary
orders in the kinematic expansion.  There are still several important steps to
take, though, before this picture can lead to a practical implementation,
including additional computations that we outlined in \subsec{nlomap}.  We list
here three topics which are natural next steps, and which we believe should be
straightforward to approach:
\begin{enumerate}
\item{ This work has only considered $q \rightarrow qg$ splittings and an
    abelian theory.  One should include the full nonabelian results and compute
    the coefficients required for gluon splitting as well.  This is required to
    properly treat color correlation corrections in a manner determined by the
    \NLOl interference pattern.  For collinear particles we expect that one can
    include the dominant part of these effects by considering nearest-neighbor
    interference since this arises from the kinematic expansion, and thus leaves
    the rest of the shower as before.  }
\item{ Only a subset of the terms required for a full NLL$_{\rm exp}$
    resummation were considered here.  We determined the LL$_{\rm
    exp}$ evolution for subleading operators, but did not carry out
    the computation of the NLL$_{\rm exp}$ evolution of the leading
    operator in a scheme that is consistent with our power corrections
    (we only considered it in $\overline {\rm MS}$).  In order for a
    consistent treatment as a probabilistic process, the real emission
    probabilities and Sudakov no-branching corrections must go hand in
    hand.  Furthermore, once these evolution factors are determined, the
    reweighting discussed in \subsec{nlomap} must be tested
    in an actual shower Monte Carlo.  }
\item{ Since soft modes in SCET can communicate between different collinear
    jets, they carry the ability to spoil their factorization.  Fortunately,
    this does not happen for their LO interactions, which yield angular ordering
    and coherent branching of soft gluons in $\sceti$.  It is open question as
    to what extent NLO soft couplings can be factorized in the shower tree and
    the necessary SCET computations were discussed but not carried out here. The
    treatment of soft NLO interactions in SCET in other contexts has always led
    to factorized structures, so we remain optimistic that such effects will be
    tractable for the shower.  }
\end{enumerate}
Future investigation of these items is well warranted.

\begin{acknowledgments}
  This work was supported in part by the NSF grant, nsf-phy/0401513, by the
  Office of Nuclear Physics of the U.S.\ Department of Energy under the Contract
  DE-FG02-94ER40818, and by a Friedrich Wilhelm Bessel award from the Alexander
  von Humboldt foundation.  I.S. thanks the Werner-Heisenberg
  Max-Planck Institute for Physics for hospitality while this work was
  completed.  The authors would like to thank Christian Bauer, Kirill Melnikov,
  Frank Tackmann, and Jesse Thaler for useful discussions.
\end{acknowledgments}

\appendix

\section{More SCET basics}
\label{app:scet}

Soft-Collinear Effective Theory describes the interactions of collinear and soft
quarks and gluons \cite{Bauer:2000ew, Bauer:2000yr, Bauer:2001ct, Bauer:2001yt}.
As we mentioned in Sec.~\ref{subsec:introSCET}, to define the collinearity of a
particle, the momentum is decomposed along two light-cone vectors, $n$ and
$\bar{n}$, with $n^2=0$, $\bar{n}^2=0$ and $n\cdot \bar{n}=2$
\begin{align}
\label{eq:decV}
p^{\mu}&= n\!\cdot\!p \frac{\bar{n}^\mu}{2}+\bar{p} \frac{n^\mu}{2}+p^\mu_\perp\, ,
\end{align} 
where $\bar{p}=\bar{n} \cdot p$. A particle is collinear to the direction $n$ if
its momentum scales as:
\begin{equation}
(n \cdot p, \,\bar{p},\, p_\perp)\sim (\lambda^2, \,1, \,\lambda)\,Q\, ,
\label{eq:powercountingSCETA}
\end{equation}
where $Q$ is the hard scale of the process, and $\lambda \ll 1$.  A particle is
soft if:
\begin{equation}
(n \cdot p, \,\bar{p},\, p_\perp)\sim (\lambda^2, \, \lambda^2, \,\lambda^2)\,Q\, .
\label{eq:powercountingSCETsoftA}
\end{equation}
We obtain SCET from QCD by expanding in powers of $\lambda$ and
integrating out modes harder than $\sim Q^2\lambda^2$.  Both
Eqs.~(\ref{eq:powercountingSCETA}) and
(\ref{eq:powercountingSCETsoftA}) imply that $p^2 = \bar{p}\, (n \cdot
p) + p_\perp ^2 \lesssim Q^2 \lambda^2$.

In addition to the expansion, we also want to divide the quark and gluon fields
into separate soft and collinear modes.  For the collinear case, the fields are
indexed by $n$, and two collinear sectors are distinct if $n_i \cdot n_j\gg
\lambda^2$. In addition, we introduce a momentum-space lattice for the 
$\mo(\lambda^0)$ and $\mo(\lambda)$ momenta in order to facilitate carrying out the
multipole expansion with respect to the $\mo(\lambda^2)$ momenta.  To divide the
QCD fields in this way, we split the momentum of a collinear particle into a
``large'' part $\tilde{p}^\mu$ and a residual one $k^\mu \sim \lambda^2$
\begin{align}
p^\mu=\tilde{p}^\mu+k^\mu \, ,\quad \mathrm{where} \;\,\,\, \tilde{p}^\mu \equiv n\!\cdot\! p \frac{n^\mu}{2} + p_\perp^\mu \, .
\end{align}
We can pull out the large momenta $\tilde{p}$ from the fermion field by the
phase redefinition
\begin{align}
\psi(x) = \sum_{\tilde{p},\, n} e^{-i \tilde{p} \cdot x}\, \psi_{n, \tilde{p}}\, .
\end{align}
For a collinear particle along $n$, $\partial^\mu \psi_{n , \tilde{p}} (x) \sim
\lambda^2$.  The four component field, $\psi_{n , \tilde{p}}$, has two large
components, $\xi_{n,\tilde{p}}$, and two small components $\xi_{\nb,\tilde{p}}$,
that can be separated using the following projectors:
\begin{align}
\psi_{n , \tilde{p}} =  \frac{\ns \nbs}{4}   \psi_{n , \tilde{p}} 
+ \frac{\nbs \ns}{4}  \psi_{n , \tilde{p}}
\equiv \xi_{n,\tilde{p}} + \xi_{\nb,\tilde{p}}.
\end{align}
These satisfy the relations,
\begin{align}
\frac{\ns \nbs}{4} \, \xi_{n,\tilde{p}}  & =\xi_{n,\tilde{p}}  \, ,  \quad \ns  \, \xi_{n,\tilde{p}} =0 \, ,   \nonumber \\
\frac{\nbs \ns}{4} \,  \xi_{\nb,\tilde{p}}& =\xi_{\nb,\tilde{p}} \, ,\quad \nbs \,  \xi_{\nb,\tilde{p}} =0\, .
\end{align}
Similarly, we can define a collinear gluon field, $A^\mu_{n,\tilde{q}}(x)$.
Pictorially, we can think of $\xi_{n, \tilde{p}}(x)$ and $A^\mu_{n,\tilde{q}}(x)$
as fields that create a particle whose three-momentum lies inside a cone with
opening angle $\sim \lambda$ about the three-direction $\vec{n}$.
$\mathcal{P}_n^\mu$ is the momentum operator that picks up the large components
of the momentum, $\mathcal{P}^\mu_n\, \xi_{n,\tilde{p}} (x)= \tilde{p}^\mu\,
\xi_{n,\tilde{p}} (x)$.  Collinear fields always appear with a sum over
$\tilde{p}$, and both label and residual momenta are separately
conserved. Therefore it is often useful to abbreviate the notation as
\begin{align}
\xi_{n}&= \sum_{\tilde{p}}  \xi_{n,\tilde{p}}\,, 
\quad A_{n}= \sum_{\tilde{q}}  A_{n,\tilde{q}}\, .
\end{align}
The SCET collinear Lagrangian, $\mathcal{L}_{n}$, describes the interaction
between the collinear fields $\xi_{n}$ and $A^\mu_{n}(x)$.  It is derived from
the QCD Lagrangian by integrating out the field, $\xi_{\nb}$. At LO, for the
kinetic and purely collinear interaction terms we have \cite{Bauer:2000yr,
  Bauer:2001ct}:
\begin{align} 
\mathcal{L}_{n}^{(0)} = \bar{\xi}_{n} \big( i n \!\cdot\!\partial + g\, n\!\cdot\!  A_n 
+ i \slashed{D}_{\,n\perp} W_n \frac{1}{\mathcal{P}_n} W^\dagger_n i \slashed{D}_{\,n\perp}  \big) \frac{\nbs}{2}\xi_{n}\, ,
\label{eq:LagSCET}
\end{align}
where we intrinsically sum over the large, label momenta, $\tilde{p}$. The
$in\cdot \partial$ derivative picks out the $\mo(\lambda^2)$ momenta.  The collinear
derivative, $D^\mu_n$, and collinear Wilson line, $W_n$, are defined as ~\cite{Bauer:2001yt}:
\begin{align}
i D^\mu_n&=\mathcal{P}_n^\mu + g A_n^\mu  \, ,   \nonumber \\
W_n(x) &= \Big[ \sum_{\rm{perms.}}  \mathrm{exp}\Big( -\frac{g}{\bar{\mathcal{P}_n}} \,\bar{n} \!\cdot\! A_n(x) \Big) \Big] \, .
\end{align}
The leading order coupling of collinear quarks to soft gluons is eikonal,
\begin{align} \label{eq:Lsoftn}
  {\cal L}_{sn}^{(0)} &= \bar\xi_n \, g\, n\cdot A_s \,\frac{\nbs}{2}\xi_{n} \,,
\end{align}
while the Lagrangian for purely soft quarks and gluons has the same form as full
QCD. The LO collinear Lagrangian for gluons has similar properties and is given
in Ref.~\cite{Bauer:2001yt}. The interactions between soft and collinear
particles, such as the one in \eq{Lsoftn}, can be removed from the Lagrangian
by the field redefinitions~\cite{Bauer:2001yt}:
\begin{align}
 & \xi_n \to Y_n\, \xi_n \,,
 & A_n^\mu & \to Y_n\, A_n^\mu\, Y_n^\dagger \,,
\end{align}
where the soft Wilson line $Y_n$ is defined in \eq{Yn}.
This causes soft interactions to be represented by Wilson lines in operators, as
in \eq{OpLOsoft}.

Now that we have split up gluons according to a momentum-space lattice, the
gauge structure of the theory has become more complex and involves global,
collinear, and soft gauge transformations.    Fortunately, with the collinear Wilson
line, it is possible to construct fermion and gluon fields that are manifestly
invariant under collinear gauge transformations.  The definitions are:
\begin{align}
\chi_{n}(x) &= W^\dagger_n (x) \xi_n (x) \, , 
& \mb^\mu_{n}(x) &=\frac{1}{g} \big[  W^\dagger_n (x)\, i D^\mu_{n} (x) W_n (x) \big] \, ,
\label{eq:collFieldsDef}
\end{align}
where the derivative in $\mb^\mu_{n}$ does not act outside of the brackets in
its definition, and we always have $\nb\cdot {\cal B}_n=0$.  In the $\nb\cdot
A_n=0$ light-cone gauge, $W_n = 1$ and $\mb^\mu_{n} \,=\, A^\mu_{n}$.  One can
construct collinear operators out of just three objects: the fermion field,
$\chi_n$, the perpendicular gluon field, $\mb_{n\perp}^\mu$, and the
perpendicular momentum operator, $\mathcal{P}_{n\perp}^\mu$. All the other
operators, like $n\!\cdot\mb_n$, or $n\!\cdot\!\partial$ can be written in terms
of these three using the equation of motions \cite{Marcantonini:2008qn}.

\section{Finite RPI}
\label{app:rpi}

\begin{figure}[t!]
\centering
\includegraphics[width=0.5\textwidth]{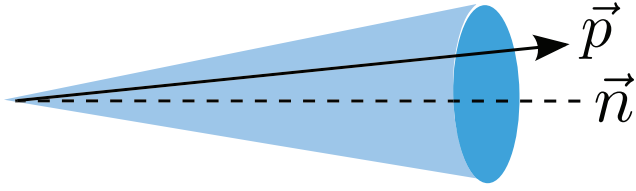}
\caption{In SCET, a particle is collinear to the direction $n$
if it is inside a cone centered in $\vec{n}$ and of opening angle $\lambda$.} 
\label{fig:singlecone}
\end{figure}

Even though SCET explicitly breaks Lorentz invariance, the symmetry
returns at each order in $\lambda$ by reparametrization invariance
(RPI).  RPI$_i$ is the version appropriate for $\sceti$.  As usual, we
define $p$ as collinear to the direction $n$ in $\sceti$ if its
components scale as $(n\!\cdot\!p, \,\bar{p},\, p_\perp)\sim
(\lambda^{2i}, \,1, \,\lambda^i)\,Q$ , where $Q$ is the hard scale and
$\lambda \ll 1$ ({\it cf.} Eq.~\ref{eq:decV}).  The vector $n$ has
physical meaning as its 3-vector subset, $\overrightarrow{n}$, is the
direction where most of the momentum is allocated. The direction
$\overrightarrow{p}$ is therefore inside a cone of opening angle
$\lambda^i$ around $\overrightarrow {n}$, ({\it cf.}
Fig~\ref{fig:singlecone}).  By contrast, $\bar{n}$ is an auxiliary
vector only needed to decompose the momentum.  The parameter $\lambda$ gives
the amount of collinearity to $n$.  The decomposition is not unique
since we can shift $n$ by an amount $\lambda$ and the particle will
still be collinear to it.  This means that if we move $n$ inside the
cone in Fig.~\ref{fig:singlecone}, $p$ is still collinear to it.  This
is called a reparametrization invariance (RPI) transformation of
type-I.  Thus, if a particle is collinear to $n$, it is also collinear
to any direction $n^\prime$ related by a type-I transformation.  To be
more formal, we can divide the space of light-cone vectors, $\{ n_i
\}$, into equivalence classes, $\{[n_i]\}$, where $[n_j]=\{ n\in [n_j]
|\, n\cdot n_j \lesssim \lambda^{2i}\}$.  The meaningful objects in
$\sceti$ are the $[n_j]$. 

By extension from SCET \cite{Bauer:2002nz}, two collinear sectors in $\sceti$, $n_1$ and $n_2$, are distinct if
\begin{align}
n_1\!\cdot\! n_2 \gg \lambda^{2i} \,,
\end{align}
Just as in regular SCET, we can write the external state with the 
$n$-label to which each particle is collinear.  For working in $\sceti$, 
we give a subscript to indicate the appropriate definition of collinearity.
For example, $| q_{n_1} \rangle_i$ is a state with one quark, collinear 
to $n_1$ that can be annihilated by any $\chi_n$ such that 
$n \cdot n_1 \,\ll\, \lambda^{2i}$, or $n \in [n_1]$.

For each $\{n,\nb \}$, the type-I RPI infinitesimal transformations  
are\footnote{Infinitesimal does not refer to the expansion in $\lambda$.}
\begin{eqnarray}\label{repinv}
\text{(I)} \left\{
\begin{tabular}{l}
$n^\mu \to n^\mu + \Delta^\mu_{n \perp}$ \\
$\nb^\mu \to \nb^\mu$
\end{tabular}
\right. \,,
\label{eq:rpiOne}
\end{eqnarray}
where $\Delta^\mu_{n \perp}\sim\lambda$ and $n\cdot\Delta_{n
  \perp}=\nb\cdot\Delta_{n \perp}=0$.  These transformations preserve the
relations $n^2=0$, $\nb^2=0$ and $n\cdot\nb=2$.  \footnote{It is also possible
  to rotate $\bar{n}\to \bar{n} +\varepsilon_\perp$ where $\varepsilon_\perp\sim
  \lambda^0$, which is a type-II RPI transformation. Finally, a type-III
  transformation takes $n\to e^{\alpha} n$ and $\nb\to e^{-\alpha}\nb$.}
 
The general problem of matching $\sceti \rightarrow \scetipone$ is our need to
rotate the direction $n$ of objects in the amplitude (such as spinors and
vectors) to $n^\prime$ that is close enough to the particle momentum such that
$p$ is collinear to $n^\prime$ in $\scetipone$.  Thus, RPI$_i$ is crucial for
matching as it determines how formerly identical $\sceti$ configurations wind up
in different $\scetipone$ terms.  Any transformation in RPI$_i$/RPI$_{i+1}$ is
therefore of consequence.  By contrast, the choice within $\scetipone$ is purely
a convention we may use to our convenience ({\it cf.} Fig.~\ref{fig:ConeC}).
For example, we can pick $n^\prime$ as that direction $n_p$ such that $p$ has
zero perpendicular momentum in the $n_p - \bar{n}$ frame:
\begin{align}
p&= \bar{p} \frac{n^\mu_p}{2} + n_p\!\cdot\!p\frac{\nb^\mu}{2}\, .
\end{align}
This is satisfied for:
\begin{align}
n_p^\mu&= n^\mu+2\, \frac{p_\perp^\mu}{\bar{p}}-\bar{n}^\mu\frac{(p_\perp)^2}{\bar{p}^2} \, , \label{nnprel}
\end{align}
with $p_\perp$ defined in the $n$-frame.  Unlike \eq{rpiOne}, this RPI$_i$
transformation is finite. It is easy to check that $n_p^2=0$, $n_p\!\cdot\!
\nb=2$ and that $p_{n_p \perp}^\mu=p^\mu-n_p\!\cdot\!p \, \bar{n}^\mu/2 -
\bar{p} \, n_p^\mu/2 =0$.

We can derive similar relations for other quantities.
To see how the quark field transforms, we use the RPI invariant fermion field \cite{Marcantonini:2008qn}:
\begin{align}
\psi_n=\Big( 1+ \frac{ \slashed{D}_n^\perp}{\nb\!\cdot\!D_n }\frac{\nbs}{2} \Big)\xi_n  \, . \label{psi}
\end{align} 
Since (\ref{psi}) is invariant under RPI, $\psi_n=\psi_{n_p}$ and we can write,
\begin{align}
\Big( 1+ \frac{ \slashed{D}_n^\perp}{\nb\!\cdot\!D_n }\frac{\nbs}{2} \Big)\xi_n
=\Big( 1+ \frac{\slashed{D}_{n_p}^\perp }{\nb\!\cdot\!D_{n_p} }\frac{\nbs}{2} \Big)\xi_{n_p}  \, . \label{psinp}
\end{align}
Multiplying (\ref{psinp}) by the projector $\ns \nbs/4$ we get the finite RPI$_i$ relation
\begin{align}
\label{eq:spinSqSn}
\xi_{n}=\frac{\ns \nbs}{4}\xi_{n_p} \, .
\end{align}
The relation (\ref{eq:spinSqSn}) is in agreement with the spinor equation (A7) in \cite{Bauer:2006mk} 
upon setting $\bar{n}_1=\bar{n}_2$.  Objects with a full Lorentz index, like $p^\mu$ or $\gamma^\mu$, 
are RPI invariant as there is no reference to the light-cone vectors $n$ and $\nb$.  Those 
in the perpendicular direction though, such as $p^\mu_\perp$ or $\gamma^\mu_\perp$, are not, as $\perp$ is 
defined with respect to $n$ and $\bar{n}$. Using the relation $\gamma_\perp^\mu=\gamma^\mu-\nb^\mu\,\ns/2-n^\mu \, \nbs/2$, 
we derive the expression
\begin{align}
\gamma_{n_p \perp}^\mu = \gamma_{\perp}^\mu - \nb^\mu\frac{\slashed{p}_\perp}{\bar{p}} - p_\perp^\mu\frac{\nbs}{\bar{p}}
+ \nb^\mu \frac{(p_\perp)^2}{\bar{p}^2}\nbs\, .
\end{align}
  
\begin{figure}[t!]
 \centering
 \includegraphics{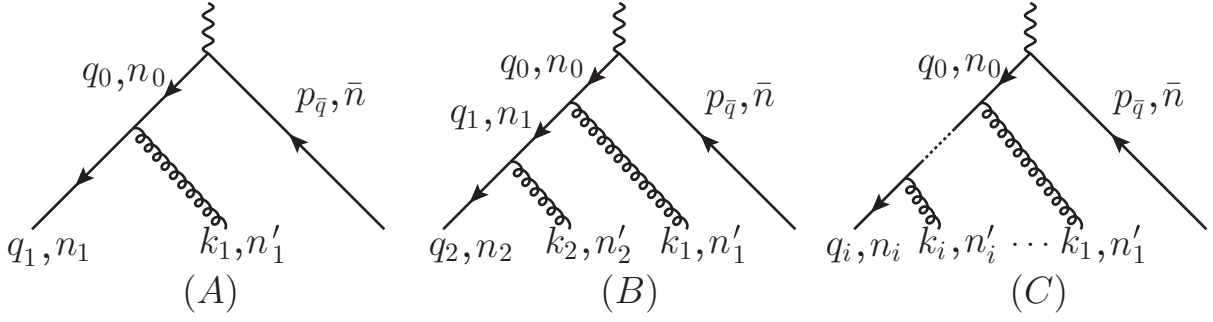}
 \caption[Kinematic variables for one-gluon emission and two-gluon
 emission] {Kinematic variables for one-gluon emission ($A$), two-gluon
 emission ($B$) and $i$-gluon emission ($C$).  The $n$'s are defined
 such that the corresponding particle's momentum has no perpendicular
 component along the directions $n-\nb$.  Note that $q_1$ is a different
 vector for single and double emissions.
 }
 \label{fig:figureKinWithNC}
 \end{figure}

We now focus on those transformations needed for one-gluon emission.  As in
\subsec{loShowerRe}, we consider the case of a virtual quark with momentum $q_0$
emitting an external gluon and quark with momentum $k_1$ and $q_1$,
respectively.  In Fig.~\ref{fig:figureKinWithNC}(A), we portray this kinematics
for one-gluon emission where the initial quark $q_0$ comes from a QCD current
$\bar{q}\gamma^\mu q$.
We call $n_0$, $n'_1$ and $n_1$ the directions where $q_0$, $k_1$ and $q_1$ zero have perpendicular component, that is:
\begin{align}
\label{eq:Rot1}
q_0&= \bar{q}_0 \frac{n_0^\mu}{2} + n_0\!\cdot\!q_0\frac{\nb^\mu}{2}\, , \nonumber \\
k_1&= \bar{k}_1 \frac{{n}^{\prime\mu}_1}{2} \, , \nonumber \\
q_1&= \bar{q}_1 \frac{n^\mu_{q_1}}{2} \,  , 
\end{align}
Using Eq.~(\ref{nnprel}), we can relate $n'_1$ and $n_1$ to $n_0$,
\begin{align}
n_1^{\prime\mu}&= n_0^\mu-2 \frac{(q_1)_{n_0 \perp}^\mu}{\bar{k}_1}-\bar{n}^\mu\frac{(q_1)_{n_0 \perp}^2}{\bar{k}_1^2}\, , \nonumber \\ \label{n1nq1def}
n_1^\mu&= n_0^\mu+2 \frac{(q_1)_{n_0 \perp}^\mu}{\bar{q}_1}-\bar{n}^\mu\frac{(q_1)_{n_0 \perp}^2}{\bar{q}_1^2}\, ,
\end{align}
where we have used the equality $(k_1)_{n_0 \perp}^\mu= -(q_1)_{n_0 \perp}^\mu$. Some useful relations are:
\begin{align}
\label{nrel}
n_1\!\cdot\!n'_1&=n_0\!\cdot\!n'_1 \frac{\bar{q}_0^2}{\bar{k}_1^2}=n_0\!\cdot\!n_1 \frac{\bar{q}_0^2}{\bar{q}_1^2}
=-2 \frac{(q_1)_{n_0 \perp}^2 \bar{q}_0^2}{\bar{q}_1^2\,\bar{k}_1^2}\, ,  \nonumber \\
n_0^\mu&=\frac{\bar{k}_1 n_1^{\prime\mu} + \bar{q}_1 n_1^\mu}{\bar{q}_0} -\bar{n}^\mu   \, (n_1\!\cdot\!n'_1 )
 \frac{\bar{q}_1 \bar{k}_1}{2\, \bar{q}_0^2}  \, , \nonumber \\
(q_1)_{n_0 \perp}^\mu&=\frac{\bar{q}_1\bar{k}_1}{\bar{q}_0} \sqrt{n_1 \cdot n'_1} \frac{v^{\mu}_1}{2}
- \bar{n}^\mu \,(n_1\!\cdot\! n'_1)\frac{\bar{q}_1 \bar{k}_1 (\bar{k}_1^2 -\bar{q}_1^2)}{4\, \bar{q}_0^3}\, , \nonumber \\
\gamma_{n_0 \perp}^\mu&=\gamma_{n_1'\perp}^\mu-\nb^\mu\frac{(\slashed{q}_1)_{n_0 \perp}}{\bar{k}_1}-(q_{1})^\mu_{n_0 \perp}\frac{\nbs}{\bar{k}_1}
+\nb^\mu\,(n_1\!\cdot\!n'_1) \frac{\bar{q}_1^2}{2 \,\bar{q}_0^2}\nbs\, ,
\end{align}
where
\begin{align}
\label{eq:v}
v^{\mu}_1 = \frac{n_1^\mu-n_1^{\prime \mu}}{\sqrt { n_1 \cdot n'_1}}\, ,
\end{align}
and $|v_1^2|=2$. Another useful relation is
\begin{align}
\label{q021E}
q_0^2&=(q_1+k_1)^2= n'_1\!\cdot\!n_1  \frac{  \bar{q}_1 \bar{k}_1 }{2}\, . 
\end{align}
We can express all quantities of interest in terms of the vectors $n'_1$, $n_1$ and the momenta $\bar{q}_1$ and $\bar{k}_1$.

In two-gluon emissions, the kinematic variables are assigned in Fig.~\ref{fig:figureKinWithNC}(B).
We define $n_0$, $n'_1$, $n'_2$ and $n_1$ as follows (note that $q_1$ and $n_1$ are different from above):
\begin{align}
\label{eq:n12p}
q_0&= \bar{q}_0 \frac{n_0^\mu}{2} + n_0\!\cdot\!q_0\frac{\nb^\mu}{2}\, , \nonumber \\
k_1&= \bar{k}_1 \frac{n_1^{\prime\mu}}{2}\, , \nonumber \\
q_1&= \bar{q}_1 \frac{n_1^\mu}{2}+n_1\!\cdot\!q_1\frac{\nb^\mu}{2} \,  ,\nonumber \\
k_2&= \bar{k}_2 \frac{n_2^{\prime\mu}}{2} \, , \nonumber \\
q_2&= \bar{q}_2 \frac{n_2^\mu}{2} \,  ,
\end{align}
Eq.~(\ref{n1nq1def}) is still valid, and we can similarly define $n'_2$ and $n_2$ as:
\begin{align}
n_2^{\prime \mu}&= n_1^\mu-2 \frac{(q_2)_{n_1 \perp}^\mu}{\bar{k}_2}-\bar{n}^\mu\frac{(q_2)_{n_1 \perp}^2}{\bar{k}_2^2}\, , \nonumber \\ 
n_2^\mu&= n_1^\mu+2 \frac{(q_2)_{n_1 \perp}^\mu}{\bar{q}_2}-\bar{n}^\mu\frac{(q_2)_{n_1 \perp}^2}{\bar{q}_2^2}\, ,
\label{eq:n2nq2def}
\end{align}
where  $(k_2)_{n_1 \perp }= -(q_2)_{n_1 \perp}$.  Also, Eq.~(\ref{nrel}) is still valid,  
and we get a new set by sending $0\rightarrow1$ and  $1\rightarrow 2$:
\begin{align}
\label{nrel2}
n_2\!\cdot\!n'_2&=n_1\!\cdot\!n'_2 \frac{\bar{q}_1^2}{\bar{k}_2^2}=n_1\!\cdot\!n_2\frac{\bar{q}_1^2}{\bar{q}_2^2}
=-2 \frac{(q_2)_{n_1 \perp}^2 \bar{q}_1^2}{\bar{q}_2^2\,\bar{k}_2^2}\, ,  \\
n_1^\mu&=\frac{\bar{k}_2 n_2^{\prime\mu} + \bar{q}_2 n_2^\mu}{\bar{q}_1} -\bar{n}^\mu   \, (n_2\!\cdot\!n'_2 )
 \frac{\bar{q}_2 \bar{k}_2}{2\, \bar{q}_1^2}  \, , \nonumber \\
(q_2)_{n_1 \perp}^\mu&= \frac{\bar{q}_2\bar{k}_2}{\bar{q}_1} \sqrt{n_2 \cdot n'_2} \frac{v^{\mu}_2}{2}
- \bar{n}^\mu \,(n_2\!\cdot\! n'_2)\frac{\bar{q}_2 \bar{k}_2 (\bar{k}_2^2 -\bar{q}_2^2)}{4\, \bar{q}_1^3}\, , \nonumber \\
\gamma_{n_1 \perp}^\mu&=\gamma_{n_2'\perp}^\mu-\nb^\mu\frac{(\slashed{q}_2)_{n_1 \perp}}{\bar{k}_2}
-(q_{2})^\mu_{n_1 \perp}\frac{\nbs}{\bar{k}_2}
+\nb^\mu\,(n_2\!\cdot\!n'_2) \frac{\bar{q}_2^2}{2 \,\bar{q}_1^2}\nbs\, , \nonumber
\end{align}
where
\begin{align}
\label{eq:v2}
v^{\mu}_2 = \frac{n_2^\mu-n_2^{\prime \mu}}{\sqrt { n_2 \cdot n'_2}}\, ,
\end{align}
and $|v_2^2|=2$.  We can write $n_1 \cdot n'_1$ and $v_1$  in terms of $n_2$, $n^\prime_1$ and $n^\prime_2$
so that once again we only need to work with external quantities:
\begin{align}
\label{eq:v1n2}
n_1 \!\cdot\! n'_1& = \frac{\bar{k}_2 (n'_2 \!\cdot\! n'_1) + \bar{q_2} (n_2 \!\cdot\! n'_1)}{\bar{q}_1}
 - (n_2 \!\cdot\!n'_2)\frac{\bar{q}_1 \bar{k}_1}{\bar{q}_0^2} \, , \\
v^{\mu}_1 &= \frac{2\,\bar{k}_2 \bar{q}_1 \,n_2^{\prime\mu} +2\, \bar{q}_1 \bar{q}_2 n_1^\mu-  \bar{k}_2 \bar{q}_2    \, (n_2\!\cdot\!n'_2 )  \,\bar{n}^\mu
-2\, (\bar{q}_1)^2n_1^{\prime \mu}}{2 (\bar{q}_1)^2\sqrt { n_1 \cdot n'_1}}\, . \nonumber 
\end{align}
Eq.~(\ref{q021E}) for two emissions is modified to:
\begin{align}
\label{q022E1}
q_0^2=(q_2+k_1+k_2)^2= n'_1\!\cdot\!n_2  \frac{  \bar{q}_2 \bar{k}_1 }{2}+ n'_2\!\cdot\!n_2  
\frac{  \bar{q}_2 \bar{k}_2 }{2}+ n'_{1}\!\cdot\!n'_2  \frac{  \bar{k}_1 \bar{k}_2 }{2}\, . 
\end{align}
Other useful relations are
\begin{align}
\label{q022E2}
\gamma_{n_0 \perp}^\mu&=\gamma_{n'_2 \perp}^\mu-\bar{n}^\mu 
\Big( \frac{(\slashed{q}_1)_{n_0 \perp}}{\bar{k}_1}  + \frac{(\slashed{q}_2)_{n_1 \perp}}{\bar{k}_2}   \Big) 
-\Big( \frac{(q_1)_{n_0 \perp}^\mu}{\bar{k}_1}  + \frac{(q_2)_{n_1
    \perp}^\mu}{\bar{k}_2}   \Big) \nbs \nonumber \\
& \qquad 
- \bar{n}^\mu \nbs \Big( \frac{(q_1)_{n_0 \perp}^2}{\bar{k}_1^2}  + \frac{(q_2)_{n_1 \perp}^2}{\bar{k}_2^2}   \Big)  ,\nonumber \\
q_1^2&=(k_2+q_2)^2=n'_2\!\cdot\!n_2 \frac{  \bar{k}_2  \bar{q}_2}{2} \, ,\nonumber \\
(k_1+q_2)^2&=n'_1\!\cdot\!n_2 \frac{  \bar{k}_1  \bar{q}_2}{2} \, .
\end{align}

For $i$ gluon emissions, Fig.~\ref{fig:figureKinWithNC}(C), $n'_k$ is
parallel to the $k$-gluon, $n_i$ parallel to the external quark, and
$n_k$ is the light cone vector such that the $k^{\rm th}$ virtual quark
has zero perpendicular momentum with respect to ($n_k$, $\nb$).  To
calculate $n_i$, $n'_i$ we can iterate the formulas above up to $i$
emissions. That is we can calculate $n_i$, $n'_i$ from $n_{i-1}$ using
Eq.~(\ref{n1nq1def}) with $0 \rightarrow( i-1)$, $1 \rightarrow i$.

\section{Matching QCD to $\scetone$}
\label{app:QCD/SCET1}

\begin{figure}[t!]
\centering
\includegraphics{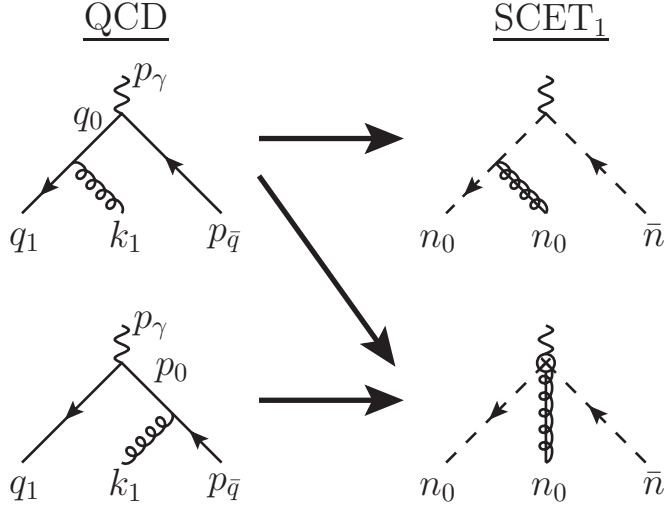}
\caption[Matching QCD to $\mathrm{SCET}_{1}$  for the two-jet configuration]
{Matching QCD to $\mathrm{SCET}_{1}$  for the two-jet configuration:
 In the first column there are the two Feynman graphs for one-gluon emission in QCD, 
labeled by the 4-momenta. In the second column there are the two   Feynman graphs in $\scetone$
 that reproduce the same amplitude in the case the quark and gluon are collinear along the direction $n_0$. 
The first graph comes from the operator $\mo^{(0)}_{1}(n_0)$ with the insertion of the $\scetone$ Lagrangian, 
the second graph comes from the operators $\mo^{(1)}_{1}(n_0,\,n_0)$ and $\mathcal{T}^{(1)}_{1}(n_0,\,n_0)$. } 
\label{fig:figureAppQS1C}
\end{figure}

To study the process of $q \,\rightarrow\, qg$ emission, we match the QCD current,
\begin{align}
J^\mu_{\rm{QCD}}=\bar{q} \gamma^\mu q \, , 
\label{eq:QCDcurret}
\end{align}
to $\mathrm{SCET}_1$ operators for a final state with a quark, antiquark, and gluon. 
The particle momenta are $q_1$ for the quark, $p_{\bar{q}}$ for the antiquark, 
and $k_1$ for the gluon, ({\it cf.} Fig.~\ref{fig:figureAppQS1C}).  We do the matching 
in the center of mass frame with 
\begin{align}
\label{eq:pgamma}
p_\gamma=q_1+p_{\bar{q}}+k_1=(Q,0,0,0)\, .
\end{align}
$\mathrm{SCET}_1$, being equivalent to the usual SCET, is formulated as an
expansion in the parameter $\lambda$.  The current in Eq.~(\ref{eq:QCDcurret})
matches onto an infinite series of SCET$_1$ operators.  We will perform the
matching up to NNLO$(\lambda)$ for one gluon emission, and focus only on the
cases when the gluon is either collinear to the quark or has its own direction.
Obtaining the result for gluon-antiquark collinearity from our work is a simple
exercise in charge conjugation.  We can construct the $\scetone$ operators out
of a few building blocks: the quark field $\chi_n$, the gluon field
$\mb_{n\perp}^\alpha$ and the perpendicular momentum operator
$\mathcal{P}_{n\perp}^\alpha$, plus Dirac structures.  $\chi_n$,
$\mb_{n\perp}^\alpha$ and $\mathcal{P}_{n\perp}^\alpha$ all scale $\sim \,
\lambda$.  The basis of $\scetone$ operators for one emission up to
NNLO($\lambda$) is \cite{Marcantonini:2008qn}: \footnote{$\mathcal{T}^{(1)}_{1}
  (n_1,n'_1)$ encodes redundant information that can be obtained with RPI. For
  example we can choose the directions $n_1$ and $n'_1$ to
  align perfectly with particle momenta such that {\it e.g.}. $\mcp_{n_1\perp}
  \mb_{n_1\perp}$ = 0. This is not possible for $\mathcal{T}_1^{(1)}(n_0,n_0)$.}
\begin{align}
\label{eq:SCET1op}
\mo^{(0)}_{1}(n_0)&= \bar{\chi}_{n_0}  \chi_{\bar{n}} \, , \nonumber \\
\mo^{(1)}_{1}(n_0,n_0) &=\,\chib_{n_0}\,g \mb^\alpha_{n_0\perp} \chi_{\nb}\, , \nonumber \\
\mathcal{T}^{(1)}_{1} (n_0,n_0)&= 
\bar{\chi}_{n_0} \left[ \mathcal {P}_{n_0 \perp}^\beta \, g \mathcal{B}_{n_0 \perp}^\alpha  \right] \chi_{\bar{n}} \, , \nonumber \\
\mo^{(1)}_1 (n_1, n'_1) &= \bar{\chi}_{n_1} g \mathcal{B}_{n'_1\perp}^\alpha \chi_{\bar{n}}\, ,
\end{align}
Following the convention of \eq{opDef},
we do not write the antiquark direction as it is always $\nb$.
$\mo^{(0)}_{1}$ is the LO operator and scales as $\lambda^2$, $\mo^{(1)}_{1}(n_0,n_0)$ and 
$\mo^{(1)}_1 (n_1, n'_1) $ 
are NLO($\lambda$) operators, scaling like $\lambda^3$, and $\mathcal{T}^{(1)}_{1} \sim \lambda^4$. 

In $\mathrm{SCET}_1$, two particles are collinear if they are inside a
cone with opening angle $\sim \lambda$, equivalently $p_1 \cdot p_2
\lesssim (Q \lambda)^2/\eta^4$.  The factor $\eta \sim 1/2$ represents 
the average $\bar{p}$ fraction taken by the daughter from the mother as discussed in Sec.~II C.
Usually, we formulate this condition with
dimensionless quantities, $n_{p_1} \cdot n_{p_2} \lesssim \lambda^2/\eta^4$,
where $n_{p_i}$ is exactly proportional to the particle momentum.  To
distinguish a ``two-jet'' from a ``three-jet'' state, we label the
external states with the direction to which the particles are
collinear.  A state $|q_{n_0} \rangle_1$ indicates a state where a
quark with momentum $q_1$ is collinear to the direction $n_0$, that is
$( \bar{q}_1, n_0\cdot q_1, (q_1 )_{n_0 \perp}) \sim (1,\lambda^2,
\lambda) Q$, and the subscript, 1, tells us the state can be
annihilated by any operator, $\chi_n$, where $n$ and $n_0$ are in the
same $\scetone$ equivalence class, $\{ [n] \}$.  As we will see, when
we match to lower-scale $\sceti$, we will change this number
appropriately.  A two-jet state with a collinear quark and gluon, and
an antiquark is given by $ | q_{n_0}\,g_{n_0}\,\bar{q}_{\nb}
\rangle_1$.  The fact that the quark and gluon share an index label
implies that $q_1 \cdot k_1 \lesssim (Q\lambda)^2/\eta^4$.  A three-jet state
is indicated by $ | q_{n_1}\,g_{n_1^\prime}\,\bar{q}_{\nb} \rangle_1$,
where each particle is collinear to a different direction.  The
operators $\mo^{(0)}_{1}(n_0)$, $\mo^{(1)}_{1}(n_0,n_0)$ and
$\mathcal{T}^{(1)}_{1} (n_0,n_0)$ can only create a two-jet state,
whereas $\mo^{(1)}_1 (n_1, n'_1)$ is for three-jets.  Multiplying the
terms in (\ref{eq:SCET1op}) by the Wilson coefficients, we have:
\begin{align}
\label{eq:JqcdS1}
 J^\mu_{\rm{QCD}}=&\,
   C^{(0)}_{1,\rm{LO}}(n_0)  \mo^{(0)}_{1}
+  C^{(1)}_{1,\,{\rm NLO}}  (n_0, n_0)\mo^{(1)}_{1}
 + C^{(1)}_{1, \mathcal{T}}(n_1, n'_1) \mathcal{T}^{(1)}_{1} \nonumber \\
 &+ C^{(1)}_{1}(n_1, n'_1) \mo^{(1)}_1 + \ldots\, ,
\end{align}
where the ellipses indicate higher order terms in $\lambda$. 
When it is unambiguous, we will only write the $n$-labels in the Wilson
coefficients, as above.
We begin by looking at two-jet operators in detail.  Here, because we are in
the center of mass frame, the two jets are back to back.  We define the
kinematics as follows, the antiquark is exactly parallel to
$\bar{n}=(1,0,0,-1)$, while the quark and the gluon are collinear to
$n_0=(1,0,0,1)$, such that $q_0=q_1+k_1$ has no component perpendicular to $n_0$
and $\nb$, and:
\begin{align}
p_{\bar{q}}^\mu&=n_0\!\cdot\!p_{\bar{q}} \frac{\bar{n}^\mu}{2}\, , \nonumber \\
q_1^\mu&=\bar{q}_1 \frac{n_0^\mu}{2} + n_0\!\cdot\! q_1 \frac{\nb^\mu}{2} + (q_1)_{n_0 \perp}^\mu \, , \nonumber \\
k_1^\mu&=\bar{q}_1 \frac{n_0^\mu}{2} + n_0\!\cdot\! k_1 \frac{\nb^\mu}{2} + (k_1)_{n_0 \perp}^\mu \, , \label{eq:SCETkin1}
\end{align}
where $(n_0\cdot q_1, \bar{q}_1,  q_{1\perp})$ and  $(n_0\cdot k_1, \bar{k}_1,  k_{1\perp})$ 
scale as $ (\lambda^2, 1, \lambda)$, and $(q_1)_{n_0\perp}^\mu=-(k_1)_{n_0 \perp}^\mu$ by momentum conservation. 
The Wilson coefficients are defined through the equation
\begin{align} \label{eq:machQS1}
\langle 0|  J^\mu_{\rm{QCD}}| q_{n_0}\,g_{n_0}\,\bar{q}_{\nb} \rangle_1 =&
   C^{(0)}_{1,\rm{LO}} (n_0,n_0) \int\!\! dx^4\langle 0| T\{ \mathcal{L}_{\scetone}(x) \mo^{(0)}_{1}  \}| q_{n_0}\,g_{n_0}\,\bar{q}_{\nb} \rangle_1 \\
& +  C^{(1)}_{1}  (n_0,n_0) \langle 0|   \mo^{(1)}_{1}| q_{n_0}\,g_{n_0}\,\bar{q}_{\nb} \rangle_1 
 + C^{(1)}_{1, \mathcal{T}}   (n_0,n_0) \langle 0|  \mathcal{T}^{(1)}_{1} | q_{n_0}\,g_{n_0}\,\bar{q}_{\nb} \rangle_1 \, . \nonumber
\end{align}
Calculating the $C$'s for this two-jet process goes as follows.
We decompose the QCD amplitude along $n_0$ and $\bar{n}$, using Eq.~(\ref{eq:SCETkin1}), and we write the QCD spinor in terms 
of the $\scetone$ spinor, \eq{QCDSCETspinor}.  Expanding in $\lambda$ up to NNLO, on the RHS we compute the amplitudes
for the three $\scetone$ terms. 
The coefficient $C^{(0)}_{1,\rm{LO} }$ was already determined from matching QCD to $\scetone$ for zero gluon emission, it is:
\begin{align}
C^{(0)}_{1,\rm{LO} }=\gamma^\mu \, .
\end{align}
The coefficients $C^{(1)}_1$ and $ C^{(1)}_{1, \mathcal{T}}$ come from solving Eq.~(\ref{eq:machQS1}) at NLO($\lambda$) and NNLO($\lambda$), respectively.
Since $ \mo^{(1)}_{1}$ and $\mathcal{T}^{(1)}_{1} $ are at different orders in $\lambda$, there are no ambiguities.

In order to do the matching, we need the relation between the QCD and SCET spinors. Using Eq.~(\ref{psi}), we can write:
\begin{align}
u(p)=\Big( 1+ \frac{p\!\!\!\slash_\perp \nbs}{2\,\bar{p}}\Big) u_n(p)\, ,
\label{eq:QCDSCETspinor}
\end{align}
where $u(p)$ is the QCD spinor and $u_n(p)$ is the $\scetone$ one.
It easy to see that the SCET spinor satisfies:
\begin{align}
\frac{\nbs \ns}{4}u_n &=0 \, , \nonumber \\
\frac{\ns \nbs}{4}u_n &=u_n \, , \nonumber \\
\sum_{s}\bar{u}_n^s u_n^s &= \bar{p} \frac{\nbs}{2} \, .
\label{eq:SCETspinorprop}
\end{align}

The QCD amplitude for $\gamma^*\rightarrow q\bar{q}g$ (shown in Fig.~\ref{fig:figureAppQS1C}) is:
\begin{align}
A^{q\bar{q}g}_{\rm{QCD}}= \bar{u}(q_{1}) i g \gamma^\alpha \frac{i \slashed{q}_{0}}{q_{0}^2}\gamma^\mu v(p_{\bar{q}})
-\bar{u}(q_{1}) i g \gamma^\mu \frac{i \slashed{p}_{0}} {p_0^ 2}\gamma^\alpha v(p_{\bar{q}}) \, . \label{eq:ampli1EmQCD}
\end{align}
Using Eqs.~(\ref{eq:SCETkin1}) \& (\ref{eq:QCDSCETspinor}) in (\ref{eq:ampli1EmQCD}) 
and expanding to NNLO in $\lambda$ we get:
\begin{align} \label{eq:ampliS1}
A^{q\bar{q}g}_{\rm{QCD}} = A^{q\bar{q}g}_{\rm{LO}}+A^{q\bar{q}g}_{\rm{NLO}}+A^{q\bar{q}g}_{\rm{NNLO}} \, , 
\end{align}
where,
\begin{align}
A^{q\bar{q}g}_{\rm{LO}}&=-g\bar{u}_{n_0} \Big[\Big( n_0^\alpha + \frac{(\slashed{q}_1)_{n_0 \perp}}{\bar{q}_1}
 \gamma_{n_0 \perp}^\alpha \Big) \frac{\bar{q}_0}{q_0^2}+\frac{\bar{n}^\alpha}{\bar{k}_1}  \Big] 
                             \gamma_{n_0 \perp}^\mu v_{\bar{n}} \, ,  \nonumber \\
A^{q\bar{q}g}_{\rm{NLO}}  &=g \frac{n_0^\mu - \nb^\mu}{Q}\bar{u}_{n_0}  \Big( \gamma^\alpha_{n_0 \perp}
 - \frac{(\slashed{k}_1)_{n_0 \perp}}{\bar{k}_1}\nb^\alpha \Big) v_{\bar{n}} \, ,\nonumber \\
A^{q\bar{q}g}_{\rm{NNLO}}   &= g\Big( \frac{1}{\bar{q}_1} +\frac{1}{\bar{k}_1} \Big)  \frac{1}{Q}\bar{u}_{n_0}  
\gamma_{n_0 \perp}^\mu (\slashed{k}_1)_{n_0 \perp}
                          \Big( \gamma^\alpha_{n_0 \perp}
                           - \frac{(\slashed{k}_1)_{n_0 \perp}}{\bar{k}_1}\nb^\alpha \Big) v_{\bar{n}} \nonumber \\
& - g\frac{2 }{\bar{q}_1 Q} \bar{u}_{n_0}(k_1)_{n_0 \perp}^\mu   \Big( \gamma^\alpha_{n_0 \perp}
 - \frac{(\slashed{k}_1)_{n_0 \perp}}{\bar{k}_1}\nb^\alpha \Big) v_{\bar{n}} \, .
 \label{eq:ANNLO}
\end{align}
We already know $C^{(0)}_{1,\rm{LO}} $, and it is easy to determine the other two Wilson coefficients to 
reproduce $A^{q\bar{q}g}_{\rm{QCD}}$, they are:
\begin{align}
C^{(1)}_{1,\,{\rm NLO}}(n_0,n_0)&= \frac{1}{Q}  (n_{0}^{\mu}-\nb^{\mu} ) \gamma_{n_0 \perp}^\alpha \, , \nonumber \\
C^{(1)}_{1, \mathcal{T}}(n_0,n_0)  &= \frac{1}{\bar{q}_1 \bar{k}_1 }   \gamma_{n_0 \perp}^\mu  \gamma_{n_0 \perp}^\beta \gamma_{n_0 \perp}^\alpha 
- \frac{2}{\bar{q}_1 Q} \,g^{\beta \mu} \gamma_{n_0 \perp}^\alpha \,,
\label{eq:CoeS1}
\end{align}
where we have used the relation $\bar{q}_1+\bar{k}_1=Q$.

For the three-jet operator $O^{(1)}_1 (n_1, n'_1) $, the matching was already
done in \cite{Marcantonini:2008qn}, but we will translate it to the notation
used here.  In this case, we need three distinct directions in SCET$_1$ to
describe the three external particles, and there is no small parameter to expand
in.  This means that the amplitude for this operator is exactly equal to the
tree-level QCD amplitude for a $q\bar{q}g$ process.  One may wonder then, why we
simply do not apply this everywhere instead of just the three-jet region.  The
answer has to do with factorization and running effects.  The RG kernels of our
two-jet operators, $\mo^{(0)}_1$, $\mo^{(1)}_1$, and $\mathcal{T}^{(1)}_1$, will
resum the large collinear logarithms of those configurations ({\it cf.}
\subsec{opRunning}).  It is for this reason that we gain by keeping track of
them as separate contributions.
  
Even though they are all in independent directions, we need only four
independent vectors to decompose the particles.  In the center of mass frame,
$q_0=q_1 + k_1$ is back to back with the antiquark, $p_{\bar{q}} \propto \nb$.
We decompose $q_0$ along ($n_0$, $\nb$) such that it has no component
perpendicular to them: $q_0= \bar{n}\cdot q_0 \, n_0 /2 + n_0\cdot q_0 \,
\bar{n} /2$.  Using Eq.~(\ref{n1nq1def}) we can define $n_1$ and $n'_1$ such
that they are parallel to $q_1$ and $k_1$, respectively and such that the quark
is decomposed along ($n_1$, $\nb$) and the gluon along ($n'_1$, $\nb$), both
without $\perp$ components.  Unlike the two-jet case, where $(q_1)_{n_0 \perp}
\lesssim \lambda$, since the quark was collinear to $n_0$, here $(q_1)_{n_0
  \perp} > \lambda$ in Eq.~(\ref{n1nq1def}).  We have:
\begin{align}
 q_1^\mu&= \bar{n} \!\cdot\! q_1 \frac{n_1^\mu}{2} \, , \\
 p_{\bar{q}}^\mu&= n_0\!\cdot\! p_{\bar{q}} \frac{\bar{n}^\mu}{2} \, , \nonumber \\
 k_1^\mu&= \bar{n} \!\cdot\! k_1 \frac{n_1^{\prime\mu}}{2} \, ,\nonumber 
 \end{align}
 where $\bar{n} \cdot q_1$, $n_0\cdot p_{\bar{q}}$ and $ \bar{n} \cdot k_1 $ are
 $\mo(Q)$, and $n_1 \cdot n'_1 >\lambda^2/\eta^4$.  With this setup
 $\mathcal{T}_1^{(1)}(n_1, n'_1)=\bar{\chi}_{n_1}
 \mathcal{P}_{n'_1}^\perp\mathcal{B}_{n'_1\perp}^\alpha \chi_{\nb}=0 $.

The matching is therefore given by:
\begin{align}
 \langle 0| J^\mu_{\rm{QCD}}| q_{n_1}\,g_{n'_1}\bar q_{\nb} \rangle_1
 = C^{(1)}_{1} (n_1, n'_1) \langle 0|\mo^{(1)}_1  
 | q_{n_1}\,g_{n'_1}\,\bar{q}_{\nb}\rangle_1 \, ,
\end{align}
and the Wilson coefficient is:
\begin{align} \label{eq:C3j}
C^{(1)}_{1} (n_1, n'_1)&= -\frac{2}{ (n_1\!\cdot\!n'_1) \bar{q}_1 \bar{k}_1} \gamma^\alpha \slashed{p}_\gamma \,\gamma_T^\mu \nonumber \\
&+ \Big[ \frac{1}{ (n\!\cdot\! p_{\bar{q}} )\bar{k}_1 } 
\Big( \gamma_T^\mu \, \slashed{p}_\gamma - \bar{q}_1\,n_{1T}^\mu \Big) 
+\frac{2  (n\!\cdot\! p_{\bar{q}} ) }{ (n_1\!\cdot\!n'_1) \bar{q}_1  \bar{k}_1} \nb_T^\mu \Big] \gamma^\alpha_{n'_1\, \perp} \, ,
\end{align}
where the subscript $T $ applied to a generic four vector $f^\mu$ means: 
$f_T^\mu \equiv f^\mu-p_\gamma^\mu \,(f\cdot p_\gamma)/p_\gamma^2$,
and $p_\gamma$ is defined in Eq.~(\ref{eq:pgamma}).

Before moving on to lower scale $\sceti$, we note that all the Wilson coefficients in SCET$_1$ are of order $\lambda^0$. 
This will change with $\scettwo$ as these factors will determine the relative importance of different contributions.
As we discussed at the very end 
of \subsec{qcdToSCET}, we do not need to compute any suppressed two-gluon operators in $\scetone$ to the order at which we are working.  
Their Wilson coefficient will be $\mo(\lambda^0)$.  Matching this contribution to a two-gluon $\scettwo$ 
operator will leave this factor unchanged as there are no further emissions from it.
The field content in $\scettwo$ will scale $\sim \lambda^8$.  As shown in \eq{PCS21E} though, LO in $\scettwo$ is at $\lambda^5$.

Lastly, we described the effects of adding running effects in \subsec{opRunning}.
In the next Appendix we will match SCET$_1$ to SCET$_2$. Before doing it we have to run the SCET$_1$ operators from 
$Q$ down to $\mu_1$, where we have the first emission:
\begin{align}
\label{eq:S1coeRun2}
C^{(0)}_{0}(n_0)&=U^{(2,0,0)}(n_0;\,Q,\mu) \,  \gamma^\mu_{n_0 \perp}\nonumber  \\
C^{(1)}_{1,\,\rm{NLO}}(n_0,n_0)&=  U^{(2,1,0)}(n_0, n_0;\,Q,\mu)  \otimes\, \frac{n_{0}^{\mu}-\nb^{\mu} }{Q} 
\gamma_{n_0 \perp}^\alpha \, , \nonumber \\
C^{(1)}_{1, \mathcal{T}}(n_0,n_0)  &= U^{(2,1,1)}_{ \mathcal{T}} (n_0,n_0;\,Q,\mu)\otimes\,\frac{1}{\bar{q}_1 \bar{k}_1 }  
\Big( \gamma_{n_0 \perp}^\mu  \gamma_{n_0 \perp}^\beta \gamma_{n_0 \perp}^\alpha 
- \frac{2}{\bar{q}_1 Q} \,g^{\mu\beta} \gamma_{n_0 \perp}^\alpha \Big)\,, \nonumber \\
C^{(1)}_{1} (n_1, n'_1)&= U^{(2,1,0)} (n_1, n'_1;\,Q,\mu) \, \Big( - \frac{2}{ (n_1\!\cdot\!n'_1) \bar{q}_1 \bar{k}_1}
 \gamma_{n_0 \perp}^\alpha \,\slashed{p}_\gamma \,\gamma_T^\mu \nonumber \\
&- \Big[ \frac{1}{(n\!\cdot\! p_{\bar{q}} )\bar{k}_1 } 
\Big( \gamma_T^\mu \, \slashed{p}_\gamma - \bar{q}_1\,n_{1T}^\mu \Big) 
+\frac{2  (n\!\cdot\! p_{\bar{q}} ) }{ (n_1\!\cdot\!n'_1) \bar{q}_1  \bar{k}_1} \nb_T^\mu \Big] \gamma_{n_0 \perp}^\alpha \Big) \, .
\end{align} 
For the definition of the running factors $U^{(i,j,k)}(Q,\mu)$ see
Eqs.~(\ref{eq:rgKernel})-(\ref{eq:runningAsSudakov}), and
(\ref{eq:kernelEq}). The convolution symbol, $\otimes$, is only relevant beyond
LL; that is beyond the level required here.

\section{Matching $\scetone$ to $\scettwo$} 
\label{app:scet1/scet2}

\subsection{One-Gluon Emission}

We now match $\scetone$ to $\scettwo$ for one and two-gluon emissions, starting
with the former.  The basis of $\scettwo$ operators necessary for the matching
up to NNLO($\lambda$) is equal to Eq.~(\ref{eq:SCET1op}), but defined in
SCET$_2$: $\mo^{(0)}_{2}(n_0)$, $\mo^{(1)}_{2}(n_0,n_0)$, $\mathcal{T}^{(1)}_{2}
(n_0,n_0)$,$\mo^{(1)}_2 (n_1, n^\prime_1) $.\footnote{ As before, we do not
  consider operators like $\mo^{(1)}_{2}(n_0,\nb)$ that describe a gluon
  collinear to the antiquark.}  In the previous section, we matched QCD to
SCET$_1$ for one emission and found either a two-jet ($ |
q_{n_0}\,g_{n_0}\,\bar{q}_{\nb} \rangle_1$) or three-jet configuration ($ |
q_{n_1}\,g_{n'_1}\,\bar{q}_{\nb} \rangle_1$), depending on the collinearity of
the external particles.  When we go to $\scettwo$, our definition of
collinearity becomes stricter.  Particles with momenta $p_1$ and $p_2$ are
collinear only if $p_1 \cdot p_2 \lesssim Q^2 \lambda^4/\eta^4$, where $\eta
\sim \frac{1}{2}$ is the average energy loss factor between mother and daughters
discussed in \subsec{sceti}.  As a result of this change, a two-jet
configuration in SCET$_1$ can be matched both onto $ |
q_{n_0}\,g_{n_0}\,\bar{q}_{\nb} \rangle_2$ and $ |
q_{n_1}\,g_{n^\prime_1}\,\bar{q}_{\nb} \rangle_2$ in SCET$_2$.  The three-jet
configuration in SCET$_1$ can, of course, only go to the three-jet state $ |
q_{n_1}\,g_{n^\prime_1}\,\bar{q}_{\nb} \rangle_2$ in SCET$_2$.  The matching is
given by
\begin{align}
\label{eq:S1S2mat0}
  J^\mu_{\rm{QCD}}  
=   & \, C^{(0)}_{1,\,\rm{LO}}(n_0) \mo^{(0)}_{1}  
 + C^{(1)}_{1,\,{\rm NLO}} (n_0,n_0)  \mo^{(1)}_{1} 
 +  C^{(1)}_{1} (n_1, n'_1)    \mo^{(1)}_1 \\
&  +   C^{(1)}_{1, \mathcal{T}} (n_0,n_0)  \mathcal{T}^{(1)}_{1}  + \cdots  \nonumber\\
 \label{eq:S1S2mat}
= & \, C^{(0)}_{2}(n_0) \mo^{(0)}_{2}   
 + C^{(1)}_{2} (n_0,n_0)  \mo^{(1)}_{2}  
 +  C^{(1)}_{2} (n_1, n'_1)    \mo^{(1)}_2  \\
&  +   C^{(1)}_{2, \mathcal{T}} (n_0,n_0)  \mathcal{T}^{(1)}_{2}  + \cdots \nonumber
 \, ,
 \end{align}
where we give the decomposition into both $\scetone$ and $\scettwo$ operators.  
The ellipses indicate higher order terms.  If we close Eq.~(\ref{eq:S1S2mat}) with the state 
$| q_{n_0}\,g_{n_0}\,\bar{q}_{\nb} \rangle_2$, we get
\begin{align}
\label{eq:S1S2coe2j}
   &  C^{(0)}_{1,\,\rm{LO}}(n_0) \int  dx^4 \langle 0|T \{ \mathcal{L}_{\scetone}(x)  \mo^{(0)}_{1}  \} | q_{n_0}\,g_{n_0}\,\bar{q}_{\nb} \rangle_2  \nonumber\\
& + C^{(1)}_{1,\,{rm NLO}} (n_0,n_0) \langle 0 |  \mo^{(1)}_{1} | q_{n_0}\,g_{n_0}\,\bar{q}_{\nb} \rangle_2
  +   C^{(1)}_{1, \mathcal{T}} (n_0,\,n_0) \langle 0 |  \mathcal{T}^{(1)}_{1}  | q_{n_0}\,g_{n_0}\,\bar{q}_{\nb} \rangle_2 \nonumber\\
 &=C^{(0)}_{2}(n_0) \int  dx^4 \langle 0|T \{ \mathcal{L}_{\scettwo}(x)  \mo^{(0)}_{2}  \} | q_{n_0}\,g_{n_0}\,\bar{q}_{\nb} \rangle_2 \nonumber\\
& + C^{(1)}_{2} (n_0,n_0) \langle 0 |  \mo^{(1)}_{2}  | q_{n_0}\,g_{n_0}\,\bar{q}_{\nb} \rangle_2 
  +   C^{(1)}_{2, \mathcal{T}} (n_0,n_0) \langle 0 |  \mathcal{T}^{(1)}_{2} | q_{n_0}\,g_{n_0}\,\bar{q}_{\nb} \rangle_2 \, .
 \end{align}
Since the structure of the operators  in Eq.~(\ref{eq:S1S2coe2j}) is the same on the LHS and RHS, we simply get:
\begin{align}
   \label{eq:S2coe2j}
 C^{(0)}_{2}(n_0)&= C^{(0)}_{1,\,\rm{LO}}(n_0) \, ,\nonumber \\
  C^{(1)}_{2} (n_0,n_0)&= C^{(1)}_{1,\,{\rm NLO}} (n_0,n_0) \, ,\nonumber \\
   C^{(1)}_{2, \mathcal{T}} (n_0,n_0) &= C^{(1)}_{1, \mathcal{T}} (n_0,n_0)  \, .
\end{align}
Acting on Eq.~(\ref{eq:S1S2mat}) with the state $ | q_{n_1}\,g_{n^\prime_1}\,\bar{q}_{\nb} \rangle_2$, we have:
\begin{align}
\label{eq:S1S2coe}
&   C^{(0)}_{1,\,\rm{LO}}(n_0) \int  dx^4 \langle 0|T \{ \mathcal{L}_{\scetone}(x) \mo^{(0)}_{1} \}   | q_{n_1}\,g_{n^\prime_1}\,\bar{q}_{\nb} \rangle_2 
 + C^{(1)}_{1,\,{\rm NLO}} (n_0,n_0) \langle 0|   \mo^{(1)}_{1} | q_{n_1}\,g_{n^\prime_1}\,\bar{q}_{\nb} \rangle_2 \nonumber\\
& +  C^{(1)}_{1} (n_1, n'_1)  \langle 0|  \mo^{(1)}_1  | q_{n_1}\,g_{n^\prime_1}\,\bar{q}_{\nb} \rangle_2
  +   C^{(1)}_{1, \mathcal{T}} (n_0,n_0) \langle 0| \mathcal{T}^{(1)}_{1} | q_{n_1}\,g_{n^\prime_1}\,\bar{q}_{\nb} \rangle_2    \nonumber \\
 &=C^{(1)}_{2} (n_1,n_1^\prime)\langle 0 | \mo^{(1)}_{2} | q_{n_1}\,g_{n^\prime_1}\,\bar{q}_{\nb} \rangle_2 \,.
 \end{align}
We decompose $C^{(1)}_{2}  (n_1,n_1^\prime)$ as 
\begin{align} \label{eq:WCS21E}
C^{(1)}_{2}(n_1,n_1^\prime) = C^{(1)}_{2,\,\rm{LO}}(n_1,n_1^\prime) +  
C^{(1) H,a}_{2,\,\rm{NLO}}(n_1,n_1^\prime)+ C^{(1)H,b}_{2,\,\rm{NLO}}(n_1,n_1^\prime) 
+ C^{(1) H}_{2,\,\rm{NNLO}}(n_1,n_1^\prime) \, ,
\end{align}
where  $C^{(1)}_{2,\,\rm{LO}}$ is the coefficient that reproduces
the first term on the LHS of Eq.~(\ref{eq:S1S2coe}), {\it etc}. 
All the SCET$_2$ coefficients in Eq.~(\ref{eq:S2coe2j}) scale as $\lambda^0$, like in SCET$_1$,
but we will see that those in Eq.~(\ref{eq:WCS21E}) scale differently,
giving the hierarchy indicated in the subscript.  We will show that:
\begin{align} \label{eq:PCS21E}
  C^{(0)}_{2} (n_0) \mo^{(0)}_{2} & \sim \lambda^{4} \, , 
  & C^{(1)}_{2,\,\rm{LO}} (n_1,n_1^\prime) \mo^{(1)}_{2} & \sim \lambda^{5} \, , \\
  C^{(1)}_{2} (n_0,n_0) \mo^{(1)}_{2} & \sim \lambda^{6} \, 
  ,&  C^{(1) H ,\,a}_{2,\,\rm{NLO}}(n_1,n_1^\prime)  \mo^{(1)}_{2} & \sim \lambda^{6} \, ,\nonumber \\
       C^{(1)}_{2, \mathcal{T}} (n_0,n_0)  \mathcal{T}^{(1)}_{2} & \sim \lambda^{8} \, ,
       &C^{(1)H ,\,b}_{2,\,\rm{NLO}}(n_1,n_1^\prime)  \mo^{(1)}_{2} & \sim \lambda^{6} \, ,\nonumber \\
 && C^{(1) H}_{2,\,\rm{NNLO}} (n_1,n_1^\prime) \mo^{(1)}_{2}  & \sim \lambda^{7} \, . \nonumber
\end{align}
In the second column we have only one operator $\mo^{(1)}_{2}  (n_1,n_1^\prime)$
and we have decomposed its coefficient according to Eq.~(\ref{eq:WCS21E}).
The matching does not conserve the power counting,
as collinear $\scetone$ fields scale as $\lambda$, but in $\scettwo$ they go as $\lambda^2$.
For example, we have that the LO operator in SCET$_1$ is $C^{(0)}_{1}  \mo^{(0)}_{1}\sim \lambda^{2}$, 
but for the  LO operator in SCET$_2$ we have $ C^{(0)}_{2} \,\mo^{(0)}_{2}   \sim \lambda^{4}$.
 
If we want to calculate a cross section for a fixed number of external
particles, then we need all the SCET$_2$ operators in Eq.~(\ref{eq:S1S2mat}).
Our interest, though, is in improving shower Monte Carlo, and so we only
calculate operators needed for that ({\it cf.} discussion in
\subsec{loShowerRe}).  To reproduce the LL emission of two gluons, the only
higher dimension operator we need is $C^{(1)}_{2,\,\rm{LO}}(n_1,n_1^\prime)
\mo^{(1)}_{2}$.  The operators $\mo^{(0)}_{2} (n_0) $ and $\mo^{(1)}_{2}
(n_0,n_0)$ only tell us about the no-branching probabilities already determined
by the one-loop cusp anomalous dimension.  For example, $\mo^{(0)}_{2} (n_0) $
describes a quark which does not emit until after the scale of matching $k_{1
  \perp}$.  For this reason, we call $C^{(1)}_{2,\,\rm{LO}}(n_1,n_1^\prime)
\mo^{(1)}_{2}$ our LO operator.  Naively, two-gluon contributions from
$\mo^{(0)}_{2} (n_0) $ and $\mo^{(1)}_{2} (n_0,\,n_0)$ are lower order at
tree-level, but this does not take into account the exponential suppression from
running.  The dominant contribution to showers comes from strong-ordering, not
``every emission as collinear as possible.''  Thus, we build our shower around
$C^{(1)}_{2,\,\rm{LO}} (n_1,n_1^\prime) \mo^{(1)}_{2}$.  The coefficients
$C^{(1), H}_{2,\,\rm{NLO}}(n_1,n_1^\prime)$ and $C^{(1),
  H}_{2,\,\rm{NNLO}}(n_1,n_1^\prime)$ give corrections for one emission.  We
therefore obtain a correction if we run a LL shower based on a matrix element
computed with one of these suppressed terms.

We now turn to calculate the terms in Eq.~(\ref{eq:WCS21E}).
in three steps: first we calculate the amplitudes in  
$\scetone$ on the LHS of (\ref{eq:S1S2coe});
second we rotate it using the finite RPI$_1$ transformations defined in App.~\ref{app:rpi}, 
so that the necessary operators overlap with SCET$_2$ states;
and third  we calculate the Wilson coefficients necessary to match the two sides. 
We do it order by order and we start by calculating the coefficient $C^{(1)}_{2,\,\rm{LO}}$. 
The first term of the LHS of (\ref{eq:S1S2coe}) is 
\begin{align}
 \label{eq:1eALOS1}
A^{q \bar{q} g}_{\rm{LO}} = U^{(2,0,0)}(n_0;\,Q,\mu_1)\,g\,\bar{\xi}_{n_0} \left( n_0^{\alpha}
 + \frac{(\slashed{q}_1)_{n_0 \perp}\gamma^{\alpha}_{n_0 \perp}}{\bar{q}_1} \right)    \gamma^{\mu}_{n_0 \perp} \xi_{\bar{n}}\, ,
\end{align}
where $U^{(2,0,0)}(n_0)$ is the running factor ({\it cf.} Eqs.~\ref{eq:rgKernel}, \ref{eq:oneLoopCusp}, and \ref{eq:kernelEq}), 
and $\mu_1\sim \lambda Q$ is at the scale of the emission.  In (\ref{eq:1eALOS1}), we have omitted the terms proportional to $\nb^\alpha$ as they are unnecessary for matching.  Gauge invariance
constrains all appearances of $\nb \cdot A_n$ to come from the Wilson lines in $\chi$ and $\mb$. 
The amplitude is written in terms of objects projected in the $n_0$ and $\nb$ directions. 
As discussed in Appendix \ref{app:rpi}, 
these directions are not suitable for a $\scettwo$ states, but we can use the formulas (\ref{eq:spinSqSn}) 
and write (\ref{eq:1eALOS1}) in terms of the directions $n_{1}$ and $n'_{1}$
where the quark and gluon have zero perpendicular component,  this gives
\beq
A^{q \bar{q} g}_{\rm{LO}} = U^{(2,0,0)}(n_0;\,Q,\mu_1)\,g\, \bar{\xi}_{n_1} \left( n_{1}^{\prime\alpha} + 2\frac{(q_1)_{n_0 \perp}^\alpha}{\bar{k}_1} 
+ \frac{(\slashed{q}_1)_{n_0 \perp}\gamma^{\alpha}_{n'_1 \perp}}{\bar{q}_1} \right)  \frac{\bar{q}_0}{q_0^2}    \gamma^{\mu}_{n_0 \perp} \xi_{\bar{n}}\, .
\label{eq:1eALOS2}
\eeq
In (\ref{eq:1eALOS2}) we have rotated the spinor in the $n_{1}$ direction, 
$\gamma_{n_0 \perp}$ in the $n'_1$ direction 
and we have written $n_{0}$ in terms of $n_1$,  $n'_{1}$ and $(q_1)_{n_0 \perp}$. 
We have dropped all the terms proportional to $\bar{n}^\alpha$ and we made use of relations $\nbs \nbs=0$
and $ \bar{\xi}_{n_{1}} \ns_{1}=0$.
Since the gluon momentum is parallel to $n_1^{\prime \mu}$, 
only the polarizations in the perpendicular direction with respect to $n_1^{\prime \mu}$ are physical, 
thus we can neglect the term proportional to $n_1^{\prime }$ in Eq.~(\ref{eq:1eALOS2}).
The $\scettwo$  amplitude
$ \langle 0|  \bar{\chi}_{n_1}g \mathcal{B}_{n'_1 \perp}^\alpha \chi_{\nb}|q_{n_1}g_{n'_1} \bar{q}_{\nb} \rangle_2$
is
\begin{align}
\label{eq:1eS2LO}
\langle 0|  \bar{\chi}_{n_1} g \mathcal{B}_{n'_1 \perp}^\alpha \chi_{\nb}|q_{n_1} g_{n'_1} \bar{q}_{\nb}\rangle_2
= g \, \bar{u}_{n_1}  \epsilon_{n'_1 \perp}^{\alpha} v_{\nb}\, ,
 \end{align}
 where in Eq.~(\ref{eq:1eS2LO}) we have explicitly written the polarization
 vector for the gluon.  From Eq.~(\ref{eq:1eALOS2}) and Eq.~(\ref{eq:1eS2LO}),
 we can see that the LO Wilson coefficient is
\begin{align}
\label{eq:CLOS2}
 C^{(1)}_{2,\,\rm{LO}}  &= U^{(2,0,0)}(n_0;\,Q,\mu_1)  
 c^{\alpha}_{\rm{LO}}(n_0) 
 \frac{\bar{q}_0}{q_0^2}  \gamma^{\mu}_{n_0 \perp} \, ,
\end{align}
where
 \begin{align}
 c^{\alpha}_{\rm{LO}}(n_0)  =   \left(  2\frac{(q_1)_{n_0 \perp}^\alpha}{\bar{k}_1} 
+ \frac{(\slashed{q}_1)_{n_0 \perp}\gamma^{\alpha}_{n'_1 \perp}}{\bar{q}_1} \right)\frac{\nbs \ns_0}{4}
    \,   \Theta_{ \delta_2} [n_1 \cdot n'_1 ] \, .
 \label{eq:Calpha}
\end{align}
The difference with \eq{loCoeff} is that we replaced $n_0$ in terms of external
vectors.  $\Theta_{ \delta_2 } [n_1 \cdot n'_1 ] $ is the phase space cutoff
that guarantees $(n_1 \cdot n'_1) \lesssim \lambda^2/\eta^4$,\footnote{The
  factor of $\eta \simeq \frac{1}{2}$ tracks the average energy loss between
  mother and daughter.  In choosing appropriate values for the parameters
  $\delta_k$ in the numerical implementation of $\Theta$ it is important to
  track these $\eta$ factors in the scaling of $n_1 \cdot n'_1$.}  we will say
more about it below.  Since this comes from matching to a SCET$_1$ operator,
$(q_1)_{n_0 \perp} \sim \lambda$ and $q_0^2 \sim \lambda^2$, thus $
C^{(1)}_{2,\,\rm{LO}}$ scales as $\lambda^{-1}$.  Using formulas (\ref{nrel}),
we can write (\ref{eq:CLOS2}) only in terms of $n_{1}$ and $n'_1$, this gives
\begin{align}
\label{eq:C1S2}
 C^{(1)}_{2,\,\rm{LO}}&= U^{(2,0,0)}(n_0;\,Q,\,\mu_1) \left( \frac{\bar{q}_1}{Q}  \sqrt{n_1 \cdot n'_1}v_1^{\alpha} 
+\frac{\bar{k}_1}{ 2Q}  \sqrt{n_1 \cdot n'_1} \slashed{v}_1   \, \gamma^{\alpha}_{n'_1 \perp}     \right)
 \frac{2 \bar{q}_0}{ (n_1\!\cdot\!n'_1)  \,\bar{q}_1 \bar{k}_1} \nonumber \\
 &\times \big( \gamma^{\mu}_{n'_1}  - \nb^\mu \frac{1}{2}\frac{\bar{q}_1}{\bar{q}_0} \sqrt{n_1 \cdot n'_1} \slashed{v}_1 \big)  \, ,
\end{align}
where $v_1^\mu$ is defined in Eq.~(\ref{eq:v}), $\bar{q}_1+\bar{k}_1=Q$. For $\mu_1$, 
as explained in \subsec{loShowerRe}, we take it at the scale of $(k_1)_{n_0\perp}$ 
as in \eq{muk2} for $k=1$.
Since $|v_1^2|=2$,  the power counting of (\ref{eq:C1S2}) is given by the scalar product $n_1 \cdot n'_1$,
that is $\mo(\lambda^2)$.
In a similar way, we can calculate  $  C^{(1) H,a}_{2,\,\rm{NLO}} $ and  $ C^{(1) H}_{2,\,\rm{NNLO}}  $.

We have done the matching  starting from the vector current $J^\mu_{\rm{QCD}}=\bar{q}\,\gamma^\mu q$.
If we had started from a general structure, $\bar{q}\, \Gamma^\mu q$,
the results (\ref{eq:CLOS2}) for $C^{(1)}_{2,\,\rm{LO}}$ would have been the same upon the substitution 
\begin{align}
\label{eq:JetSub}
 \gamma^\mu_{n_0 \perp} \rightarrow  \Gamma^\mu\, .
\end{align}
We can obtain $ C^{(1)}_{2,\,\rm{LO}}(n_1, n'_1) \mo^{(2)}_{2}$ from the SCET$_1$
operator $\bar{\chi}_{n_0} \Gamma^\mu \chi_{\bar{n}}$ by running down from $Q$ to
$\mu_1$, multiplying by the factor $U^{(2,0,0)}(n_0;\,Q,\,\mu_1)$, and subsequently 
using the replacement rule
\begin{align}
\label{eq:repRuleS2}
(\chib_{n_{0}})_i \rightarrow  \,
( c^{\alpha}_{\rm{LO}}(n_0))_{ji} (\bar{\chi}_{n_{1}})_j \,g\mb_{\alpha}^{n'_1 \perp}\, .
\end{align}
The coefficients $C^{(1)H,\,a}_{2,\,\rm{NLO}}$, $C^{(1)H,\,b}_{2,\,\rm{NLO}}$,
$C^{(1)H}_{2,\,\rm{NNLO}}$, however, are sensitive to the particular QCD
current.  This is why we refer to them as hard-scattering corrections, denoted
by the superscript, $H$.

For the NLO($\lambda$) and NNLO($\lambda$) amplitudes in the second and third
line of the LHS of Eq.~(\ref{eq:S1S2coe}) we have
\begin{align}
\label{eq:1eANLOS1}
A^{q\bar{q}g}_{\rm{NLO}}&= U^{(2,1,0)}(n_0, n_0;\,Q,\,\mu_1)
\frac{n_0^\mu - \bar{n}^\mu}{  Q}g\bar{u}_{n_0} \gamma_{n_0 \perp}^\alpha v_{\bar{n}}
=\frac{n_0^\mu - \bar{n}^\mu}{ Q}g\bar{u}_{n_1} \gamma_{n'_1 \perp}^\alpha  v_{\bar{n}} \,, \nonumber\\
A^{q\bar{q}g}_{\rm{NNLO}}&= U^{(2,1,1)}(n_0, n_0;\,Q,\,\mu_1) \nonumber \\
& \times \Big( \frac{1}{\bar{q}_1} +\frac{1}{\bar{k}_1} \Big)  \frac{1}{Q}g\bar{u}_{n_0}  \gamma_{n_0 \perp}^\mu (\slashed{k}_1)_{n_0 \perp}
                           \gamma^\alpha_{n_0 \perp} v_{\bar{n}}
- \frac{2 }{\bar{q}_1 Q} g\bar{u}_{n_0} (k_1)_{n_0 \perp}^\mu    \gamma^\alpha_{n_0 \perp} v_{\bar{n}} \nonumber \\
                                        &=U^{(2,1,1)}(n_0, n_0;\,Q,\,\mu_1) \nonumber \\
&      \times                                     \Big( \frac{1}{\bar{q}_1} +\frac{1}{\bar{k}_1} \Big) \frac{1}{Q}g \bar{u}_{n_1}  
\gamma_{n_0 \perp}^\mu (\slashed{k}_1)_{n_0 \perp}
                           \gamma^\alpha_{n'_1 \perp} v_{\bar{n}}
- \frac{2 }{\bar{q}_1 Q} g\bar{u}_{n_1}  (k_1)_{n_0 \perp}^\mu    \gamma^\alpha_{n'_1 \perp} v_{\bar{n}} \,.
\end{align}
The SCET$_2$ coefficients needed to reproduce the amplitudes in
Eq.~(\ref{eq:1eANLOS1}) are:
\begin{align}
\label{eq:C2S2}
C^{(1)H,a}_{2,\,\rm{NLO}}   & =   U^{(2,1,0)}(n_0, n_0;\,Q,\,\mu_1) \otimes c^{H,a}_{2,\,\rm{NLO}}(n_0, n_0), \\
 C^{(1)H}_{2,\,\rm{NNLO}}  &=U^{(2,1,1)}_{\mathcal{T}}(n_0, n_0;\,Q,\,\mu_1) \otimes
 c^{H}_{2,\,\rm{NNLO}}(n_0, n_0), \nonumber  
\end{align}
where
\begin{align}
c^{H,a}_{2,\,\rm{NLO}}(n_0, n_0) & =  U^{(2,1,0)}(n_0, n_0;\,Q,\,\mu_1)
 \frac{n_0^\mu - \bar{n}^\mu}{ Q}   \gamma_{n_0 \perp}^\alpha  \Theta_{ \delta_2} [n_1 \cdot n'_1]  \\
&=  U^{(2,1,0)}(n_0, n_0;\,Q,\,\mu_1)\frac{1}{Q}\bigg[\frac{\bar{k}_1
  n_1^{\prime\mu} + \bar{q}_1 n_1^\mu}{\bar{q}_0} 
 - \Big( 1+\frac{\bar{q}_1 \bar{k}_1}{2\, \bar{q}_0^2} (n_1\!\cdot\!n'_1 ) \Big)
  \nb^\mu \bigg]
\nonumber\\
&\qquad
\times \gamma_{n'_1 \perp}^\alpha
  \Theta_{ \delta_2} [n_1 \cdot n'_1 ]  \, ,\nonumber \\
 c^{H}_{2,\,\rm{NNLO}}(n_0, n_0)  &= U^{(2,1,1)}_{\mathcal{T}}(n_0, n_0;\,Q,\,\mu_1)\!
   \bigg[ \Big( \frac{1}{\bar{q}_1} +\frac{1}{\bar{k}_1} \Big)  \frac{1}{Q} \gamma_{n_0 \perp}^\mu (\slashed{k}_1)_{n_0 \perp}
                           \gamma^\alpha_{n'_1 \perp} 
                  \!     \!- \!\frac{2 }{\bar{q}_1 Q}   (k_1)_{n_0 \perp}^\mu
                  \! \gamma^\alpha_{n'_1 \perp}  \bigg]
   \nonumber\\
 &\qquad\times
                 \Theta_{ \delta_2} [n_1 \cdot n'_1 ]  \nonumber  \\
                           &= - U^{(2,1,1)}_{\mathcal{T}}(n_0, n_0;\,Q,\,\mu_1)
                           \Big(  \frac{1}{2 \,Q} \Big(\gamma^\mu_{n'_1 \perp} \sqrt{n_1 \cdot n'_1} \slashed{v}_1
                           + \nb^\mu \frac{\bar{q}_1}{Q} (n_1 \cdot n'_1) \Big)\gamma^\alpha_{n'_1 \perp}\nonumber \\
                          &\ \ +\frac{\bar{k}_1}{ Q^2} \Big(   \sqrt{n_1 \cdot n'_1} v_1^\mu 
- \bar{n}^\mu \,(n_1\!\cdot\! n'_1)\frac{  (\bar{k}_1^2 -\bar{q}_1^2)}{2\, Q^2} \Big)  \gamma^\alpha_{n'_1 \perp} \Big) 
 \Theta_{ \delta_2} [n_1 \cdot n'_1 ]  \,. \nonumber
\end{align}
The coefficients scale $C^{(1)H,a}_{2,\,\rm{NLO}} \,\sim\, \lambda^0$ and $
C^{(1)H}_{2,\,\rm{NNLO}} \sim \lambda$.  As discussed below \eq{S1coeRun}, we
have a convolution because SCET fields collinear to the same direction can
exchange longitudinal momentum during the running.  However, this convolution is
only needed beyond the LL level considered here.

For the coefficient $C^{(1)H,b}_{2,\,\rm{NLO}}$, the matching comes from the
SCET$_1$ three-jet operator where $n_1 \cdot n'_1 \,\sim\,
\lambda^0/\eta^4$.\footnote{With our conventions where $n_i\cdot\nb=2$, two well
  seperated directions $n_1$ and $n_2$ really do give $n_1\cdot n_2\sim 16$.}
Since the $n$-labels in $C^{(1)}_{1}(n_1, n'_1)$ are already parallel to the
external particles, we can simply write:
\begin{align} 
\label{eq:S23j}
C^{(1)H,b}_{2,\,\rm{NLO}}(n_1, n'_1) = C^{(1)}_{1}(n_1, n'_1)  \tilde{\Theta}_{ \delta_2} [n_1 \cdot n'_1 ] \, ,
\end{align}
where $\tilde{\Theta}_{\delta_2}  [n_1 \cdot n'_1 ]$ only has support for $(n_1 \cdot n'_1)>\lambda/\eta^4$, 
where it is equal to 1.
Knowing that for this term, $n_1 \cdot n'_1 \sim \lambda^0/\eta^4$,  $C^{(1)H,b}_{2,\,\rm{NLO}}(n_1, n'_1)$ scales
$\sim \lambda^0$, and
\begin{align}
\label{eq:CS23j}
C^{(1)H,b}_{2,\,\rm{NLO}}(n_1, n'_1)  \mo^{(1)}_{2} \sim \lambda^{6} \, .
\end{align}
In keeping with our conventions, we keep track of dependence on $\eta \sim
\frac{1}{2}$ for our $\Theta$ functions and their dot product arguments, where
the various $2^n$ factors affect where the step function turns over.  We do not
include them in the power counting for operators, where $\lambda$ parametrizes
strong-ordering and the deviations from it.  Accounting for $\eta$ here is
certainly possible, but in the end we always will compare amplitudes with the
same number of external particles, so $\eta$ factors from operators will not
play any role.

The operator $\mathcal{O}_2^{(1)}(n_1, n'_1)$
only knows that  $n_1\cdot n'_1 > \lambda^4/\eta^4$, it is not able to distinguish its two-jet contributions,
\eq{C1S2} and \eq{C2S2}, from its three-jet one, \eq{S23j}.  
This information must then be in the Wilson coefficients, and we have put it
in the functions $\Theta$ and $\tilde{\Theta}$, first described in \subsec{loShowerRe}. 
We can think of $\Theta_{ \delta_2} [x ] $ as usual theta function:
$\Theta_{ \delta_2} [x ]= \theta[\delta_2 - x]$ 
and $\tilde{\Theta}_{ \delta_2} [x ] =1-\Theta_{ \delta_2} [x]$,
but for integrating phase space, this can lead to numerical problems.
Instead, we can use a smoother theta function, such as the following, plotted in Fig.~\ref{PlotTheta-3JB}
\begin{equation}
\Theta_{\Lambda, a}(x) =
\left\{
\begin{array}{ll}
0 & \text{if }  x< \Lambda -a \\
-\frac{\text{Sign}(x-\Lambda)}{2} e^{2+\frac{2 a \text{Sign}(x-\Lambda) }{(x-\Lambda) - a \, \text{Sign}(x-\Lambda) }}
+\frac{\text{Sign}(x-\Lambda)+1}{2}
& \text{if } \Lambda-a<  x< \Lambda +a \\
1 & \text{if } x> \Lambda +a \\
\end{array} \\
\right. \, ,
\label{eq:thetaEx}
\end{equation}
The parameter $\Lambda$ determines where the function switches from 0
to 1, and $a$ governs how fast it does it.  For the SCET$_2$
coefficients, we have $\Lambda \simeq \delta_2$.  In order to have
$n_1 \cdot n'_1 \lesssim \lambda^2/\eta^4$, we need $ \lambda^2/\eta^4
\ll \delta_2 < 1/\eta^{4} $, so we choose $\delta_2 = \lambda/\eta^4 $.  When we
go down to lower SCET$_i$, in general the Wilson coefficient has to
encode the that either $ n_i \cdot n_j \,\leq\, \lambda^{2( i-1)}/\eta^4$ or
$ n_i \cdot n_j \,>\, \lambda^{2( i-1)}/\eta^4$, in order to do so, we will
use $ \Theta_{\delta _i}$ where $\delta _i = \lambda^{2 i-3}/\eta^4$.  To see
how this $\Theta$ works, we look at the amplitude squared up to
NLO($\lambda$).\footnote{We perform some trivial azimuthal integrals
in order to eliminate some terms that will drop out of typical
observables.  Also, by NLO($\lambda$) corrections for amplitudes
squared, we mean suppressed by two powers of $\lambda$.  Since there
are no odd powers of $\lambda$ in the expansion, this means
NLO($\lambda$) for the cross section.}  The LO amplitude squared is
\begin{align}
\label{eq:AmSqLO}
|A^{q\bar{q}g}|_{\rm{LO}}^2&=|C^{(1)}_{2,\rm{LO}}(n_1,n^\prime_1)|^2
G(q_1,k_1,k_2,p_{\bar{q}}) \, , 
\end{align}
where
\begin{align}
G(q_1,k_1,k_2,p_{\bar{q}})&= \; {}_2\langle q_{n_1}\,g_{n^\prime_1}\,\bar{q}_{\nb} |   \mo^{(1) \dagger}_{1} (n_1,n'_1) |0\rangle 
\langle 0| \mo^{(1)}_{1} (n_1,n'_1)| q_{n_1}\,g_{n^\prime_1}\,\bar{q}_{\nb} \rangle_2 \, .
\end{align}
The NLO($\lambda$) amplitude squared is
\begin{align}
\label{eq:AmSqNLO}
|A^{q\bar{q}g}|_{\rm{NLO}}^2 &= |A^{q\bar{q}g}|_{\rm{NLO},\,2-jet}^2+|A^{q\bar{q}g}|_{\rm{NLO},\,3-jet}^2\, , 
\end{align}
where
\begin{align}
\label{eq:AmSqNLO2}
|A^{q\bar{q}g}|_{\rm{to \,NLO},\,2-jet}^2 &= ( C^{(1)\dagger}_{2,\rm{LO}}(n_1,n^\prime_1) 
C^{(1), H}_{2,\rm{NNLO}}(n_1,n^\prime_1) \nonumber \\
& +C^{(1), H\dagger}_{2,\rm{NNLO}}(n_1,n^\prime_1)
C^{(1)}_{2,\rm{LO}}(n_1,n^\prime_1) \nonumber \\
&+|C^{(1), Ha}_{2,\rm{NLO}}(n_1,n^\prime_1)|^2  )
G(q_1,k_1,k_2,p_{\bar{q}}) \, , \nonumber \\
|A^{q\bar{q}g}|_{\rm{NLO},\,3-jet}^2&=|C^{(1), Hb}_{2,\rm{NLO}}(n_1,n^\prime_1)|^2  
G(q_1,k_1,k_2,p_{\bar{q}}) \, .
\end{align} 
In Fig.~\ref{Plot:Plot_A22jvsQCDB} we plot the ratios  
$|A^{q\bar{q}g}|_{\rm{LO}}^2/|A^{q\bar{q}g}|_{\rm{QCD}}^2$  
and $(|A^{q\bar{q}g}|_{\rm{LO}}^2+|A^{q\bar{q}g}|_{\rm{NLO},\,2-jet}^2)/|A^{q\bar{q}g}|_{\rm{QCD}}^2$ versus $(k_1)_{n_0\perp}$,  
We note that including NLO($\lambda$) corrections extends the region where tree-level SCET$_2$ and QCD agree.
In Fig.~\ref{Merge2J-3JB}, we plot the the merging of the  two-jet and and three-jet amplitude squared using the theta function.
Although we have not undertaken any systematic study of how our phase space cutoff enters observables, we take 
Fig.~\ref{Merge2J-3JB} as visual evidence of minimal sensitivity.
\begin{figure}[ht!]
\centering
\includegraphics[width=0.8\textwidth]{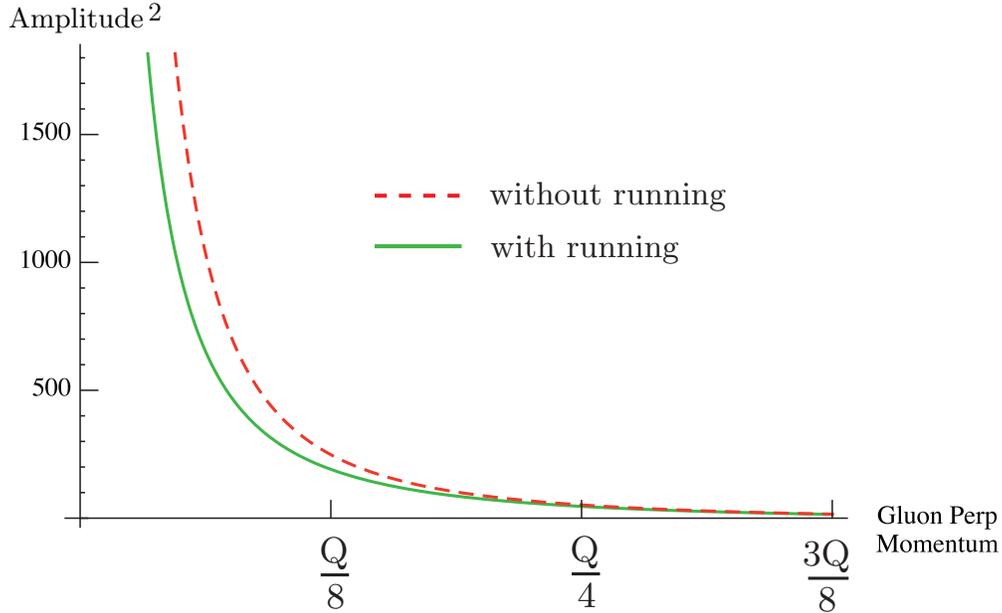}
\caption[Plot of the SCET$_2$ amplitude squared up to NLO]
{ Plot of the SCET$_2$ amplitude squared up to NLO, $|A^{q\bar{q}g}|_{\rm{LO}}^2+|A^{q\bar{q}g}|_{\rm{NLO}}^2$,  
with (green) and without (red) running factors versus $(k_1)_{n_0\perp}$ for $\bar{k}_1/\bar{q}_0=0.4$.} 
\label{A2running}
\end{figure} 
Lastly, in Fig.~\ref{A2running} we plot $|A^{q\bar{q}g}|_{\rm{LO}}^2+|A^{q\bar{q}g}|_{\rm{NLO}}^2$ with and without
running factors.  As expected, the latter is suppressed relative to the former.

\subsection{Two-Gluon Emission}

\begin{figure}[t!]
\centering
\includegraphics{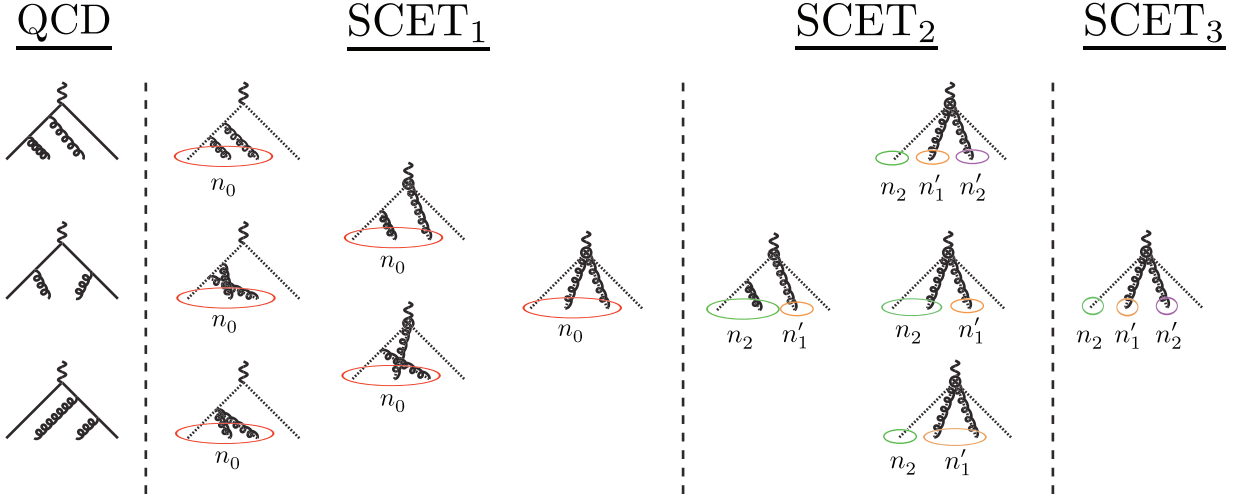}
\caption{Matching SCET$_1$ to SCET$_2$ to SCET$_3$ for two emissions
to the two-jet configuration in SCET$_1$.  We organize here by column number:
(1) QCD Feynman diagrams; (2) SCET$_1$ diagrams
from the operator $\mo^{(0)}_1(n_0)$; (3) SCET$_1$
diagrams from the operators $\mo^{(1)}_1(n_0, n_0)$ and $
\mathcal{T}^{(1)}_{1} (n_0, n_0) $; (4) SCET$_1$
diagram from the operator $\mo^{(2)}_1(n_0, n_0,n_0)$, this operator
contributes only at N$^3$LO to the SCET$_2$ matching; (5) 
SCET$_2$ diagram from the operator $\mo^{(1)}_1(n_2,
n^\prime_1)$; (6) SCET$_2$ diagrams from the
operators $\mo^{(2)}_2(n_2, n^\prime_1, n^\prime_2)$,
$\mo^{(2)}_2(n_2, n'_1, n_2)$ and $\mo^{(2)}_2(n_2, n^\prime_1,
n^\prime_1)$; (7) SCET$_3$ diagram from the
operator $\mo^{(2)}_3(n_1, n^\prime_1, n^\prime_2)$. }
\label{fig:figureS1S2S3twoemiC}
\end{figure}

We show the Feynman diagrams corresponding to the operators needed for matching 
two-gluon emission in Fig.~\ref{fig:figureS1S2S3twoemiC}.
As discussed at the very end of of \subsec{qcdToSCET} and in \app{QCD/SCET1}, we
do not need the two-gluon, $\scetone$ operator, $\mo_1^{(2)}(n_0,n_0,n_0)$ at
this order.  Thus, the ones in (\ref{eq:SCET1op}) are sufficient.

The $\scettwo$ basis has the following two gluon operators:
\begin{align} 
\label{eq:S2BBa} \mo^{(2)}_2(n_2,n'_1,n_2)=\bar{\chi}_{n_2}g \mb_{n'_1\perp}^\alpha g \mb_{n_2\perp}^\beta \chi_{\nb} \, ,  \\
 \mo^{(2)}_2(n_2,n'_1,n'_1)=\bar{\chi}_{n_2} g\mb_{n'_1\perp}^\alpha  g\mb_{n'_1\perp}^\beta \chi_{\nb} \, , \nonumber  \\
   \mo^{(2)}_2(n_2,n'_1,n'_2)=\bar{\chi}_{n_2}g \mb_{n'_1\perp}^\alpha g \mb_{n'_2\perp}^\beta \chi_{\nb} \, , \nonumber \\
 \mo^{(2)}_2(n_0,n_0,n_0)=\bar{\chi}_{n_0}g \mb_{n_0\perp}^\alpha g \mb_{n_0\perp}^\beta \chi_{\nb} \, . \nonumber
 \end{align}
The last operator in (\ref{eq:S2BBa}) is not necessary for
the matching at NNLO($\lambda$).  It can only be closed with states $ |q_{n_0}
g_{n_0} g_{n_0} \bar{q}_{\nb} \rangle_2$ having
both gluons collinear in $\scettwo$.  Its coefficient can only come from the
SCET$_1$ operator $ \mo^{(2)}_1(n_0,n_0,n_0)$.  Any contribution involving
$\scetone$ Lagrangian emission that matches to a higher-dimension operator
in $\scettwo$ will necessarily have some partons in different $\scettwo$
directions, {\it e.g.} $(n_0,n_0,n'_1)$.
Since $C^{(2)}_1(n_0,n_0,n_0) \sim \lambda^0$, and the matching does not change this, 
$\mo^{(2)}_2(n_0,n_0,n_0)$ contributes at N$^3$LO.  The Wilson coefficients of the operators
(\ref{eq:S2BBa}) are defined such that
\begin{align}
\label{eq:S1S2match0}
J^\mu_{\rm{QCD}} = 
&  \,C^{(0)}_{1,\,\rm{LO}}(n_0) \mo^{(0)}_{1} + C^{(1)}_{1}(n_0,n_0) \mo^{(1)}_{1}
+ C^{(1)}_{1, \mathcal{T}}  \mathcal{T}^{(1)}_{1} \\
&+C^{(1)}_{1}(n_1,n^\prime_1) \mo^{(1)}_{1} + \cdots  \nonumber \\
=&\,C^{(0)}_{2}(n_0) \mo^{(0)}_{2} + C^{(1)}_{1}(n_0,n_0)
\mo^{(2)}_{2} + C^{(2)}_{2, \mathcal{T}}(n_0, n_0)
\mathcal{T}^{(2)}_{2} \nonumber \\ 
& +C_{2}^{(1)} (n_1, n^\prime_1)
\mo^{(1)}_2  + C_{2}^{(2)} (n_2,n'_1,n_2) \mo^{(2)}_2 \nonumber \\ 
&+C_{2}^{(2)} (n_2,n'_1,n'_1) \mo^{(2)}_2 + C_{2}^{(2)}(n_2,n'_1,n'_2)\mo^{(2)}_2 + \cdots \nonumber
\end{align}
where we have written the QCD current in terms of SCET$_1$ and SCET$_2$ operators.  The
ellipses indicate higher order terms.
  
We divide the subleading Wilson coefficients in two categories: jet-structure and
hard-scattering, labeling their contributions with the superscripts
$J$ and $H$.  As mentioned previously, the latter come from suppressed
operators in the QCD $\rightarrow \scetone$ matching and depend on the
details of the hard partons' creation.  The former are subleading
terms from the $\scetone$ Lagrangian that correct \eq{loArrow} as
we match to lower-scale theories.  They are completely independent of
the initial hard process.

We have seen in the previous section that the LO single gluon
coefficient$\times$operator is
$C^{(1)}_{2}(n_1,n'_1) \mo^{(1)}_2 \sim \lambda^5$, Eq.~(\ref{eq:PCS21E}).  We are interested in calculating
the amplitude squared to NLO($\lambda$).  We therefore only need to calculate those
NNLO($\lambda$) contributions that can interfere with the LO amplitude. 
These operators are of the form $\mo^{(2)}_2 (n_2,n_2,n'_1)$, as the others in 
\eq{S2BBa} are not strongly-ordered.\footnote{In principle, we also have
$\mathcal{T}^{(2)}_2 (n_2,n_2,n'_1)$.  However, the field content alone makes
this $\lambda^{10}$, but all our correction operators have Wilson coefficients
at $\mo(\lambda^{-2})$, so its contribution is beyond NNLO($\lambda$).}

We now calculate the coefficients in (\ref{eq:S2BBa}), 
starting with $C^{(2)}_{2} (n_2,n'_1,n'_2)$, which we decompose as:
\begin{align} 
\label{eq:coeJH1}
C^{(2)}_{2} (n_2,n'_1,n'_2)&=C^{(2) J}_{2,\,\rm{NLO}}(n_2,n'_1,n'_2)
+  C^{(2)  H}_{2} (n_2,n'_1,n'_2)\,,
\end{align}
where
\begin{align}
 &C^{(2) J}_{2,\,\rm{NLO}}(n_2,n'_1,n'_2)  \langle 0|\mo^{(2)}_2 |q_{n_2} g_{n_1^\prime} g_{n_2^\prime} \bar{q}_{\nb} \rangle_2,
 \label{eq:S2CoeA}
 =  \\
  &C^{(0)}_{1,\,\rm{LO}} (n_0)\int \! \! dx_1 dx_2 \langle 0| T \{ \mathcal{L}_{\scetone}(x_1) \mathcal{L}_{\scetone}(x_2) \mo^{(0)}_{1} (n_0, n_0) \}  
   |q_{n_2} g_{n_1^\prime} g_{n_2^\prime} \bar{q}_{\nb} \rangle_2 \,  \nonumber 
\end{align}
and
\begin{align}
 & C^{(2) H}_{2}(n_2,n'_1,n'_2)  \langle 0|\mo^{(2)}_2 |q_{n_2} g_{n_1^\prime} g_{n_2^\prime} \bar{q}_{\nb} \rangle_2 
 \label{eq:S2CoeA1}
 =  \\
  & + C^{(1)}_{1} (n_0, n_0) \int \! \! dx\langle 0|   T \{ \mathcal{L}_{\scetone}(x) \mo^{(1)}_{1} \}  
  |q_{n_2} g_{n_1^\prime} g_{n_2^\prime} \bar{q}_{\nb} \rangle_2  \nonumber \\
& + C^{(1)}_{1, \mathcal{T}}(n_0, n_0) \int \! \! dx  \langle 0| T \{ \mathcal{L}_{\scetone}(x) \mathcal{T}^{(1)}_{1}  \}
 |q_{n_2} g_{n_1^\prime} g_{n_2^\prime} \bar{q}_{\nb} \rangle_2    \nonumber \\
& + C^{(1)}_{1}(n_1, n^\prime_1)  \int \! \! dx\langle 0|   T \{ \mathcal{L}_{\scetone}(x) \mo^{(1)}_{1} \}  
  |q_{n_2} g_{n_1^\prime} g_{n_2^\prime} \bar{q}_{\nb} \rangle_2 \, . \nonumber
\end{align}
We decompose $C^{(2)H}_{2} (n_2,n'_1,n'_2)$ as
\begin{align} 
 \label{eq:coeH1}
 C^{(2)  H}_{2} (n_2,n'_1,n'_2)&= C^{(2) H,\, a}_{2,\,\rm{NNLO}} (n_2,n'_1,n'_2)
 +C^{(2) H}_{2,\, \rm{N^3LO}}(n_2,n'_1,n'_2) +  C^{(2) H,\,b}_{2,\,\rm{NNLO}} (n_2,n'_1,n'_2)\,,
\end{align}
where  $C^{(2) H,\, a}_{2,\,\rm{NNLO}} (n_2,n'_1,n'_2)$ is the coefficient  
that reproduces the the second line in Eq.~(\ref{eq:S2CoeA1}), {\it etc.}

Since $\mo^{(2)}_2(n_2,n'_1,n'_2)$ does not interfere with the LO
operator, we only need the coefficient, $C^{(2)J}_{2,\,\rm{NLO}}
(n_2,n'_1,n'_2)$.  We also calculate $C^{(2)H,\,a}_{2,\,\rm{NNLO}}
(n_2,n'_1,n'_2)$ though, because it will be useful later.  We prove below
that these coefficients and their corresponding operators are of order
$\lambda^6$ and $\lambda^7$, respectively (in Eq.~\ref{eq:PCS21E}, we
show that LO is at $\lambda^5$).  $C^{(2)J}_{2,\,\rm{NLO}}(n_2,n'_1,n'_2)$ and
$C^{(2)H,\,a}_{2,\,\rm{NNLO}} (n_2,n'_1,n'_2)$ come from two-jet
operators in SCET$_1$.  Thus, they both contain factors of
$\Theta_{\delta_2}[n_2 \cdot n'_1] \Theta_{\delta_2}[n_2 \cdot n'_2]
\Theta_{\delta_2}[n'_2 \cdot n'_1]$.  We first described these phase
space cutoffs in \subsec{loShowerRe}, and made use of them in previous
section on single-gluon matching.  The subscript, $\delta_2$,
constrains the argument to be $\lesssim \lambda/\eta^4$.

\begin{figure}[t!]
\centering
\includegraphics{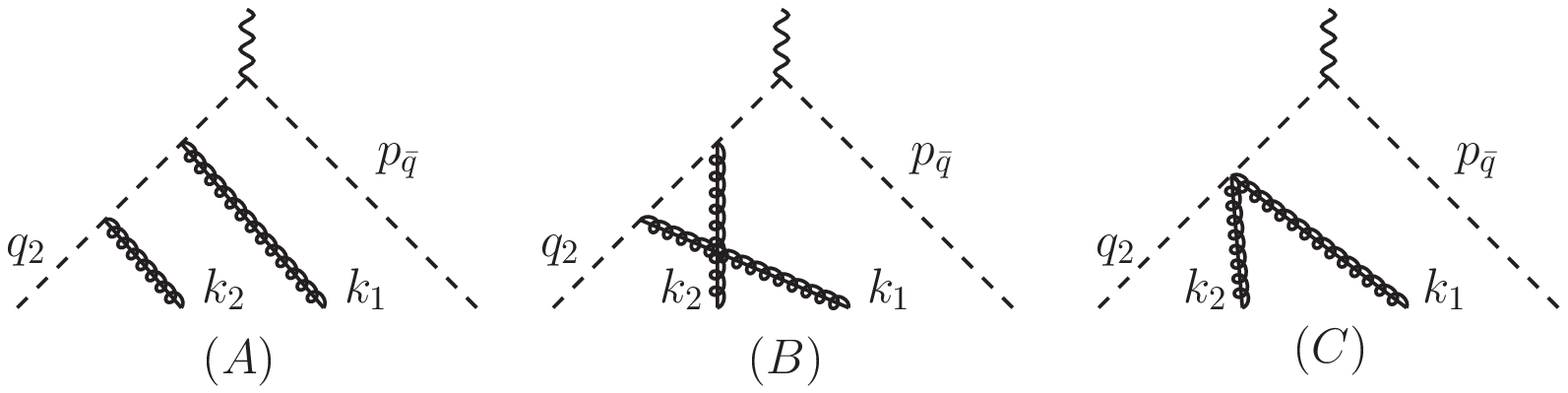}
\caption{Amplitudes for two emissions in $\scetone$ from the operator $ \mo^{(0)}_{1}$.} 
\label{fig:figureABCC}
\end{figure}

To calculate the coefficients, we proceed as with
one-gluon emission: on the LHS of Eqs.~(\ref{eq:S2CoeA}) and
(\ref{eq:S2CoeA1}) we calculate the SCET$_1$ amplitude and rotate it
along the directions $n_2$, $n'_1$, $n'_2$ where the quark and the two
gluons are aligned using the finite RPI$_1$ described in App.(\ref{app:rpi}); 
on the RHS we write the SCET$_2$ amplitude and
calculate the Wilson coefficient necessary for the matching.  We
decompose the SCET$_1$ amplitude:
\begin{align}
\label{Am2ETh}
 A^{q\bar{q}gg}_{\rm{NLO}}  = C_{1}^{(0)}(n_0)
  \int\!\! dx_1 dx_2 \langle 0| T \{ \mathcal{L}_{\scetone}(x_1) \mathcal{L}_{\scetone}(x_2) \mo^{(0)}_{1} \}  
    |q_{n_2} g_{n_1^\prime} g_{n_2^\prime} \bar{q}_{\nb} \rangle_2 \, , 
\end{align}
in 
\begin{align}
 A^{q\bar{q}gg}_{\rm{NLO}}= A^{q\bar{q}gg}_{\mathrm{NLO},\, A }+ A^{q\bar{q}gg}_{\mathrm{NLO},\,B}+ A^{q\bar{q}gg}_{\mathrm{NLO},\,C} \, ,
\end{align}
where $A$, $B$, $C$ correspond to the three graphs in  Fig.~\ref{fig:figureABCC}.  
Using the SCET$_1$ Feynman rules, we have:
\begin{align}
\label{Am2ES1}
A^{q\bar{q}gg}_{\mathrm{NLO},\,A}&=U^{(2,0,0)}(n_0;\,Q,\mu_1)\, g^2\,\bar{u}_{n_0}  
\Big[ n_0^\beta + \gamma^\beta_{n_0 \perp} \frac{(\slashed{q}_1)_{n_0 \perp}}{\bar{q}_{1}} 
+ \frac{(\slashed{q}_2)_{n_0 \perp}}{\bar{q}_{2}} \gamma^\beta_{n_0 \perp}  \Big]  \\
& \times\Big[ n_0^\alpha+ \frac{(\slashed{q}_1)_{n_0 \perp}}{\bar{q}_{1}} \gamma^\alpha_{n_0 \perp}  \Big]
 \frac{\bar{q}_{1}}{q_{1}^2}  \frac{\bar{q}_{0}
}{q_{0}^2}  \gamma^\mu_{n_0 \perp}  v_{\bar{n}}\, , \nonumber \\
A^{q\bar{q}gg}_{\mathrm{NLO},\,B}&= U^{(2,0,0)}(n_0;\,Q, \mu_1)\, g^2\,\bar{u}_{n_0}  
\Big[ n_0^\alpha + \gamma^\alpha_{n_0 \perp} \frac{   (\slashed{q}_2  +\slashed{k}_1)_{n_0 \perp} }{\bar{q}_2+\bar{k}_1} 
  + \frac{(\slashed{q}_2)_{n_0 \perp}}{\bar{q}_{2}} \gamma^\alpha_{n_0 \perp}  \Big] 
  \nonumber \\ 
& \times  \Big[ n_0^\beta+ \frac{  (\slashed{q}_2  +\slashed{k}_1)_{n_0 \perp}}{\bar{q}_2+\bar{k}_1} \gamma^\beta_{n_0 \perp}  \Big] \frac{\bar{q}_2+\bar{k}_1}{(q_2+k_1)^2} \frac{\bar{q}_{0}
}{q_{0}^2} \gamma^\mu_{n_0 \perp}  v_{\bar{n}} \, , \nonumber \\
A^{q\bar{q}gg}_{\mathrm{NLO},\,C}&= U^{(2,0,0)}(n_0;\,Q, \mu_1)\, g^2\,\bar{u}_{n_0} 
\Big[ \frac{1}{\bar{q}_{2}+\bar{k}_1} \gamma_{n_0 \perp}^\alpha \gamma_{n_0 \perp}^\beta 
+ \frac{1}{\bar{q}_2+\bar{k}_2}  \gamma_{n_0 \perp}^\beta \gamma_{n_0 \perp}^\alpha \Big] \frac{\bar{q}_{0}
}{q_{0}^2} \gamma^\mu_{n_0 \perp} v_{\bar{n}} \, , \nonumber
\end{align}
where $q_1=q_2+k_2$ and $q_0=q_2+k_1+k_2$.  As before, we do not write terms
with $\bar{n}^\alpha$ and $\bar{n}^\beta$, as they are not necessary for the
matching because the operator $\bar{n}\!\cdot\! A_n$ is constrained by gauge
invariance to be only in Wilson lines.  Now we rotate the amplitude
(\ref{Am2ES1}) to the directions $n_{2}$ and $n'_1$ and $n'_2$ parallel to the
quark and the two gluons, as described in Eq.~(\ref{eq:n12p})
{\allowdisplaybreaks
\begin{align}
 \label{ampliAS2}
A^{q\bar{q}gg}_{\mathrm{NLO},\,A}&=U^{(2,0,0)}(n_0;\,Q, \mu_1) \,g^2\,\bar{u}_{n_2} 
 \Big[\frac{\bar{q}_2}{\bar{q}_1} \sqrt{n_2 \!\cdot\! n'_2}\, v_2^\beta   
  +\frac{\bar{k}_2}{\bar{q}_1}  \sqrt{n_2 \!\cdot\! n'_2} \,\frac{\slashed{v}_2 }{2} \gamma^\beta_{n'_2 \perp} \Big] \\ 
&\times \Big[\frac{\bar{q}_1}{\bar{q}_0} \sqrt{n_1 \!\cdot\! n'_1}\, v_1^\alpha 
  +\frac{\bar{k}_1}{\bar{q}_0}  \sqrt{n_1 \!\cdot\! n'_1} \,\frac{\slashed{v}_1 }{2} \gamma^\alpha_{n'_1 \perp} \Big]  
\frac{\bar{q}_1}{q_1^2}\frac{\bar{q}_0}{q_0^2}\,  \gamma^\mu_{n_0 \perp} v_{\bar{n}}\, ,\nonumber   \\  
A^{q\bar{q}gg}_{\mathrm{NLO},\,B}&=U^{(2,0,0)}(n_0;\,Q, \mu_1) \,g^2\,\bar{u}_{n_2} 
 \Big[\frac{\bar{q}_1}{\bar{q}_0} \sqrt{n_1 \!\cdot\! n'_1}\, v_1^\alpha
   +\frac{\bar{k}_2}{\bar{q}_1}  \sqrt{n_2 \!\cdot\! n'_2} \,\frac{\slashed{v}_2 }{2} \gamma^\alpha_{n'_1 \perp} \nonumber  \\ 
  & + \frac{\bar{k}_1}{\bar{q}_0} \sqrt{n_1 \!\cdot\! n'_1} \frac{\slashed{v}_1}{2}\gamma^\alpha_{n'_1 \perp}
   + \frac{\bar{q}_2 \, \bar{k}_2}{\bar{q}_1(\bar{q}_2 + \bar{k}_1)}  \sqrt{n_2 \!\cdot\! n'_2} 
   \,\gamma^\alpha_{n'_1 \perp} \frac{\slashed{v}_2 }{2}  \nonumber \\
        & - \frac{\bar{k}_1 \bar{k}_2}{\bar{q}_0(\bar{q}_2 + \bar{k}_1)}  \sqrt{n_1 \!\cdot\! n'_1} 
   \,\gamma^\alpha_{n'_1 \perp} \frac{\slashed{v}_1 }{2} \Big]
 \Big[\frac{\bar{q}_2}{\bar{q}_1} \sqrt{n_2 \!\cdot\! n'_2}\, v_2^\beta   
  -\frac{\bar{k}_1}{\bar{q}_0}  \sqrt{n_1 \!\cdot\! n'_1}\, v_1^\beta  \nonumber \\
 &+   \frac{\bar{q}_2 \, \bar{k}_2}{(\bar{q}_2+\bar{k}_1) \bar{q}_1}  \sqrt{n_2 \!\cdot\! n'_2} 
 \frac{\slashed{v}_2}{2}  \gamma^\beta_{n'_2 \perp}
 -   \frac{\bar{k}_2\, \bar{k}_1 }{(\bar{q}_2+\bar{k}_1)\bar{q}_0}  \sqrt{n_1 \!\cdot\!  n'_1} \,
\frac{\slashed{v}_1 }{2} \gamma^\beta_{n'_2 \perp}\Big]   
\frac{\bar{q}_2+\bar{k}_1}{(q_2+k_1)^2}\frac{\bar{q}_0}{q_0^2}
\gamma^\mu_{n_0 \perp} v_{\bar{n}} \, ,   \nonumber \\
A^{q\bar{q}gg}_{\mathrm{NLO},\,C}&=U^{(2,0,0)}(n_0;\,Q, \mu_1) \,g^2\, \bar{u}_{n_2}
 \Big[ \frac{1}{\bar{q}_{2}+\bar{k}_1} \gamma^\alpha_{n'_1 \perp}  \gamma^\beta_{n'_2 \perp}  
 + \frac{1}{\bar{q}_2+\bar{k}_2} \gamma^\beta_{n'_2 \perp} \gamma^\alpha_{n'_1 \perp} \Big]
   \frac{\bar{q}_0}{q_0^2}\,  \gamma^\mu_{n_0 \perp} v_{\bar{n}} \, , \nonumber
\end{align}
}
and $n_1 \cdot n'_1$ is defined in terms of $n_2$, $n'_1$ and $n'_2$ in
Eqs.~(\ref{eq:v1n2}),
 
The vectors $v_1$ and $v_2$ are defined in Eqs.~(\ref{eq:v}), (\ref{eq:v2}) and
(\ref{eq:v1n2}).  The values of $q_0^2$, $q_1^2$ and $(q_1+k_2)^2$ are given in
Eqs.~(\ref{q022E1}) and (\ref{q022E2}).  As with the one-gluon emission, we can
neglect the terms with $n_1^{\prime\alpha}$ and $n_2^{\prime\beta}$ as they are
orthogonal to the $\mb_{n'_1 \perp}^\alpha$ and $\mb_{n'_2 \perp}^\beta$ fields.
The SCET$_2$ amplitude for $ \langle 0| \mathcal{O}_{2}^{(2)}(n_2, n'_1,n'_2)
|q_{n_2} g_{n'_1} g_{n'_2} \bar{q}_{\nb}\rangle$ is:
\begin{align}
\label{eq:S2A2g}
 \langle 0|  \bar{\chi}_{n_2} g \mathcal{B}_{n'_2 \perp}^\alpha g \mathcal{B}_{n'_1 \perp}^\beta \chi_{\nb}
  |q_{n_2} g_{n'_1}  g_{n'_2} \bar{q}_{\nb}\rangle
 = g^2 \bar{u}_{n_2}    \epsilon_{n'_1 \perp}^{\alpha}   \epsilon_{n'_2 \perp}^{\beta}v_{\nb}\, .
\end{align}
In Eq.~(\ref{eq:S2A2g}) we have explicitly written the polarization vectors of
the external gluons.  For the jet-structure corrections, we get:
\begin{align}
\label{eq:C21p2p}
C^{(2) J}_{2,\,\rm{NLO}}(n_2,n'_1,n'_2)= U^{(2,0,0)}(n_0;\,Q, \mu_1) \,
d^{J\,\alpha\beta}_1(n_2,n'_1,n'_2)
   \Theta_{\delta_2}[n_2 \!\cdot\! n'_1] 
  \Theta_{\delta_2}[n_2 \!\cdot\! n'_2] 
  \Theta_{\delta_2}[n'_2 \!\cdot\! n'_1]  \, ,
\end{align} 
where
\begin{align}
\label{eq:AppD1}
d^{J\,\alpha\beta}_1(n_2,n'_1,n'_2)=
 d^{J\,\alpha\beta}_{1,A} (n_2,n'_1,n'_2)
+d^{J\,\alpha\beta}_{1,B} (n_2,n'_1,n'_2)
+d^{J\,\alpha\beta}_{1,C} (n_2,n'_1,n'_2) \, ,
\end{align} 
with
\begin{align}
\label{CA}
 d^{J\,\alpha\beta}_{1,A} (n_2,n'_1,n'_2)
&=\Big[\frac{\bar{q}_2}{\bar{q}_1} \sqrt{n_2 \!\cdot\! n'_2}\, v_2^\beta   
  +\frac{\bar{k}_2}{\bar{q}_1}  \sqrt{n_2 \!\cdot\! n'_2} \,\frac{\slashed{v}_2 }{2} \gamma^\beta_{n'_2 \perp} \Big]  \\ 
&\times \Big[\frac{\bar{q}_1}{\bar{q}_0} \sqrt{n_1 \!\cdot\! n'_1}\, v_1^\alpha 
  +\frac{\bar{k}_1}{\bar{q}_0}  \sqrt{n_1 \!\cdot\! n'_1} \,\frac{\slashed{v}_1 }{2} \gamma^\alpha_{n'_1 \perp} \Big]  
\frac{\bar{q}_1}{q_1^2}\frac{\bar{q}_0}{q_0^2}\,  \gamma^\mu_{n_0 \perp} \, ,\nonumber   \\ %
d^{J\,\alpha\beta}_{1,B} (n_2,n'_1,n'_2)
&= \Big[\frac{\bar{q}_1}{\bar{q}_0} \sqrt{n_1 \!\cdot\! n'_1}\, v_1^\alpha
   +\frac{\bar{k}_2}{\bar{q}_1}  \sqrt{n_2 \!\cdot\! n'_2} \,\frac{\slashed{v}_2 }{2} \gamma^\alpha_{n'_1 \perp} \nonumber  \\ 
  & + \frac{\bar{k}_1}{\bar{q}_0} \sqrt{n_1 \!\cdot\! n'_1} \frac{\slashed{v}_1}{2}\gamma^\alpha_{n'_1 \perp}
   + \frac{\bar{q}_2 \, \bar{k}_2}{\bar{q}_1(\bar{q}_2 + \bar{k}_1)}  \sqrt{n_2 \!\cdot\! n'_2} 
   \,\gamma^\alpha_{n'_1 \perp} \frac{\slashed{v}_2 }{2}  \nonumber \\
        & - \frac{\bar{k}_1 \bar{k}_2}{\bar{q}_0(\bar{q}_2 + \bar{k}_1)}  \sqrt{n_1 \!\cdot\! n'_1} 
   \,\gamma^\alpha_{n'_1 \perp} \frac{\slashed{v}_1 }{2} \Big] \nonumber \\
 &\times \Big[\frac{\bar{q}_2}{\bar{q}_1} \sqrt{n_2 \!\cdot\! n'_2}\, v_2^\beta   
  -\frac{\bar{k}_1}{\bar{q}_0}  \sqrt{n_1 \!\cdot\! n'_1}\, v_1^\beta  \nonumber \\
 &+   \frac{\bar{q}_2 \, \bar{k}_2}{(\bar{q}_2+\bar{k}_1) \bar{q}_1}  \sqrt{n_2 \!\cdot\! n'_2} 
 \frac{\slashed{v}_2}{2}  \gamma^\beta_{n'_2 \perp}
 -   \frac{\bar{k}_2\, \bar{k}_1 }{(\bar{q}_2+\bar{k}_1)\bar{q}_0}  \sqrt{n_1 \!\cdot\!  n'_1} \,
\frac{\slashed{v}_1 }{2} \gamma^\beta_{n'_2 \perp}\Big]   \nonumber \\
& \times    \frac{\bar{q}_2+\bar{k}_1}{(q_2+k_1)^2}\frac{\bar{q}_0}{q_0^2}
\gamma^\mu_{n_0 \perp} \, , \nonumber \\ 
d^{J\,\alpha\beta}_{1,C} (n_2,n'_1,n'_2)
&=  \Big[ \frac{1}{\bar{q}_{2}+\bar{k}_1} \gamma^\alpha_{n'_1 \perp}  \gamma^\beta_{n'_2 \perp}  
 + \frac{1}{\bar{q}_2 + \bar{k}_2} \gamma^\beta_{n'_2 \perp} \gamma^\alpha_{n'_1 \perp} \Big]
   \frac{\bar{q}_0}{q_0^2}\,  \gamma^\mu_{n_0 \perp}\,  .
\end{align}
The $\Theta$ functions in Eq.~(\ref{eq:C21p2p}) 
show that $C^{(2) J}_{2,\,\rm{NLO}}(n_2,n'_1,n'_2)$
comes from the two-jet SCET$_1$ operators.  To examine the power
counting of $C^{(2) J}_{2,\,\rm{NLO}}(n_2,n'_1,n'_2)$, we have to
consider that this coefficient comes from matching SCET$_1$ to
SCET$_2$ in the region where $n_2 \cdot n'_1\sim n_2 \cdot n'_2 \sim
n'_1 \cdot n'_2 \sim \lambda^2/\eta^4$, thus we have
\begin{align}
C^{(2) J}_{2,\,\rm{NLO}}(n_2,n'_1,n'_2)\sim \lambda^{-2}\, ,
\end{align} 
and since this multiplies $\mo^{(2)}_2 \,\sim\, \lambda^8$, by comparison with 
\eq{PCS21E} we see that we get an NLO($\lambda$) contribution.

\begin{figure}[t!]
\centering
\includegraphics{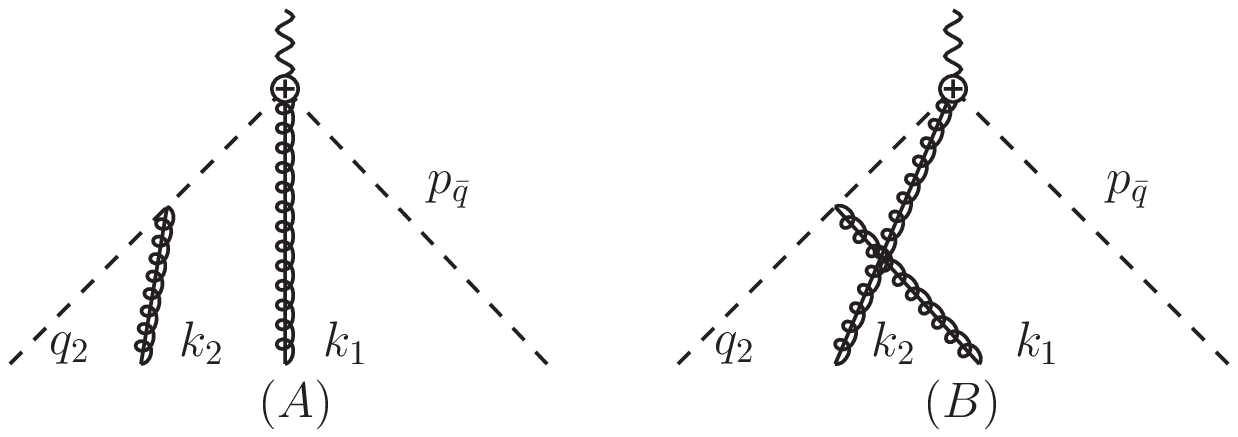}
\caption{Amplitudes for two emissions in $\scetone$ from the operator $ \mo^{(1)}_{1}$.} 
\label{fig:figureKinABC}
\end{figure}

We proceed similarly to calculate the coefficient  $C^{(2), H,\,a}_{2,\,\rm{NNLO}}(n_2,n'_1,n'_2)$
and show that it is $\mo(\lambda^{-1})$.  We decompose the  SCET$_1$ amplitude:
\begin{align}
 A^{q\bar{q}gg}_{\rm{NNLO}}   = C_{1,\,{\rm NLO}}^{(1)}(n_0, n_0) \int\!\! dx \langle 0| T \{ \mathcal{L}_{\scetone}(x)  \mo^{(1)}_{1} \} 
    |q_{n_2} g_{n_1^\prime} g_{n_2^\prime} \bar{q}_{\nb} \rangle_2 \nonumber \, , 
\end{align}
in 
\begin{align}
 A^{q\bar{q}gg}_{\rm{NNLO}}= A^{q\bar{q}gg}_{\mathrm{NNLO},\, A }+ A^{q\bar{q}gg}_{\mathrm{NNLO},\,B} \, ,
\end{align}
where $A$, $B$ correspond to the two graphs in  Fig.~\ref{fig:figureKinABC}.  
We have:
\begin{align}
\label{ampliAS2NLO}
A^{q\bar{q}gg}_{\mathrm{NNLO},\,A}&=- U^{(2,1,0)}(n_0, n_0;\,Q, \mu_1)
 \, g^2 \,\bar{u}_{n_2} \Big[\frac{\bar{q}_2}{\bar{q}_1}
 \sqrt{ n_2 \!\cdot\! n'_2}\, v_2^\beta \\ &
 +\frac{\bar{k}_2}{\bar{q}_1} \sqrt{ n_2 \!\cdot\! n'_2}
 \,\frac{\slashed{v}_2 }{2} \gamma^\beta_{n'_2 \perp} \Big]
 \gamma^{\alpha}_{n'_1 \perp} \frac{\bar{q}_1}{q_1^2}\, \frac{n_0^\mu -
 \nb^\mu}{Q} v_{\bar{n}}\, ,\nonumber \\
A^{q\bar{q}gg}_{\mathrm{NNLO},\,B}&=-U^{(2,1,0)}(n_0, n_0;\,Q, \mu_1) \, g^2 \,\bar{u}_{n_2} 
 \Big[\frac{\bar{q}_1}{\bar{q}_0} \sqrt{ n_1 \!\cdot\! n'_1}\, v_1^\alpha
   +\frac{\bar{k}_2}{\bar{q}_1}  \sqrt{ n_2 \!\cdot\! n'_2} \,\frac{\slashed{v}_2 }{2} \gamma^\alpha_{n'_1 \perp} \nonumber  \\ 
  & + \frac{\bar{k}_1}{\bar{q}_0} \sqrt{ n_1 \!\cdot\! n'_1} \frac{\slashed{v}_1}{2}\gamma^\alpha_{n'_1 \perp}
   + \frac{\bar{q}_2 \, \bar{k}_2}{\bar{q}_1(\bar{q}_2 + \bar{k}_1)}  \sqrt{ n_2 \!\cdot\! n'_2} 
   \,\gamma^\alpha_{n'_1 \perp} \frac{\slashed{v}_2 }{2}  \nonumber \\
        & - \frac{\bar{k}_1 \bar{k}_2}{\bar{q}_0(\bar{q}_2 + \bar{k}_1)}  \sqrt{ n_1 \!\cdot\! n'_1} 
   \,\gamma^\alpha_{n'_1 \perp} \frac{\slashed{v}_1 }{2} \Big]
   \gamma^{\beta}_{n'_1 \perp} \frac{\bar{q}_1}{q_1^2}\, 
 \frac{n_0^\mu - \nb^\mu}{Q} v_{\bar{n}}\, , \nonumber
\end{align}
where in Eq.~(\ref{ampliAS2NLO}) we have already rotated the amplitude
to the directions $n_{2}$, $n'_1$ and $n'_2$.
From Eqs.~(\ref{eq:S2A2g}) and (\ref{ampliAS2NLO}) 
we can see that the  Wilson coefficient $C^{(2) H,\,a}_{2,\,\rm{NNLO}}(n_2,n'_1,n'_2)$ is
\begin{align}
\label{eq:S2J1a}
C^{(2) H,\,a}_{2,\,\rm{NNLO}}(n_2,n'_1,n'_2)= &U^{(2,1,0)}(n_0, n_0;\,Q, \mu_1)  d_{1}^{H\,\alpha\beta}(n_2,n'_1,n'_2)
 \nonumber \\
& \times \Theta_{\delta_2}[n_2 \!\cdot\! n'_1] 
  \Theta_{\delta_2}[n_2 \!\cdot\! n'_2] 
  \Theta_{\delta_2}[n'_2 \!\cdot\! n'_1]\, ,
\end{align} 
where
\begin{align}
d_{1}^{H\,\alpha\beta}(n_2,n'_1,n'_2)
 = d_{1, A}^{H\,\alpha\beta}(n_2,n'_1,n'_2)
+ d_{1, B}^{H\,\alpha\beta}(n_2,n'_1,n'_2)
\end{align}
with
\begin{align}
\label{eq:factNLO}
d_{1, A}^{H\,\alpha\beta}(n_2,n'_1,n'_2)
 &= \Big[\frac{\bar{q}_2}{\bar{q}_2+\bar{k}_2} \sqrt{ n_2 \!\cdot\! n'_2}\, v_2^\beta  \\
  &  +\frac{\bar{k}_2}{\bar{q}_2+\bar{k}_2}  \sqrt{ n_2 \!\cdot\! n'_2} \,\frac{\slashed{v}_2 }{2} \gamma^\beta_{n'_2 \perp} \Big] 
    \gamma^{\alpha}_{n'_1 \perp} 
       \frac{\bar{q}_1}{q_1^2}\, 
 \frac{n_0^\mu - \nb^\mu}{Q} \, ,\nonumber   \\ 
 d_{1, B}^{H\,\alpha\beta}(n_2,n'_1,n'_2)
 &=  \Big[\frac{\bar{q}_1}{\bar{q}_0} \sqrt{ n_1 \!\cdot\! n'_1}\, v_1^\alpha
   +\frac{\bar{k}_2}{\bar{q}_1}  \sqrt{ n_2 \!\cdot\! n'_2} \,\frac{\slashed{v}_2 }{2} \gamma^\alpha_{n'_1 \perp} \nonumber  \\ 
  & + \frac{\bar{k}_1}{\bar{q}_0} \sqrt{ n_1 \!\cdot\! n'_1} \frac{\slashed{v}_1}{2}\gamma^\alpha_{n'_1 \perp}
   + \frac{\bar{q}_2 \, \bar{k}_2}{\bar{q}_1(\bar{q}_2 + \bar{k}_1)}  \sqrt{ n_2 \!\cdot\! n'_2} 
   \,\gamma^\alpha_{n'_1 \perp} \frac{\slashed{v}_2 }{2}  \nonumber \\
        & - \frac{\bar{k}_1 \bar{k}_2}{\bar{q}_0(\bar{q}_2 + \bar{k}_1)}  \sqrt{ n_1 \!\cdot\! n'_1} 
   \,\gamma^\alpha_{n'_1 \perp} \frac{\slashed{v}_1 }{2} \Big]  
    \gamma^{\beta}_{n'_1 \perp} \frac{\bar{q}_1}{q_1^2}\, 
 \frac{n_0^\mu - \nb^\mu}{Q}\, . \nonumber
\end{align}
To get the power counting of $C^{(2) H,\,a}_{2,\,\rm{NNLO}}(n_2,n'_1,n'_2)$, 
as in the previous case, we have to consider that the matching 
is done in a region where $n_2 \cdot n'_1\sim n_2 \cdot n'_2 \sim n'_1 \cdot n'_2 \sim \lambda^2/\eta^4$.
This implies:
\begin{align}
C^{(2) H,\,a}_{2,\,\rm{NNLO}}(n_2,n'_1,n'_2)\sim \lambda^{-1}\, ,
\end{align} 
which justifies its labeling as NNLO($\lambda$).


We now turn to calculate the coefficient $C^{(2)}_{2}(n_1,n'_1,n'_1)$.  
We will proceed as above.
We decompose $C^{(2)}_{2}(n_1,n'_1,n'_1)$ as:
\begin{align}
C^{(2)}_{2}(n_2,n'_1,n'_1)=C^{(2) J}_{2,\,\rm{ NLO}}(n_2,n'_1,n'_1) 
+C^{(2) H}_{2}(n_2,n'_1,n'_1)\, ,
\end{align}
where
\begin{align}
   \label{eq:2HC0}
& C^{(2)J}_{2,\,\rm{ NLO}}(n_2,n'_1,n'_1)  \langle 0|\mo^{(2)}_2(n_2,n'_1,n'_1) 
|q_{n_2} g_{n'_1} g_{n_1^\prime} \bar{q}_{\nb} \rangle_2 \\
& =  C^{(0)}_{1,\,\rm{LO}}(n_0) \int \! \! dx_1 dx_2 
\langle 0|  T \{ \mathcal{L}_{\scetone}(x_1) \mathcal{L}_{\scetone}(x_2) \mo^{(0)}_{1} (n_0) \} 
|q_{n_2} g_{n'_1} g_{n_1^\prime} \bar{q}_{\nb} \rangle_2 \nonumber \, ,
\end{align}
and
\begin{align}
\label{eq:2HC}
 &  C^{(2) H}_{2}(n_2,n'_1,n'_1)  \langle 0|\mo^{(2)}_2
|q_{n_2} g_{n'_1} g_{n_1^\prime} \bar{q}_{\nb} \rangle_2 \nonumber \\
&  = C^{(1)}_{1,\,{\rm NLO}}(n_0,n_0) \int \! \! dx \langle 0|   T \{ \mathcal{L}_{\scetone}(x) \mo^{(1)}_{1} \} 
|q_{n_2} g_{n'_1} g_{n_1^\prime} \bar{q}_{\nb} \rangle_2   \nonumber\\ 
& + C^{(1)}_{1, \mathcal{T}}(n_0,n_0) \int \! \! dx \langle 0| T \{ \mathcal{L}_{\scetone}(x) \mathcal{T}^{(1)}_{1}   \}
|q_{n_2} g_{n'_1} g_{n_1^\prime} \bar{q}_{\nb} \rangle_2    \, .
\end{align}
We further set:
\begin{align}
C^{(2)H}_{2}(n_2,n'_1,n'_1)=C^{(2) H}_{2,\, \rm{NNLO}}(n_2,n'_1,n'_1)
+C^{(2) H}_{2,\, \rm{N^3LO}}(n_2,n'_1,n'_1)\, ,
\end{align}
where $C^{(2) H}_{2,\, \rm{NLO}}(n_2,n'_1,n'_1)$ is the coefficient
of the contribution that reproduces the second line in \eq{2HC}, {\it etc.}
We will only calculate $C^{(2) J}_{2,\, \rm{NLO}}(n_2,n'_1,n'_1)$ and show that
it scales as $\lambda^{-2}$. This is the only operator of this form that we need to
calculate the amplitude squared at NLO($\lambda$). 

To calculate the amplitude on the RHS of Eq.~(\ref{eq:2HC0}), we can
use Eqs.~(\ref{ampliAS2}), which are written in terms of $n_2$, $n'_1$
and $n'_2$ that are parallel to the external particles, and take the
limit $n'_2 \cdot n'_1\rightarrow \lambda^4/\eta^4$.  In this
case the two gluons are collinear in SCET$_2$.  Thus, we can define
$C^{(2) J}_{2,\, \rm{NLO} }(n_2,n'_1,n'_1)$ as
\begin{align}
\label{eq:C11p1pa}
C^{(2) J}_{2,\, \rm{NLO} }(n_2,n'_1,n'_1)&= 
  U^{(2,0,0)}(n_0;\,Q, \mu_1) 
d^{J\, \alpha\beta}_2 (n_2,n'_1,n'_1)\Theta_{\delta_2}[n_2 \cdot n'_1]  \, ,
\end{align}
where
\begin{align}
d^{J\,\alpha\beta}_2 (n_2,n'_1,n'_1)&= 
\label{eq:C11p1pb} 
\displaystyle\lim_{n'_2 \cdot n'_1\rightarrow \lambda^4/\eta^4} d^{\alpha\beta}_1 (n_2,n'_1,n'_2) 
 \\
&= \Bigg(
\Big[\frac{\bar{q}_2}{\bar{q}_1} \sqrt{ n_2 \!\cdot\! n'_2}\, v_2^\beta   
  +\frac{\bar{k}_2}{\bar{q}_1}  \sqrt{ n_2 \!\cdot\! n'_2} \,\frac{\slashed{v}_2 }{2} \gamma^\beta_{n'_2 \perp} \Big] \nonumber \\ 
&\times \Big[\frac{\bar{q}_1}{\bar{q}_0} \sqrt{ n_1 \!\cdot\! n'_1}\, v_1^\alpha 
  +\frac{\bar{k}_1}{\bar{q}_0}  \sqrt{ n_1 \!\cdot\! n'_1} \,\frac{\slashed{v}_1 }{2} \gamma^\alpha_{n'_1 \perp} \Big]  
\frac{\bar{q}_1}{q_1^2}\frac{2\bar{q}_0}{\bar{q}_2 \, \bar{k}_1 (n_2 \!\cdot\! n'_1) 
+ \bar{q}_2 \, \bar{k}_2 (n_2 \!\cdot\! n'_2)} \nonumber   \\ 
&+ \Big[\frac{\bar{q}_1}{\bar{q}_0} \sqrt{ n_1 \!\cdot\! n'_1}\, v_1^\alpha
   +\frac{\bar{k}_2}{\bar{q}_1}  \sqrt{ n_2 \!\cdot\! n'_2} \,\frac{\slashed{v}_2 }{2} \gamma^\alpha_{n'_1 \perp} \nonumber  \\ 
  & + \frac{\bar{k}_1}{\bar{q}_0} \sqrt{ n_1 \!\cdot\! n'_1} \frac{\slashed{v}_1}{2}\gamma^\alpha_{n'_1 \perp}
   + \frac{\bar{q}_2 \, \bar{k}_2}{\bar{q}_1(\bar{q}_2 + \bar{k}_1)}  \sqrt{ n_2 \!\cdot\! n'_2} 
   \,\gamma^\alpha_{n'_1 \perp} \frac{\slashed{v}_2 }{2}  \nonumber \\
        & - \frac{\bar{k}_1 \bar{k}_2}{\bar{q}_0(\bar{q}_2 + \bar{k}_1)}  \sqrt{ n_1 \!\cdot\! n'_1} 
   \,\gamma^\alpha_{n'_1 \perp} \frac{\slashed{v}_1 }{2} \Big] 
 \times \Big[\frac{\bar{q}_2}{\bar{q}_1} \sqrt{ n_2 \!\cdot\! n'_2}\, v_2^\beta   
  -\frac{\bar{k}_1}{\bar{q}_0}  \sqrt{ n_1 \!\cdot\! n'_1}\, v_1^\beta  \nonumber \\
 &+   \frac{\bar{q}_2 \, \bar{k}_2}{(\bar{q}_2+\bar{k}_1) \bar{q}_1}  \sqrt{ n_2 \!\cdot\! n'_2} 
 \frac{\slashed{v}_2}{2}  \gamma^\beta_{n'_2 \perp}
 -   \frac{\bar{k}_2\, \bar{k}_1 }{(\bar{q}_2+\bar{k}_1)\bar{q}_0}  \sqrt{ n_1 \!\cdot\!  n'_1} \,
\frac{\slashed{v}_1 }{2} \gamma^\beta_{n'_2 \perp}\Big]   \nonumber \\
& \times    \frac{\bar{q}_2+\bar{k}_1}{(q_2+k_1)^2}\frac{2\bar{q}_0}{\bar{q}_2 \, \bar{k}_1 (n_2 \!\cdot\! n'_1) 
+ \bar{q}_2 \, \bar{k}_2 (n_2 \!\cdot\! n'_2)}
 \nonumber \\ 
&+  \Big[ \frac{1}{\bar{q}_{2}+\bar{k}_1} \gamma^\alpha_{n'_1 \perp}  \gamma^\beta_{n'_2 \perp}  
 + \frac{1}{\bar{q}_2 + \bar{k}_2} \gamma^\beta_{n'_2 \perp} \gamma^\alpha_{n'_1 \perp} \Big]
   \frac{\bar{q}_0}{\bar{q}_2 \, \bar{k}_1 (n_2 \!\cdot\! n'_1) 
+ \bar{q}_2 \, \bar{k}_2 (n_2 \!\cdot\! n'_2)}\, 
\Bigg) \nonumber \\
& \times \gamma^\mu_{n_0 \perp} \Big |_{n'_1=n'_2} \, . \nonumber 
\end{align}
In Eqs.~(\ref{eq:C11p1pb}) there is a difference
in the notation between the LHS and RHS. On the LHS, we have labeled the
quark with $n_2$ and the two gluons with $n'_1$ because the
coefficient (\ref{eq:C11p1pa}) is for the operator
$\mo^{(2)}_2(n_2,n'_1,n'_1)$, where the gluons are collinear. On the
RHS, $n_2$, $n'_1$ and $n'_2$ are the directions parallel to the
quarks and the two gluons as defined in Eqs.~(\ref{n1nq1def}) and
(\ref{eq:n2nq2def}).  We encode that the two gluons are collinear using
the $\Theta$ function on the RHS of Eq.~(\ref{eq:C11p1pb})
with $\delta_3=\lambda^{3}/\eta^4$.  It restricts that $n'_1 \cdot n'_2
\lesssim \lambda^4/\eta^4$.  On the RHS of Eq.~(\ref{eq:C11p1pb}) we could
decompose $n'_2$ in terms of $n'_1$ and avoid inserting the $\Theta$,
but it is convenient to leave $n'_2$ explicit
because it will make the matching easier to SCET$_3$.  We notice
that the RHS of Eq.~(\ref{eq:C11p1pb}) is just equal to the
coefficient $C^{(2) J}_{2,\,\rm{NLO}}(n_2,n'_1,n'_2)$ defined in
Eq.~(\ref{eq:C21p2p}) with the substitution $q_0^2 \rightarrow
\bar{q}_2 \bar{k}_1 (n_2 \!\cdot\! n'_1)/4 + \bar{q}_2 \bar{k}_2 (n_2
\!\cdot\! n'_2)/4$.  Knowing that $n'_1 \cdot n'_2 \sim \lambda^4/\eta^4$,
$n_1 \cdot n'_2 \sim \lambda^2/\eta^4$ and $n_1 \cdot n'_1 \sim \lambda^2/\eta^4$,
it is easy to check that Eq.~(\ref{eq:C11p1pb}) scales as
$\lambda^{-2}$.  The information that $C^{(2) J}_{2,\,\rm{LO}}(n_2,n'_1,n'_1)$
comes from a two-jet SCET$_1$ operator, 
is encoded in the $\Theta$-functions of \eq{C11p1pa}.


For the coefficient $C^{(2)}_{2}(n_2,n'_1,n_2)$, we decompose it as:
\begin{align}
C^{(2)}_{2}(n_2,n'_1,n_2)=C^{(2) J}_{2,\,\rm{NLO}} (n_2,n'_1,n_2)+C^{(2) H}_{2}(n_2,n'_1,n_2) \, ,
\end{align}
where 
\begin{align}
\label{eq:CS2n1n1np1}
& C^{(2) J}_{2,\,\rm{NLO}}(n_2,n'_1,n_2)  \langle 0|\mo^{(2)}_2 
|q_{n_2}  g_{n_1^\prime}g_{n_2} \bar{q}_{\nb} \rangle_2 \\
& =  C^{(0)}_{1,\,\rm{LO}}(n_0) \int \! \! dx_1 dx_2 \langle 0|  
T \{ \mathcal{L}_{\scetone}(x_1) \mathcal{L}_{\scetone}(x_2) \mo^{(0)}_{1}  \} 
|q_{n_2}  g_{n_1^\prime}g_{n_2} \bar{q}_{\nb} \rangle_2 \nonumber \\
 &  -C^{(1)}_{2\, ,\rm{LO}}(n_1, n'_1) \int \! \! dx \langle 0|T \{ \mathcal{L}_{\scettwo}(x)\mo^{(1)}_{2}\}
|q_{n_2}  g_{n_1^\prime}g_{n_2} \bar{q}_{\nb} \rangle_2  \, ,\nonumber
\end{align}
and 
\begin{align}
\label{eq:CS2n1n1np1a}
& C^{(2)H}_{2}(n_2,n'_1,n_2)  \langle 0|\mo^{(2)}_2
|q_{n_2}  g_{n_1^\prime}g_{n_2} \bar{q}_{\nb} \rangle_2 \\
 & =  C^{(1)}_{1,\,{\rm NLO}}(n_0, n_0) \int \! \! dx \langle 0|   T \{ \mathcal{L}_{\scetone}(x) \mo^{(1)}_{1} \} 
|q_{n_2}  g_{n_1^\prime}g_{n_2} \bar{q}_{\nb} \rangle_2 \nonumber\\
&  -C^{(1)}_{2\, ,\rm{NLO}}(n_2, n'_1) \int \! \! dx \langle 0|T \{ \mathcal{L}_{\scettwo}(x)\mo^{(1)}_{2}\}
|q_{n_2}  g_{n_1^\prime}g_{n_2} \bar{q}_{\nb} \rangle_2  \nonumber\\ 
& + C^{(1)}_{1, \mathcal{T}}(n_0, n_0) \int \! \! dx \langle 0| 
T \{ \mathcal{L}_{\scetone}(x) \mathcal{T}^{(1)}_{1}   \}
|q_{n_2}  g_{n_1^\prime}g_{n_2} \bar{q}_{\nb} \rangle_2   \nonumber\\ 
&  -C^{(1)}_{2\, ,\rm{NNLO}}(n_2, n'_1) \int \! \! dx \langle 0|T \{ \mathcal{L}_{\scettwo}(x)\mo^{(1)}_{2}\}
|q_{n_2}  g_{n_1^\prime}g_{n_2} \bar{q}_{\nb} \rangle_2  \, , \nonumber
\end{align}
We write $C^{(2) H }_{2}(n_2,n_2,n'_1)$ as
\begin{align}
\label{eq:CoeMis}
C^{(2)}_{2}(n_2,n'_1,n_2) &=
C^{(2) H}_{2,\,\rm{NNLO}} (n_2,n'_1,n_2) 
+C^{(2) H}_{2,\,\rm{N^3LO}} (n_2,n'_1,n_2) \, ,
\end{align}
where $C^{(2) H}_{2,\,\rm{NNLO}} (n_2,n'_1,n'_2)$ is the coefficient of
the contribution that reproduces the the second and third line in the
Eq.~(\ref{eq:CS2n1n1np1a}), and $C^{(2) H}_{2,\,\rm{N^3LO}}$ the fourth
and fifth line.  As for the previous cases, the coefficient $C^{(2)
J}_{2,\,\rm{NLO}} (n_2,n'_1,n_2)$ scales as $\lambda^{-2}$, $C^{(2)
H}_{2,\,\rm{NNLO}} (n_2,n'_1,n_2)$ as $\lambda^{-1}$ and $C^{(2)
H}_{2,\,\rm{N^3LO}} (n_2,n'_1,n_2)$ as $\lambda^{0}$.  
Since $\mo^{(2)}_2(n_2,n'_1,n_2)$ interferes with the LO
operator, $\mo^{(1)}_2(n_1,n'_1)$, to have the amplitude squared up to
NLO($\lambda$) we need both $C^{(2) J}_{2,\,\rm{NLO}}$ and $C^{(2)
H}_{2,\,\rm{NNLO}}$.  We start with $C^{(2) J}_{2,\,\rm{NLO}}$.  To
calculate the amplitude in the second line in
Eq.~(\ref{eq:CS2n1n1np1}), we use Eq.~(\ref{ampliAS2}) and take the
limit $n_2 \cdot n'_2\rightarrow \lambda^4/\eta^4$ with $n_2 \cdot n'_1 \sim
n'_1 \cdot n'_2 \sim \lambda^2/\eta^4$. (We could alternatively take
the limit $n_2 \cdot n'_1\rightarrow \lambda^4/\eta^4$ with $n_2 \cdot n'_2
\sim n'_2 \cdot n'_1 \sim \lambda^2/\eta^4$.) It is easy to check that
\begin{align}
\label{eq:RelAS2}
& \displaystyle\lim_{n_2 \cdot n'_2\rightarrow \lambda^4/\eta^4} 
d^{J\, \alpha\beta}_{1,A} (n_2,n'_1,n'_2)  \langle 0 | \mo^{(2)}_2 
|q_{n_2}  g_{n_1^\prime}g_{n_2} \bar{q}_{\nb} \rangle_2 
\nonumber \\
& \ \ = C^{(1)}_{2\, ,\rm{LO}}(n_2,n'_1) \int \! \! dx \langle 0|T \{ \mathcal{L}_{\scettwo}(x)\mo^{(1)}_{2} \}
|q_{n_2}  g_{n_1^\prime}g_{n_2} \bar{q}_{\nb} \rangle_2 \, .
\end{align}
With Eq.~(\ref{eq:RelAS2}), we can write $C^{(2) J}_{2\, ,\rm{NLO}}(n_2,n'_1,n_2) $ as
\begin{align}
\label{eq:Cn2n2n1p}
C^{(2) J}_{2,\,\rm{NLO}} (n_2,n'_1,n_2) = 
  U^{(2,0,0)}(n_0;\,Q, \mu_1) \,
d^{J\, \alpha\beta}_3 (n_2,n'_1,n_2) \Theta_{\delta_2}[n_2 \cdot n'_1]
\end{align}
where
\begin{align}
\label{eq:ApMS}
d^{J\,\alpha\beta}_3 (n_2,n'_1,n_2)&=
 \displaystyle\lim_{n_2 \cdot n'_2\rightarrow \lambda^4/\eta^4} 
(d^{J\, \alpha\beta}_{1 B} (n_2,n'_1,n'_2)
+d^{J\, \alpha\beta}_{1 C} (n_2,n'_1,n'_2))  \\  
&= \Big(
 \Big[\frac{\bar{q}_1}{\bar{q}_0} \sqrt{ n_1 \!\cdot\! n'_1}\, v_1^\alpha
  + \frac{\bar{k}_1}{\bar{q}_0} \sqrt{ n_1 \!\cdot\! n'_1} \frac{\slashed{v}_1}{2}\gamma^\alpha_{n'_1 \perp}
 \nonumber  \\ 
        & - \frac{\bar{k}_1 \bar{k}_2}{\bar{q}_0(\bar{q}_2 + \bar{k}_1)}  \sqrt{ n_1 \!\cdot\! n'_1} 
   \,\gamma^\alpha_{n'_1 \perp} \frac{\slashed{v}_1 }{2} \Big] \nonumber \\
 &\times  \Big[  - \frac{\bar{k}_1}{\bar{q}_0}  \sqrt{ n_1 \!\cdot\! n'_1}\, v_1^\beta    
 -  \frac{\bar{k}_2\, \bar{k}_1 }{(\bar{q}_2+\bar{k}_1)\bar{q}_0}  \sqrt{ n_1 \!\cdot\!  n'_1} \,
\frac{\slashed{v}_1 }{2} \gamma^\beta_{n'_2 \perp}\Big]   \nonumber \\
&+ \Big[ \frac{1}{\bar{q}_{2}+\bar{k}_1} \gamma^\alpha_{n'_1 \perp}  \gamma^\beta_{n'_2 \perp}  
 + \frac{1}{\bar{q}_2 + \bar{k}_2} \gamma^\beta_{n'_2 \perp} \gamma^\alpha_{n'_1 \perp} \Big] \gamma^\mu_{n_0 \perp}  \Big)\nonumber\\
  &\times \frac{2 \bar{q}_0}{ \bar{k}_1 ( \bar{k}_2 (n'_2 \!\cdot\! n'_1) + \bar{q}_2 (n_2 \!\cdot\! n'_1) )}  
   \Big |_{n_2=n'_2} \nonumber \, .
\end{align}
The scaling of the dot products of $n$'s in this configuration make
the coefficient $C^{(2) J}_{2,\,\rm{NLO}} (n_2,n_2,n'_1)
\sim\lambda^{-2}$.  As previously for $C^{(2) J}_{2,\,\rm{NLO}}
(n_1,n'_1,n'_1) $, we prefer leaving (\ref{eq:Cn2n2n1p}) in terms
of $n_2$, $n^\prime_1$ and $n^\prime_2$.  To calculate
$C^{(2) H}_{2\, ,\rm{NNLO}}(n_2,n_2,n'_1) $ we proceed in the same
way.  We have
\begin{align}
\label{eq:C2H221p}
&C^{(2) H}_{2\, ,\rm{NNLO}}(n_2,n_2,n'_1)= U^{(2,1,0)}(n_0, n_0;\,Q, \mu_1) d^{H\, \alpha\beta}_{3}(n_2,n'_1,n_2)
\Theta_{\delta_2}[n_2 \cdot n'_1] \, ,
\end{align}
where
\begin{align}
d^{H\, \alpha\beta}_{3} (n_2,n'_1,n_2) &= \lim_{n_2 \cdot n'_2\rightarrow \lambda^4/\eta^4} 
d^{H\, \alpha\beta}_{1 B} (n_2,n'_1,n'_2)  \nonumber \\
 & = \Big[\frac{\bar{q}_1}{\bar{q}_0} \sqrt{ n_1 \!\cdot\! n'_1}\, v_1^\alpha 
   + \frac{\bar{k}_1}{\bar{q}_0} \sqrt{ n_1 \!\cdot\! n'_1} \frac{\slashed{v}_1}{2}\gamma^\alpha_{n'_1 \perp}
    \nonumber \\
        & - \frac{\bar{k}_1 \bar{k}_2}{\bar{q}_0(\bar{q}_2 + \bar{k}_1)}  \sqrt{ n_1 \!\cdot\! n'_1} 
   \,\gamma^\alpha_{n'_1 \perp} \frac{\slashed{v}_1 }{2} \Big]  
    \gamma^{\beta}_{n'_1 \perp} \frac{\bar{q}_1}{q_1^2}\, 
 \frac{n_0^\mu - \nb^\mu}{Q} \Big |_{n_2=n'_2} .
\label{eq:d3hSubtrLimit}
\end{align}
In Eq.~(\ref{eq:C2H221p}) we use the fact that,
\begin{align}
& \displaystyle\lim_{n_2 \cdot n'_2\rightarrow \lambda^4/\eta^4} 
d^{H\, \alpha\beta}_{1,\,A}(n_2,n'_1,n'_2) \langle 0 |
\mo^{(2)}_2 |q_{n_2} g_{n_1^\prime} g_{n_2} \bar{q}_{\nb} \rangle_2 \nonumber \\
& \ \ = C^{(1)}_{2\, ,\rm{NLO}} \int \! \! dx \langle 0|T \{ \mathcal{L}_{\scettwo}(x)\mo^{(1)}_{2} \}
 |q_{n_2} g_{n_1^\prime} g_{n_2} \bar{q}_{\nb} \rangle_2 \, .
\end{align}
In Eqs.~(\ref{eq:ApMS}, \ref{eq:d3hSubtrLimit}) there is again a difference in the
notation between the LHS and RHS similar to \eq{C11p1pb}.
Since $C^{(2)J}_{2,\,\rm{LO}}(n_2,n'_1,n_2)$ and $C^{(2) H}_{2\,
,\rm{NNLO}}(n_2,n'_1,n_2)$ come from SCET$_1$ two-jet operators,
we include the appropriate $\Theta$-functions in
Eqs.~(\ref{eq:Cn2n2n1p}, \ref{eq:C2H221p}).

We have that all the NLO($\lambda$) terms for two gluon matching come from the
SCET$_1$ operator, $\mo^{(0)}_1(n_0)$, and are jet-structure
corrections.  At NNLO($\lambda$) we have only hard corrections.  Before matching
SCET$_2$ to SCET$_3$, we have to insert in the coefficients the
SCET$_2$ running factors. Below we list all the needed SCET$_2$
coefficients to NNLO($\lambda$) that we have calculated with the appropriate RG
kernels.  From the matching of one-gluon emission, we have:
\begin{align}
\label{eq:S2SumA}
C^{(1)}_{2,\,\rm{LO}}(n_1, n'_1)= & \, U^{(2,1,0)}(n_1, n'_1;\,\mu_1,\mu)
U^{(2,0,0)}(n_0;\,Q,\mu_1)  
 c_{\rm{LO}}(n_0) 
 \frac{\bar{q}_0}{q_0^2}  \gamma^{\mu}_{n_0 \perp} \, , \\
  C^{(1)H,a}_{2,\,\rm{NLO}}  (n_1, n'_1)= &\,   U^{(2,1,0)}(n_1, n'_1;\,\mu_1,\mu)
U^{(2,1,0)}(n_0, n_0;\,Q,\,\mu_1) \otimes c^{H,a}_{2,\,\rm{NLO}}(n_0, n_0) \, , \nonumber\\
  C^{(1)H}_{2,\,\rm{NNLO}}  (n_1, n'_1)= & \, U^{(2,1,0)}(n_1, n'_1;\,\mu_1,\mu) 
U^{(2,1,0)}(n_0, n_0;\,Q,\,\mu_1) \otimes c^{H}_{2,\,\rm{NNLO}}(n_0, n_0) \, , \nonumber\\
 C^{(1)H,b}_{2,\,\rm{NLO}} (n_1, n'_1)= & \,  U^{(2,1,0)}(n_1, n'_1;\,\mu_1,\mu)
  C^{(1)}_{1}(n_1, n'_1)  \tilde{\Theta}_{ \delta_2} [n_1 \cdot n'_1]  \, , \nonumber
\end{align}
where the coefficient in (\ref{eq:S2SumA}) without the $\scettwo$ RG-kernel is defined
in Eq.~(\ref{eq:CLOS2}), the second and third in Eqs.~(\ref{eq:C2S2}),
and the last in (\ref{eq:S23j}).  From the matching of two-gluon
emission we have the coefficients:
\begin{align}
\label{eq:S2SumE}
C^{(2) J}_{2\, ,\rm{NLO}}(n_2,n'_1,n'_2)&= \, 
U^{(2,1,0)}(n_2,n'_1,n'_2;\,\mu_1, \mu)\,
U^{(2,0,0)}(n_0;\,Q, \mu_1) \, \\
& \times d^{J}_1(n_2,n'_1,n'_2)
   \Theta_{\delta_2}[n_2 \!\cdot\! n'_1] 
  \Theta_{\delta_2}[n_2 \!\cdot\! n'_2] 
  \Theta_{\delta_2}[n'_2 \!\cdot\! n'_1]\, , \nonumber \\ 
C^{(2) J}_{2,\,\rm{NLO}}(n_2,n'_1,n'_1) &= \,
 U^{(2,1,0)}(n_2,n'_1,n'_1;\,\mu_1,\mu)\otimes U^{(2,0,0)}(n_0;\,Q, \mu_1) \nonumber\\
& \times d^{J}_2 (n_2,n'_1,n'_1)
\Theta_{\delta_2}[n_2 \cdot n'_1] \Theta_{\delta_2}[n_2 \cdot n'_2] \, , \nonumber \\
C^{(2) J}_{2,\,\rm{NLO}}(n_2,n'_1,n_2)&=\,
U^{(2,1,0)}(n_2,n'_1,n_2;\,\mu_1,\mu) \otimes
U^{(2,0,0)}(n_0;\,Q, \mu_1) \nonumber \\
& \times d^{J}_3 (n_2,n'_1,n_2)
 \Theta_{\delta_2}[n_2 \cdot n'_2]\Theta_{\delta_2}[n'_1 \cdot n'_2 ]\, , \nonumber \\
C^{(2) H}_{2\, ,\rm{NNLO}}(n_2,n'_1,n_2)&= 
\, U^{(2,1,0)}(n_2,n'_1,n_2;\,\mu_1,\mu) \otimes
 U^{(2,1,0)}(n_0, n_0;\,Q, \mu_1) \nonumber  \\
 & \otimes d^{H}_{3}  (n_2,n'_1,n_2)
\Theta_{\delta_2}[n_2 \cdot n'_2]\Theta_{\delta_2}[n'_1 \cdot n'_2] \nonumber \, ,
\end{align}
where the coefficients without $\scettwo$ running are defined in
Eqs.~(\ref{eq:C21p2p}, \ref{eq:C11p1pa}, \ref{eq:Cn2n2n1p},
\ref{eq:C2H221p}).  The RG kernels are given in
Eqs.~(\ref{eq:rgKernel}, \ref{eq:oneLoopCusp}, and \ref{eq:kernelEq}).
As discussed below \eq{S1coeRun}, we have a convolution because SCET
fields collinear to the same direction can exchange longitudinal
momentum during the running.

\section{Matching $\scettwo$ to $\scetthree$, $\scetn$}
\label{app:scet2/scet3}

We match SCET$_2$ to SCET$_3$ before proceeding to the general case and listing
a set of master operators for $\scetn$ The SCET$_3$ operators necessary for
matching up to two-gluon emission are: $\mo^{(0)}_{3}(n_0)$,
$\mo^{(1)}_{3}(n_0,n_0)$, $\mo^{(1)}_3 (n_1, n'_1)$, $\mo^{(2)}_3(n_2,
n'_1,n'_2)$, $\mo^{(2)}_3(n_2, n_2, n_1^\prime)$, $\mo^{(2)}_3(n_2, n_1^\prime,
n_1^\prime)$.  We have seen that to describe the parton shower for one emission,
we only need the coefficient of the SCET$_2$ operator,
$\mo^{(1)}_{2}(n_1,n^\prime_1)$.  Similarly, in SCET$_3$ we need the coefficient
of the operator $\mo^{(2)}_3(n_2, n_1^\prime, n_2^\prime)$.  We can follow the
same steps from \app{scet1/scet2} to calculate the Wilson coefficients
$C^{(2)}_3(n_2, n_1^\prime,n_2^\prime)$.  In this way, it is not difficult to
show that
\begin{align}
C^{(2)}_3(n_2, n_1^\prime, n_2^\prime)&=  \, C^{(2)}_{3, \,\rm{LO}} \nonumber \\
&+C^{(2)H,a}_{3, \,\rm{NLO}}
+C^{(2)H,b}_{3, \,\rm{NLO}}
+C^{(2)J\,a}_{3, \,\rm{NLO}}
+C^{(2)J\,b}_{3, \,\rm{NLO}}
+C^{(2)J\,c}_{3, \,\rm{NLO}} \nonumber \\
&+C^{(2)H,a}_{3, \,\rm{NNLO}}
+C^{(2)H,b}_{3, \,\rm{NNLO}}\, ,
\end{align}
where
{\allowdisplaybreaks
\begin{align}
\label{eq:CoeS3a}
C^{(2)}_{3, \,\rm{LO}} (n_2, n'_1, n'_2) &=C^{(1)}_{3, \,\rm{LO}}(n_2,  n_2^\prime)C^{(1)}_{2, \,\rm{LO}}(n_1, n_1^\prime) \, , \\
C^{(2)H,\,a}_{3, \,\rm{NLO}}(n_2, n'_1, n'_2)&=C^{(1)}_{3, \,\rm{LO}}(n_2,  n_2^\prime)C^{(1)H,a}_{2, \,\rm{NLO}}(n_1, n_1^\prime)\,, \nonumber \\
C^{(2)H,\,b}_{3, \,\rm{NLO}}(n_2, n'_1, n'_2)&=C^{(1)}_{3, \,\rm{LO}}(n_2,  n_2^\prime)C^{(1)H,b}_{2, \,\rm{NLO}}(n_1, n_1^\prime)\,, \nonumber\\
C^{(2)J,\,1}_{3, \,\rm{NLO}}(n_2, n'_1, n'_2)&=  C^{(2) J}_{2, \,\rm{NLO}}(n_2, n'_1, n'_2) 
 \tilde{\Theta}_{\delta_3}[ n_2\!\cdot\! n_2^\prime] 
\tilde{\Theta}_{\delta_3}[n_2\!\cdot\! n_1^\prime] 
\tilde{\Theta}_{\delta_3}[n_2^\prime\!\cdot\! n'_1]  \,, \nonumber\\ 
C^{(2)J,\,2}_{3, \,\rm{NLO}}(n_2, n'_1, n'_2)  &= C^{(2) J}_{2, \,\rm{NLO}} (n_2, n'_1, n_1^\prime )
 \tilde{\Theta}_{\delta_3}[ n_2\!\cdot\! n_2^\prime] 
\tilde{\Theta}_{\delta_3}[n_2\!\cdot\! n_1^\prime]
\Theta_{\delta_3}[n_2^\prime\!\cdot\! n_1^\prime] \,, \nonumber \\
C^{(2)J,\,3}_{3, \,\rm{NLO}}(n_2, n'_1, n'_2)&= C^{(2) J}_{2, \,\rm{NLO}} (n_2, n'_1, n_2) 
\Theta_{\delta_3}[ n_2\!\cdot\! n_2^\prime]  
\tilde{\Theta}_{\delta_3}[n_2\!\cdot\! n_1^\prime] 
\tilde{\Theta}_{\delta_3}[n_2^\prime\!\cdot\! n_1^\prime]\, , \nonumber\\
C^{(2)H,\,a}_{3, \,\rm{NNLO}}(n_2, n'_1, n'_2)&=C^{(1)}_{3, \,\rm{LO}}(n_2,  n_2^\prime)C^{(1)}_{2, \,\rm{NNLO}}(n_1, n_1^\prime)\, , \nonumber \\
C^{(2)H,\,b}_{3, \,\rm{NNLO}}(n_2, n'_1, n'_2)&=C^{(2) H}_{2, \,\rm{NNLO}}(n_2, n'_1, n_2)
\Theta_{\delta_3}[ n_2\!\cdot\! n_2^\prime] 
\tilde{\Theta}_{\delta_3}[n_2\!\cdot\! n_1^\prime]
\tilde{\Theta}_{\delta_3}[n_2^\prime\!\cdot\! n_1^\prime]\, , \nonumber
\end{align}
}
and
\begin{align}
C^{(1)}_{3, \,\rm{LO}}(n_2,  n_2^\prime) = \left(  2\frac{(q_2)_{n_1 \perp}^\beta}{\bar{k}_2} 
+ \frac{(\slashed{q}_2)_{n_0 \perp}\gamma^{\beta}_{n'_2 \perp}}{\bar{q}_2} \right)
 \frac{\bar{q}_1}{q_1^2}\frac{\nbs n\!\!\!\slash_1}{4}  \,  \Theta_{\delta_3}[ n_2\!\cdot\! n_2^\prime]  \, .
\end{align}
On the LHS of the equations in the first, second and third line of
(\ref{eq:CoeS3a}) we can write $n_1$ in terms of $n_2$, $n'_2$ and
$n'_1$ using the formulas in (\ref{nrel2}). The SCET$_2$ coefficients
$C^{(2)J}_{2, \,\rm{NLO}}$ and $C^{(2)H}_{2, \,\rm{NNLO}}$ are
defined in Eqs.~(\ref{eq:S2SumE}).  $C^{(1)H,a}_{2, \,\rm{NLO}}(n_1, n_1^\prime),\,
C^{(1)H,b}_{2, \,\rm{NLO}}(n_1, n_1^\prime),$ and $C^{(1)}_{2, \,\rm{NNLO}}(n_1, n_1^\prime)$   
are given in \eq{S2SumA}, and $C^{(1)}_{2, \,\rm{LO}}(n_1, n_1^\prime)$ in \eq{loRep}.
As with any $\sceti \rightarrow \scetipone$ matching, we encode the 
definition of collinearity from the higher scale theory in the lower
one by $\Theta$ functions ({\it cf.} discussion in \subsec{loShowerRe}).  
Some of the $\scettwo$ coefficients above already contained such factors as a 
result of matching to $\scetone$.  In \eq{CoeS3a}, we write out the new ones that 
appear with $\Theta_{\delta_3}$, with $\delta_3 = \lambda^3/\eta^4$ according to our 
usual convention. Since all the coefficients above multiply $\mo^{(2)}_3$,
the scaling of contributions comes from them alone, with 
$C^{(2)}_{3, \,\rm{LO}} \sim \lambda^{-3}$, the NLO terms $\sim \lambda^{-2}$, 
and NNLO going as $\lambda^{-1}$.

At LO, the contribution in SCET$_3$ is given by the replacement procedure on the 
LO contribution in SCET$_2$, $C^{(1)}_{2, \,\rm{LO}}(n_1,n'_1)
\mo^{(1)}_{2}$.  We multiply it by the running function
$U^{(1)}(n_1, n'_1;\,\mu_1,\,\mu)$ and apply the replacement:
\begin{align}
\label{eq:repRuleS3}
 (\chib_{n_{2}})_i \rightarrow  \,
 ( c^{\alpha}_{\rm{LO}}(n_1))_{ji} (\bar{\chi}_{n_{1}})_j g \mb_{\alpha}^{n'_1 \perp}\, ,
 \end{align}
 where $c^{\alpha}_{\rm{LO}}(n_1)$ is
 \begin{align}
  c^{\alpha}_{\rm{LO}}(n_1)  =   \left(  2\frac{(q_2)_{n_1 \perp}^\alpha}{\bar{k}_2} 
+ \frac{(\slashed{q}_2)_{n_1 \perp}\gamma^{\alpha}_{n'_2 \perp}}{\bar{q}_2} \right)\frac{\nbs \ns_{1}}{4}
    \,   \Theta_{ \delta_3} [n_2 \cdot n'_2 ] \, .
\end{align}

Eq.~(\ref{eq:repRuleS3}) has the same structure as Eq.~(\ref{eq:repRuleS2}).
If we go on with the matching down to SCET$_N$, we find that the LO  
result would be given by applying the above replacement $N-1$ times.
At SCET$_N$ we could match everything to the operator $\mo^{(N-1)}_{N}(n_{N-1}, n'_1,\dots,n'_{N-1})$,
and the LO coefficient is
\begin{align}
C^{(N-1)}_{N,\,\rm{LO}} &= \prod_{k=1}^{N-1} U^{(2,k-1,0)}(n_{k-1}, n'_1,\dots,n'_{k-1};\mu_{k-1},\,\mu_k) 
c^{\alpha_k}_{\rm{LO}}(n_{k-1}) \Gamma^{\mu} \,, 
\label{eq:loMaster}
\end{align}
with $\mu_k \sim (k_k)_{n_{k-1} \perp}$ given in \eq{muk2} and 
\begin{align}
  c^{\alpha}_{\rm{LO}}(n_k)  =   \left(  2\frac{(q_{k+1})_{n_0 \perp}^\alpha}{\bar{k}_{k+1}} 
+ \frac{(\slashed{q}_{k+1})_{n_k \perp}\gamma^{\alpha}_{n'_{k+1} \perp}}{\bar{q}_{k+1}} \right)\frac{\nbs \ns_k}{4}
    \,   \Theta_{ \delta_k} [n_{k+1} \cdot n'_{k+1} ] \, ,
\label{eq:littlecDef}
\end{align}
where $\delta_k= \lambda^{2k-3}/\eta^4$.

At NLO($\lambda$), we have two kinds of corrections: hard-scattering and jet-structure. 
We notice that the NLO($\lambda$) hard-scattering terms in SCET$_3$ are just given by those
in SCET$_2$ with the application of the replacement rule (\ref{eq:repRuleS3}).  If we
go on with the matching down to SCET$_N$, we find that we get NLO($\lambda$) hard-scattering
by applying the above replacement rule $N-2$
times to the SCET$_2$ hard-scattering operators.  Thus, we can consider
this as a correction to the matrix elements that we pass to a LL shower:
\begin{align}
\label{eq:hardMasterNLO}
C^{(N-1) H}_{N,\rm{NLO}} &= \big( C^{(1)H,a}_{2, \,\rm{NLO}}(n_1, n_1^\prime)
+C^{(1)H,b}_{2, \,\rm{NLO}}(n_1, n_1^\prime) \big) \\
&\times \Big(\prod_{k=2}^{N-1} U^{(2,k-1,0)}(n_{k-1}, n'_1,\dots,n'_{k-1};\mu_{k-1} \, ,\mu_k) 
c^{\alpha_k}_{\rm{LO}}(n_{k-1})\Big)\,. \nonumber
\end{align}

This approach also works for hard-scattering at NNLO($\lambda$).  Since we did not get
$C^{(2)H,\,b}_{3, \,\rm{NNLO}}$ from a replacement rule, it contains one less
factor of $c^{\alpha_k}_{\rm{LO}}$.
\begin{align}
C^{(N-1) H}_{N,\rm{NNLO}} &= \,C^{(1)H,a}_{2, \,\rm{NNLO}}(n_1, n_1^\prime)
 \left(\prod_{k=2}^{N-2} U^{(2,k-1,0)}(\mu_{k-1},\,\mu_k) 
c^{\alpha_k}_{\rm{LO}}(n_{k-1})\right) \nonumber \\
 &+ C^{(1)H,b}_{2, \,\rm{NNLO}}(n_2, n_1^\prime,n_2^\prime)
  \left(\prod_{k=3}^{N-3} U^{(2,k-1,0)}(\mu_{k-1},\,\mu_k)
c^{\alpha_k}_{\rm{LO}}(n_{k-1})\right) \,, 
\label{eq:hardMasterNNLO}
\end{align}
where the coefficients $C^{(1)H,a}_{2, \,\rm{NNLO}}(n_1, n_1^\prime)$
and $C^{(1)H,b}_{2, \,\rm{NNLO}}(n_2, n_1^\prime,n_2^\prime)$ are
defined in Eqs.~(\ref{eq:S2SumE}).

The NLO($\lambda$) jet-structure corrections in $\scetthree$ are given by
$C^{(2)J,\,I}_{3,\,\rm{NLO}}(n_2, n'_1, n'_2) \mo^{(2)}_3$, where $I=\{1,2,3\}$, are
given by the LO SCET$_1$ operator $\bar{\chi}_{n_0} \gamma^\mu
\chi_{\nb}$ in three steps:  First, we multiply it by the running
factor $U^{(1)}(n_0;\,Q,\,\mu_1)$, second, we apply the replacements
\begin{align}
\label{eq:repRuleS3a}
 (\chib_{n_{2}})_i \rightarrow & \, 
( h^{\alpha \beta}_I)_{ji}(n_2, n'_1, n'_2) (\bar{\chi}_{n_{1}})_j \,g \mb_{\alpha}^{n'_1 \perp} g 
\mb_{\beta}^{n'_2 \perp}\, ,
\end{align}
where
\begin{align}
\label{eq:hsc}
 h^{\alpha \beta}_1 (n_2, n'_1, n'_2)&= %
\,  d_1^{\alpha \beta}(n_2, n'_1, n'_2)
 \tilde{\Theta}_{\delta_3}[ n_2\!\cdot\! n_2^\prime] 
\tilde{\Theta}_{\delta_3}[n_2\!\cdot\! n_1^\prime] 
\tilde{\Theta}_{\delta_3}[n_2^\prime\!\cdot\! n'_1]  \, ,  \\
 h^{\alpha \beta}_2 (n_2, n'_1, n'_2)&=  %
  \, d_2^{\alpha \beta}(n_2, n'_1, n'_1)
 \tilde{\Theta}_{\delta_3}[ n_2\!\cdot\! n_2^\prime] 
\tilde{\Theta}_{\delta_3}[n_2\!\cdot\! n_1^\prime]
\Theta_{\delta_3}[n_2^\prime\!\cdot\! n_1^\prime] \, , \nonumber\\
 h^{\alpha \beta}_3 (n_2, n'_1, n'_2)&= %
\, d_3^{\alpha \beta}(n_2, n'_1, n_2) 
 \Theta_{\delta_2}[ n_2\!\cdot\! n_2^\prime] 
\tilde{\Theta}_{\delta_3}[n_2\!\cdot\! n_1^\prime]
\tilde{\Theta}_{\delta_3}[n_2^\prime\!\cdot\! n_1^\prime]\, . \nonumber
\end{align} 
The $d^{\alpha \beta}_I$ coefficients are defined in Eqs.(\ref{eq:AppD1}, \ref{eq:C11p1pb}, \ref{eq:ApMS}).
Third, we multiply the operators that come from applying  Eqs.~(\ref{eq:hsc})
by the second running factor. This depends on the SCET$_2$ operator
so each replacement rule (\ref{eq:repRuleS3a}) is followed by a different factor:
$ h^{\alpha \beta}_1 $ by $U^{(2,2,0)}(n_2, n'_1, n'_2;\,\mu_1,\mu_2)$, 
$ h^{\alpha \beta}_2 $ by $U^{(2,2,0)}(n_2,n'_1,n'_1;\,\mu_1, \mu_2)$ and
$ h^{\alpha \beta}_3 $by $U^{(2,2,0)}(n_2,n'_1,n_2;\,\mu_1,\, \mu_2)$.
Since these corrections are independent of the initial hard process,
we would encounter the same calculations we have done just now for
SCET$_1$ to SCET$_3$,  at any matching SCET$_i$ to SCET$_{i+2}$.
Thus, the NLO($\lambda$) jet-structure coefficients for the SCET$_N$ operator are:
\begin{align}
\label{eq:CNLOjet}
C^{(N-1) J}_{N,\,\rm{NLO}} =  \sum_{l=1}^{N-2} C^{(N) J}_{N,\,\rm{NLO}} (l) \, ,
\end{align}
where
\begin{align}
\label{eq:masterJet}
C^{(N-1) J}_{N,\,\rm{NLO}} (l)&= \sum_{I=1}^3 \Big[ \Big(
\prod_{k=1}^{l-1} U^{(2,k-1,0)}(n_{k-1}, n'_1,\dots,n'_{k-1};\mu_{k-1} ,\,\mu_k)
c^{\alpha_k}_{\rm{LO}}(n_{k-1})\Big)\\
& \times U^{(l+1)}_I(\mu_{l} ,\,\mu_{l+1})
\otimes h^{\alpha \beta}_I (n_{l+1}, n'_{l}, n'_{l+1})\nonumber \\ \nonumber&
\times \Big( \prod_{k=l+1}^{N-1} U^{(2,k-1,0)}(n_{k-1}, n'_1,\dots,n'_{k-1};\mu_{k-1},\,\mu_k)
c^{\alpha_k}_{\rm{LO}}(n_{k-1})\Big) \Big] \Gamma^{\mu} \, , \nonumber
\end{align}
with
\begin{align}
U^{(l+1)}_1(\mu_{l} ,\,\mu_{l+1})&= U^{(2,l+1,0)}(n_{l+1}, n'_1,\dots,n'_{l},n'_{l+1};\mu_{l},\,\mu_{l+1}) \, ,\\
U^{(l+1)}_2(\mu_{l} ,\,\mu_{l+1})&= U^{(2,l+1,0)}(n_{l+1}, n'_1,\dots,n'_{l},n'_{l};\mu_{l},\,\mu_{l+1}) \, , \\ \nonumber
U^{(l+1)}_3(\mu_{l} ,\,\mu_{l+1})&= U^{(2,l+1,0)}(n_{l+1}, n'_1,\dots,n'_{l},n_{l+1};\mu_{l},\,\mu_{l+1}) \, , \nonumber
\end{align}
and 
\begin{align}
h^{\alpha \beta}_1 (n_{l+1}, n'_{l}, n'_{l+1})&= %
\,  d_1^{\alpha \beta}(n_{l+1}, n'_{l}, n'_{l+1})
 \tilde{\Theta}_{\delta_{l+1}}[ n_{l+1}\!\cdot\! n_{l+1}^\prime] 
\tilde{\Theta}_{\delta_{l+1}}[n_{l+1}\!\cdot\! n_{l}^\prime] 
\tilde{\Theta}_{\delta_{l+1}}[n_{l+1}^\prime\!\cdot\! n'_1]  \, , \nonumber  \\
 h^{\alpha \beta}_1 (n_{l+1}, n'_{l}, n'_{l})&=  %
  \, d_2^{\alpha \beta}(n_{l+1}, n'_{l}, n'_{l+1})
 \tilde{\Theta}_{\delta_{l+1}}[ n_{l+1}\!\cdot\! n_{l+1}^\prime] 
\tilde{\Theta}_{\delta_{l+1}}[n_{l+1}\!\cdot\! n_{l}^\prime]
\Theta_{\delta_{l+1}}[n_{l+1}^\prime\!\cdot\! n_{l}^\prime] \, , \nonumber\\
 h^{\alpha \beta}_3(n_{l+1}, n'_{l}, n_{l+1})&= %
\, d_3^{\alpha \beta}(n_{l+1}, n'_{l}, n'_{l+1})
 \Theta_{\delta_{l+1}}[ n_{l+1}\!\cdot\! n_{l+1}^\prime] 
\tilde{\Theta}_{\delta_{l+1}}[n_{l+1}\!\cdot\! n_{l}^\prime]
\tilde{\Theta}_{\delta_{l+1}}[n_{l+1}^\prime\!\cdot\! n_{l}^\prime]\, .
\end{align}
The coefficients $d_I^{\alpha \beta}$ here are equal to the coefficients 
$d^{\alpha \beta}_I$ defined in Eqs.(\ref{eq:AppD1}, \ref{eq:C11p1pb}, \ref{eq:ApMS})
upon the substitution $(n_2, n'_1, n'_2)\rightarrow (n_{l+1}, n'_{l}, n'_{l+1})$ and $\delta_3 \rightarrow \delta_{l+1}$.

\section{$\mo(\alpha_s^2)$ Correction to Splitting Function}
\label{app:nloSplit}

\begin{figure}[t!]
\centering
\includegraphics{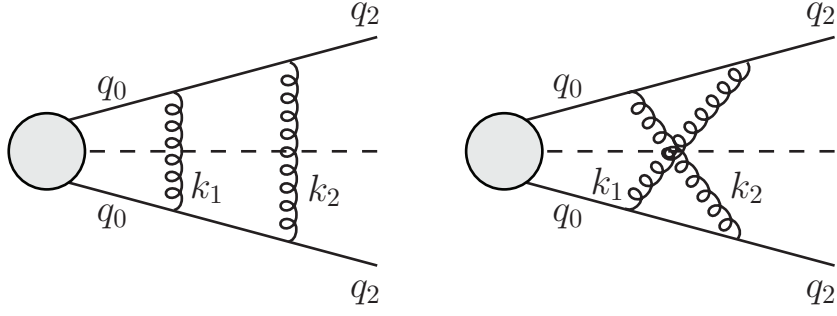}
\caption{Two distinct real emission contributions to $\nlosplit$ drawn as amplitudes squared.  
They are referred to as the box ({\bf L}) and crossed ({\bf R}) contributions.} 
\label{fig:boxCross}
\end{figure}

One of the cross-checks on our results is the rederivation of (the abelian part
of) the $\mo(\alpha_s^2)$ correction to the $q \rightarrow qg$ splitting
function, $\nlosplit$.  This follows from obtaining the NLO($\lambda$)
correction to two-gluon emission.  For comparison, we have chosen the classic
result of Curci {\it et al.}~\cite{Curci:1980uw}.  The full expression for
$\nlosplit$ involves many real and virtual contributions.  Here we will only
explicitly calculate the $\sim \, C_F^2$ component of $\nlosplit$ and show it
agrees.  (Obtaining the full result requires additional
non-abelian diagrams.)  Ref.~\cite{Curci:1980uw} splits the abelian, two-gluon,
real emission contributions to $\nlosplit$ into two topologically inequivalent
diagrams, the box and crossed graphs, \fig{boxCross}.  We calculated each of
these individually.

\begin{figure}[t!]
\centering
\includegraphics{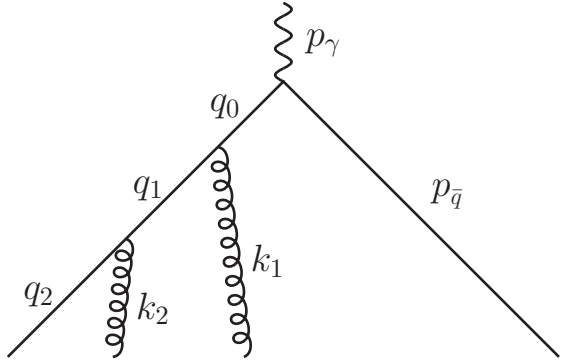}
\caption{Kinematics for double gluon emission.  This particular diagram corresponds to the ``A'' graph of Fig.~(\ref{fig:figureABCC}).} 
\label{fig:doubleGluKinRep}
\end{figure}

The $\scetone$ amplitude contains three graphs for two-gluon emission.  These are shown in \fig{figureABCC}, 
and we give the corresponding amplitudes in Eqs.~(\ref{Am2ES1}).  In order to obtain $\nlosplit$, we will 
need to square the amplitudes and partially integrate over phase space.  
Thus, we need to choose an explicit kinematics.  
We redraw, in Fig.~(\ref{fig:doubleGluKinRep}), our vector labels for two-gluon emission.
We choose a somewhat nonstandard assignment for our variables.  This is to aid
in the comparison with \cite{Curci:1980uw}.  The final state parton shower
occurs for timelike virtual particles, and momentum fractions decrease the
farther we are from the initial hard scattering.  By contrast,
\cite{Curci:1980uw} considered a DIS-type process where the shower is spacelike.
Since the radiation in that case comes from initial states, the momentum
fractions decrease toward the hard interaction.  Only at LO in $\alpha_s$ are
the spacelike and timelike splitting functions equal, by the Gribov-Lipatov
relation \cite{Gribov:1972rt}.  At higher orders, this gets violated, but there
is a straightforward conversion procedure, detailed in \cite{Curci:1980uw,
  Stratmann:1996hn}.  We, however, choose our kinematics such that our variable
relations are equivalent to those for a spacelike process.  For example,
$\nlosplit$ is a function of $x \, \equiv \, \bar{q}_0/\bar{q}_2$.  In a
spacelike process, $x \in [0,1]$.  Rather than convert our answer, we will also
define $x$ as above, even though this means for us $x \in [1,\infty)$.  Other
integration variables will have their ranges shifted so that they have the same
relation with $x$ as in DIS, and thus they enter into our expression in the same
way.  Lastly, we do not do the phase space integration for $q_2$.  While this
{\it is} necessary for the timelike splitting function, the analogous particle
for a spacelike process is a fixed initial state.  Thus, for comparison
purposes, we can leave it undone.  Our vectors are as follows (note that this is
a different frame from the one used previously for matching):
%
\begin{eqnarray}
q_2 &=& \left\{ p,0,0,p \right\} \nonumber \\
k_1 &=& \left\{ -z_1 p - \frac{k^2_{1\perp}}{4p z_1}, \, k_{1\perp} \cos(\phi_1), 
\, k_{1\perp} \sin(\phi_1), -z_1 p + \frac{k^2_{1\perp}}{4p z_1}  \right\}  \nonumber \\
k_2 &=& \left\{ -z_2 p - \frac{k^2_{2\perp}}{4p z_2}, \, k_{2\perp} , \, 0, -z_2 p + \frac{k^2_{2\perp}}{4p z_2}  \right\} 
\nonumber \\
q_0 &=& \left\{ x\, p + \frac{q_0^2 + |\overrightarrow{{\bf k_1}}_{\perp} + \overrightarrow{{\bf k_2}}_{\perp} |^2}{4p x}, 
\overrightarrow{{\bf k_1}}_{\perp} + \overrightarrow{{\bf k_2}}_{\perp},
x\, p - \frac{q_0^2 + |\overrightarrow{{\bf k_1}}_{\perp} + \overrightarrow{{\bf k_2}}_{\perp} |^2}{4p x}   \right\}.
\label{eq:doubleKin}
\end{eqnarray}
%
Before proceeding, we wish to note some things about our assignment.  First of all, while it is 
redundant to include $q_0 = k_1 + k_2 + q_2$, we will integrate over $d^4 q_0$ and wanted to 
present our parametrization.  We see that $x = 1 - z_1 - z_2$.  This is consistent with
the spacelike case, but here, $z_1, z_2 \in (-\infty, 0]$, hence the minus signs in $k_1$ and $k_2$.  
Additionally, only the relative azimuthal angle between $k_1$ and $k_2$ is physical.  Thus, to simplify 
our formulas, we fix $k_2$ in the $x-z$ plane.

As a last step before squaring and integrating, we will introduce our measure
and integral parametrization.  While one could integrate the full final state
phase space including the antiquark, we instead exploit the factorization of the
the cross section into a hard interaction ${\cal H}$, a radiation-function
${\cal K}$, and fragmentation functions $q_{B,\, F}(x)$ which determine how the
partons arrange themselves into hadrons. Schematically, $\sigma = \mathcal{H}
\otimes (\mathcal{K}_{\rm LO}(x,q^2)\,+\mathcal{K}_{J,\,{\rm NLO}}(x,q^2)\, +
\ldots ) \otimes \Pi\: q_{B,\, F}(x) = {\cal H}\otimes ({\cal R}_{\rm LO} +
{\cal R}_{\rm NLO}+\ldots)$.  For our computations we need only integrate the
phase space for ${\mathcal R}$, and it will remain independent of the details of
${\cal H}$.  Taking $d \,\equiv\, 4+\epsilon$:
\begin{eqnarray}
 \mathcal{R}_{\rm LO} &=& \sum_i \frac{2}{\bar{q}_i} \int \prod_{j=1}^{i} 
\frac{\slashed{d}^{d-1}k_j}{z_j}\, \slashed{d}^d q_0\, dq^2\,{\rm PP}
\left[ | C^{(i-1)}_{i,\,{\rm LO}} \langle 0 | \mo^{(i-1)}_i | q(i-1)g\bar{q}
  \rangle |^2 \right] \nonumber  \\ 
&&\times \delta(x - \bar{q}_0/\bar{q}_i) \delta(q^2 - (q_i + \sum_{j=1}^i k_j)^2)\, (2\pi)^d
\delta^{(d)}(q_0 - \sum_{j=1}^i k_j)), \nonumber \\
\mathcal{R}_{J,\,{\rm NLO}} &=& \sum_i \frac{2}{\bar{q}_i} \int \prod_{j=1}^{i} 
\frac{\slashed{d}^{d-1}k_j}{z_j}\, \slashed{d}^d q_0\, dq^2\,{\rm PP}\left[ | C^{(i-1)J}_{i,\,{\rm NLO}} 
\langle 0 | \mo^{(i-1)}_i | q(i-1)g\bar{q} \rangle |^2 \right] \nonumber \\ 
&&\times \delta(x - \bar{q}_0/\bar{q}_i) \delta(q^2 - (q_i + \sum_{j=1}^i k_j)^2)\, (2\pi)^d 
\delta^{(d)}(q_0 - \sum_{j=1}^i k_j)),
\label{eq:jDef}
\end{eqnarray}
and the $q_i$ phase space and spin-sum are moved into $\mathcal{H}$.  We define
$z_j$ analogously to Eqs.~(\ref{eq:doubleKin}).  The setup we describe in the
body of the paper uses Wilsonian cutoffs in phase space, both to keep the
contributions of different operators distinct via $\Theta$'s and to cutoff soft
and collinear divergences via some shower resolution parameter which keeps
configurations outside of nonperturbative regimes.  In the shower language the
$\Pi\, q_{B,F}(x)$ term in ${\cal R}$ signifies the hadronization model and may
depend on more than just $x$ variables, and the ${\cal K}$ term signifies the
infrared finite fully differential shower computations.  In \eq{jDef} we are
integrating over $\perp$-momenta to carry out the perturbative comparison with
Curci {\it et al}. Here we are implicitly in the $\overline{\rm MS}$ scheme, and it is
the perturbative IR divergences in ${\cal R}$ that get absorbed by $q_{B,\,
  F}(x)$.  The $\mathcal{R}$ terms that we need consist of only the
$1/\epsilon^2$ and $1/\epsilon$ portions of the corresponding operator
expectation values.  The non-pole contributions from $C^{(i-1)}_{i,\,{\rm LO}}$
and $C^{(i-1)J}_{i,\,{\rm NLO}}$ (Eqs.~\ref{eq:loMaster} and
\ref{eq:masterJet}), along with higher-order corrections are in higher order
terms in the ${\cal R}$ functions.  The hard-scattering corrections are in
$\mathcal{H}$.  The reason we extract only the pole terms is that these are
precisely what give the expression for $\losplit$ and $\nlosplit$.  In addition to 
selecting the pole part, we also
define ${\rm PP}$ to remove those portions of the matrix element which enter
into $\mathcal{H}$ such as the final quark spin-sum, current $\Gamma$, and
antiquark quantities.

In this $\overline{\rm MS}$ factorization scheme, we need to define our correction
operator differently than in Apps.~\ref{app:scet1/scet2} and
\ref{app:scet2/scet3}.  Since $\nlosplit$ requires the calculation of
two-gluon emission, we find it simplest here to calculate in $\scetthree$ where only
$C^{(2)J,\,1}_{3, \,\rm{NLO}}$ in \eq{CoeS3a} contributes. This corresponds to
taking limits such that only its $\Theta$-function equals one, while the other
jet-structure coefficients are zero.  Since we integrate it over all of phase
space, which includes the strongly-ordered limit, we need to subtract the LO
contribution.  This just comes from $C^{(2)}_{3,\, {\rm LO}}(n_2,n'_1,n'_2)
\mo^{(2)}_3$, but we take care to only remove the pole parts consistent with
$\overline{\rm MS}$.  We can thus write the subtraction as:
\begin{align}
\mathcal{R}_{J,\,{\rm NLO}}^{q\rightarrow qgg} 
\,&=\, \int d\Pi_{k_1,\, k_2,\, q_0} {\rm PP} \bigg[
|C^{(2)J,\,1}_{3, \,\rm{NLO}}(n_2,n'_1,n'_2) \langle 0| \mo^{(2)}_3 |qgg\bar{q}
\rangle|^2
\nonumber\\
&\qquad  - \left(
|{C^{(2)}_{3,\,{\rm LO} }}(n_2,n'_1,n'_2) \langle 0| \mo^{(2)}_3 \Gamma^{\mu}  
|qgg\bar{q} \rangle |^2 \right)_{\overline{\rm MS}} \bigg],
\end{align}
where $C^{(2)J,\,1}_{3, \,\rm{NLO}}$ is evaluated such that $\Theta = 1$ over all of phase space.
We will describe the subtraction portion in detail below, but first we concentrate on the correction term.

By fixing the virtuality of $q_0^2 \equiv q^2$, we can obtain an expression
without having to know its exact limits, which will depend on the details of the
hard scattering.  For $\nlosplit$, one only needs to calculate one-loop
corrections to single emission and tree-level double emission, and we now
specialize to the latter case.  We perform the $d$-dimensional integration
over $d^d q_0$ and rewrite the integral in terms of $k_{1\perp}$ and
$k_{2\perp}$ dependent functions with $z_{1,2}$-dependent coefficients.  Using
the same parametrization as Ref.~\cite{Ellis:1996nn}, we can write:
\begin{align}
\label{eq:ABCDj}
\mathcal{R}^{q\rightarrow qgg}_{J,\,{\rm NLO}} &= \frac{1}{(16\pi^2)^2} \int \, dq^2\, \frac{dz_1}{z_1}\, \frac{dz_2}{z_2}\, 
\frac{d^{d-2}{\bf k_1}_{\perp}}{\pi}\, \frac{d^{d-2}{\bf k_2}_{\perp}}{\pi}\, \delta(1 - x - z_1 - z_2)  \\
\times & \delta \left(q^2 - (a_1 \, {\bf k_1}_{\perp}^2 + a_2 \, {\bf k_2}_{\perp}^2 - 
{\bf k_1}_{\perp} \cdot {\bf k_2}_{\perp} \, ) \right) \nonumber \\
\times & \frac{1}{q^4} \left( A(z_1, z_2) + B(z_1, z_2) \frac{{\bf k_1}_{\perp} \cdot {\bf k_2}_{\perp}}{{\bf k_1}_{\perp}^2} + 
C(z_1, z_2) \frac{{\bf k_1}_{\perp} \cdot {\bf k_2}_{\perp}}{{\bf k_2}_{\perp}^2} \right. \nonumber \\
& \left. + \; D(z_1, z_2) \frac{({\bf k_1}_{\perp} \cdot {\bf k_2}_{\perp})^2}{{\bf k_1}_{\perp}^2 {\bf k_2}_{\perp}^2} 
+ \; E(z_1,z_2) \, \frac{{\bf k_1}_{\perp}^2}{{\bf k_2}_{\perp}^2} \; 
+ \; F(z_1,z_2) \, \frac{{\bf k_2}_{\perp}^2}{{\bf k_1}_{\perp}^2}   \right)
 - \left[{\rm LO} \right],&  
\end{align}
where $a_1=-(1-z_2)/z_1$ and $a_2=-(1-z_1)/z_2$.
The functions $A, \, B, \, C, \, D$ are defined in \cite{Ellis:1996nn}, and
their corresponding ${\bf k_i}_{\perp}$ integrals are finite.  We can check the
intermediate step of their integration with \cite{Ellis:1996nn}.  The terms in
our $q^2$ $\delta$-function have a relative sign compared to theirs, as our
$q^2>0$.  As a computational aside, we found it easiest to pass to a change of
variables: ($u \,\equiv\, k_{1\perp} k_{2\perp},$ $w \equiv\,
k_{1\perp}/k_{2\perp}$).  Then the $\delta$-function just enforces:
\beq
u \,=\, u_0 \,\equiv\, \frac{q^2 \, w}{a_1 w^2 + a_2 - 2 \, w \cos(\phi_1)}.
\label{eq:uDelta}
\eeq
Performing all but the $dz_i$ integrals in $\mathcal{R}$, we get Table \ref{tbl:singlePole}, 
which corresponds to \cite{Ellis:1996nn}'s Table 5.
\begin{center}  
\begin{table}[ht!]        
  \begin{tabular}{|c|l|}
  \hline
    Function of ${\bf k_i}_{\perp}$             & Contribution to $\mathcal{R}$ multiplying  \\
    in integrand of equation (\ref{eq:ABCDj}) & $\frac{q^2}{(16\pi^2)^2 x} \int dz_1 \, dz_2 \, \delta(1-z_1-z_2-x)$   \\ \hline
    1   & $A(z_1,z_2)$ \\ \hline
    $\frac{{\bf k_1}_{\perp} \cdot {\bf k_2}_{\perp}}{{\bf k_1}_{\perp}^2}$     & $-\frac{z_2}{1-z1} B(z_1,z_2)$ \\ \hline
    $\frac{{\bf k_1}_{\perp} \cdot {\bf k_2}_{\perp}}{{\bf k_2}_{\perp}^2}$     & $-\frac{z_1}{1-z2} C(z_1,z_2)$ \\ \hline
    $\frac{({\bf k_1}_{\perp} \cdot {\bf k_2}_{\perp})^2}{{\bf k_1}_{\perp}^2 {\bf k_2}_{\perp}^2}$ & 
    $\left( 1 + \frac{x}{2 z_1 z_2} \ln \left[ \frac{x}{(1-z_1)(1-z_2)} \right] \right) D(z_1,z_2)$ \\ \hline
  \end{tabular}
 \caption{Purely finite contributions to $\mathcal{R}$}
\label{tbl:singlePole}
\end{table}
\end{center}
We thus reproduce the earlier result.

The $E, \; F$ functions multiply integrals that lead to single $\epsilon$ poles
after the $dk_{i\perp}$ integrals (and double poles after integrating $q^2$),
and so we must be more careful in treating them.  These double poles correspond
to the LO contribution, which we are explicitly subtracting as it does not
contribute to $\nlosplit$.  We discuss the subtraction in detail below \eq{loArrowSubt}.  For
now we concentrate on the divergent integrals multiplying $E$ and $F$.  When we
did our computations for Table (\ref{tbl:singlePole}), we were helped by the
finiteness of the expressions under the $dk_{i\perp}$ integration.  We could
thus take $\epsilon \rightarrow$ 0 for these terms, which greatly simplifies
their integrals.  By contrast, we will need to keep the $\epsilon$-dependence of
the $E, \; F$ terms, which results in an intractable computation.  To get around
this, one can introduce subtraction functions, which simply reproduce the
$\epsilon$ poles (these are merely a computational aid and are not related to
the subtraction of LO).  We will need to take care that they do not remove any
finite pieces.  Secondly, since their full contribution to $\mathcal{R}$ is
$\propto 1/\epsilon^2$, we will need to include for $E$ and $F$ any terms
$\propto \, \epsilon$ that multiply $\frac{{\bf k_1}_{\perp}^2}{{\bf
    k_2}_{\perp}^2}$ or $\frac{{\bf k_1}_{\perp}^2}{{\bf k_2}_{\perp}^2}$.
These arise from doing Dirac algebra in $d$-dimensions.

To do the integrals in $\mathcal{R}$ which multiply $E$ and $F$, we will change
variables to $u, \, w$, and perform the $u$ integration as well as the trivial
$\phi_2$ azimuthal one.  We get for this contribution to $\mathcal{R}$:
\begin{align}
\mathcal{R}|_{E, \, F} &= \frac{1}{(16\pi^2)^2} \frac{2}{\pi} \int \, dq^2 \, \frac{dz_1}{z_1} \frac{dz_2}{z_2} 
\, d\phi_1 \, dw \, \delta(1-x-z_1-z_2) \nonumber \\
&\times \left( \frac{w\, u_0^{2+\epsilon}}{2\, q^2} E(z_1,z_2) \,+\,\frac{u_0^{2+\epsilon}}{2\,w\, q^2} F(z_1,z_2)  
\right) \frac{1}{q^4}, 
\label{eq:jEF}
\end{align}
where $u_0$ is defined by equation (\ref{eq:uDelta}). 
We only need the leading poles in $\epsilon$, and so rather than performing the
$w$ and $\phi_1$ integrals for the functions multiplying $E, \, F$, we will
define subtraction functions to reproduce the poles of $\frac{w\,
  u_0^{2+\epsilon}}{2\, q^2},\, \frac{u_0^{2+\epsilon}}{2\,w\, q^2}$,
respectively:
\begin{align}
\mathcal{S}_E &= \frac{q^2}{2\, a_1^2} \frac{w^{-\epsilon}}{(w+1)} \,, \nonumber \\
\mathcal{S}_F &= \frac{q^2}{2\, a_2^2} \frac{w^{\epsilon}}{(w+w^2)}.
\label{eq:subtrTerms}
\end{align}
Integrating these in $w$ gives us a pure $1/\epsilon$ term.  Subtracting them from the functions in equation 
(\ref{eq:jEF}): 
\begin{eqnarray}
\mathcal{A}_E &\equiv \frac{w\, u_0^{2+\epsilon}}{2\, q^2} &= \frac{q^2\, w^3}{2(a_2 + a_1 w^2 - 2 w \cos(\phi_1))^2} 
\left(\frac{w\, q^2}{a_2 + a_1 w^2 - 2 w \cos(\phi_1)}\right)^{\epsilon}\,, \nonumber \\
\mathcal{A}_F &\equiv \frac{u_0^{2+\epsilon}}{2\,w\, q^2}  &= \frac{q^2}{2\, w(a_2 + a_1 w^2 - 2 w \cos(\phi_1))^2} 
\left(\frac{w\, q^2}{a_2 + a_1 w^2 - 2 w \cos(\phi_1)}\right)^{\epsilon}
\label{eq:addTerms}
\end{eqnarray}
leads to finite integrals, allowing us to pass to the $\epsilon \rightarrow$ 0
limit prior to integration, making the calculation tractable.  After integrating
$w$ and $\phi_1$, we want the $\epsilon^{-1,\, 0}$ pieces as these turn into the
single and double poles upon doing the $q^2$ integral and contribute to
$\mathcal{R}_{J,\, {\rm NLO}}$.  The $\epsilon^0$ piece has one contribution
besides that from $(\mathcal{A}_{E,F} \,-\, \mathcal{S}_{E,F})|_{\epsilon=0}$
($\mathcal{S}_{E,F}$ contributes a pure 1/$\epsilon$ pole).  Our $w$ integration
goes from 0 to $\infty$, and we obtained $\mathcal{S}_{E,F}$ by expanding
$\mathcal{A}_{E,F}$ in the appropriate $w\rightarrow 0,\infty$ limit to pick up
the pole, while carefully regulating the other integration limit so as not to
contribute its own spurious divergence or any subleading terms.  However, we see
that in equation (\ref{eq:addTerms}), taking these limits actually results in
factors $(a_1 w)^{-\epsilon}$ and $(w/a_2)^{\epsilon}$.  Expanding the
$a_i^{\pm\epsilon}$ to LO in $\epsilon$ does not affect $\mathcal{S}_{E,F}$.
Nonetheless, since the subtraction functions have $1/\epsilon$ poles, including
the NLO part of the $\epsilon$-expansion will yield an $\epsilon^0$ contribution.  This
${\cal O}(\epsilon^0)$ term is not in $\mathcal{A}_{E,F}|_{\epsilon=0}$ since
they send $u_0^\epsilon \,\rightarrow\,$ 1.  Thus, we have the following
addition to the contributions from the integration of $\mathcal{R}|_{E, \, F}$:
\begin{align}
\mathcal{B}_E &= -\epsilon \ln(a_1) \frac{q^2}{2\, a_1^2}
\frac{w^{-\epsilon}}{(w+1)} 
  \,, \nonumber \\
\mathcal{B}_F &= -\epsilon \ln(a_2) \frac{q^2}{2\, a_2^2} \frac{w^{\epsilon}}{(w+w^2)}.
\label{eq:bTerms}
\end{align}
In the end, our $\epsilon^{-1,\, 0}$ contributions after $w$ and $\phi_1$
\begin{table}[t!]        
  \begin{tabular}{|c|l|}
  \hline
  Function of ${\bf k_i}_{\perp}$             & Contribution to $\mathcal{R}$ multiplying  \\
  in integrand of equation (\ref{eq:ABCDj}) & $\frac{q^2}{(16\pi^2)^2 x} \int dz_1 \, dz_2 \, \delta(1-z_1-z_2-x)$   \\ \hline
  $\frac{{\bf k_1}_{\perp}^2}{{\bf k_2}_{\perp}^2}$   & $\left[ \frac{2\,x\, z_1}{(1-z_2)^2 z_2\, \epsilon} 
    \left( 1 -\epsilon \ln\left[ -\frac{1-z_2}{z_1} \right] \right) \right.$ \\
    &  $\left. \;+\; \frac{z_1}{z_2 (1-z_2)^2} \left( z_1 z_2 + x\, \left( \ln \left[ \frac{z_2 (1-z_2)}{z_1 x} \right] -1 \right) 
    \right) \right] E(z_1, z_2) $\\ \hline
  $\frac{{\bf k_2}_{\perp}^2}{{\bf k_1}_{\perp}^2}$   & $\left[ \frac{2\,x\, z_2}{(1-z_1)^2 z_1\, \epsilon} 
    \left( 1 -\epsilon \ln\left[ -\frac{1-z_1}{z_2} \right] \right) \right.$ \\
    &  $\left. \;+\; \frac{z_2}{z_1 (1-z_1)^2} \left( z_1 z_2 + x\, \left( \ln \left[ \frac{z_1 (1-z_1)}{z_2 x} 
      \right] -1 \right) \right) \right] F(z_1, z_2) $\\ \hline
  \end{tabular}
  \caption{Contributions to $\mathcal{R}|_{E, \, F}$}
\label{tbl:doublePole}
\end{table}
integration come from: $\mathcal{S}_{E,F} \,+\, \mathcal{B}_{E,F} \,+\,
(\mathcal{A}_{E,F} \,-\, \mathcal{S}_{E,F})|_{\epsilon=0}$.  For integrating the
first two terms, we leave the full $\epsilon$ dependence as this was tractable.
Collecting everything, we can obtain the counterpart to Table
\ref{tbl:singlePole} for $E, \; F$, (Table \ref{tbl:doublePole}).  
%
%

Having set up this much of the integration, we can take the amplitude squared
from the process of interest and decompose it in terms of the $A(z_1,z_2),\,
B(z_1,z_2),\, etc.$ basis.  We then simply have to read off the results from
Tables \ref{tbl:singlePole} and \ref{tbl:doublePole}, and perform the $z_{1,\,
  2}$ integrals.  One of these is made trivial by the remaining $x$-dependent
$\delta$-function.  As mentioned at the beginning of this Appendix,
\cite{Curci:1980uw} recognizes two topologically distinct contributions, which
we shall refer to as box and crossed ({\it cf.}~\fig{boxCross}), because of
their appearance as cut two-loop diagrams.  We can identify them in our
calculation by their color structures ($C_F^2$ and $C_F^2 - \frac{1}{2}C_F\,
C_A$, respectively).  In fact, we can already calculate the entire crossed
contribution as it only involves terms from Table \ref{tbl:singlePole}, having
no double pole contribution to $\mathcal{R}$ and thus requiring no subtraction of LO.  
Determining the box graph, however, involves treating the LO subtraction properly.

As this subtraction is one of the more subtle points of the computation, we will present it in some detail.  
Its handling is tied up with what one means precisely by a ``subleading splitting function.''  At LO in $\alpha_s$, 
the definition is clear.  The same splitting function that gives us the probability for a 1 $\rightarrow$ 2 radiation 
also determines the running of parton densities:
\beq
Q^2 \frac{\partial}{\partial Q^2} f(x,Q^2) \,=\, \int^1_x \frac{dz}{z} \, P_{qq} 
\left( \frac{x}{z},\alpha_s(Q^2)  \right) \, f(z,\, Q^2),
\label{eq:pdfRun}
\eeq
where the $\mo(\alpha_s)$ part of $P_{qq}$, $P_{qq}^{(0)}$ is given by
Eq.~(\ref{eq:ap}).  To determine $\nlosplit$, we have had to calculate a 1
$\rightarrow$ 3 splitting, thus the probabilistic interpretation in terms of
radiation is nontrivial as it involves a mix of 1 $\rightarrow$ 2 and 1
$\rightarrow$ 3 processes.  At the level of Eq.~(\ref{eq:pdfRun}) though, we see
that we are just correcting PDF evolution.  In addition to the real-emission
calculation that we are pursuing, one can alternatively determine $P_{qq}$ from
the anomalous dimension of certain twist-2 operators \cite{Floratos:1977au,
  Floratos:1978ny}.  Ref. \cite{Curci:1980uw} made a comparison to this approach
and found agreement to $\mo(\alpha_s^2)$.  Since $\nlosplit$ is thus a two-loop
object, it has the scheme dependence one would expect at this order, and so we
need to make sure that we compute in the same scheme, which is why we do our LO
subtraction in $\overline{\rm MS}$.  In SCET, one could attempt the same
cross-check from a straightforward two-loop calculation after fixing to one's
renormalization scheme of choice.

We will now show how to subtract the LO portion in the calculation of
$\mathcal{R}_{J,\,{\rm NLO}}^{q\rightarrow qgg}$. We get a double collinear pole
associated with the strongly-ordered emission of two gluons.  We want to write
this as removing the emission coming from our LO operator, $C^{(2)}_{3,\,{\rm
    LO}} \mo^{(2)}_{3}$.  As with any subtraction scheme, while the pole is
unambiguous, we need to make sure to remove the appropriate finite pieces.  We
note that $c^{\alpha}_{\rm LO}$ defined by Eq.~(\ref{eq:loCoeff}) contains \NLOl
pieces (in $\scetthree$ power counting) which come from the offshellness of the
intermediate quark.  It is true that the LO replacement rule,
Eq.~(\ref{eq:loArrow}), gives only the splitting function times the logarithmic,
collinear divergence.  Nonetheless, the Wilson coefficients given by
Eq.~(\ref{eq:loOp}) for offshell quarks have additional terms.  From the point
of view of amplitude matching, this poses no problem.  However, if we want to
copy \cite{Curci:1980uw}'s scheme, then we can only subtract poles associated
with the pure LO result after integration.  As an operator subtraction in
$\scetthree$, this means we need to change $C^{(2)}_{3,\, {\rm LO}}$.  In order
to recover the correct splitting function with no NLO contribution, we will need
to project the offshell quark momentum to an onshell one with the same
$\bar{p}$-fraction.  This alone, though, does not specify the spatial
orientation of the vector and will not necessarily kill the subleading terms.
To do that, we write the replacement rule, but in the limit that the offshell
quark's daughters are exactly collinear with it.  Equivalently, if we are in the
frame determined by $\bar{n} = \{1,0,0,-1 \}$, we can project the quark momentum
along $n = \{1,0,0,1 \}$, {\it i.e.} $q_i \,\rightarrow\, \frac{\bar{q}_i}{2} n
= q'_i$.  Since the replacement rule also makes reference to the quark's
parent's momentum, we also need to project it to what it would be if it had
emitted an onshell quark with $q'_i$.  Thus, $q_{i-1} \,\rightarrow\, k_i + q'_i
= q'_{i-1}$.  In the end, this changes our replacement rule coefficient for the
$j^{\rm th}$ quark to:
\begin{align}
{c'_{\rm LO }}^{\alpha_{j+1}} & =  \frac{\bar{q}_j}{{q'_j}^2} \left( n_{j'}^{\alpha} 
+ \frac{(\slashed{q}'_{j+1})_{n_{j'}\perp}\:\gamma^{\alpha}_{n'_{j+1}\perp}}{\bar{q}_{j+1}} \right) 
\frac{\nbs \ns_{j}}{4}  \,,
\label{eq:loArrowSubt}
\end{align}
where $q_j^{\prime\mu} = \bar q_j\, n_{j'}^\mu/2$. Thus $c_{\rm LO}'$ has the
same form as $c_{\rm LO}$ but with a different orientation for its momenta.
This changes the expression for $C^{(2)}_{3,\,{\rm LO}}\mo^{(2)}_3$ to involve
$c'_{\rm LO}$ instead of $c_{\rm LO}$ ({\it cf.} Eq.~\ref{eq:loOp})

After the $dq^2$ integration, the $1/\epsilon$ term in $\mathcal{R}_{J,\,{\rm
    NLO}}^{q\rightarrow qgg}$ will allow us to read off $\nlosplit$.  As a
reminder, we need this subtraction operator because our \NLOl term,
$C^{(2)J,\,1}_{3, \,\rm{NLO}}(n_2,n'_1,n'_2) \mo^{(2)}_{3}$ is supported over
all of phase space, and thus contains LO portions.  We therefore have
\begin{align}
\mathcal{R}_{J,\,{\rm NLO}}^{q\rightarrow qgg} \,&=\, \int d\Pi_{k_1,\, k_2,\, q_0} \mathrm{PP} \bigg[
|C^{(2)J,\,1}_{3, \,\rm{NLO}}(n_2,n'_1,n'_2) \langle 0| \mo^{(2)}_3 |qgg\bar{q}
\rangle|^2
\nonumber \\
 &\qquad - \left(
|{c'_{\rm LO }}^{\alpha_{1}} \, {c'_{\rm LO }}^{\alpha_{2}} \langle 0| \mo^{(2)}_3 \Gamma^{\mu}  
|qgg\bar{q} \rangle |^2 \right)_{\overline{\rm MS}} \bigg] \,.
\end{align}
The $\overline{\rm MS}$ indicates that we are only subtracting pole parts of the
LO contribution with no finite pieces.  However, there is still an ambiguity
over {\it which} pole parts we subtract, since the LO contribution has a double
pole from its two collinear divergences, but we are at some liberty to decide
which single pole parts we remove as well.  As we expect, this subtraction
operator squared takes the form of a convolution of two splitting functions:
\begin{align}
\int d\Pi
\left( | {c'_{\rm LO }}^{\alpha_{1}} \, {c'_{\rm LO }}^{\alpha_{2}}\langle 0| \mo^{(2)}_3 \Gamma^{\mu}  
|q\bar{q}gg \rangle |^2 \right)_{\rm \overline{MS}} &= 2 \int \slashed{d}^4q_2\, \delta(q_2^2)\, dq^2\, dy\, x\, p \, \left(1-y 
\right)^{\frac{\epsilon}{2}} (q^2)^{-1+\epsilon/2} \frac{\alpha^2}{2\pi^2} \nonumber \\
&\times \frac{1}{y} P^{(0)}_{d,qq}(y) \frac{P^{(0)}_{4,qq}(x/y)}{\epsilon} 
{\rm Tr}\left[\bar{q}_2 \frac{\slashed{n}_2}{2} \Omega \Omega^\dagger  \right],
\label{eq:splitConv}
\end{align}
where the trace contains those terms that get passed to the hard function, $\mathcal{H}$,
along with the $q_2$ phase space by the projector PP.  This includes
the final quark spin-sum and phase space, the current $\Gamma$ which is a spectator for both LO and 
jet-structure corrections, and quantities related to the antiquark ({\it cf.} Eq.~\ref{eq:loAmpFact}).  
What may seem surprising is that the two splitting functions live in different dimensions.  The reason
for this particular scheme for regulating phase space has to do with the alternate, two-loop method
for calculating $\nlosplit$, which was the original approach.  For that result, in ${\rm \overline{MS}}$
we would subtract a pure pole counterterm, regulate the loop integral in $d$-dimensions, and leave external
particles in 4d.  Since the phase space integrals are related to loops by cuts, we see above that our $y$-integral
is, in fact, in $d$-dimensions, but the splitting involving two external particles is left simply in four.  

Looking at the $\scetone$ diagrams for the process (Fig.~\ref{fig:figureABCC}), the amplitude
${c'_{\rm LO }}^{\alpha_{1}} \, {c'_{\rm LO }}^{\alpha_{2}}\langle 0| \mo^{(2)}_3 \Gamma^{\mu} 
|qgg\bar{q} \rangle$ comes from a subset of diagrams $A^2$ and $B^2$.  The expression for subtraction is 
thus:
\begin{align}
\mathrm{PP} \Big[\int d\Pi
\left( | {c'_{\rm LO }}^{\alpha_{1}} \, {c'_{\rm LO }}^{\alpha_{2}}\langle 0| \mo^{(2)}_{\rm S_3} \Gamma^{\mu}  
|q\bar{q}gg \rangle |^2 \right)_{\rm \overline{MS}}\Big] &= 
\int dq^2\, d z_1 x\, p \, \left(\frac{z_1}{x+z_1} \right)^{\frac{\epsilon}{2}} \nonumber \\
\times \, \frac{(q^2)^{-1+\epsilon/2}}{\epsilon} \frac{\alpha^2}{2\pi^2}\, \frac{1}{x+z_1} 
\Biggr[ \frac{1+\bigl( \frac{x}{x+z_1} \bigr)^2}{\frac{x}{x+z_1}-1} & 
+ \frac{\epsilon}{2} \left(1-\frac{x}{x+z_1} \right) \Biggr] \left( \frac{1+(x+z_1)^2}{x+z_1 - 1} \right)\Big] \nonumber \\
\,+\,& z_1 \leftrightarrow z_2,
\label{eq:subtrTerm}
\end{align}
where we now act with $\mathrm{PP}$, dropping the trace from \eq{splitConv} and keeping only those 
terms needed for the computation of $\mathcal{R}$ and
$\nlosplit$.  We can note several things about this expression.  For concreteness, we discuss the $z_1$-dependent term corresponding 
to graph $A^2$, \fig{figureABCC}.  The $\bar{p}$ fraction of $q_0$ relative to $q_1$ is $x/(x+z_1)$, and that of $q_1$ to $q_2$ 
is $x+z_1$, in terms of the variables in \eq{splitConv}, $y' \,=\, x/(x+z_1)$.  
Performing the integrals leads to double and single poles.  For later use, we write down the result of 
doing the $dq^2, \, dz_i$ integrals, where one of latter is trivial since we have 
$\delta(1-x-z_1-z_2)$ sitting inside the phase space integral ({\it cf.} Eq.~\ref{eq:ABCDj}).
\begin{align}
\int d\Pi \,
\mathrm{PP} \left[ \left( | {c'_{\rm LO }}^{\alpha_{1}} \, {c'_{\rm LO }}^{\alpha_{2}}\langle 0| \mo^{(2)}_3 \Gamma^{\mu}  
|q\bar{q}gg \rangle |^2 \right)_{\rm \overline{MS}} \right] &= \nonumber \\
         2\,  x\, p \, \frac{\alpha^2}{2\pi^2} \left( 
   \frac{1}{\epsilon^2} \frac{2}{x-1} \left[ \left(-2 \left(2
   \left(x^2+1\right) \log (\lambda) 
   +(x-1)^2\right) \right.\right.\right. & \nonumber \\ 
   + 
   4 \left(x^2+1\right) \log
   (x-1)- & 
   \left.\left. \left(x^2-1\right) \log
   (x)\right) \right] \nonumber \\  
   + \frac{2}{\epsilon}\frac{1}{(x-1)^2 x} 
   \left[ 
   2 x (x-1) \left(2
   \left(x^2+1\right) 
   \left(\text{Li}_2(1-x)-\text{Li}_2 \left(\frac{1}{x}\right)\right) \right.\right.
   & \left. +\left(x^2-1\right) 
   \text{Li}_2\left(\frac{x-1}{x}\right)\right) \nonumber \\
   +x \left(-2 
   \left(x^2+1\right) (x-1) \log
   ^2(\lambda )+4
   \left(\left(x^2+1\right) \left(\log
   \left(\frac{x-1}{x}\right) \right. \right. \right. &
         \left. 
   +x \log
   \left(\frac{x}{x-1}\right)\right) \nonumber \\
   -  \left. \left.
   (x-1)^3\right) \log (\lambda)-\left(3 x^2+5\right) (x-1) \log^2(x) 
   +2 (x-1)^2 \log (x)\right) & \nonumber \\
   +6x \left(x^2+1\right) (x-1) \log^2(x-1)-2 x (x-1)^3 & \nonumber \\
   -2 x (x+1)
   (x-1)^2 \log (x-1) \log(x)
         ((x-1)^2 x) \biggr] \biggr) &
         \label{eq:subtrInt}
\end{align}
where we have done the $dz_i$ integrals between $1-x+\lambda$ and $-\lambda$ to regulate soft divergences.  
All $\lambda$-dependence cancels out of the final answer, which gives us a consistency check on the scheme.

Before comparing $\nlosplit$, we can check our setup with $\losplit$, by looking at the $\mo(\alpha_s)$
contribution to $\mathcal{R}_{\rm LO}$  We see that \cite{Curci:1980uw} gets the following contribution:
\beq
\losplit \,=\, \left( \frac{\alpha_s}{2\pi} \right)\frac{2}{\epsilon}\frac{1+x^2}{1-x}.
\label{eq:curciLO}
\eeq
Calculating in $\scetone$, we get the following amplitude squared:
\beq
A_{q\rightarrow qg} \,=\, \frac{\bar{q}_0}{q_0^2} \left( \frac{2 n \cdot k_1}{\bar{k}_1} + 
\frac{2 k_{1\perp} \cdot q_{1 \perp}}{\bar{q}_1 \bar{k}_1} - \frac{q_{1\perp}^2}{\bar{q}_1^2} \right)
{\rm Tr}[\bar{q}_0 \frac{\slashed{n}_0}{2} \Omega \Omega^\dagger ]
\eeq
With our definition of $\mathcal{R}_{\rm LO}$ in Eq.~(\ref{eq:ABCDj}), we get:
\beq
\losplit \,=\, \left( \frac{\alpha_s}{2\pi} \right)\frac{2}{\epsilon}\frac{1+x^2}{x-1}.
\label{eq:ourLO}
\eeq
The overall minus sign between Eqs.~(\ref{eq:curciLO}) and (\ref{eq:ourLO}) is due to the difference 
between the spacelike and timelike processes.  It arises in the $dz_i$ integral.  Even though the $z_i$ 
dependence is the same in the two calculations, and the integration limits are the same, 0 and $1-x$.  
For us, $1-x \,<\, 0$, but in \cite{Curci:1980uw}, it is positive.  

We will compare the different contributions to double emission separately.  In $\scetone$, the C graph in 
Fig.~\ref{fig:figureABCC} will give box and crossed terms when interfered with itself and the A and B ones.  
We identify the crossed contribution by inserting the color structure and taking those terms 
proportional to $C_F^2 - \frac{1}{2}C_F\, C_A$.  As mentioned above, it only contains the integrals 
in Table \ref{tbl:singlePole}.  In terms of its notation, we have:
\begin{center}  
\begin{table}[ht!]        
  \begin{tabular}{|c|l|}
  \hline
    Function defined in Eq.~(\ref{eq:ABCDj})          &  Value in crossed diagram \\ \hline
    $A(z_1,z_2)$ & $-\frac{16 x \left(x^2+x z_1+(z_1-1)z_1+1\right)}{z_1(x+z_1-1)}$ \\ \hline
    $B(z_1,z_2)$ & $\frac{8 \left(x^2 (z_1-2)-x z_1+z_1-1\right)}{x+z_1-1}$ \\ \hline
    $C(z_1,z_2)$ & $\frac{8 \left(x \left(x^2+(x-1)z_1+2\right)+z_1\right)}{z_1}$ \\ \hline
    $D(z_1,z_2)$ & $16 \left(x^2+1\right)$ \\ \hline
  \end{tabular}
 \caption{Contributions to crossed amplitude squared diagram}
\label{tbl:abcdX}
\end{table}
\end{center}
The box contribution additionally contains the functions in Table \ref{tbl:doublePole}, though we are 
only interested in the finite parts.  Their $z_i$ dependence is:
\begin{center}  
\begin{table}[ht!]        
  \begin{tabular}{|c|l|}
  \hline
    Function defined in Eq.~(\ref{eq:ABCDj})          &  Value in box diagram \\ \hline
    $A(z_1,z_2)$ & $12 x^2+8 x z_1+8 (z_1-1)z_1+12$ \\ \hline
    $B(z_1,z_2)$ & $\frac{8 (z_1-1)\left(x^2+(z_1-2)z_1+2\right)}{x+z_1-1}$ \\ \hline
    $C(z_1,z_2)$ & $\frac{8 (x+z_1) \left(2 x^2+2 x z_1+z_1^2+1\right)}{z_1}$ \\ \hline
    $D(z_1,z_2)$ & 0 \\ \hline
    $E(z_1,z_2)$ & $4 \left[\frac{\left(2 x^4+6 x^3z_1+x^2 \left(7z_1^2+2\right)+2 x \left(2z_1^3+z_1\right)+z_1^4+z_1^2\right)}{z_1^2}\right] $ \\
                                                 &      $+ 4\epsilon \left[ \frac{\left(x^2 (x+z_1-1)^2+z_1^2\left((x+z_1-1)^2+x+z_1\right)+x z_1(x+z_1-1)^2\right)}{z_1^2}\right]$             \\      \hline
    $F(z_1,z_2)$ & $4\left[\frac{\left(z_1^2-2z_1+2\right)\left(x^2+(z_1-1)^2\right)}{(x+z_1-1)^2}\right]$ \\
                                                 &      $+ 4\epsilon \left[ \frac{\left(x^2((z_1-1) z_1+1)+x(z_1-1) ((z_1-2)z_1+2)+(z_1-1)^2((z_1-1)z_1+1)\right)}{(x+z_1-1)^2}\right]$ \\ \hline
  \end{tabular}
 \caption{Contributions to box amplitude squared diagram}
\label{tbl:abcdB}
\end{table}
\end{center}
For the crossed contribution, we perform the multiplication in Table \ref{tbl:singlePole} with the functions 
defined in Table \ref{tbl:abcdX} and integrate $dz_1$, having already done the trivial $dz_2$ integral.  We 
again use a cutoff to avoid soft divergences, thus its range is between $1-x+\lambda$ and $-\lambda$.  In the end, 
we obtain:
\beq
P^{(1)}_{qq\, {\rm crossed}} \,=\, \left( \frac{\alpha_s}{2\pi} \right)^2 \left[ \left(\frac{1+x^2}{x-1} \right)\left(4 \ln(x-1) - \ln^2(x) -\ln(\lambda) \right) 
-2(x+1)\ln(x) \right].
\label{eq:cross}
\eeq
The $\lambda$-dependent pieces will cancel against those from the box contribution.  The other terms agree 
with \cite{Curci:1980uw} up to the previously discussed minus sign, and wherever $\ln (1-x)$ appears in the 
spacelike calculation, we get $\ln (x-1)$.  Since our integrand and integration region are real, the imaginary 
pieces generated by $\ln (1-x)$ when making $x>1$ all must cancel.  

The box calculation proceeds similarly using the functions defined in
Tables \ref{tbl:singlePole} and \ref{tbl:abcdB} We also include the terms proportional to $E(z_1,z_2)$ 
and $F(z_1,z_2)$ and we have subtracted the appropriate contribution, \eq{subtrInt} from that
given by $C^{(2)J,1}_{3,\,\mathrm{NLO}}$. Doing all this, we get:
\beq
P^{(1)}_{qq\, {\rm box}} \,=\, \left( \frac{\alpha_s}{2\pi} \right)^2 \left[ \left(\frac{1+x^2}{x-1} \right)\left(\ln(\lambda) - \ln(x-1) \right) 
+2(2x-1)\ln(x) \right].
\label{eq:box}
\eeq
The soft divergent pieces cancel against the crossed contribution, and once again we agree with 
\cite{Curci:1980uw} up to an overall sign, and the continuation $\ln (1-x) \,\rightarrow\, \ln(x-1)$.  

\begin{figure}[ht!]
\centering
\includegraphics{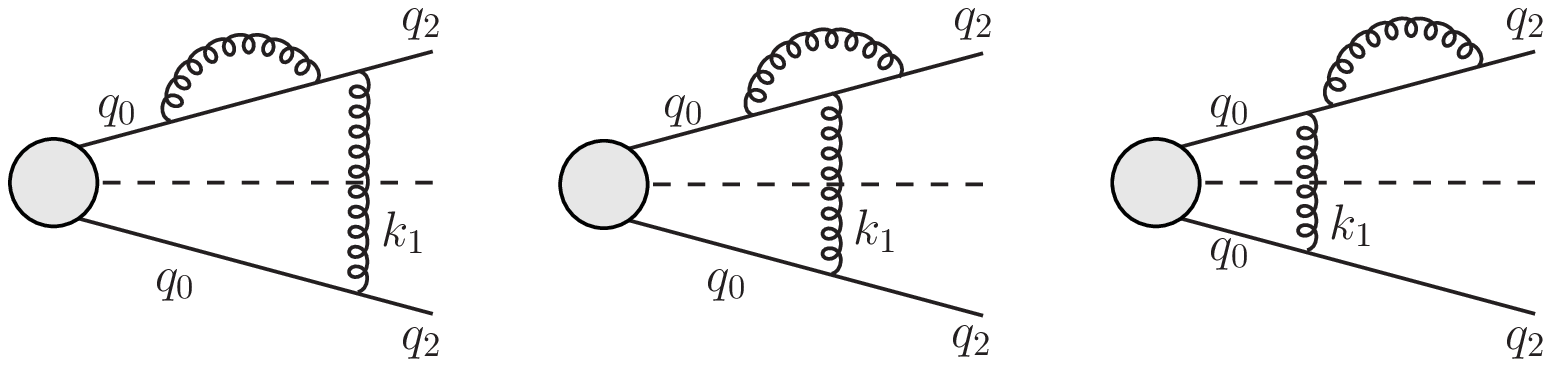}
\caption{Single emission, one-loop contributions to $\nlosplit$.}
\label{fig:oneLoopConts}
\end{figure}

In addition to these real emission contributions to the $C_F^2$ portion of $\nlosplit$, there are also 
single-emission, one-loop diagrams, shown in \fig{oneLoopConts}.
We can account for their contributions in SCET easily.  
We have already derived the tree-level expression for single emission (Eqs.~\ref{eq:curciLO} and \ref{eq:ourLO}).  
Furthermore, both the quark wavefunction renormalization and the vertex renormalization 
are the same in SCET as in QCD \cite{Bauer:2000yr}.  Thus, we recover the entire, gauge-invariant, 
$\propto \, \alpha_s^2 \, C_F^2$ contribution to the splitting function, in
agreement with Ref.~\cite{Curci:1980uw},
\begin{align}
P_{qq\,{\rm abelian}}^{(1)} &=\, C_F^2 \, \frac{\alpha_s^2}{2\pi} 
  \left[ (1-x)\ln(x) -\frac{3}{2}\frac{1+x^2}{1-x}\ln(x) 
  -  2\frac{1+x^2}{1-x} \ln(x)\, \ln(1-x) \right. \nonumber \\
  & \left. -\, \frac{1}{2} (1+x)\ln^2(x) - 5 (1-x) - \frac{5}{2} (1+x) \ln(x)
  \right]
  \,.
\label{eq:nlosplitOfX}
\end{align}
Here we have written $P_{qq}^{(1)}$ with its usual sign conventions for
spacelike evolution.

\bibliographystyle{physrev4}
\bibliography{shower}

\providecommand{\href}[2]{#2}\begin{thebibliography}{10}

\bibitem{Mangano:2002ea}
M.~L. Mangano, M.~Moretti, F.~Piccinini, R.~Pittau, and A.~D. Polosa,
\newblock JHEP {\bf 07}, 001 (2003),
  [\href{http://arXiv.org/abs/hep-ph/0206293}{hep-ph/0206293}].

\bibitem{Gleisberg:2008fv}
T.~Gleisberg and S.~Hoche,
\newblock JHEP {\bf 12}, 039 (2008),
  [\href{http://arXiv.org/abs/0808.3674}{0808.3674}].

\bibitem{Papadopoulos:2006mh}
C.~G. Papadopoulos and M.~Worek,
\newblock \href{http://arXiv.org/abs/hep-ph/0606320}{hep-ph/0606320}.

\bibitem{Kilian:2007gr}
W.~Kilian, T.~Ohl, and J.~Reuter,
\newblock \href{http://arXiv.org/abs/0708.4233}{0708.4233}.

\bibitem{KeithEllis:2009bu}
R.~Keith~Ellis, K.~Melnikov, and G.~Zanderighi,
\newblock Phys. Rev. {\bf D80}, 094002 (2009),
  [\href{http://arXiv.org/abs/0906.1445}{0906.1445}].

\bibitem{Berger:2009ep}
C.~F. Berger {\em et~al.},
\newblock Phys. Rev. {\bf D80}, 074036 (2009),
  [\href{http://arXiv.org/abs/0907.1984}{0907.1984}].

\bibitem{Bredenstein:2010rs}
A.~Bredenstein, A.~Denner, S.~Dittmaier, and S.~Pozzorini,
\newblock \href{http://arXiv.org/abs/1001.4006}{1001.4006}.

\bibitem{Anastasiou:2003ds}
C.~Anastasiou, L.~J. Dixon, K.~Melnikov, and F.~Petriello,
\newblock Phys. Rev. D {\bf 69}, 094008 (2004),
  [\href{http://arXiv.org/abs/hep-ph/0312266}{hep-ph/0312266}].

\bibitem{Catani:2010en}
S.~Catani, G.~Ferrera, and M.~Grazzini,
\newblock \href{http://arXiv.org/abs/1002.3115}{1002.3115}.

\bibitem{Anastasiou:2005qj}
C.~Anastasiou, K.~Melnikov, and F.~Petriello,
\newblock Nucl. Phys. {\bf B724}, 197 (2005),
  [\href{http://arXiv.org/abs/hep-ph/0501130}{hep-ph/0501130}].

\bibitem{Catani:2007vq}
S.~Catani and M.~Grazzini,
\newblock Phys. Rev. Lett. {\bf 98}, 222002 (2007),
  [\href{http://arXiv.org/abs/hep-ph/0703012}{hep-ph/0703012}].

\bibitem{GehrmannDeRidder:2007bj}
A.~Gehrmann-De~Ridder, T.~Gehrmann, E.~W.~N. Glover, and G.~Heinrich,
\newblock Phys. Rev. Lett. {\bf 99}, 132002 (2007),
  [\href{http://arXiv.org/abs/0707.1285}{0707.1285}].

\bibitem{GehrmannDeRidder:2007hr}
A.~Gehrmann-De~Ridder, T.~Gehrmann, E.~W.~N. Glover, and G.~Heinrich,
\newblock JHEP {\bf 12}, 094 (2007),
  [\href{http://arXiv.org/abs/0711.4711}{0711.4711}].

\bibitem{Weinzierl:2008iv}
S.~Weinzierl,
\newblock Phys. Rev. Lett. {\bf 101}, 162001 (2008),
  [\href{http://arXiv.org/abs/0807.3241}{0807.3241}].

\bibitem{Weinzierl:2009ms}
S.~Weinzierl,
\newblock JHEP {\bf 06}, 041 (2009),
  [\href{http://arXiv.org/abs/0904.1077}{0904.1077}].

\bibitem{Sjostrand:2006za}
T.~Sj{\"o}strand, S.~Mrenna, and P.~Skands,
\newblock JHEP {\bf 05}, 026 (2006),
  [\href{http://arXiv.org/abs/hep-ph/0603175}{hep-ph/0603175}].

\bibitem{Sjostrand:2007gs}
T.~Sj{\"o}strand, S.~Mrenna, and P.~Skands,
\newblock Comput. Phys. Commun. {\bf 178}, 852 (2008),
  [\href{http://arXiv.org/abs/arXiv:0710.3820}{arXiv:0710.3820}].

\bibitem{Corcella:2000bw}
G.~Corcella {\em et~al.},
\newblock JHEP {\bf 01}, 010 (2001),
  [\href{http://arXiv.org/abs/hep-ph/0011363}{hep-ph/0011363}].

\bibitem{Bahr:2008pv}
M.~Bahr {\em et~al.},
\newblock Eur. Phys. J. C {\bf 58}, 639 (2008),
  [\href{http://arXiv.org/abs/arXiv:0803.0883}{arXiv:0803.0883}].

\bibitem{Bauer:2000ew}
C.~W. Bauer, S.~Fleming, and M.~E. Luke,
\newblock Phys. Rev. D {\bf 63}, 014006 (2000),
  [\href{http://arXiv.org/abs/hep-ph/0005275}{hep-ph/0005275}].

\bibitem{Bauer:2000yr}
C.~W. Bauer, S.~Fleming, D.~Pirjol, and I.~W. Stewart,
\newblock Phys. Rev. D {\bf 63}, 114020 (2001),
  [\href{http://arXiv.org/abs/hep-ph/0011336}{hep-ph/0011336}].

\bibitem{Bauer:2001ct}
C.~W. Bauer and I.~W. Stewart,
\newblock Phys. Lett. B {\bf 516}, 134 (2001),
  [\href{http://arXiv.org/abs/hep-ph/0107001}{hep-ph/0107001}].

\bibitem{Bauer:2001yt}
C.~W. Bauer, D.~Pirjol, and I.~W. Stewart,
\newblock Phys. Rev. D {\bf 65}, 054022 (2002),
  [\href{http://arXiv.org/abs/hep-ph/0109045}{hep-ph/0109045}].

\bibitem{Bauer:2006mk}
C.~W. Bauer and M.~D. Schwartz,
\newblock Phys. Rev. {\bf D76}, 074004 (2007),
  [\href{http://arXiv.org/abs/hep-ph/0607296}{hep-ph/0607296}].

\bibitem{Bauer:2006qp}
C.~W. Bauer and M.~D. Schwartz,
\newblock Phys. Rev. Lett. {\bf 97}, 142001 (2006),
  [\href{http://arXiv.org/abs/hep-ph/0604065}{hep-ph/0604065}].

\bibitem{Lonnblad:2001iq}
L.~Lonnblad,
\newblock JHEP {\bf 05}, 046 (2002),
  [\href{http://arXiv.org/abs/hep-ph/0112284}{hep-ph/0112284}].

\bibitem{Catani:2001cc}
S.~Catani, F.~Krauss, R.~Kuhn, and B.~R. Webber,
\newblock JHEP {\bf 11}, 063 (2001),
  [\href{http://arXiv.org/abs/hep-ph/0109231}{hep-ph/0109231}].

\bibitem{Caravaglios:1998yr}
F.~Caravaglios, M.~L. Mangano, M.~Moretti, and R.~Pittau,
\newblock Nucl. Phys. {\bf B539}, 215 (1999),
  [\href{http://arXiv.org/abs/hep-ph/9807570}{hep-ph/9807570}].

\bibitem{Catani:1991hj}
S.~Catani, Y.~L. Dokshitzer, M.~Olsson, G.~Turnock, and B.~R. Webber,
\newblock Phys. Lett. {\bf B269}, 432 (1991).

\bibitem{Gleisberg:2003xi}
T.~Gleisberg {\em et~al.},
\newblock JHEP {\bf 02}, 056 (2004),
  [\href{http://arXiv.org/abs/hep-ph/0311263}{hep-ph/0311263}].

\bibitem{Marchesini:1991ch}
G.~Marchesini {\em et~al.},
\newblock Comput. Phys. Commun. {\bf 67}, 465 (1992).

\bibitem{Mueller:1981ex}
A.~H. Mueller,
\newblock Phys. Lett. {\bf B104}, 161 (1981).

\bibitem{Ermolaev:1981cm}
B.~I. Ermolaev and V.~S. Fadin,
\newblock JETP Lett. {\bf 33}, 269 (1981).

\bibitem{Bassetto:1982ma}
A.~Bassetto, M.~Ciafaloni, G.~Marchesini, and A.~H. Mueller,
\newblock Nucl. Phys. {\bf B207}, 189 (1982).

\bibitem{Dokshitzer:1982xr}
Y.~L. Dokshitzer, V.~S. Fadin, and V.~A. Khoze,
\newblock Zeit. Phys. {\bf C15}, 325 (1982).

\bibitem{Dokshitzer:1982ia}
Y.~L. Dokshitzer, V.~S. Fadin, and V.~A. Khoze,
\newblock Z. Phys. {\bf C18}, 37 (1983).

\bibitem{Frixione:2002ik}
S.~Frixione and B.~R. Webber,
\newblock JHEP {\bf 06}, 029 (2002),
  [\href{http://arXiv.org/abs/hep-ph/0204244}{hep-ph/0204244}].

\bibitem{Nason:2004rx}
P.~Nason,
\newblock JHEP {\bf 11}, 040 (2004),
  [\href{http://arXiv.org/abs/hep-ph/0409146}{hep-ph/0409146}].

\bibitem{Catani:1996vz}
S.~Catani and M.~H. Seymour,
\newblock Nucl. Phys. {\bf B485}, 291 (1997),
  [\href{http://arXiv.org/abs/hep-ph/9605323}{hep-ph/9605323}].

\bibitem{Catani:2002hc}
S.~Catani, S.~Dittmaier, M.~H. Seymour, and Z.~Trocsanyi,
\newblock Nucl. Phys. {\bf B627}, 189 (2002),
  [\href{http://arXiv.org/abs/hep-ph/0201036}{hep-ph/0201036}].

\bibitem{Schumann:2007mg}
S.~Schumann and F.~Krauss,
\newblock JHEP {\bf 03}, 038 (2008),
  [\href{http://arXiv.org/abs/0709.1027}{0709.1027}].

\bibitem{Dinsdale:2007mf}
M.~Dinsdale, M.~Ternick, and S.~Weinzierl,
\newblock Phys. Rev. {\bf D76}, 094003 (2007),
  [\href{http://arXiv.org/abs/0709.1026}{0709.1026}].

\bibitem{Nagy:2007ty}
Z.~Nagy and D.~E. Soper,
\newblock JHEP {\bf 09}, 114 (2007),
  [\href{http://arXiv.org/abs/0706.0017}{0706.0017}].

\bibitem{Nagy:2008ns}
Z.~Nagy and D.~E. Soper,
\newblock JHEP {\bf 03}, 030 (2008),
  [\href{http://arXiv.org/abs/0801.1917}{0801.1917}].

\bibitem{Nagy:2008eq}
Z.~Nagy and D.~E. Soper,
\newblock JHEP {\bf 07}, 025 (2008),
  [\href{http://arXiv.org/abs/0805.0216}{0805.0216}].

\bibitem{Soper:2008zp}
D.~E. Soper and Z.~Nagy,
\newblock \href{http://arXiv.org/abs/0805.4371}{0805.4371}.

\bibitem{Robens:2010zr}
T.~Robens and C.~H. Chung,
\newblock \href{http://arXiv.org/abs/1001.2704}{1001.2704}.

\bibitem{Gustafson:1987rq}
G.~Gustafson and U.~Pettersson,
\newblock Nucl. Phys. {\bf B306}, 746 (1988).

\bibitem{Andersson:1989ki}
B.~Andersson, G.~Gustafson, and L.~Lonnblad,
\newblock Nucl. Phys. {\bf B339}, 393 (1990).

\bibitem{Pettersson:1988zu}
U.~Pettersson,
\newblock LU-TP-88-5.

\bibitem{Lonnblad:1992tz}
L.~Lonnblad,
\newblock Comput. Phys. Commun. {\bf 71}, 15 (1992).

\bibitem{Giele:2007di}
W.~T. Giele, D.~A. Kosower, and P.~Z. Skands,
\newblock Phys. Rev. {\bf D78}, 014026 (2008),
  [\href{http://arXiv.org/abs/0707.3652}{0707.3652}].

\bibitem{Larkoski:2009ah}
A.~J. Larkoski and M.~E. Peskin,
\newblock Phys. Rev. {\bf D81}, 054010 (2010),
  [\href{http://arXiv.org/abs/0908.2450}{0908.2450}].

\bibitem{Bauer:2008qh}
C.~W. Bauer, F.~J. Tackmann, and J.~Thaler,
\newblock JHEP {\bf 12}, 010 (2008),
  [\href{http://arXiv.org/abs/0801.4026}{0801.4026}].

\bibitem{Bauer:2008qj}
C.~W. Bauer, F.~J. Tackmann, and J.~Thaler,
\newblock JHEP {\bf 12}, 011 (2008),
  [\href{http://arXiv.org/abs/0801.4028}{0801.4028}].

\bibitem{Catani:1990rr}
S.~Catani, B.~R. Webber, and G.~Marchesini,
\newblock Nucl. Phys. {\bf B349}, 635 (1991).

\bibitem{Jadach:2009gm}
S.~Jadach and M.~Skrzypek,
\newblock Acta Phys. Polon. {\bf B40}, 2071 (2009),
  [\href{http://arXiv.org/abs/0905.1399}{0905.1399}].

\bibitem{Skrzypek:2009jk}
M.~Skrzypek and S.~Jadach,
\newblock \href{http://arXiv.org/abs/0909.5588}{0909.5588}.

\bibitem{Jadach:2010ew}
S.~Jadach, M.~Skrzypek, A.~Kusina, and M.~Slawinska,
\newblock \href{http://arXiv.org/abs/1002.0010}{1002.0010}.

\bibitem{Bonciani:2003nt}
R.~Bonciani, S.~Catani, M.~L. Mangano, and P.~Nason,
\newblock Phys. Lett. B {\bf 575}, 268 (2003),
  [\href{http://arXiv.org/abs/hep-ph/0307035}{hep-ph/0307035}].

\bibitem{Frixione:2007vw}
S.~Frixione, P.~Nason, and C.~Oleari,
\newblock JHEP {\bf 11}, 070 (2007),
  [\href{http://arXiv.org/abs/0709.2092}{0709.2092}].

\bibitem{Curci:1980uw}
G.~Curci, W.~Furmanski, and R.~Petronzio,
\newblock Nucl. Phys. {\bf B175}, 27 (1980).

\bibitem{Bauer:2002nz}
C.~W. Bauer, S.~Fleming, D.~Pirjol, I.~Z. Rothstein, and I.~W. Stewart,
\newblock Phys. Rev. D {\bf 66}, 014017 (2002),
  [\href{http://arXiv.org/abs/hep-ph/0202088}{hep-ph/0202088}].

\bibitem{Bauer:2003mga}
C.~W. Bauer, D.~Pirjol, and I.~W. Stewart,
\newblock Phys. Rev. {\bf D68}, 034021 (2003),
  [\href{http://arXiv.org/abs/hep-ph/0303156}{hep-ph/0303156}].

\bibitem{Beneke:2002ph}
M.~Beneke, A.~P. Chapovsky, M.~Diehl, and T.~Feldmann,
\newblock Nucl. Phys. {\bf B643}, 431 (2002),
  [\href{http://arXiv.org/abs/hep-ph/0206152}{hep-ph/0206152}].

\bibitem{Beneke:2002ni}
M.~Beneke and T.~Feldmann,
\newblock Phys. Lett. {\bf B553}, 267 (2003),
  [\href{http://arXiv.org/abs/hep-ph/0211358}{hep-ph/0211358}].

\bibitem{Marcantonini:2008qn}
C.~Marcantonini and I.~W. Stewart,
\newblock \href{http://arXiv.org/abs/arXiv:0809.1093}{arXiv:0809.1093}.

\bibitem{Amati:1980ch}
D.~Amati, A.~Bassetto, M.~Ciafaloni, G.~Marchesini, and G.~Veneziano,
\newblock Nucl. Phys. {\bf B173}, 429 (1980).

\bibitem{Bassetto:1984ik}
A.~Bassetto, M.~Ciafaloni, and G.~Marchesini,
\newblock Phys. Rept. {\bf 100}, 201 (1983).

\bibitem{Brown:1990nm}
N.~Brown and W.~J. Stirling,
\newblock Phys. Lett. {\bf B252}, 657 (1990).

\bibitem{Dokshitzer:1992ip}
Y.~L. Dokshitzer, G.~Marchesini, and G.~Oriani,
\newblock Nucl. Phys. {\bf B387}, 675 (1992).

\bibitem{Manohar:2006nz}
A.~V. Manohar and I.~W. Stewart,
\newblock Phys. Rev. D {\bf 76}, 074002 (2007),
  [\href{http://arXiv.org/abs/hep-ph/0605001}{hep-ph/0605001}].

\bibitem{Frixione:1995ms}
S.~Frixione, Z.~Kunszt, and A.~Signer,
\newblock Nucl. Phys. {\bf B467}, 399 (1996),
  [\href{http://arXiv.org/abs/hep-ph/9512328}{hep-ph/9512328}].

\bibitem{Ellis:1991qj}
R.~K. Ellis, W.~J. Stirling, and B.~R. Webber,
\newblock Camb. Monogr. Part. Phys. Nucl. Phys. Cosmol. {\bf 8}, 1 (1996).

\bibitem{Bauertalk}
C.~W. Bauer, F.~J. Tackmann, and J.~Thaler,
\newblock talk by C.W.~Bauer at SCET 2010,
  http://wwwth.mppmu.mpg.de/members/scet2010/.

\bibitem{Peskin:1995ev}
M.~E. Peskin and D.~V. Schroeder,
\newblock Reading, USA: Addison-Wesley (1995) 842 p.

\bibitem{Gribov:1972rt}
V.~N. Gribov and L.~N. Lipatov,
\newblock Sov. J. Nucl. Phys. {\bf 15}, 675 (1972).

\bibitem{Stratmann:1996hn}
M.~Stratmann and W.~Vogelsang,
\newblock Nucl. Phys. {\bf B496}, 41 (1997),
  [\href{http://arXiv.org/abs/hep-ph/9612250}{hep-ph/9612250}].

\bibitem{Ellis:1996nn}
R.~K. Ellis and W.~Vogelsang,
\newblock \href{http://arXiv.org/abs/hep-ph/9602356}{hep-ph/9602356}.

\bibitem{Floratos:1977au}
E.~G. Floratos, D.~A. Ross, and C.~T. Sachrajda,
\newblock Nucl. Phys. {\bf B129}, 66 (1977).

\bibitem{Floratos:1978ny}
E.~G. Floratos, D.~A. Ross, and C.~T. Sachrajda,
\newblock Nucl. Phys. {\bf B152}, 493 (1979).

\end{thebibliography}

\end{document}